\documentclass[twoside,12pt]{report}
\usepackage{epsfig}
\usepackage{amssymb}
\usepackage{graphicx}
\usepackage[dvips]{color}
\usepackage{amsfonts}
\def\gsim{\mathrel{\rlap{\lower4pt\hbox{\hskip1pt$\sim$}}
    \raise1pt\hbox{$>$}}}         
\def\lsim{\mathrel{\rlap{\lower4pt\hbox{\hskip1pt$\sim$}}
    \raise1pt\hbox{$<$}}}         
\newcommand{\be}{\begin{equation}}
\newcommand{\ee}{\end{equation}}
\newcommand{\bea}{\begin{eqnarray}}
\newcommand{\eea}{\end{eqnarray}}
\newcommand{\nn}{\nonumber}
\newcommand{\lb}{\label}
\newcommand{\ga}{\gamma}
\newcommand{\qd}{\quad}
\newcommand{\qqd}{\qquad}
\newcommand{\ind}[2]{^{#1}_{\mbox{\scriptsize #2}}}
\newcommand{\al}[2]{\alpha\ind{#1}{#2}}

\begin{document}
\pagestyle{myheadings}

\parindent 0mm
\parskip 6pt

\begin{figure}[!ht]
\vspace{-4.5truecm}
\centerline{
\includegraphics{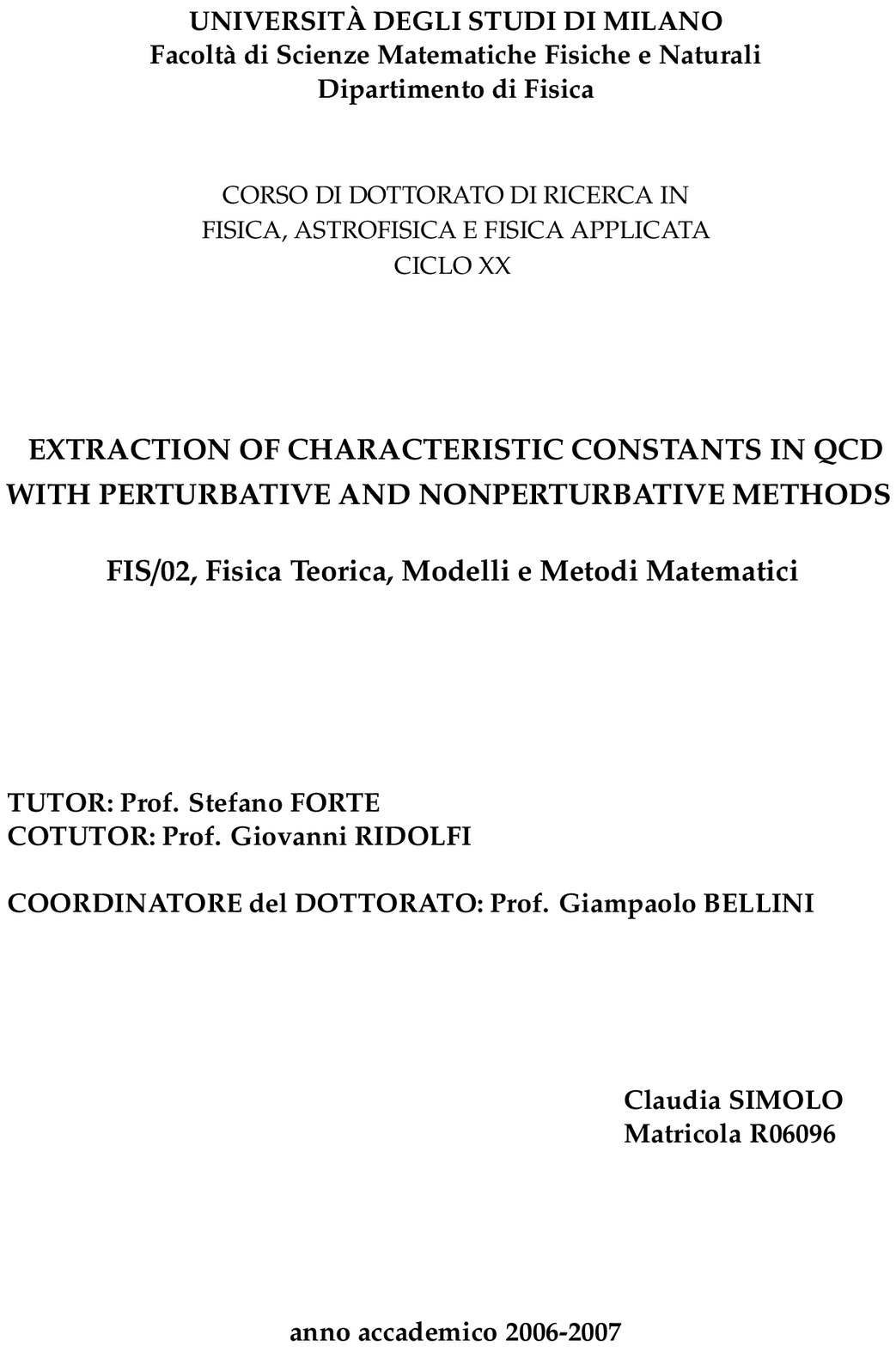}} 
\end{figure}
\vspace{5truecm}

\begin{figure}[!ht]
\vspace{-4.5truecm}
\centerline{
\includegraphics{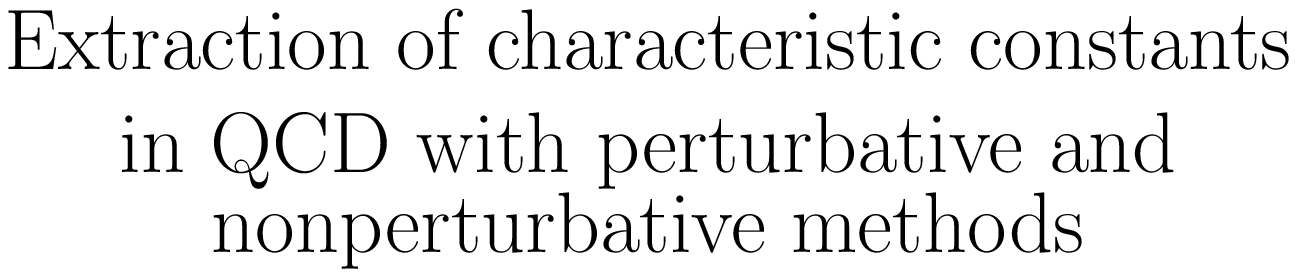}} 
\end{figure}
\vspace{5truecm}

\tableofcontents

\part{On the QCD coupling at low energy scales}
\chapter{The QCD coupling in Perturbation Theory}

The Renormalization Group method is an inherent part of the theoretical 
description of strong interaction processes. As known, the values of the 
QCD coupling extracted from high-energy data are in good agreement with
the theoretical running coupling $\alpha_s(Q^2)$ as derived by the RG
equations and properly normalized at the $Z$ boson mass, thus providing 
a powerful test of asymptotic freedom~\footnote{For a summary of the high energy 
mesurements of $\alpha_s(Q^2)\,$ see~\cite{bethke,pdg}.}.\\
On the other hand, in the low-energy domain, the running coupling $\alpha_s(Q^2)$ 
develops, at any loop level, unphysical (so-called Landau) singularities that 
contradict 
the general principles of the local Quantum Field Theory and severely 
complicate the theoretical analysis of hadron dynamics.\\ 
A reasonable prescription to get rid of the Landau singularities in the 
QCD coupling is given by the Analytic Perturbation
Theory (APT) approach. In this framework, the analycity properties are 
recovered by imposing
$\alpha_s(Q^2)$ to satisfy a dispersion relation with only the unitary 
cut on the time-like
axis, while preserving the asymptotic freedom constraint.\\
This prescription has been exploited in the framework of a second order 
Bethe-Salpeter-like (BS) formalism adjusted for QCD in recent 
works~\cite{BMP}, for the calculation of the meson spectrum in the light and 
heavy quark sectors. It has been there enphasized the relevance of the
infrared evolution of the APT coupling in the few hundreds MeV 
region for a reasonable reproduction of ground state mesons involving light 
and strange quarks, 
at variance with a naive truncation prescription.
This feature strongly supports the APT approach within the BS framework.\\
On the other hand, in the present work~\cite{Baldicchi:2007ic,Baldicchi:2007zn}, 
the calculations of the meson 
spectrum 
within the BS formalism have been exploited in order to obtain an experimental 
determination 
of the strong coupling $\alpha_s^{\rm exp}(Q^2)$ below 1~GeV, by comparison with 
the data.\\
The results turn out to be in satisfying agreeement with
the APT coupling and its evolution in the 200-500~MeV region. Furthermore, below this
scale, some hints on the deep IR behavior are further provided, albeit affected
by large theoretical errors.

In Sec.~1 the Renormalization Group method in QCD is briefly reviewed, and the
state-of-the-art on the infrared behavior of the QCD-invariant coupling sketched,
whereas in Sec.~2 the basis of the Analytic Perturbation Theory approach are 
outlined. 
Sec.~3 is devoted to the determination of the experimental values of the QCD coupling
from the meson spectrum, the analysis method and the discussion of the results.\\
Finally, some technical material is exposed in the Appendices. A short description
of the derivation of the second order BS formalism is given in App.~A, whereas 
numerical tables in App.~B display all the results in detail.


\section{The Renormalization Group method in QCD}
\vspace{-0.3truecm}

In the framework of Quantum Field Theory, 
an arbitrary change in the renormalization prescription basically
results in a repa\-rame\-trization of the theory by a suitable
redefinition of couplings and masses, so as to leave physical
quantities unaffected. The behavior of relevant
Green functions under a change in the extra mass scale $\mu$ introduced
by renormalization is specified by a set of
continuous transformations, properly expressed by the differential
equations of the renormalization group (RG)~\footnote{For an
overview on history and applications see for 
instance~\cite{Sh2001} and references therein.}. This leads to the notion
of effective couplings and masses.\\
Choosing a renormalization scheme (RS) which performs all
subtractions off shell, mass parameters can be treated on the same
foot as couplings.\\
In the ``minimal subtraction scheme'' (MS) 
the renormalization factors exhibit no mass-dependence and are of the
general form
\be
Z_{j}(\varepsilon,\alpha_s)=1+\sum_{n=1}^{\infty}\varepsilon^{-n}Z^{(n)}_{j}(\alpha_s)\,,
\label{zalpha}
\ee
where $\alpha_s=g^2/4\pi$ is the strong coupling and $\varepsilon$ is the 
deviation from
the physical space-time dimension, $d=4-\varepsilon\,$. 
The so-called ``modified minimal substraction scheme'' ($\overline{\rm MS}$) 
amounts to subtracting the whole $(\ln 4\pi-\gamma_{E})$-term together with 
the poles, and it has become a common standard in QCD. 
In MS-like schemes the arbitrary mass scale $\mu$ arises to keep the 
coupling dimensionless in $4-\varepsilon$ dimensions, and the
renormalized objects turn out to be scale dependent.\\
In transition from one parametrization
$(\mu_0^2,\alpha_s(\mu_0^2),m(\mu_0^2))$ to a different one 
$(\mu^2,\alpha_s(\mu^2),m(\mu^2))$
Green functions are connected by finite renormalization transformations,
and, in order to have a unique theory
under an infinitesimal shift of the scale, they 
must satisfy the (inhomogeneous) RG equation
\be
\mathbb{R}\,\Gamma(p_{i};\alpha_{\mu},m_{\mu},\mu^2)=
\gamma_\Gamma(\alpha_{\mu})\Gamma(p_{i};\alpha_{\mu},m_{\mu},\mu^2),
\label{RGequ}
\ee
where $p_i$ are relevant momenta and the shorthand $\alpha_{\mu}$
stands for $\alpha_s(\mu^2)$ (similarly for $m_{\mu}$).
The RG operator is the total derivative over the scale
parameter, namely
\be
\mathbb{R}\equiv
\mu^2\frac{\partial}{\partial\mu^2}+\beta\frac{\partial}{\partial\alpha_{\mu}}
+\gamma_{m}m_{\mu}\frac{\partial}{\partial m_{\mu}}\,,
\label{Rdf}
\ee
where the sum over $n_{f}$ quark flavors in the mass-term is
understood, and the universal functions
\be
\beta(\alpha_{\mu})\equiv\mu^2\frac{\partial\alpha_{\mu}}{\partial\mu^2}\,,\quad
\gamma_{m}(\alpha_{\mu})\equiv\frac{\mu^2}{m_{\mu}}\frac{\partial
m_{\mu}}{\partial\mu^2}\quad\textrm{and}\quad
\gamma_\Gamma(\alpha_{\mu})\equiv\mu^2\frac{\partial}{\partial\mu^2}\ln
Z_\Gamma
\label{intr1}
\ee
have been defined. The RG coefficients $\beta, \gamma_{m}$ and 
$\gamma_\Gamma$ are computed in perturbation theory (PT)
with the aid of the relative counterterms. Generally speaking, they are
gauge-dependent finite functions of couplings and masses, once the
physical limit $\varepsilon\rightarrow 0$ is performed; actually, in
any mass-independent RS (such as minimal subtraction prescriptions or trivially in 
massless theories), the RG coefficients are mass-independent and  
function of $\alpha_s$ alone.

As far as the QCD coupling is concerned, the 
beta function in the first of Eqs.\ (\ref{intr1}) 
has the usual PT power expansion 
\be
\beta(\alpha_s)=-\beta_0\alpha_s^2 -\beta_1 \alpha_s^3 - 
\beta_2\alpha_s^4 + {\mathcal O}(\alpha_s^5)\,,
\lb{intr3}
\ee
where $\beta_l$ are the l-loop coefficients. 
At 1-loop level (i.e. keeping only the first term in 
(\ref{intr3})) solution to the first of Eqs.\ (\ref{intr1}) reads 
\be
\alpha_s(\mu^2)={\alpha_s(\mu_0^2) \over 1+ 
\beta_0 \,\alpha_s(\mu_0^2)\ln(\mu^2/\mu_0^2)}=
\alpha_s(\mu_0^2) \sum_{n=0}^\infty \left (-  
\beta_0\, \alpha_s(\mu_0^2)\ln {\mu^2 \over \mu_0^2}
\right)^n \,,
\lb{intr4}
\ee
namely, while the theory does not predict the actual size of $\alpha_s\,$,
its scale evolution is completely known. Indeed, $\alpha_s\,$
is not an observable by itself, but plays the role of the effective
expansion parameter for physical quantities, and any determination of the 
strong coupling is 
then extracted from the measurements of observables for which PT
predictions exist. Conventionally the mass of the $Z_0$ boson, 
$M_Z\sim 91.2$ GeV, 
is used as a reference scale to the world average 
of all (high energy) determinations, which is currently  
$\alpha_s(M_Z^2)=0.1176(20)$~\cite{pdg,bethke}.\\
Defining the overall scale parameter
\be
\Lambda^2 = \mu_0^2 \exp \left [-{1 \over \beta_0}
{1 \over \alpha_s(\mu_0^2)} \right ]\,,
\lb{intr5}
\ee
Eq.\ (\ref{intr4}) becomes 
\be
\alpha_s(\mu^2)= {1 \over \beta_0 \ln(\mu^2/\Lambda^2)}\,.
\lb{intr6}
\ee
Since $4\pi\beta_0=(11-2n_f/3)$ is a
positive number for all presently known flavours, $\alpha_s(\mu^2)$
decreases to zero as $\mu^2\to0$ (asymptotic freedom), this property being not
spoiled by higher loop corrections.

Given the arbitrariness of the renormalization scale $\mu\,$, the best choice 
is a matter of expediency. The commonly adopted prescription is to fix $\mu$ 
at the physical scale of the process at hand. This can be motivated by a 
glance to the homogeneous RG equation satisfied by any (dimensionless) 
observable $G$,
namely~\footnote{Actually quark masses have been neglected, assuming that
both the physical and the renormalization scales $Q^2$ and $\mu^2$ are larger than
any other relevant mass scale. If non negligible, quark masses
give rise to an additional term in Eq.\ (\ref{intr7}), 
similarly to Eqs.\ (\ref{RGequ}) and (\ref{Rdf}).}
\be   
\mu^2{d \over d \mu^2} G = \left ( \mu^2{\partial \over \partial
  \mu^2} + \beta (\alpha_s){\partial \over \partial\alpha_s}\right ) G = 0 
\lb{intr7}
\ee
where $G$ is typically function of the ratio~\footnote{Actually, according 
to the process, more than one scale can be involved, and this can be spacelike 
($q^2<0$ and $Q^2=-q^2$) or timelike ($q^2>0$ and $s=q^2$).} $Q^2/\mu^2$ 
and of the renormalized coupling $\alpha_s(\mu^2)\,$, and  
is given in PT by a power series in $\alpha_s(\mu^2)\,$
\be
G = g_0\left({Q^2 \over \mu^2}\right) + 
g_1\left({Q^2 \over \mu^2}\right)
{\alpha_s(\mu^2) \over \pi} + 
g_2\left({Q^2 \over \mu^2}\right)
\left ({\alpha_s(\mu^2)\over \pi}\right )^2 
+ {\mathcal O}(\alpha_s^3)\,.
\lb{intr8}
\ee
Eq.\ (\ref{intr7}) clearly states the scale invariance of
any physical observable, that is fulfilled only in infinite order. 
In any truncated order, however, the cancellation of the scale
among the coupling and the coefficients is not complete, that is, 
a residual scale and scheme dependence still persists, and the
degree of cancellation improves with the inclusion of higher orders.\\
The solution of Eq.\ (\ref{intr7}) for the PT coefficients $g_j$ exhibits a 
logarithmic dependence on $Q^2/\mu^2\,$ (starting from the order $\alpha_s^2$), 
that can be controlled by setting $Q^2=\mu^2\,$. Then Eq.\ (\ref{intr8}) takes 
the form
\be
G(q) = \overline g_0+\overline g_1 {\alpha_s(Q^2) 
 \over \pi} + 
\overline g_2
\left ({\alpha_s(Q^2)\over \pi}\right )^2 
+ {\mathcal O}(\alpha_s^3)\,, 
\lb{intr9}
\ee
and the energy dependence of $G$ translates into the energy
dependence of the effective parameter $\alpha_s(Q^2)\,$. A change in 
the value of $\mu^2$ clearly amounts to a reorganization of the PT
expansion.

The RG improved perturbative QCD yields a consistent picture of high energy
strong interaction processes~\footnote{Some care is required
for timelike observables since the PT coefficients, in this case, are modified
by non negligible corrections, proportional to powers of $\beta_0 \pi\,$, due 
to the analytic continuation fron the space-like to the time-like domain (see
Sec.~1.6).} from a few GeV 
up to a few hundred GeV. On the other hand, the 
existence of infrared singularities
in the RG-invariant coupling $\alpha_s(Q^2)$ spoils the theoretical 
analysis of low energy hadron dynamics. 
Indeed, at 1-loop level $\alpha_s(Q^2)$, as given by Eq.\ (\ref{intr6}), has 
a pole-type
singularity at $Q^2=\Lambda^2\,$, and the effect of including higher order 
contributions 
does not overcome the problem and only amounts to modifying 
the singularity structure. 
On the contrary, physical observables are known to be analytic in the entire $q^2$ 
complex plane aside from a cut on the real positive (time-like) axis, and 
their exact expressions should be free of these singularities.\\
Many strategies have been suggested to get rid of unphysical singularities
within PT calculations, so as to handle the strong interaction processes at 
low energies. Some of such methods, reviewed e.g. in Ref.~\cite{Prosperi:2006hx}, 
originate in the general properties of the perturbative power series for the QCD 
observables in the framework of the RG formalism, such as the arbitrariness of
the renormalization scheme and scale. One can 
remind, for instance, the Brodsky-Lepage-Mackenzie criterion~\cite{Brodsky:1982gc}, 
the ``Optimal conformal mapping method''~\cite{Fischer} 
(see also Ref.~\cite{Maxwell}), and the so-called ``Fastest Apparent Convergence''  
technique~\cite{Grunberg:1980ja}. The latter, e.g. amounts to setting the scale so 
that all the higher (than LO) coefficients vanish. It is related to  
the {\it effective charges},~\cite{dokshitzer2}-\cite{ALEPH},
i.e. alternative definitions of the QCD coupling, straightforwardly related 
to a physical 
observable, either of spacelike or time-like argument.\\ 
In this context it is also worth mentioning the ``Optimized Perturbation Theory'' 
approach~\cite{stevenson,higgs}, with the aim of improving the convergence of PT 
expansions by a proper choice of the scale and of RS parameters (such as $\beta_2$ 
and $\beta_3$), on the basis of a minimum sensitivity criterion.\\
However, all these definitions can be related to each other by referring back 
to the conventional ${\rm \overline {MS}}$. 
Most of these techniques yield theoretical predictions 
on the low energy behavior of the RG-invariant coupling which are in qualitative  
agreement, suggesting in all cases an IR finite limit.\\ 
Furthermore, recent results of lattice simulations testify to the absence of spurious 
IR singularities in the QCD coupling~\cite{dvFourier03,Lattice}.

The ghost-pole issue gives rise to severe complications, in particular, 
as far as the bound states problem is concerned, since the characteristic 
scale $Q$ involved (i.e., the momentum transfer in the quark-antiquark 
interaction for mesons) is typically below 1~GeV, according to the 
state and the mass of quarks involved.\\
Among the many potential inspired attempts to modify the expression of
$\alpha_s$ in the low energy domain, there exist either trivial tricks
as the trucation prescription~\footnote{This simply amounts to assuming 
$\alpha_s(Q^2)$ frozen at a maximum finite value in the IR region.}, as well as 
more sophisticated models. For instance, one can remind the 1-loop definition 
of the QCD coupling 
in connection with the momentum 
representation of the Cornell-like potential~\cite{richardson}
\be
\alpha_{\rm V}(Q^2) = {1 \over \beta_0 \ln (1 + Q^2/
    \Lambda^2)}\,,\quad
 \tilde V(Q) = - {1 \over 2 \pi^2}{4 \over 3}{\alpha_{\rm V}(Q^2) \over 
    Q^2}\,;
\label{richar}
\ee
as $Q\to\infty\,$, the usual 1-loop coupling (\ref{intr6})  is recovered 
from the first of Eqs.\ (\ref{richar}), whereas as $Q \to 0$ the latter diverges as 
$\Lambda^2/\beta_0 Q^2\,$, incorporating non-perturbative effects~\footnote{This 
can be 
seen by comparing the second of (\ref{richar}) with the usual Cornell potential
in the momentum space representation
\be
 V(Q) =- {1 \over 2 \pi^2}{4 \over 3}{\alpha_{\rm s}(\mu^2) \over 
    Q^2} - {1 \over \pi^2} {\sigma \over Q^4} \,, 
\label{fourier}
\ee
and identifying the quantity $2\Lambda^2/3\beta_0\,$ with the string tension $\sigma$
that controls the confining part of the potential.}. Estimation of higher order 
corrections
and further developments along this line can be found in~\cite{buchmuller}.
On the other hand, a number
of analyses seems rather consistent with an IR finite coupling 
and a suitable modification of the quark-antiquark potential to encode confinement 
(see e.g.~\cite{eichten,isgur,Ball:1995ni,Badalian:2007km}).

A reliable algorithm to get rid of Landau singularities is
provided by the Analytic Perturbation Theory (APT) approach~\cite{ShSol96-7},
which represents a next step, after RG summation, in improving the 
perturbative results. The APT method, as discussed to some extent in Sec.~2, is 
 based on 
the causality condition which imposes the analyticity constraint on the RG-invariant 
coupling on the whole cut complex plane ${\mathcal C}-\{q^2>0\}\,$. 
PT is then used to evaluate the spectral function, i.e. the discontinuity
of $\alpha_s(Q^2)$ across the cut. At 1-loop, e.g., it reads 
\be
{\rm Im}\,[\alpha_s(-\sigma)]={\pi\beta_0\alpha^2_s(\mu_0^2)
\over \left [1 + \beta_0   
\alpha_s(\mu_0^2) \ln(\sigma/\mu_0^2) \right ]^2
+[\pi\beta_0\alpha_s(\mu_0^2)]^2}\,,
\lb{intr10}
\ee
where Eq.\ (\ref{intr4}) has been used.
Indeed, within the causal approach, Landau singularities are suppressed at 
their very roots, and the IR-safe QCD coupling possesses a universal
finite limit as $Q^2\to0\,$.

In what follows the basis of the derivation of the QCD coupling from
the Renormalization Group equations are reviewed, both in the space-like
and time-like domains.

\section{QCD $\beta$ function}

The RG equation for the effective coupling is actually solved in PT 
with the aid of the relation
\be
\beta(\alpha_s)=-\alpha_s\frac{\mu\partial}{\partial\mu}\ln Z_{\alpha}
\label{beta1}
\ee
where the limit $\varepsilon\rightarrow 0$ is understood.
Thus one needs to compute the renormalization factor for the coupling 
$Z_\alpha$ and this can be accomplished in several ways. One 
can start from the quark-gluon vertex $Z_{\bar q qg}$ 
together with renormalization factors of quark and gluon 
propagators $Z_{q}$ and $Z_{g}$ to obtain 
$Z_{\alpha}=Z_{\bar q qg}^{2}Z_{q}^{-2}Z_{g}^{-1}$, but other 
choices, as ghost-ghost-gluon vertex or trilinear and quartic 
gluon interactions, equally work. In a minimal subtraction
prescription the four-dimensional $\beta$-function is entirely
specified by the residuum of the simple pole in the
$\varepsilon$-expansion of $Z_{\alpha}$ (\ref{zalpha})
\be
\beta(\alpha_s)=\frac{1}{2}\alpha_s^2 \frac{d}{d\alpha_s}
Z_{\alpha}^{(1)}(\alpha_s)\,.
\lb{beta2}
\ee
Then the MS-coefficients in Eq.\ (\ref{intr3}) reads~\cite{vanRitbergen:1997va}
\bea
&&\beta_{0}={1 \over 4\pi} \left[11-\frac{2}{3}n_{f}\right]\nonumber\\
&&\beta_{1}={1 \over (4\pi)^2} \left[102-\frac{38}{3}n_{f}\right]\nonumber\\
&&\beta_{2}={1 \over (4\pi)^3} \left[\frac{2857}{2}
   -\frac{5033}{18}n_{f}+\frac{325}{54}n_{f}^2\right]\nonumber\\
&&\beta_{3}={1 \over (4\pi)^4} \left[\Big(\frac{149753}{6}+
   3564\zeta_{3}\Big)-\Big(\frac{1078361}{162}+
  \frac{6508}{27}\zeta_{3}\Big)n_{f}\right . \nonumber\\
&&\qquad+\left . \Big(\frac{50065}{162}+\frac{6472}{81}\zeta_{3}\Big)n_{f}^2
  +\frac{1093}{729}n_f^3\right]
\lb{beta-coeffs}
\vspace{-0.5truecm}
\eea
where $\zeta_\nu$ is the Riemann zeta-function, $\zeta_3\simeq1.202057$. 
The coefficients $\beta_{j}$ generally depend on the RS,
whereas the first two are universal within massless schemes.
Moreover, in the $\rm{MS}$-scheme the $\beta$-function is 
gauge-independent in any order~\cite{Caswell:1974cj}, and in an
arbitrary mass-dependent scheme this feature is preserved only 
in the first order.\\ 
As already noted, the universal 1-loop coefficient~\cite{Gross:1973id} 
has a positive sign provided there is a small enough number of quark 
fields $(n_f\le33/2)$; thus the theory is asymptotically free, that 
is, the $\beta$-function has a stable UV fixed point as its argument 
decreases to zero. Indeed this coefficient is the sum of two 
contributions, the relevant one with respect to asymptotic freedom 
property being the first, which arises from pure gauge field  
effects i.e. from the nonlinear Yang-Mills interaction terms. The 
2-loop coefficient has been computed for the first time  
in~\cite{Caswell:1974gg} and is positive up to $n_f=8\,$.\\ 
Higher order approximations are scheme-sensitive, and it is common 
practice to perform computation within minimal subtraction prescriptions, 
in which the $\beta$-function is unchanged. The first 
calculation of 3-loop coefficient is due to~\cite{Tarasov:1980au}, 
where the ghost-ghost-gluon vertex in the Feynman gauge was used, and
it turns out to be negative for $6\le n_f\le40\,$.  
In a more recent work \cite{Larin:1993tp} the quark-gluon vertex 
was used instead, providing an independent check in an arbitrary 
covariant gauge of the previous result and its gauge-independence. 
Finally, the original 4-loop calculation \cite{vanRitbergen:1997va}
has been performed using the ghost-ghost-gluon vertex in a arbitrary
covariant gauge, and for a generic semi-simple compact Lie symmetry 
group. The result turns out to be gauge-independent, and involves 
higher order group invariants 
such as quartic Casimir operators; specialized to the standard 
SU(3) symmetry, the 4-loop coefficient is a positive number 
for every positive $n_f$ (see also \cite{Czakon:2004bu}). 
Finally note that (except for $\beta_2$) all four coefficients 
are positive up to $n_f=6\,$.

\section{RG-invariant coupling}

The formal solution of the first of Eq.\ (\ref{intr1}) is easily worked out
\be
\ln {\mu^2 \over \mu_0^2 }= \int_{\alpha_s(\mu_0^2)}^{\alpha_s(\mu^2)}
{d \alpha_s \over \beta(\alpha_s)}\,,
\lb{RGScoupl}
\ee
and yields the evolution of the effective coupling as a function
of two dimensionless arguments $t=\mu^2/\mu_0^2$ and 
$\alpha_0=\alpha_s(\mu_0^2)$, where $\mu_0^2$ can be viewed as a 
fixed reference scale and $\mu^2$
as a sliding one. It is worth noting that by differentiating 
Eq.\ (\ref{RGScoupl}) with respect to $\alpha_0$ one has
\be
\left(t\frac{\partial}{\partial t}-\beta(\alpha_0)\frac{\partial}{\partial\alpha_0}
\right)\alpha_s(t,\alpha_0)=0\,,
\label{rginv}
\ee
namely the homogeneous RG equation satisfied by the effective coupling
(or invariant charge).\\
The exact 1-loop solution (\ref{intr4}) or (\ref{intr6}) is 
obtained by retaining only the first 
term in (\ref{intr3}). In Eq.\ (\ref{intr6}), in particular, 
the dimensional scale $\Lambda$ keeps track of the 
initial parametrization $(\mu_0,\alpha_s(\mu_0^2))$, and it 
is scale-invariant; its value is not predicted by the theory 
but must be extracted from a measurement of $\alpha_s$ at a 
given reference scale. 
Emergence of a scale parameter, sometimes referred to as dimensional 
transmutation, breaks naive scale invariance of the massless theory, 
and it is commonly believed to be associated with the typical hadron 
size i.e. to the energy scale where confinement effects set in. 
Roughly speaking, $\Lambda$ is the scale at which the (1-loop) coupling 
diverges (Landau pole), and perturbation 
theory becomes meaningless. Further, $\Lambda$ is scheme-dependent and receives 
further corrections at each loop level (but for simplicity the 
same notation is used throughout).

At the 2-loop level the integration of (\ref{RGScoupl}) leads to a 
transcendental equation, that is, starting from the 2-loop 
approximation to the $\beta$-function in (\ref{intr3}), 
one has 
\be
\ln\frac{\mu^2}{\mu_0^2}=\textrm{C}+\frac{1}{\beta_0\alpha_s}+B_1\ln\alpha_s 
-B_1\ln\left(1+\frac{\beta_1}{\beta_0}\alpha_s\right)
\lb{RG2loop-int}
\ee
where $B_1=\beta_1/\beta_0^2\,$ and the constant term $C$ from the 
lower end points can be again reabsorbed into the 
$\Lambda$-parametrization, with the commonly adopted prescription 
(see e.g.\cite{Bardeen:1978yd,Collins})
\be
\ln\frac{\Lambda^2}{\mu_0^2}=\textrm{C}-B_1\ln\beta_0\,,
\lb{lambdafixing}
\ee
which fixes a specific choice for $\Lambda\,$. Thus one gets the 
2-loop implicit solution 
\be
\ln\frac{\mu^2}{\Lambda^2}=\frac{1}{\beta_0\alpha_s}-B_1\ln\left(B_1+
\frac{1}{\beta_0\alpha_s}\right)\,,
\lb{RG2loop-impl}
\ee
from which the 2-loop scaling parameter is immediately read. 
To achieve an explicit expression for the running coupling 
at this level one should resort to the many-valued Lambert function 
$W(\zeta)$ implicitly defined by the equation
\be
W(\zeta)\exp\left[W(\zeta)\right]=\zeta\,.
\lb{Lamb}
\ee
%
The function $W(\zeta)$ has an infinite number of branches 
$W_k(\zeta)\,$ $k=0,\pm1,\pm2\dots$ 
such that $W_n^*(\zeta)=W_{-n}(\zeta^*)\,$
(for more details see~\cite{Corless:1996}). 
The exact solution to Eq.\ (\ref{RG2loop-impl})  
for $0\le n_f\le8$ reads~\footnote{Note that if $9\le n_f\le16$
the principal branch $W_0$ is involved, but here and throughout the 
discussion is focused on $n_f\le6\,$.}\ (see e.g. 
\cite{Grunberg98})
\be
\alpha_{ex}^{(2)}(z)=-\frac{1}{\beta_0B_1}\frac{1}{1+W_{-1}(\zeta)}
\qquad\zeta=-\frac{1}{eB_1}\left(\frac{1}{z}\right)^{1/B_1}
\lb{RG2loop-ex}
\ee
where $z=\mu^2/\Lambda^2\,$, and $W_{-1}(\zeta)$ is the ``physical'' 
branch of the Lambert function, since it defines a real values 
function for $\zeta\in(-e^{-1},0)\,$ which fulfills the asymptotic 
freedom constraint. As $\varepsilon\to0^+\,$ the asymptotic relations
\be
W_{-1}(-\varepsilon)=\ln\varepsilon+O(\ln|\ln\varepsilon|)
\lb{lamb1}
\ee
\be
W_{-1}\left(-\frac{1}{e}+\varepsilon\right)=-1-\sqrt{2e\varepsilon}+O(\varepsilon)
\lb{lamb2}
\ee
hold. Outside this region of the real axis $W_{-1}(-\varepsilon)$
takes on complex values. 
Though not easy for practical aims, Eq.\ (\ref{RG2loop-ex}) yields 
the most accurate expression for investigating the IR behavior of 
the running coupling, since it has not been derived by means of deep 
perturbative approximations (aside from the truncation of the 2-loop 
$\beta$ function).\\
Actually, a frequently used 2-loop approximate solution, known as the 
\emph{iterative} solution~\cite{ShSol96-7}, is obtained 
starting from Eq.\ (\ref{RG2loop-int}) together with a single iteration 
of the 1-loop formula (\ref{intr6}), that is  
\be
\alpha_{it}^{(2)}(z)=\frac{\beta_0^{-1}}{\ln z+B_1\ln(1+B_1^{-1}\ln z)}\,,
\lb{2loopECit}
\ee
where $z=\mu^2/\tilde\Lambda^2\,$, and $\tilde\Lambda$ now defined 
in (\ref{RG2loop-int}) by
\be
\ln\frac{\tilde\Lambda^2}{\mu_0^2}=\textrm{C}-B_1\ln\frac{\beta_1}{\beta_0}
\lb{delta2}\,.
\ee
This definition is related to the standard one (\ref{lambdafixing})
by the factor
\be
\ln(\Lambda/\tilde{\Lambda})=\frac{1}{2}B_1\ln{B_1}\,.
\lb{conv1Lambda}
\ee  
However, the commonly used 2-loop solution is an asymptotic formula 
which strictly relies on the smallness of $\alpha_s\,$ for fairly 
large $\mu^2$, since it amounts to solving Eq.\ (\ref{RG2loop-int}) (with the
choice (\ref{lambdafixing})) where 
the last term in the r.h.s. has been neglected. Again, after one iteration 
of the 1-loop formula, the result is then re-expanded in powers of $1/L$, 
where $L=\ln z$ and $z=\mu^2/\Lambda^2\,$ as before
\be
\alpha_s^{(2)}(z)=\frac{1}{\beta_0\ln z}\,\left[1-\frac{\beta_1}{\beta_0^2}
\frac{\ln\left(\ln z\right)}{\ln z}\right]\,.
\lb{2loopEC}
\ee
Eq.(\ref{2loopEC}) is known as the standard 2-loop running coupling 
and works only in the deep UV regime, i.e. for $L\gg 1\,$. In Fig.~1(a)
the fractional differences of the two approximate formulas Eqs.~(\ref{2loopECit})
and~(\ref{2loopEC}) with respect to the exact 2-loop coupling 
Eq.~(\ref{RG2loop-ex}) are displayed.

Under the same assumptions of Eq.~(\ref{2loopEC}) one can easily derive 
the 3 and 4-loop asymptotic formulas. Starting with the approximate 
implicit solution at the 4-loop level, namely 
%
\be
\ln\frac{\mu^2}{\mu_0^2}=\textrm{C}+\frac{1}{\beta_0\alpha_s}+
\frac{\beta_1}{\beta_0^2}
\ln\alpha_s+\frac{\beta_2\beta_0-\beta_1^2}{\beta_0^3}\alpha_s
+\frac{\beta_1^3-2\beta_0\beta_1\beta_2+\beta_0^2\beta_3}{2\beta_0^4}\alpha_s^2
+O(\alpha_s^3)\,,
\lb{RG4loop-int}
\ee
the 4-loop coupling in the standard form of an
expansion in inverse powers of the logarithm $L$ for $L\gg1$ (see 
e.g.~\cite{Chetyrkin:1997sg}) reads
\bea
&&\alpha_s^{(4)}(\mu^2)=\frac{1}{\beta_0\,L}\,
\left\{1-\frac{\beta_1}{\beta_0^2}\frac{\ln L}{L}+
\frac{1}{\beta_0^2 L^2}\left[\frac{\beta_1^2}{\beta_0^2}
\left(\ln^2L-\ln L-1\right)+\frac{\beta_2}{\beta_0}\right]+\right.\nn\\
&&\left.\frac{1}{\beta_0^3L^3}\left[\frac{\beta_1^3}{\beta_0^3}\left(
-\ln^3L+\frac{5}{2}\ln^2L+2\ln L -\frac{1}{2}\right)-3\frac{\beta_1\beta_2}
{\beta_0^2}\ln L+\frac{\beta_3}{2\beta_0}\right]\right\}\,.
\lb{4loopEC}
\eea
The 4-loop solution (\ref{4loopEC}) turns out to be nearly indistinguishable 
from the 3-loop curve (see Fig.~1(b)).
In Eq.~(\ref{4loopEC}) the 1-loop solution (\ref{intr6}) has been 
emphasized, and being the leading UV term in (\ref{4loopEC}), it 
defines the asymptotic behavior of the RG-invariant coupling.  
On the other hand, the 2 and 3-loop asymptotic expressions are 
easily read from Eq.\ (\ref{4loopEC}) by 
keeping only the first two and three terms respectively.\\
%
\begin{figure}[t]
\begin{picture}(150,230)
 \put(-10,235){\mbox{\epsfig{file=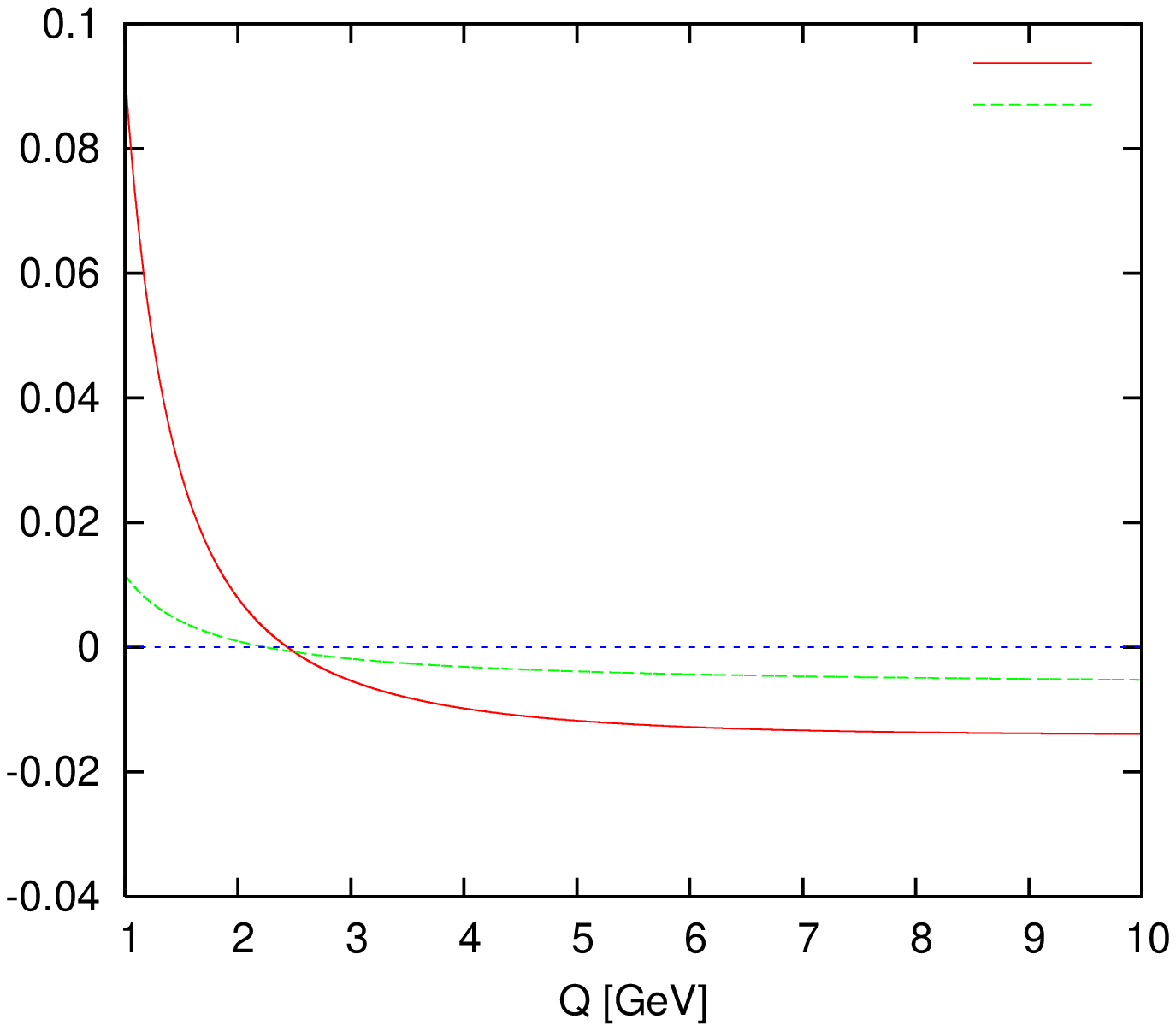,height=8.5cm,width=6.5cm,angle=-90}}}
\put(140,219){ {\tiny $1-\alpha_{ex}/\alpha_s$} }  
 \put(140,211){ {\tiny $1-\alpha_{ex}/\alpha_{it}$} } 
\put(240,50){\mbox{\epsfig{file=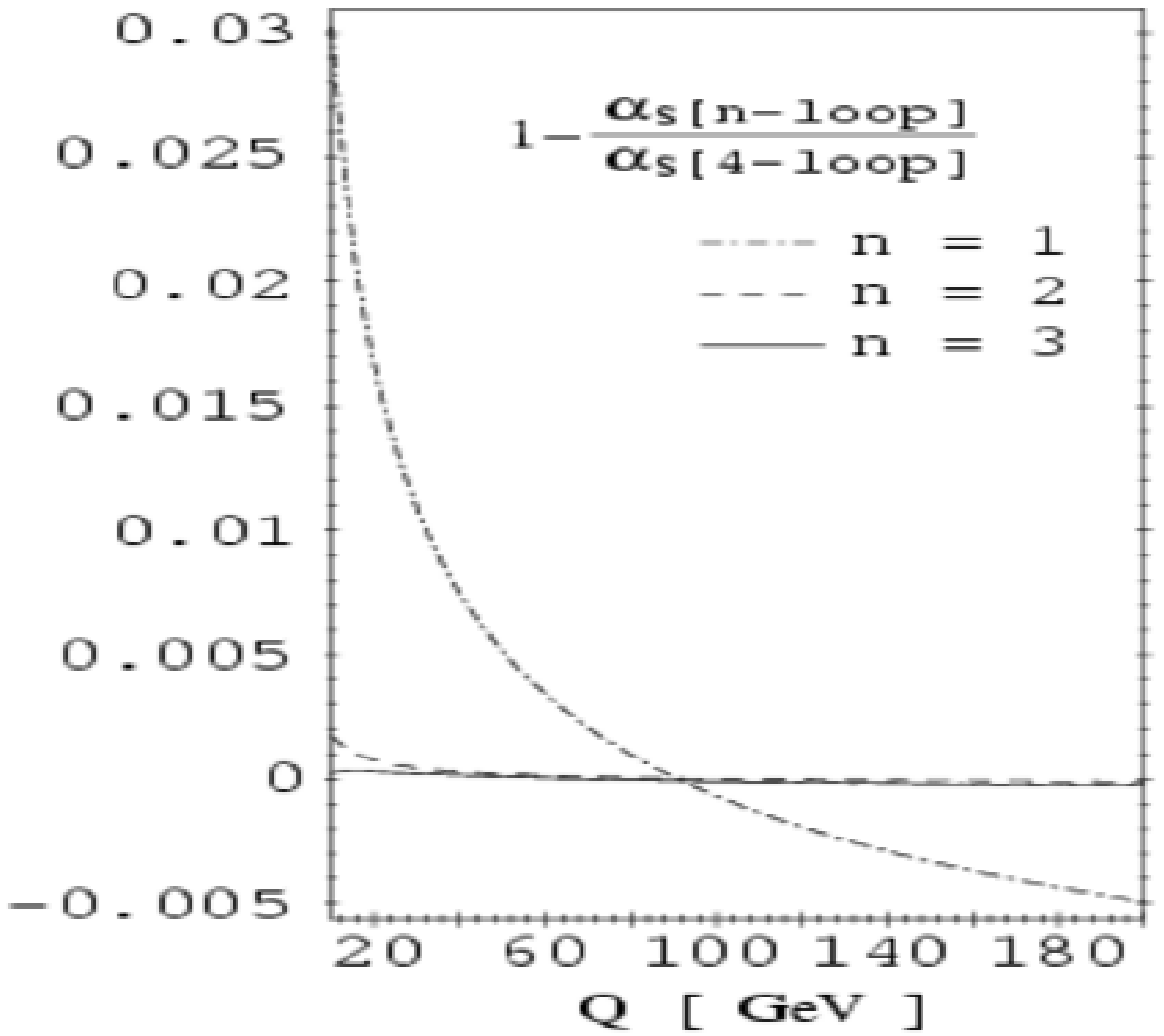,height=6.5cm,width=4.8cm}}}
\put(20,50){ {\footnotesize (a)} } 
\put(273,50){ {\footnotesize (b)}} 
\end{picture}
\vspace{-1.9truecm}
\caption{\footnotesize (a) Fractional difference between 
Eqs.\ (\ref{RG2loop-ex}), (\ref{2loopEC}) and (\ref{2loopECit}) 
(solid and dashed line respectively), with $\Lambda=350\,$MeV and $n_f=4\,$. 
(b) From~\cite{bethke}: fractional difference between 
the 4-loop and the 1-, 2-, 3-loop approximations to Eq.\ (\ref{4loopEC}), 
with $n_f=5$ and normalization conditions at $\alpha_s(M_Z^2)=0.119\,$.
\vspace{-0.3truecm}} 
\label{fig:als-asord}
\end{figure}
Exact integration of the truncated 4 or
3-loop $\beta$-function leads to a more involved structure 
than Eq.\ (\ref{RG4loop-int}). 
Nevertheless, in~\cite{Grunberg98} a useful 
solution has been still worked out at 3-loop level via the 
real branch $W_{-1}(\zeta)$ of the Lambert function together with 
the Pade' Approximant of the related $\beta$-function. Moreover, 
in \cite{Kourashev:1999ye} a reliable approximation to the higher-loop  
coupling has been suggested, via a power expansion in the 
2-loop exact coupling~(\ref{RG2loop-ex}), of the form
\be
\alpha_s^{(k)}(\mu^2)=\sum_{n\ge1} p_n^{(k)}
\left[\alpha_{ex}^{(2)}(\mu^2)\right]^n
\lb{alpha^k}
\ee
with $k\ge3$ the loop order, and $p_n^{(k)}$ proper functions of the
coefficients $\beta_j$. Comparison~\cite{Kurashev:2003pt} of these 
multi-loop approximants with the relative asymptotic 
formulas, Eq.(\ref{4loopEC}) and the 3-loop analogue, reveals 
the better agreement of the former ones with the higher-loop exact 
coupling numerically estimated (i.e. starting from the exact 
implicit solution), even at rather low energy scales (see also 
\cite{Magradze:2005ab}).

Finally, as far as the definition of the scaling constant $\Lambda$ 
is concerned, few comments are in order. As already  
pointed out, starting from the 2-loop level an arbitrary 
constant has to be fixed for $\Lambda$ to be uniquely defined; 
beside the commonly accepted convention~(\ref{lambdafixing}) 
(or~(\ref{delta2})), other prescriptions 
have been proposed \cite{Radyushkin:1982kg} in order to optimize the 
$1/L$-expansion, while~(\ref{lambdafixing}) 
does remain the preferred one as no further terms of order 
$1/L^2$ appear in the 2-loop asymptotic formula (\ref{2loopEC}). 
Thus, at higher-loop levels the scaling parameter is analogously  
related to the initial parameterization, and the 4-loop (and 3-loop) 
value reproducing the world average $\alpha_s(M_Z^2)=0.119$ is 
$\Lambda^{(n_f=5)}_{\overline{\mathrm{MS}}}=220$ MeV~\cite{bethke}.\\
Furthermore, with regard to the scheme-dependence of the coupling and
the scale parameter, restricting the discussion to mass-independent 
RS (as MS-like schemes or trivially any prescription in the 
massless theory), the renormalized couplings 
 in two such different schemes can be related 
perturbatively at any fixed scale $\mu\,$
\be
\alpha_s'=\alpha_s\left[1+v_1\frac{\alpha_s}{\pi}+
v_2\left(\frac{\alpha_s}{\pi}\right)^2
+\dots\,\right]\,.
\label{alphaconv}
\ee
Then, it can be verified that the first two coefficients of 
the relative $\beta$-functions do not change as the 
renormalization prescription is changed, while, for instance,
the 3-loop ones are related by 
$\beta'_2=\beta_2-v_1\beta_1+v_2\beta_0-\beta_0v_1^2\,$. 
As a result, the runing coupling at each loop-level (e.g. 
Eq.\ (\ref{4loopEC})) depends on the RS, through the 
coefficients $\beta_j$ with $j\ge2$ and the initial 
parameterization. The latter obviously amounts 
to suitably adjusting the $\Lambda\,$,  
and the relation is given by~\cite{Gonsalves79}
\be
\ln\left(\Lambda'/\Lambda\right)=\frac{v_1}{2\pi\beta_0}\,,  
\label{conv2lambda}
\ee
which works through all orders (e.g. 
$\Lambda_{\mathrm{\overline{MS}}}/\Lambda_{\mathrm{MS}}=
\exp{[(\ln4\pi-\gamma_e)/2]}\simeq2.66$).

\section{Threshold matching}

Quark mass effects reveal themselves in 
explicit corrections within higher order perturbation theory 
(see e.g.~\cite{Bernreuther:1997jn}), and in the energy 
dependence of the effective (running) quark masses as a 
result of the RG improvement~\cite{Vermaseren:1997fq}.
A direct effect of the quark masses 
on the evolution of the coupling is through 
the dependence of the $\beta$ coefficients on the number 
$n_f\,$ of active quarks, i.e. satisfying $m_f\ll\mu\,$, 
where $\mu$ is the renormalization scale and $m_f$ the 
$\rm{\overline{MS}}$ quark mass (the definition can be also 
formulated in terms of the pole mass $M_f\,$ that can be 
related to the former~\cite{Melnikov:2000qh}).\\ 
Then, in order to have a continuous coupling on the whole 
domain, the two functions $\alpha_s^{(n_f)}(\mu^2)$ and 
$\alpha_s^{(n_f-1)}(\mu^2)$ must be matched at each heavy quark 
mass scale $\mu^{(n_f)}=O(m_f)\,$. As a result the scaling  
parameter $\Lambda$ depends also on $n_f\,$ (see 
e.g.~\cite{Larin:1994va,Chetyrkin:1997sg}).\\
The most straightforward way is to impose continuity of 
$\alpha_s$ by means of the matching condition 
$\alpha_s^{(n_f-1)}(m^2_f)=\alpha_s^{(n_f)}(m^2_f)$, 
which works up to next-to-leading order. At 1-loop, 
it yields
\be
\Lambda^{(n_f)}=\Lambda^{(n_f-1)}\left[\frac{\Lambda^{(n_f-1)}}{m_f}\right]
^{2/(33-2n_f)}\,,
\lb{Lmatch}
\ee
that can be extended up to 2-loop, and exhibits explicit 
dependence on $m_f\,$. Since trivial matching does 
not generally hold in higher orders within $\overline{\rm{MS}}$ 
scheme, a more accurate formula is required in this case; 
specifically 
to obtain the global evolution of the 4-loop coupling the 
proper matching condition reads \cite{Chetyrkin:1997sg}
\be
\alpha_s^{(n_f-1)}=\alpha_s^{(n_f)}\left[1+
k_2\left(\frac{\alpha_s^{(n_f)}}{\pi}\right)^2 +
k_3\left(\frac{\alpha_s^{(n_f)}}{\pi}\right)^3\right]
\lb{L3match}
\ee
with
\be
k_2=\frac{11}{72}\,,\qquad k_3=\frac{564731}{124416}-
\frac{82043}{27648}\zeta_3-\frac{2633}{31104}(n_f-1)
\lb{k_j} 
\ee
if $\mu^{(n_f)}=m_f\,$ is exactly assumed. With this convention, 
Eq.\ (\ref{L3match}) yields the relationship between 
$\Lambda^{(n_f-1)}$ and $\Lambda^{(n_f)}$ in the 
$\overline{\rm{MS}}$ scheme. 
One can equally fix $\mu^{(n_f)}=M_f\,$, that amounts to a 
proper adjustment of the coefficients in (\ref{L3match}); for 
instance~\cite{bethke} starting with (\ref{4loopEC}) and 
$\Lambda^{(n_f=5)}=220\,$MeV, the values  
$\Lambda^{(n_f=4)}=305\,$MeV and $\Lambda^{(n_f=3)}=346\,$MeV are
obtained, with thresholds fixed at the pole masses $M_b=4.7\,$GeV
and $M_c=1.5\,$GeV.\\   
Note finally that Eq.~(\ref{L3match}) spoils 
continuity of $\alpha_s\,$; therefore, 
one can resort to a more sophisticated  
technique~\cite{Shirkov:1990vw}, which relies upon mass-dependent 
RS and yields a smooth transition across thresholds.

\section{Landau singularities}

As already noted, the 1-loop coupling (\ref{intr6}) is affected by a 
spacelike pole at $\Lambda$ with residue $1/\beta_0\,$. 
Adding multi-loop corrections does not overcome the
hurdle; rather, the singularity structure
is more involved due to the log-of-log dependence, 
so that a branch cut adds on to the 1-loop single pole in the IR 
domain of the spacelike axis.\\ 
Moreover, at a given loop level, Landau singularities sensibly 
depend on the approximation used. For instance, the 2-loop 
iterative formula (\ref{2loopECit}) has a pole at $z=1$ 
($\mu=\tilde\Lambda\,$, see Eqs.\ (\ref{delta2}) and (\ref{conv1Lambda}))  
with residue $1/(2\beta_0)\,$, and a cut for $0<z<\exp{(-B_1)}$ due 
to the double logarithm. On the other hand, in the 2-loop
asymptotic expression Eq.~(\ref{2loopEC}), the singularity 
in $z=1$ becomes stronger, namely 
\be
\alpha_s^{(2)}(z)\simeq -\frac{B_1}{\beta_0}\frac{\ln(z-1)}{(z-1)^2}
\qquad z=\mu^2/\Lambda^2\,\to1
\lb{IR2loop}
\ee
and the cut now runs from 0 to 1.
Analogously, the 3- and 4-loop asymptotic solutions, as given by
Eq.\ (\ref{4loopEC}), behave as 
\be
\alpha_s^{(3)}(z)\simeq \frac{B^2_1}{\beta_0}\frac{\ln^2(z-1)}{(z-1)^3}\,,
\qquad\alpha_s^{(4)}(z)\simeq -\frac{B^3_1}{\beta_0}\frac{\ln^3(z-1)}{(z-1)^4}
\qquad z\to1\,,
\lb{IR3loop}
\ee
and are equally affected by an unphysical cut. 
However, the cumbersome singularities structure of the leading 
Landau singularity, and of the unphysical cut as well, are an 
artifact of the UV approximations introduced. Therefore to deal 
with the IR behavior of the invariant coupling  
it is necessary to face with the exact solution, e.g.   
Eq.\ (\ref{RG2loop-ex}) at 2-loop; clearly it is singular when 
$W_{-1}(\zeta)=-1$ that is at $z=B_1^{-B_1}\,$ 
($\mu^2=B_1^{-B_1}\Lambda^2\,$), corresponding to the branch 
point $\zeta=-1/e$ of the Lambert function. In the neighborhood
of this point, due to the asymptotic behavior (\ref{lamb2}) of 
$W_{-1}(\zeta)\,$, one has 
\be
\alpha_{ex}^{(2)}(z)\simeq\frac{1}{\beta_0}\sqrt{\frac{B_1^{-B_1-1}}{2}}
\left[z-B_1^{-B_1}\right]^{-1/2}\,,
\lb{IR2loop-ex}
\ee
i.e., an integrable singularity (the factor $B_1^{-B_1}$ 
in front of $\Lambda$ in the singular point can be reabsorbed 
into a proper redefinition of the integration constant, through
Eq.~(\ref{conv1Lambda})).\\
A more detailed investigation about the IR singularity structure 
of higher-order approximations of the QCD coupling has been performed 
in Ref.~\cite{Magradze:2005ab}, on the basis 
of Eq.\ (\ref{alpha^k}), where, in particular, the location of Landau 
singularities as a function of $n_f\,$ are studied.

\section{Time-like coupling}

The evolution of the QCD coupling has been derived above
in the space-like region. 
However, in order to parametrize observables of time-like arguments, 
one needs an effective coupling defined on the whole real axis.\\
While this poses no special problems in high energy processes, 
at any finite energy the issue of which should be the most 
suitable parameter in the s-channel must be carefully considered. 
The standard practice is to 
merely take over to the time-like domain the space-like coupling  
at any loop level, regardless of the crossing between two 
disconnected regions, thus exporting the IR singular 
structure in the s-channel.
Nevertheless, from many early works based upon analysis of 
$e^+e^-$-annihilation into hadrons (see~\cite{Moorhouse:1976qq} and
references therein),
it is known that this should not be the 
case except far in the asymptotic regime. This is because of the 
appearance of not negligible corrections ($\pi^2$-terms) in the  
higher order coefficients of the $\alpha_s(s)$-expansions, 
due to analytic continuation from space-like to time-like axis. 
The problem is related to the IR non analyticity of the invariant
coupling, and it receives a satisfactory answer 
in the framework of Analytic Perturbation Theory, in which 
the space-like and time-like channels are treated in a unified description
(see Sec.~2).\\
Referring for definiteness to the $e^+e^-$-annihilation ratio into 
hadrons
\be
R(s)=\frac{\sigma(e^+e^-\to\mathrm{hadrons})}{\sigma(e^+e^-\to\mu^+\mu^-)}\,,
\lb{Rdef}
\ee
where $s=q^2>0\,$, one can start with the hadronic contribution to 
the photon 
polarization tensor in (space-like) momentum space
\be
\Pi_h^{\mu\nu}(q)=(g^{\mu\nu}q^2-q^\mu q^\nu)\Pi_h(-q^2)
\propto\int d^4x e^{iqx}<0|T(j^\mu(x)j^\nu(0))|0>
\lb{Pdef}
\ee
$j^\mu$ being the quark electromagnetic current operator. 
As is known, exploiting the analytical properties  of the two-point 
correlation function (\ref{Pdef}) in the cut complex plane 
$\mathbb{C}-\{q^2>0\}$ , $R(s)$ can be straightforwardly related 
to the discontinuity of $\Pi_h(-q^2)$ across the cut
\be
R(s)=\frac{1}{2\pi i}\lim_{\varepsilon\to0}\left[\Pi_h(-s+i\varepsilon)-
\Pi_h(-s-i\varepsilon)\right]\,.
\lb{Rcut}
\ee
In order to compute 
$\Pi_h(-q^2)$ on the space-like axis ($q^2<0$) 
one formally works with its first logarithmic 
derivative (thus avoiding subtraction constants), the Adler 
$D$-function \cite{Adler:1974gd}
\be
D(-q^2)=-q^2\frac{d\Pi_h(-q^2)}{dq^2}\,.
\lb{Adl}
\ee
The RG improved PT expansion of the $D$-function reads 
\be
D_{\rm{PT}}(Q^2)=3\sum_f Q_f^2\left[1+\frac{\alpha_s(Q^2)}{\pi}+
d_2\left(\frac{\alpha_s(Q^2)}{\pi}\right)^2+
d_3\left(\frac{\alpha_s(Q^2)}{\pi}\right)^3 +\dots\,\right]
\lb{D_ptb}
\ee
where $Q^2=-q^2\,$, $Q_f$ are the quark charges, and the coeffcients in the 
$\overline{\mathrm{MS}}\,$ scheme read~\cite{Chetyrkin:1979bj} 
\bea
&&d_2\,\simeq\,1.986-0.115n_f\qquad\nn\\ 
&&d_3\,\simeq\,18.244-4.216n_f+0.086n_f^2-1.24\left(3\sum_fQ_f^2\right)^{-1}
\left(\sum_fQ_f\right)^2\,.\qquad
\lb{D_coefs}
\eea
Then, by integration, $\Pi_h(Q^2)$ is readily obtained~\footnote{In the 
massless case the cut spreads over the 
whole positive axis, while taking into account quark masses the 
cut starts at the two-pion threshold $4m_\pi^2\,$.}.   
The result for $R(s)$ is then usually recasted as a series 
in the effective parameter $\alpha_s(s)$, naively obtained 
by specular reflection, that is by replacing the space-like argument 
$Q^2=-q^2>0$ with the time-like one $s=q^2>0\,$. 
Starting from $\mathcal{O}(\alpha_s^3)$ the series displays corrections 
proportional to $\pi-\,$powers, as a drawback of the analytic 
continuation of the hadronic tensor.
Then the ordinary perturbative expansion for $R(s)$  
(e.g.~\cite{Kataev:1995vh}) reads
\be
R_{\rm{PT}}(s)= 3\sum_f Q_f^2 \left[1+\frac{\alpha_s(s)}{\pi}+
r_2\left(\frac{\alpha_s(s)}{\pi}\right)^2 +
r_3\left(\frac{\alpha_s(s)}{\pi}\right)^3 +\dots\,\right]\,
\lb{R_ptb}
\ee 
\be
r_2=d_2\,;\qquad r_3=d_3-\delta_3\,,\quad \delta_3=\frac{\pi^2b_0^2}{48}
\lb{R_coefs}
\ee
with $d_2$ and $d_3$ the same as in (\ref{D_ptb}), and 
$b_j=(4\pi)^{j+1}\beta_j\,$. 
The correction $\delta_3$ gives to 
the $\mathcal{O}(\alpha_s^3)$ coefficient a strongly negative contribution 
for each $n_f$ (roughly $\delta_3 \simeq14.3$ for $n_f=4$). Higher order 
$\pi^2$-terms were analyzed in~\cite{Bjorken:1989xw} and show a remarkable
growth; for instance, the fourth order correction is
\be
\delta_4\equiv d_4-r_4=\frac{\pi^2b_0^2}{16}\left(r_2+\frac{5b_1}{24b_0}\right)
\lb{delta4}
\ee
with $\delta_4\simeq120$ for $n_f=4$. 
A similar treatment also holds for other 
s-channel observables~\cite{Kataev:1995vh}, such as the normalized rate 
for $\tau$-decay into hadrons. The effects of analytic 
continuation then make the perturbative expansions in the time-like region 
different from Euclidean ones.\\
Since the $\pi^2$-terms play a dominant role in higher order 
coefficients, expansion (\ref{R_ptb}) works only asymptotically at 
large $s$, that is when the smallness of $\alpha_s(s)$ scales 
down these large contributions. Actually, as already noted in pioneer  
works~\cite{Pennington:1981cw,Pennington:1983rz}, 
the expression of $R(s)$, resulting from Eq.\ (\ref{Rcut})  
exhibits no natural 
expansion parameter (since both real and imaginary parts of 
$\alpha_s(Q^2)$ enter into the expression of $R$), and its choice 
is essentially a matter of expediency.\\
Alongside less meaningful attempts, one can remind e.g. the  
analysis performed in Ref.~\cite{Pennington:1981cw} of the quantity
\be
|\alpha_s(-s)|=\frac{1}{\beta_0}\left[\frac{1}{\ln^2(s/\Lambda^2)+
\pi^2}\right]^{1/2}
\lb{penn}
\ee
as expansion parameter for $R(s)\,$. Among the main features, 
Eq.~(\ref{penn}) turns out to be IR finite, thus avoiding the hurdle 
of Landau singularities on the time-like domain, whereas 
asymptotically, i.e. for $s\gg\Lambda^2 e^{\pi}$, it reads 
\be
|\alpha_s(-s)|=\frac{1}{\beta_0}\frac{1}{\ln(s/\Lambda^2)}\left[1
-\frac{\pi^2}{2}\frac{1}{\ln^2(s/\Lambda^2)}+\dots\,\right]\,,
\ee
resembling the UV behavior of the related space-like coupling 
(\ref{intr6}). However, Eq.~(\ref{penn})  
cannot entirely sum up the $\pi^2$-terms.

In order to deal with these corrections, a somewhat different 
approach, known as RKP (Radyushkin-Krasnikov-Pivovarov) procedure, 
has been suggested~\cite{Radyushkin:1982kg, Krasnikov:1982fx} (see 
also~\cite{Jones:1995rd}), which is based upon the analytical 
properties of the polarization tensor $\Pi_h(-q^2)$ and of the 
related $D$-function (\ref{Adl}), that can be summarized by the dispersion 
relations, respectively 
\be
\Pi_h(-q^2)=\int_0^\infty ds\frac{R(s)}{s-q^2}\,\,,
\lb{dispPi}
\ee
\be
D(-q^2)=-q^2\int_0^\infty ds\frac{R(s)}{(s-q^2)^2}
\lb{dispAdl}
\ee
with $R(s)$ given by (\ref{Rcut}), and 
$q^2$ lying in $\mathbb{C}-\{q^2=s>0\}\,$. The key point here is the 
inverse of Eq.\ (\ref{dispAdl}), given by the contour integral
\be
R(s)= \frac{i}{2\pi}\int_{s-i\varepsilon}^{s+i\varepsilon}\frac{dq^2}{q^2}D(-q^2)\,,
\lb{invsAdl}
\ee
to be computed along a path in the analyticity region for 
the $D$-function. Eq.\ (\ref{invsAdl}) can be then generalized to an 
integral transformation, mapping a space-like argument function into 
a time-like one
\be
R(s)=\Phi\left[D(-q^2)\right]\,.
\lb{phi}
\ee
This can be straightforwardly applied to the perturbative expansion 
(\ref{D_ptb}), provided that the integration contour is always kept 
far enough from IR space-like singularities. This yields $R(s)$ as 
an expansion into the images of $\alpha_s(Q^2)$ and of its 
powers, through the map $\Phi\,$~\cite{Radyushkin:1982kg}
\be
R(s)=3\sum_f Q_f^2\left\{1+\sum_{n\ge1}d_n\Phi\left[\left(\frac{\alpha_s(Q^2)}
{\pi}\right)^n\right]\right\}\,,
\lb{R_rkp}
\ee
where the coefficients $d_n$ are the same as in Eq.\ (\ref{D_ptb}). 
Eq.\ (\ref{R_rkp}) is to be compared with the standard 
perturbative expansion (\ref{R_ptb}); here the $\pi^2$-terms are 
entirely summed up, with the drawback that, within this framework, 
there is no uniquely defined expansion parameter. However, it is 
useful to work out its behavior to $\mathcal{O}(\alpha_s)$  
for $R(s)$, namely
\be
\tilde{\alpha}^{(1)}(s)\equiv\Phi\left[\alpha_s^{(1)}(Q^2)\right]
=\frac{1}{\beta_0}\left\{\frac{1}{2}-\frac{1}{\pi}\arctan
\left[\frac{\ln\left(s/\Lambda^2\right)}{\pi}\right]\right\}
\lb{alpha_RKP}
\ee
easily obtained by applying the integral transformation (\ref{invsAdl}) 
to the 1-loop space-like coupling (\ref{intr6}). The related 
time-like coupling (\ref{alpha_RKP}) is free of unphysical 
singularities at low $s\,$ (as Eq.\ (\ref{penn})), and for 
$s\gg\Lambda^2e^{\pi}$ can be expanded into powers of 
$\pi/\ln(s/\Lambda^2)$ 
\be
\tilde{\alpha}^{(1)}(s)= 
\frac{1}{\pi\beta_0}\left[\frac{\pi}{\ln(s/\Lambda^2)}-\frac{1}{3}\frac{\pi^3}
{\ln^3(s/\Lambda^2)}+\,\dots\,\right]\,,
\lb{UValpha_RKP}
\ee
Then Eq.\ (\ref {UValpha_RKP}) can be recasted as a power series in the
1-loop space-like coupling 
\be
\tilde{\alpha}^{(1)}(s)=\alpha_s^{(1)}(s)\left[1-
\frac{\pi^2b_0^2}{48}\left(\frac{\alpha_s^{(1)}(s)}{\pi}\right)^2+\,
\dots\,\right]\,,
\lb{UValpha_RKP2}
\ee
emphasizing that the space-like and time-like couplings differ 
at 3-loop level. 
The comparison between Eq.\ (\ref{UValpha_RKP2}) and Eqs.\ (\ref{R_ptb}) and 
(\ref{R_coefs}) makes it clear the RKP resummation of the 
$\pi^2$-terms into the time-like coupling (\ref{alpha_RKP}).
The main shortcoming of this recipe is that by applying  
to Eq.\ (\ref{alpha_RKP}) the inverse transformation $\Phi^{-1}\,$, i.e. 
relation (\ref{dispAdl}), the original input 
(\ref{intr6}) is not recovered. Obviously, this is because integral transformations
(\ref{invsAdl}) and its inverse are well-behaved as long as the 
integrand possesses the correct analytical properties in the cut 
complex plane; actually this is not the case for the space-like 
coupling (\ref{intr6}) and its higher-loop approximations, and 
one is then forced to compute the integral along a path large 
enough to avoid the IR space-like singularities.


\chapter{The APT theoretical scheme}

\section{The ``Euclidean'' ghost-free coupling}

The analytic approach constitutes the next step, after the
RG-summation, in improving the perturbative results. Specifically, in
addition to the property of renormalizability, this method retains a
general feature of local QFT, the property of causality. The analytic
approach has first been devised in the context of Quantum
Electrodynamics~\cite{AQED}, and then extended to the QCD case about ten
years ago~\cite{ShSol96-7}. The basic merits of the analytic approach to
QCD are the absence of unphysical singularities of the invariant charge
and the enhanced stability of outcoming results with respect to both
higher loop corrections and the choice of renormalization
scheme~\cite{ShSol98-9}. Besides, this method enables one to process the
spacelike and timelike data in a congruent way~\cite{MS:97}. A fresh
review of the analytic approach to QCD and its applications can be found
in the paper~\cite{APT-07}.

Usually, in the framework of RG-improved perturbation theory a QCD
observable $D(Q^2)$ of a single argument $Q^2= -q^2 \ge 0$ (spacelike
momentum transfer squared) can be represented as a power series in the
strong coupling~$\alpha_s(Q^2)$ (cf. Eq.\ (\ref{D_ptb})):
\be
\label{DPert}
D\ind{}{\tiny{PT}}(Q^2) = 1 + \sum_{n \ge 1}
d_{n}\left[ \alpha_s(Q^2) \right]^{n},
\ee
where $d_n$ are the relevant perturbative coefficients. However, in
the IR domain this expansion becomes inapplicable due to the spurious
singularities of the running coupling~$\alpha_s(Q^2)$.
For example, the 1-loop expression (\ref{intr6})
%
%
possesses both the physical cut along the negative real semiaxis
$Q^2\le 0$ and the unphysical pole at~$Q^2=\Lambda^2$.\\
In the framework of the APT, the power series~(\ref{DPert}) for an
``Euclidean'' observable is replaced~\cite{dv99} by the nonpower
expansion
\be
\lb{DAPT}
D\ind{}{\tiny{APT}}(Q^2) = 1 + \sum_{n \ge 1}\,,
d_{n}\,{\cal A}_n(Q^2)
\ee
over the set of functions
\be
\lb{AAPT}
{\cal A}_{n}(Q^2) = \int\limits_{0}^{\infty}
\frac{\rho_{n}(\sigma)}{\sigma + Q^2}\, d\sigma.
\ee
Here, the spectral function $\rho_{n}(\sigma)$ is defined as the 
discontinuity of the relevant power of the perturbative 
coupling~$\alpha_s(Q^2)$ across the physical cut, namely
\be
\label{RhoDef}
\rho_{n}(\sigma) = \frac{1}{\pi}\,\mbox{Im}\!
\left[\alpha_s(-\sigma - i \varepsilon)\right]^{n}.
\ee
The APT representation~(\ref{DAPT}) for a QCD observable $D(Q^2)$ is
free of spurious singularities. Besides, it displays a better
stability, in comparison with the perturbative
parameterization~(\ref{DPert}), with respect to both, the higher-loop
corrections and the choice of the renormalization scheme (see
Ref.~\cite{APT-07} for details). The first-order function
${\cal A}_1(Q^2)$~(\ref{AAPT}) plays the role of the effective
Euclidean QCD coupling at a given loop level:
\be
\lb{ARC}
\alpha_E(Q^2) \equiv {\cal A}_{1}(Q^2) =
\int\limits_{0}^{\infty}
\frac{\rho_{1}(\sigma)}{\sigma + Q^2}\,d\sigma.
\ee
The argument $Q^2=-q^2>0$ now runs over the whole space-like axis,
that is, $\alpha_E(Q^2)$ is free of any space-like unphysical
singularities by construction; moreover, due to the asymptotically 
free nature of the perturbative input the spectral integral
(\ref{ARC}) needs no subtractions.

In the 1-loop case this equation can be integrated
explicitly~\cite{ShSol96-7}
\be
\lb{ARC1L}
\alpha_E^{(1)}(Q^2) = \frac{1}{\beta_{0}}\!\left[\frac{1}
{\ln(Q^2/\Lambda^2)}+\frac{\Lambda^2}{\Lambda^2-Q^2}\right]\!.
\ee
The analytically generated 
non-perturbative contribution in Eq.~(\ref{ARC1L}) subtracts 
the pole in a minimal way, yielding a ghost-free behavior 
which avoids any adjustable parameter. Furthermore, for $Q^2>\Lambda^2$
this non-perturbative term in Eq.~(\ref{ARC1L}) can be rewritten as
\be
\Delta_p^{(1)}(Q^2)=-\frac{1}{\beta_0}\sum_1^{\infty}
\left(\frac{\Lambda^2}{Q^2}\right)^n\,, 
\lb{npt1} 
\ee
namely, as a series of power-suppressed corrections to the 1-loop
perturbative coupling (\ref{intr6}). Since 
the pure perturbative 
contribution dominates in the deep UV region, the asymptotic
freedom constraint of the invariant coupling is preserved.\\
On the other hand, Eq.\ (\ref{ARC1L}) 
exhibits the infrared finite limit 
$\alpha_E^{(1)}(0)=1/\beta_0$ (roughly 1.4, 
if $n_f=3$). 
This value turns out to be independent of the QCD scale $\Lambda$, and  
it is also universal with respect to higher-loop corrections, 
warranting a remarkably stable IR behavior of (\ref{ARC1L}).\\ 
Obviously the scaling parameter has now to be re-defined as
\be
\Lambda^2=\mu_0^2\exp\left[-\phi\left(\beta_0\,\alpha_s(\mu_0^2)
\right)\right]\,,
\lb{lambdaSH1} 
\ee 
where the function $\phi$ is the formal inverse of 
(\ref{ARC1L}), that is, it satisfies the equation
\be
\frac{1}{\phi(x)}+\frac{1}{1-\exp\phi(x)}=x\,. 
\lb{phiSH} 
\ee
with $x=\beta_0\,\alpha_E^{(1)}$ and $Q^2/\Lambda^2=\exp\phi(x)\,$.
In a similar way, the beta-function for the 1-loop coupling (\ref{ARC1L})  
reads~\cite{ShSol96-7}
\be
\beta^{(1)}_E(x)=-\frac{1}{\phi^2(x)}+\frac{\exp\phi(x)}
{\left[\exp\phi(x)-1\right]^2}
\lb{anbeta}
\ee
with $\phi$ again given by Eq.~(\ref{phiSH}). Despite the implicit form 
of (\ref{anbeta}), the symmetry property 
$\beta_E^{(1)}(x)=\beta_E^{(1)}(1-x)$ reveals  
the existence of a IR fixed point at $x=1$, corresponding to 
the IR finite limit (see also \cite{Magradze:1999um}).

\section{Two-loop and higher orders}

Actually, as discussed in Sec.\ 1.3, the 
RG equation for the invariant coupling has no simple 
exact solution beyond the 1-loop level, and asymptotic expressions 
are commonly used (Eq.\ (\ref{4loopEC})), with a  
cumbersome IR nonanalytical structure. As a result, the 
spectral functions (\ref{RhoDef}) become rather involved.\\
At the 2-loop level, as already pointed out, to get the most accurate result  
in the IR domain, one needs to start with the exact RG solution 
as given by Eq.\ (\ref{RG2loop-ex}). In the framework of APT, this amounts to consider  
the discontinuity of (\ref{RG2loop-ex}) across the time-like  
cut~\cite{Magradze:2000hz,Magradze:1999um}
\be
\rho_{ex}^{(2)}(\sigma)=\frac{1}{\pi}\mathrm{Im}\left[\alpha_{ex}^{(2)}(-\sigma)
\right]\,,
\lb{exspc2}
\ee
and the exact 2-loop analytic coupling then reads
\be
\alpha^{(2)}_E(z)=\int_{-\infty}^{\infty}dt\,
\frac{e^t}{e^t+z}\,\bar{\rho}_{ex}^{(2)}(t)\,
\lb{exan2}
\ee
where $t=\ln(\sigma/\Lambda^2)$ and 
$z=Q^2/\Lambda^2\,$.
In spite of its accuracy, Eq.~(\ref{exan2}) cannot be easily
handled, and a slightly simpler expression can be obtained
starting from the 2-loop iterative solution (\ref{2loopECit})
which turns out to be a better approximation to the exact
solution than the usual Eq.\ (\ref{2loopEC}), as far as the 
IR domain is concerned (see Sec.~1.3). The related spectral density 
is then given by~\cite{ShSol96-7}
\bea
\rho_{it}^{(2)}(\sigma)=\frac{1}{\pi}\mathrm{Im}\,\alpha_{it}^{(2)}(-\sigma)=
\frac{1}{\pi\beta_0}\frac{I(t)}{I^2(t)+R^2(t)}
\qquad t=\ln(\sigma/\Lambda^2)\qquad\qquad\qquad\lb{spect2}\\
I(t)=\pi+B_1\arccos\frac{B_1+t}{\sqrt{(B_1+t)^2+\pi^2}}\,,\,\,\,
R(t)=t+B_1\ln\frac{\sqrt{(B_1+t)^2+\pi^2}}{B_1}\nn
\eea
that can be integrated numerically. Moreover, 
since the singularity structure of Eq.~(\ref{2loopECit}) 
is entirely removed by analytization, 
the iterative 2-loop analytic coupling can be straightforwardly
obtained by merely adding to (\ref{2loopECit}) two compensating terms,  
that cancel respectively the pole and the cut~\cite{ShSol98-9}
\bea
&&\alpha_{E,it}^{(2)}(z)=\alpha_{it}^{(2)}(z)+\Delta_p^{(2)}
+\Delta_c^{(2)}
\lb{SH2}\\
&&\Delta^{(2)}_p(z)= \frac{1}{2\beta_0}\frac{1}{1-z}\nn\\
&&\Delta^{(2)}_c(z)=
\frac{1}{\beta_0}\int_0^{\exp{(-B_1)}}\frac{d\xi}{\xi-z}\,
\frac{B_1}{\left[\ln\xi+B_1\ln\left(-1-B_1^{-1}\ln\xi\right)\right]^2+
\pi^2 B_1^2}\nn\,
\eea
with $\xi=\sigma/\Lambda^2$. From Eqs.~(\ref{SH2}) one can readily verify 
the IR limit $\alpha_{E,it}^{(2)}(0)=1/\beta_0$; for general arguments 
concerning the universality of the freezing value through all orders see
for instance~\cite{ShSol98-9,Milton:1997mi} and~\cite{Alekseev:2002zn}.\\  
An useful property of the 2-loop coupling (\ref{SH2}) is the possibility
to estimate its non-perturbative UV tail by expanding the two compensating terms  
into a power series of $1/z=\Lambda^2/Q^2$ for large $z\,$~\cite{Alekseev:2000}, 
in a way similar to Eq.~(\ref{npt1}), 
namely
\bea
&&\Delta^{(2)}_p(z)+\Delta^{(2)}_c(z)=\frac{1}{\beta_0}
\sum_{n=1}^{\infty}\frac{c_n}{z^n}\lb{alek2}\\
&&\nn\\
&& c_n=-\frac{1}{2}-\int_0^\infty
d\xi\,\frac{\exp{\left[-nB_1(1+\xi)\right]}}
{\left(1+\xi-\ln\xi\right)^2 +\pi^2}\nn\,\,.
\eea
If $n_f=6\,$ is taken, the first coefficient, e.g., in Eq.~(\ref{alek2}) 
turns out to be $c_1\simeq-0.54\,$. 
A numerical comparison between Eqs.\ (\ref{SH2}) and (\ref{exan2}), 
performed in \cite{Magradze:1999um} (see Tab.1),
reveals that they differ of about $1.8\%$ in the IR region.

\begin{figure}[t]
\includegraphics[width=14.cm,height=12.5cm]{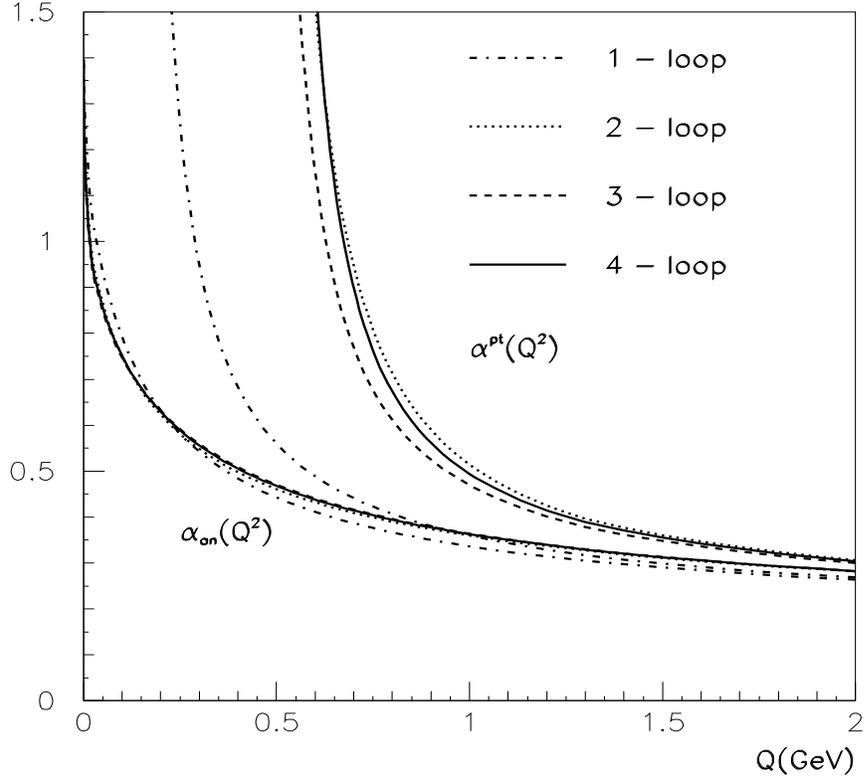} 
\caption{\footnotesize Analytic coupling taken from~\cite{Alekseev:2002zn} 
up to 4-loop compared to perturbative counterparts (\ref{4loopEC}).}
\label{alphas2}
\vspace{-0.5truecm}
\end{figure}

Increasing difficulties arise when dealing with even higher-loop levels. 
Indeed, the exact 3- and 4-loop spectral densities~(\ref{RhoDef}) have a rather cumbersome 
structure in terms of the Lambert function. Neverthless, an extensive numerical 
study of the analytic coupling~(\ref{ARC}) and its ``effective
powers''~(\ref{AAPT}) up to 4-loop can be found in
Ref.~\cite{Kurashev:2003pt}. This analysis reveals 
that the exact 3-loop spectral density can be approximated to a high accuracy
($\sim 1\%$)  
by the discontinuity of the 3-loop perturbative coupling (as deduced by 
Eq.\ (\ref{4loopEC})) across the physical cut~\cite{Nesterenko:2003xb}:
\be
\rho_1^{(3)}(\sigma)=\frac{1}{\beta_0}\frac{1}{(t^2+\pi^2)^3}
\left[t(3\pi^2-t^2)J(t)-(3t^2-\pi^2)R(t)\right].
\lb{sp3}
\ee
In this equation $t=\ln(\sigma/\Lambda^2)$,
\bea
&& J(t)=2t-B_1[tG_1(t)+G_2(t)]+B_1^2G_1(t)[2G_2(t)-1],\lb{sp3t}\\
&& R(t)=t^2-\pi^2-B_1[tG_2(t)-\pi^2G_1(t)]+ \nonumber\\
&& \qquad\quad B_1^2[G_2^2(t)-\pi^2G_1^2(t)-G_2(t)-1]+B_2,\nn\\
&& G_1(t)=\frac{1}{2}-\frac{1}{\pi}\arctan\!\left(\frac{t}{\pi}\right)\,,\quad
G_2(t)=\frac{1}{2}\ln(t^2+\pi^2)\nn
\eea
with $B_j=\beta_j/\beta_0^{j+1}$ and $\beta_j$ the
expansion coefficients (\ref{beta-coeffs}) of $\beta$~function .
The 4-loop analogue of Eqs.\ (\ref{sp3}) and (\ref{sp3t}) turns
out to be more involved but of small usefulness, being the difference 
with respect to the 3-loop approximation negligible~\cite{APT-07}. In
Fig.~\ref{alphas2} the analytic coupling up to 4-loop is compared
to the perturbative counterpart.\\
One can then argue that the 3-loop analityc coupling obtained
by integration of the spectral density (\ref{sp3}), namely
\be
\alpha_E^{(3)}(Q^2)=
\int\limits_{0}^{\infty}
\frac{\rho_1^{(3)}(\sigma)}{\sigma + Q^2}\,d\sigma\,,
\lb{ARC3L}
\ee
is a satisfactory 
improvement of the 1-loop result (\ref{ARC1L})
for all practical aims. By normalizing Eq.\ (\ref{ARC3L}) at the Z~boson
mass to the world average value $\alpha_s(M_Z^2) = 0.1176 \pm
0.0020\,$~\cite{pdg}, one finds $\Lambda_{n_f=5}^{(3)}\simeq 236\,$MeV.
Evolution at the crossing of heavy quark thresholds by
continuous matching conditions then gives 
$\Lambda_{n_f=3}^{(3)}\simeq 417\,$MeV.
Eq.~(\ref{ARC3L}) with the spectral density given by Eqs.~(\ref{sp3}) 
and~(\ref{sp3t}) is actually the expression used in the following analysis
(see Sec.~3) for the comparison with the experimental values of the QCD
coupling extracted from the Bethe-Salpeter approach.

Moreover, it should be noted that an integral
representation, similar to Eq.\ (\ref{SH2}) (and also Eq.~(\ref{alek2})), 
of the non-perturbative 
terms accounting for the divergencies of the 3-loop asymptotic solution~(\ref{4loopEC})
(that is the leading singularity in $z=1$ (\ref{IR3loop}), and the IR 
log-of-log generated cut) also exists~\cite{Alekseev:2000,Alekseev:2002zn}.
It turns out~\cite{Alekseev:2002zn} that the asymptotic behavior of $\alpha_E^{(3)}(Q^2)$
defined by Eq.~(\ref{ARC3L}), i.e. in the large $Q^2$ limit, is given by
\bea
&&\bar{\alpha}^{(3)}_{E}(z)\simeq\alpha_s^{(3)}(z)+\frac{1}{\beta_0} c_1
\frac{\Lambda^2}{Q^2}\lb{alek3}\\
&&\qquad c_1=-1+B_1(1-\gamma_{\rm E})-\frac{B_1^2}{2}\left[B_2-\frac{\pi^2}{6}+
(1-\gamma_{\rm E})^2\right]\,.\nn
\eea
Here $\gamma_{\rm E}$ is 
the Euler constant, $B_2=\beta_0\beta_2/\beta_1^2\,$,  
$\alpha_s^{(3)}(z)$ is the perturbative counterpart as given 
by (\ref{4loopEC}), and $c_1\simeq-0.52\,$ if $n_f=6$ (see 
Ref.~\cite{Alekseev:2002zn} for details).

Note finally that in Refs.~\cite{Kurashev:2003pt,Magradze:2005ab} 
high-accuracy 3- and 4-loop 
analytic coupling of the form
\be
\alpha_E^{(k)}(Q^2)=\sum_{n\ge1} p_n^{(k)}
\left[\alpha_E^{(2)}(Q^2)\right]^n
\lb{alpha^k_an}
\ee
has been also worked out, by exploiting the multi-loop approximation 
(\ref{alpha^k}) as an input in the dispersion
relation (\ref{ARC}). Here $\alpha_E^{(2)}$ is given by 
Eq.\ (\ref{exan2}) and the coefficients are 
the same as in (\ref{alpha^k}). Numerical values at low scales can be 
found in~\cite{Kurashev:2003pt}.\\
For practical applications, one can even resort
to simple expressions of the form of Eq.~(\ref{ARC1L}), with a special
model argument. According to Ref.~\cite{APTApprox}, the 3-loop
Euclidean coupling can be approximated with reasonable accuracy by
the ``1-loop-like'' model
\be
\lb{ARCApprox}
\alpha^{(3)}_{\rm{appr}}(Q^2) =
\frac{1}{\beta_{0}}\!\left[\frac{1}{L_2(Q^2)}
+\frac{1}{1-\exp[L_2(Q^2)]}\right],
\ee
where
\be
\lb{L2Appr}
L_2(Q^2) = \ln\!\left(\frac{Q^2}{\Lambda^2}\right) +
B_1\,\ln\sqrt{\ln^2\!\left(\frac{Q^2}{\Lambda^2}\right) + 2\pi^2},
\quad B_1=\frac{\beta_{1}}{\beta_{0}^2}\,.
\ee
The relative difference between the exact 3-loop
analytic coupling (numerically evaluated via the Lambert function) 
and its approximation~(\ref{ARCApprox}) is less than $3\%$ for $Q \ge
500\,$MeV (see Table~1 in Ref.~\cite{APTApprox} for the details).
However, this error could reach 5-10$\%$ below $500\,$MeV
and it is then unuseful in the deep IR domain. In this region 
the more precise expression~(\ref{ARC3L}) should be used. 
Moreover, its accuracy in the asymptotic region breaks down when taking 
into account flavor thresholds. Therefore, 
it has been suggested~\cite{Bakulev:2004cu} to use 
Eqs.\ (\ref{ARCApprox}) and~(\ref{L2Appr}) provided that the scaling constant and the 
coefficient $B_1$ are replaced by adjustable parameters (see tab. III 
in Ref.~\cite{Bakulev:2004cu}). This yields an accuracy within 
$1\%$ in the whole space-like region down to $Q \ge 1\,$GeV.\\

\section{The ``Minkowskian'' domain in APT}

As far as the definition of a reasonable expansion parameter
in the time-like domain ($q^2=s>0$) is concerned, it should be noted that such a 
definition naturally arises in a self-consistent way within the 
framework of APT~\cite{Milton:1997us,Solovtsov:1997at}. 
Further, it can be regarded as the final step in the RKP-resummation 
for $\pi^2$-terms outlined in Sec\ 1.6.\\
As already noted, Eq.\ (\ref{dispAdl}) and its formal inverse (\ref{invsAdl}) 
can be generalized to the proper tool for relating s- and t-channel
observables.
In the APT scheme, a standard power expansion for a space-like 
observables (\ref{DPert}) is replaced by a nonpower one (\ref{DAPT}). 
A time-like observables, in turn, within this framework, can be also 
re-expressed as the nonpower expansion
\be
R_{\tiny{\rm APT}}(s)=1+\sum_{n\ge1}d_n\mathfrak{A}_{n}(s) 
\lb{R_APT}
\ee 
over the set of functions
\be 
\mathfrak{A}_{n}(s)=\int_s^{\infty}\frac{d\sigma}
{\sigma}\,\rho_n(\sigma)\,,
\lb{U_k} 
\ee 
where the n-th spectral density $\rho_n$ is again given by 
Eq.\ (\ref{RhoDef}). This recipe  
is manifestly quite analogous to the RKP non-power expansion 
(\ref{R_rkp}). 
The key point here~\cite{Shirkov:2000qv} is 
that, due to the forced analyticity of the coupling and its 
analytized powers (\ref{AAPT}), the two sets (\ref{AAPT}) and 
(\ref{U_k}) are put into a one-to-one relation 
by the linear integral transformations (\ref{dispAdl}) and 
(\ref{invsAdl}), namely
\be 
\mathcal A_n(Q^2)=\mathbf{D}[\mathfrak{A}_n(s)]\,,\quad\textrm{and}\quad
\mathfrak{A}_n(s)=\mathbf{R}[\mathcal A_n(Q^2)]\,.
\lb{A-U}
\ee 
This eventually yields a closed theoretical scheme for representing 
observables of any real argument, both space-like and time-like 
(for technicalities see~\cite{Shirkov:2001sm}).\\
\begin{figure}
\begin{center}
\includegraphics[width=33.5pc]{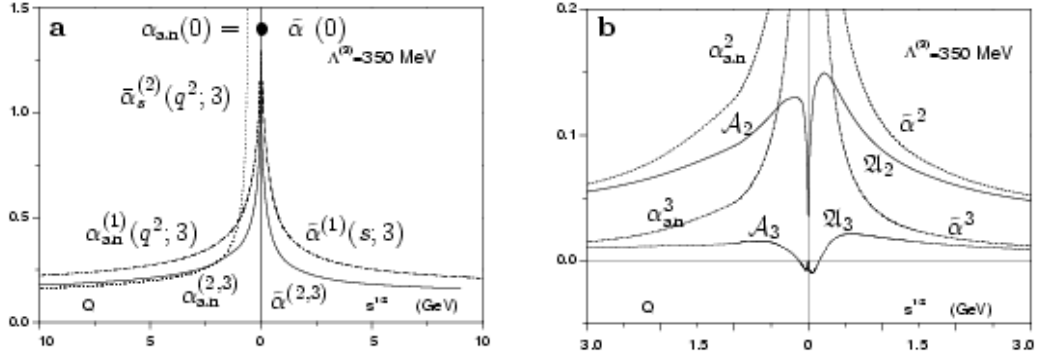} 
\end{center}
\caption{\footnotesize (a) Space-like and time-like global
analytic couplings in a few GeV domain with $n_f=3$ and $\Lambda=350\,$MeV; 
(b) ``Distorted mirror symmetry'' for global expansion
functions, corresponding to exact 2-loop solutions.} \label{fAA} 
\end{figure}
The main features of the two functional sets are illustrated in
Fig.\ \ref{fAA} taken from~\cite{Shirkov:2001sm}. For $n\ge2$ both sets
$\mathcal A_{n}$ and $\mathfrak{A}_{n}$ start with an IR zero and oscillate in the 
IR domain around $n-1$ zeros; furthermore, they all display the UV 
asymptotic behavior $1/\ln^n z$, resembling the corresponding powers
$\alpha_{\rm s}^n(z)\,$. The recursion relations~\cite{APTApprox}
\be
\frac{1}{n}\frac{d\mathcal{A}_n^{(l)}(Q^2)}{d\ln Q^2}=
-\sum_{j=1}^{l}\beta_{j-1}\mathcal{A}_{n+j}^{(l)}(Q^2)\,,\qquad
\frac{1}{n}\frac{d\mathfrak{A}_n^{(l)}(s)}{d\ln s}=
-\sum_{j=1}^{l}\beta_{j-1}\mathfrak{A}_{n+j}^{(l)}(s)\,,
\lb{rec_AU}
\ee
where $l$ is the loop level, allow one to relate different analytized 
powers within each set, albeit  
explicit expressions are available in a simple form only 
at 1-loop (see e.g.~\cite{APTApprox}). 
Nonethless, the two sets (\ref{AAPT}) and (\ref{U_k})  
have been numerically evaluated for $n=1,2,3$ in~\cite{Magradze:2000hz} 
up to 3-loop, via the Lambert function (see also~\cite{Kurashev:2003pt}
and~\cite{Magradze:2005ab}). 
Furthermore, a generalization of APT to fractional powers of $\alpha_s$ was 
also implemented
in Ref.~\cite{Bakulev:2005fp}.

As in the space-like domain, the first function of Eq.\ (\ref{U_k})
plays the role of the effective coupling of time-like argument, namely
\be
\alpha_M(s)=\int_s^{\infty}\frac{d\sigma}{\sigma}\,
\rho_1(\sigma)\,.
\lb{a_s}
\ee
Note also that $\alpha_M(s)$ can be equally defined by the 
differential equation~\cite{dokshitzer2}
\be
s {d \over ds} \alpha_M(s)\,
=\,-\rho_1(s) \qquad {\rm with}\qquad  \tilde 
\alpha(\infty) = 0 \,,
\lb{dks6}
\ee
as can be verified by differentiating the second of
Eq.\ (\ref{A-U}) with $n=1\,$, i.e.~\footnote{Note that this definition 
for the time-like coupling had been suggested also 
in~\cite{Jones:1995rd,Milton:1997us}.}
\be
\alpha_M(s)=\frac{i}{2\pi}\int_{s-i\varepsilon}^{s+i\varepsilon}
\frac{dQ^2}{Q^2}\,\alpha_E(Q^2)\,.
\lb{int-r}
\ee
Moreover, Eq.\ (\ref{dks6}) 
emphasizes the straightforward relation between the ``time-like 
$\beta$-function'' and the spectral density, thus reviving an old 
hypothesis due to Schwinger~\cite{Schwinger:1975th}.\\ 
At 1-loop from Eq.\ (\ref{a_s}) one finds again Eq.\ (\ref{alpha_RKP}); 
however this now leads by inversion of Eq.\ (\ref{int-r}) to the starting 
space-like coupling (\ref{ARC1L}), being the analytic properties 
preserved within this framework.\\
%
%
The main feature~\cite{Milton:1997us} of Eqs.~(\ref{a_s}) 
and~(\ref{ARC}) is the common IR freezing value 
$\tilde{\alpha}(0)=\alpha_E(0)=1/\beta_0$, independent of the 
loop level and of any adjustable parameter (see Fig.~4(a)). 
Moreover, they exhibit a 
similar leading UV behavior, constrained by asymptotic freedom.
Nevertheless this approximate ``mirror symmetry'' is broken in the 
intermediate energy region~\footnote{In Ref.~\cite{Milton:1998wi} an argument,
on the basis of causality principle, 
against a possible exact symmetry ruling the ``t-s dual'' couplings, 
has been developed.}, the discrepancy being about $9\%$ at 1-loop, 
and slightly less at a higher loop level 
(see Ref.~\cite{Milton:1997us} for numerical comparisons).

A last remark concerns thresholds effects, that can be included 
within the APT algorithm~\cite{Shirkov:2000qv} 
by modifying the $n$-th spectral density (\ref{RhoDef}) discontinuously at the 
heavy quark thresholds $m_f$, namely
\be
\rho_n(\sigma)=\rho_n(\sigma,3)+\sum_{n_f\ge4}\theta(\sigma-m^2_f)
\left[\rho_n(\sigma,n_f)-\rho_n(\sigma,n_f-1)\right]\,, 
\lb{ro_kf}
\ee 
descending from the trivial matching condition (Sec.~1.4).  
The global functions resulting from densities (\ref{ro_kf}), 
$\mathcal A_{n}$ and $\mathfrak{A}_{n}\,$, can be obtained from the 
local ones with $n_f$ fixed, by adding specific shift constants 
$c_n(n_f)\,$, not negligible in the $n_f=3,4$ region (technical
details are given in~\cite{Shirkov:2000qv}; for instance, in 
both the t- and s-channels it turns out $c_1(3)\simeq0.02\,$).\\  
The main tests of APT being obviously at low
and intermediate scales, a number of applications to 
specific processes, both in the space-like and time-like (low and 
high energy) domains, have been quite recently performed (see 
for instance, Refs.~\cite{Milton:1998cq,APTApprox,Solovtsov:1997at,ShSol98-9,Shirkov:2001sm,Bakulev:2004cu}).
As a result, the main 
advantages of the APT approach are 
better convergence properties of the ghost-free non-power 
expansion with respect to the usual power one, and a reduced scheme 
and loop-dependence.\\
Finally, transition from Euclidean (space-like momentum) to the distance 
picture has been also 
developed in~\cite{Shirkov:2002td}, involving a suitable modified sine-Fourier 
transformation, consistently with the APT logic.

\section{Modifications of APT and the massive case}

Different strategies to incorporate analyticity into the
RG formalism, or even to implement the above device, exist as well. 
A number of approaches are based on non-perturbative constraints either
on the invariant coupling (see e.g. Refs.~\cite{Cvetic,Schrempp:2001ir,Alekseev:2004vx}) 
or on the RG $\beta$~function (see, e.g., 
Refs.~\cite{Nesterenko:2001xa,Nesterenko:2003xb,Raczka} and 
also~\cite{Krasnikov:1995is,Grunberg98}).\\ 
Among the many attempts to cure the Landau ghost problem it is remarkable  
the existence of models that suggest an IR enhancement of the QCD coupling, 
invoking analycity as well. The most 
attractive feature is supposed to be a straightforward relation
with quark confining potential (see Sec.~1.1). As an example, it can be reminded  
the ``synthetic coupling'' model recently developed  
in~\cite{Alekseev:2004vx}, which amounts to modify 
the analytically improved coupling (\ref{ARC}) by additional 
non-perturbative pole-type terms; at 1-loop it reads
\be
\alpha_{\rm{syn}}(Q^2)=\frac{1}{\beta_0}\left[\frac{1}{\ln(Q^2/\Lambda^2)}
+\frac{\Lambda^2}{\Lambda^2-Q^2}+\frac{c\Lambda^2}{Q^2}+
\frac{(1-c)\Lambda^2}{Q^2+m_g^2}\right]\,.
\lb{a_syn}
\ee
where $m_g=\Lambda/\sqrt{c-1}\,$. The IR 
slavery, due to the pole term at $Q^2=0\,$, is controlled by one dimensionless 
parameter $c\in(1,\infty)$, which relates the 
scaling constant $\Lambda$ to the string tension $\sigma$ 
of potential models (cf. Eq.~(\ref{richar})). The pole term at 
$Q^2=-m_g^2<0$ corresponds to a non vanishing ''dynamical'' 
gluon mass, while leaving the 
analytical structure of eq.(\ref{ARC1L}) along the 
space-like axis unchanged. The value of the $m_g$ parameter 
has been estimated~\cite{Alekseev:2004vx} as $400-600\,$MeV.
Eq.~(\ref{a_syn}) can be derived, analogously to 
(\ref{ARC1L}), from a dispersion relation with a 
spectral density of the form (\ref{RhoDef}) plus 
two $\delta$-terms properly accounting for the poles. Along 
with the IR enhancement as $1/Q^2\,$, 
reproducing the linear confining part of the potential 
(\ref{richar}), construction of (\ref{a_syn}) is mainly 
motivated by the UV asymptotic behavior of its 
non-perturbative contribution, of the form $1/(Q^2)^3\,$,   
decreasing faster than (\ref{npt1}) as $Q^2\to\infty\,$
(see also~\cite{Webber:1998um}).

Actually, of some relevance to the following analysis (see Sec.~3), 
in particular, as far as the deep IR behavior of the QCD 
coupling is concerned (i.e. below 200$\,$MeV), is the ''massive'' modification of the APT
approach recently developed in~\cite{MAPT}.
In order to handle the QCD observables which do not satisfy the
integral representation of the form of Eq.~(\ref{AAPT}), the APT 
formalism has to be modified appropriately, by taking into account
the effect of a non-vanishing mass threshold $m$ in the dispersion
relations. In this case the set of APT expansion functions~(\ref{AAPT}) 
should be replaced by the set of ``massive'' ones with an 
adjustable parameter $m$
\begin{equation}
\label{AMAPT}
\textsf{A}_{n}(Q^2, m^2) = \frac{Q^2}{Q^2 + 4 m^2}
\int\limits_{4 m^2}^{\infty} \rho_{n}(\sigma)\, \frac{\sigma -
4 m^2}{\sigma + Q^2}\, \frac{d\sigma}{\sigma}\,.
\end{equation}
The APT expansion (\ref{DAPT}) should be then replaced by
\begin{equation}
\label{DMAPT}
D\ind{}{{\tiny MAPT}}(Q^2, m^2) = \frac{Q^2}{Q^2 +
4m^2} + \sum_{n \ge 1} d_{n}\, \textsf{A}_{n}(Q^2, m^2)\,.
\end{equation}
%
%
%
Obviously, in the massless limit Eqs.~(\ref{DMAPT}) and~(\ref{AMAPT})
coincide with the expressions~(\ref{DAPT}) and~(\ref{AAPT}),
respectively.\\ 
Similarly to the case of Eq.~(\ref{ARC}) in the ``massless'' APT, the
first-order function $\textsf{A}_{1}(Q^2, m^2)$~(\ref{AMAPT}) plays
the role of an effective ``massive'' running coupling at the relevant
loop level, namely
\begin{equation}
\label{MARC}
\alpha(Q^2, m^2) \equiv \textsf{A}_{1}(Q^2, m^2) =
\frac{Q^2}{Q^2 + 4 m^2} \int\limits_{4 m^2}^{\infty}
\rho_{1}(\sigma) \, \frac{\sigma - 4 m^2}{\sigma + Q^2}\,
\frac{d\sigma}{\sigma}.
\end{equation}
It is worthwhile to note that irrespective of the loop level this
coupling possesses the universal IR limiting value $\alpha(Q^2, m^2)
\to 0$ at $Q^2=0\,$, in qualitative agreement with some results from lattice
simulations~\cite{Lattice,dvFourier03} (see Ref.~\cite{MAPT} for details).
Moreover, as discussed in Sec.~3, the IR behavior of the QCD coupling, in
particular below $200\,$MeV, extracted from the meson spectrum in BS formalism,
can be reasonably described within 
this ``massive'' modification of APT, that is, by Eq.~(\ref{MARC}) if a 
proper value of the adjustable parameter $m$ is chosen.

\chapter{Bound states approach}

\section{QCD coupling from the meson spectrum}

Below 1~GeV, as discussed in Sec.~1, 
the running coupling $\alpha_s$ is affected at any loop level by unphysical
singularities that make the RG-improved pQCD useless.\\
In order to shed some light on the QCD coupling below this scale,  
one needs experimental information to be extracted with the aid of
a suitable theoretical framework. The latter can be provided
by the Bethe-Salpeter (BS) formalism developed in~\cite{BMP} for 
the study of the meson spectrum,  
since the scale involved (the momentum transfer
in the $q\bar q$ interaction) is typically below 1~GeV. 
Many relativistic formalisms for the analysis of 
the meson properties have been developed in the context of QCD, 
that take confinement into account and evaluate the meson (and baryon) spectrum in
the light and heavy quark sectors. 
Among the most recent works, one should remind 
for instance Refs.~\cite{Hecht:2000xa,Jugeau:2003df,Badalian:2007km,Ebert:2005ha,
Billo:2006zg,Dosch:1987sk,Baker:1994nq} and references therein.

In the present work the theoretical results on the meson spectrum within the
framework of a Bethe-Salpeter (BS) formalism adjusted for QCD, have been
exploited with the aim of extracting ``experimental'' values of the running  
coupling $\alpha_{s}^{\rm exp}(Q^2)$ below 1~GeV by comparison with the
data.\\ 
The {\it second order} BS formalism,  
developed in~\cite{BMP} and applied to the calculation of
a rather complete quarkonium spectrum in Refs.~\cite{BP,Baldicchi:2004wj},
is essentially derived from the QCD Lagrangian taking advantage of a Feynman-Schwinger 
representation for the solution of the iterated Dirac equation in an external
field. Confinement is encoded through an ansatz on the Wilson
loop correlator (see Ref.~\cite{BMP} for details); indeed the expression 
$i\ln W$ is written as the
sum of a one-gluon exchange~(OGE) and an area terms
\begin{equation}
 i\ln W = (i\ln W)_{\rm OGE}+ \sigma S \,.
\label{eq:wilson}
\end{equation}
By means of a three dimensional reduction, the original BS
equation takes the form of the eigenvalue equation for a squared
bound state mass
\begin{equation}
    M^2 = M_0^2 + U_{\rm OGE}+U_{\rm Conf} \,,
\label{eq:M2}
\end{equation}
where $M_0$ is the kinematic term
$ M_0 = w_1+w_2 = \sqrt{m_{1}^2 + {\bf k}^2} + \sqrt{m_{2}^2 + {\bf
     k}^2}$,
${\bf k}$ being the c.m.\ momentum of the quark, $m_1$ and $m_2$
the quark and the antiquark constituent masses, and $U=U_{\rm
OGE}+U_{\rm Conf}$ the resulting potential.\\
As a consequence of ansatz (\ref{eq:wilson}), the perturbative part of
the potential $ U_{\rm OGE}$ turns out to be proportional to
$ \alpha_s(Q^{2})\,$, that should be regularized with a proper prescription.

\begin{figure}[htbp!]
 \begin{picture}(1100,500)
 \put(10,300){\includegraphics[height=.30\textheight,
                      width=.32\textheight]{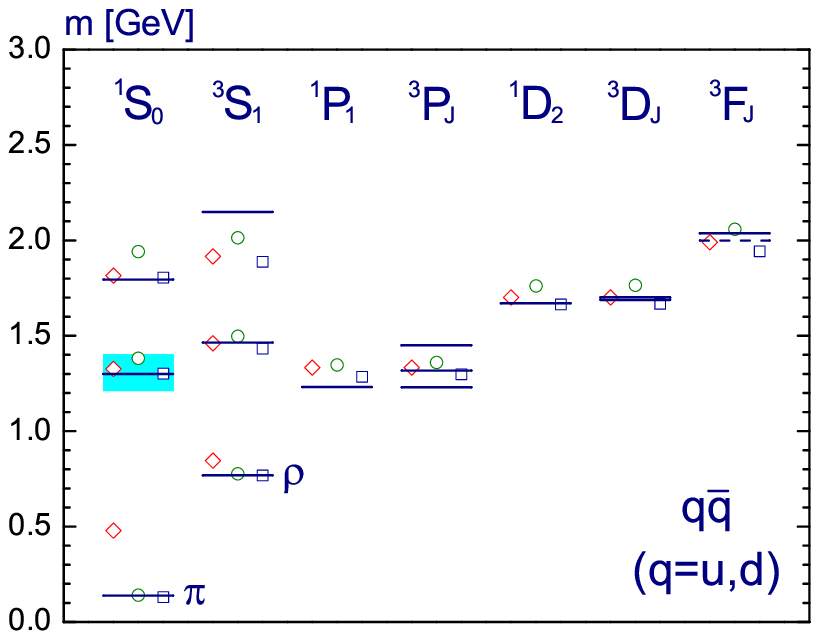}}
 \put(220,300){\includegraphics[height=.30\textheight,
                      width=.32\textheight]{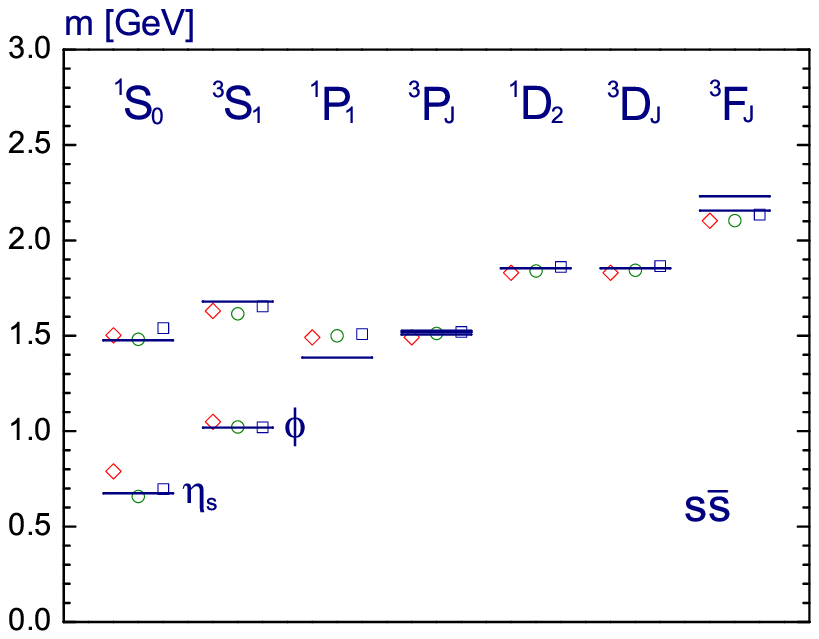}}
 \put(10,100){\includegraphics[height=.30\textheight,
                      width=.32\textheight]{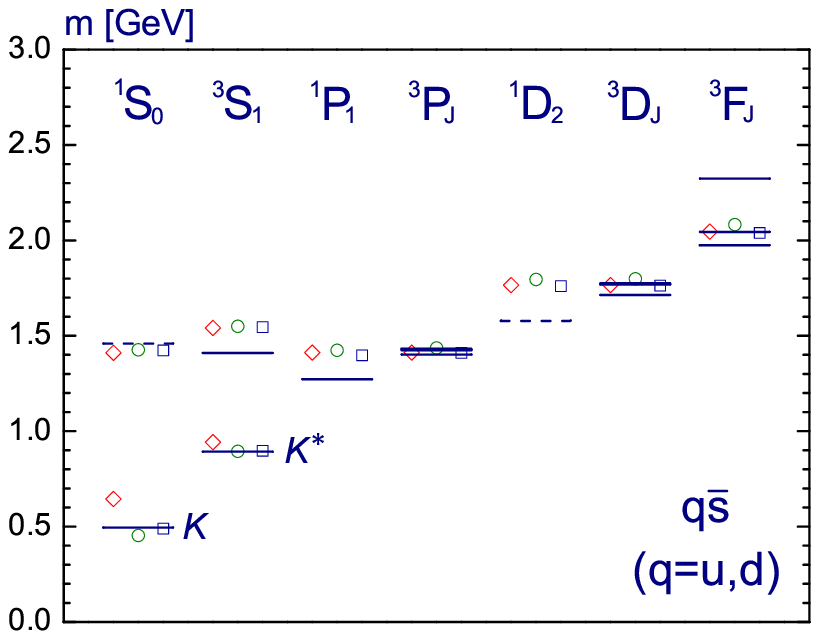}}
 \put(220,100){\includegraphics[height=.30\textheight,
                      width=.32\textheight]{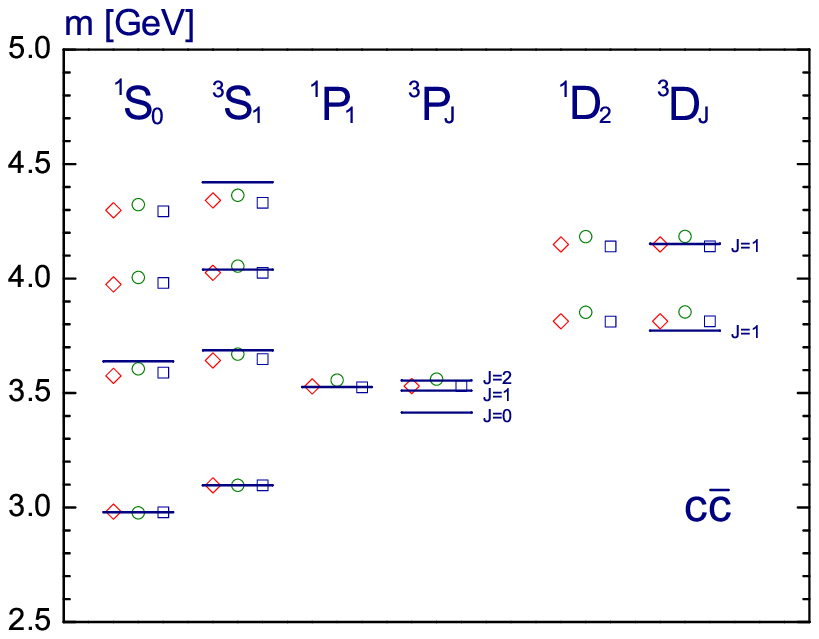}}
 \end{picture}
\vspace{-2.5truecm}
\caption{
\footnotesize
Quarkonium spectrum, three different calculations.
Diamonds refer to the truncation prescription for the running coupling, squares
and circles refer to the calculation with the 1-loop APT coupling
(\ref{ARC1L}) and two different expressions for running constituent masses
of light quarks, a solution of the Dyson-Schwinger equation and a 
phenomenological function of the c.m. quark momentum respectively. 
Horizontal lines represent experimental data.}
\label{spect}
\end{figure}

Calculations have been performed in Refs.~\cite{BP,Baldicchi:2004wj} by
using both a frozen and the 1-loop analytic coupling 
$\alpha^{(1)}_E(Q^{2})$ as given by Eq.\ (\ref{ARC1L}) with an effective scaling parameter
$\Lambda_{n_f=3}^{(1,{\rm eff})}\simeq 200\,$MeV.\\
The results of the two sets of calculations are relatively similar
for the heavy-heavy quark states. However, for the $1$S
states involving light and strange quarks, quite different results
have been obtained in the two cases. In the case of a frozen
coupling the $\pi$ and $K$ masses turn out to be too high, irrespective 
of how small the light quark mass is taken (see Fig.~\ref{spect});
e.g., if the light and the strange quark masses are fitted to the $\rho$
and the $\phi$ masses, one finds $m_\pi\sim 500\,$MeV and $m_K\sim 700\,$MeV,
respectively. On the contrary, if appropriate values for the quark
masses are chosen, the $\pi,\,\,\rho,\,\,K,\,\,K^*,\,\,\phi$
masses can be rather well reproduced when the analytic coupling
$\alpha_{E}^{(1)}(Q^{2})$ is used. This occurrence
strongly supports the use of the APT prescription, outlined in Sec.~2, within 
BS framework.\\
Indeed, the combined BS-APT theoretical scheme provides a substantial agreement 
of the calculated spin averaged c.o.g.\ masses with the data throughout the whole
calculable spectrum, and in particular the correct splittings
$1^3S_1$-$1^1S_0$ is well reproduced at least in the light-light,
light-strange and heavy-heavy sectors.\\

In this work a thorough analysis has been undertaken from the reversed point
of view, that is by comparing
theoretical results of the meson spectrum within the BS approach, obtained for a definite 
choice of the parameters, with the
results of a similar calculation performed by means of a fixed value
of~$\alpha_s$, for every quark-antiquark state. In what follows 
$\alpha_s^{\rm th}$ denotes the value that reproduces the
same theoretical result as by using 
$ \alpha_E^{(1)}(Q^2)\,$, whereas 
$\alpha_s^{\rm exp}$ the value that correctly reproduces
the experimental mass. The value $\alpha_s^{\rm th}$ is clearly
an intermediate step object, 
used to identify an effective $Q$ pertaining to each 
meson state, to be understood as the argument of
the related ``experimental'' coupling $\alpha_s^{\rm exp}(Q^2)$.\\
Since only the leading perturbative contribution in the BS kernel has
been included, a rough estimate of NLO effects on the extracted 
$\alpha_{s}^{\rm exp}$ values leads to a relative theoretical error
which spans from $20\%$ to much less than $1\%$ throughout the
spectrum, according to the quark masses involved. Furthermore,
since coupling among different quark-antiquark channels has not been
taken into account within BS formalism, the theoretical masses are expected to reproduce
the experimental ones roughly within the half width $\Gamma/2$ of the state.
If relevant, the experimental error, related to the uncertainty of the experimental masses, 
is 
added to the theoretical one.\\
The results are twofold. On the one hand, the 3-loop
APT coupling, normalized at the world average
$\alpha_s(M_Z^2)=0.1176(20)\,$ and evolved across heavy quark thresholds,
reasonably fits $\alpha_{s}^{\rm exp}(Q^2)$ from
1~GeV down to 200~MeV, quantitatively confirming the relevance of
the APT approach to IR phenomena down to 200~MeV. On the other hand,
below this scale, with the limitation due to the
large errors, the experimental points give a wee hint
about the vanishing of $\alpha_{s}(Q^2)$ as $Q\to0$. This could correlate 
with some results from lattice simulations \cite{Lattice},
and can be theoretically discussed in
the framework of a recent ``massive'' modification \cite{MAPT} of
APT (see Sec.~2.4) which takes into account effects of a finite
threshold in the dispersion relation~\footnote{It should be noted,
however, that the existence of a non vanishing
finite limit lower than the universal Shirkov-Solovtsov freezing
value, as suggested in~\cite{Badalian:2007km}, 
is still consistent with the above results.}.\\

\section{BS-model for quarkonium states}

As mentioned, in~\cite{BP,Baldicchi:2004wj} the meson spectrum is
obtained by solving the eigenvalue equation for the squared mass
operator~(\ref{eq:M2}), where the perturbative and the confinement parts 
of the potential are respectively
\bea
& &\langle {\bf k} \vert U_{\rm OGE}  \vert {\bf k}^\prime \rangle=\nn\\
\nn\\
& &  {4\over 3} {\alpha_s({\bf Q}^2) \over \pi^2}
     \sqrt{(w_1+w_2) (w_1^\prime + w_2^\prime)
     \over w_1 w_2 w_1^\prime w_2^\prime}
     \Bigg[ - \frac{1}{{\bf Q}^2}
     \bigg( q_{10} q_{20} + {\bf q}^2 - { ( {\bf Q}\cdot {\bf q})^2 \over
{\bf Q}^2 } \bigg) \nonumber \\
& & + {i\over 2 {\bf Q}^2} {\bf k}\times {\bf k}^\prime \cdot ({\bf
     \sigma}_1 + {\bf \sigma}_2 ) + {1\over 2 {\bf Q}^2 } [ q_{20}
(\alpha_1 \cdot {\bf Q}) - q_{10} (\alpha_2\cdot {\bf Q}) ]+
    \nonumber \\
& & + {1\over 6} {\bf \sigma}_1 \cdot {\bf \sigma}_2 + {1\over 4}
    \left ( {1\over 3} {\bf \sigma}_1 \cdot {\bf \sigma}_2 -
    { ( {\bf Q}\cdot \sigma_1)
    ( {\bf Q}\cdot {\bf \sigma}_2) \over {\bf Q}^2 } \right )
    + {1\over 4 {\bf Q}^2 } ( \alpha_1 \cdot {\bf Q}) ( \alpha_2 \cdot
{\bf Q}) \Bigg ]\qquad
\lb{upt}
\eea
and
\be
\langle {\bf k} \vert
U_{\rm Conf}  \vert {\bf k}^\prime \rangle ={\sigma\over ( 2 \pi)^3}
\sqrt{(w_1+w_2) (w_1^\prime + w_2^\prime)
     \over w_1 w_2 w_1^\prime w_2^\prime} \int d^3{\bf r}\,
     e^{i {\bf Q}\cdot{\bf r}}
J^{\rm inst}({\bf r}, {\bf q}, q_{10}, q_{20})
\lb{ucf}
\ee
with
\bea
&& J^{\rm inst}({\bf r}, {\bf q}, q_{10}, q_{20})= { r \over
q_{10}+q_{20}}
   \left[ q_{20}^2 \sqrt{q_{10}^2-{\bf q}^2_\perp} +
q_{10}^2 \sqrt{q_{20}^2 - {\bf q}_\perp^2}\right. +\qquad\nonumber\\
& & \qquad\left. + {q_{10}^2 q_{20}^2 \over \vert
{\bf q}_\perp \vert}
   \left(\arcsin{\vert{\bf q}_\perp \vert \over q_{10} }
   + \arcsin{\vert {\bf q}_\perp\vert \over q_{20}}\right)\right]
   \nonumber\\
& & \qquad - {1\over r}  \left[ {q_{20} \over
    \sqrt{q_{10}^2-{\bf q}^2_\perp}}
   ( {\bf r} \times {\bf q}\cdot \sigma_1 + i q_{10} ({\bf r}\cdot
\alpha_1))\right. \nonumber \\
& & \qquad\left.+ {q_{10} \over \sqrt{q_{20}^2 -
{\bf q}^2_\perp}}
   ( {\bf r}\times {\bf q} \cdot
   \sigma_2 - i q_{20} ( {\bf r}\cdot{\bf \alpha}_2)) \right]\,.
   \label{eq:uconf1}
\eea
Here $ \alpha_j^k $ denote the usual Dirac matrices $\gamma_j^0
\gamma_j^k$, $\sigma_j^k$ the $4 \times 4$ Pauli matrices $ \left( \matrix
{\sigma_j^k & 0 \cr 0 & \sigma_j^k}\right ) $ and
${\bf q} ={ {\bf k}+ {\bf k}^\prime \over 2}\,, \quad {\bf Q}= {\bf k} -
{\bf k}^\prime \,, \quad q_{j0}= {w_j+w_j^\prime \over 2}$, $m_{1}$ and
$m_{2}$ are constituent masses.\\
Eqs.~(\ref{upt}-\ref{eq:uconf1}) have been derived from the ansatz (\ref{eq:wilson})
and a 3-dimensional reduction of a Bethe-Salpeter like equation (see
Refs.~\cite{BMP,BP,Baldicchi:2004wj} and App.~A for the details). Actually
in the calculation  of Refs.~\cite{BP,Baldicchi:2004wj} only the center of
gravity (c.o.g.) masses of the fine multiplets were considered as a rule,
and the spin dependent terms in (\ref{upt}-\ref{eq:uconf1}) (spin-orbit
and tensorial terms) were neglected with the exception of the hyperfine
separation term in (\ref{upt}), proportional to ${1\over 6} {\bf \sigma}_1
\cdot {\bf \sigma}_2\,$. Within this limitation, a generally good
reproduction of the spectrum was obtained for appropriate values of the
parameters, as shown in Fig.~\ref{spect}. Here the results of three sets of calculations
are displayed. Diamonds refer to the usual perturbative 1-loop coupling
(to be replaced in Eq.\ (\ref{upt})), frozen at a maximum value $H\,$,
which has been taken as an additional adjustable parameter. Squares and
circles both refer to the 1-loop APT coupling~(\ref{ARC1L}) with
$\Lambda\simeq 200\,$MeV. For light quarks a running constituent mass was
used too.\footnote{Circles refer to a phenomenological running mass as
function of the c.m.\ quark momentum $m_u^2=m_d^2 = 0.17 |{\bf k}| - 0.025
|{\bf k}|^2 + 0.15 |{\bf k}|^4$. Squares refer to a running constituent
mass resulting from a solution of the Dyson-Schwinger equation with an analytic RG
running current mass (see App.~A for some details).}\\
It should be stressed that only with the choice (\ref{ARC1L}) the $1^1S_0$
state has been correctly reproduced when light and strange quarks were
involved, as in the case of $\pi$ and the $K$ mesons.

\begin{figure}[th]
\centerline{\includegraphics[width=110mm]{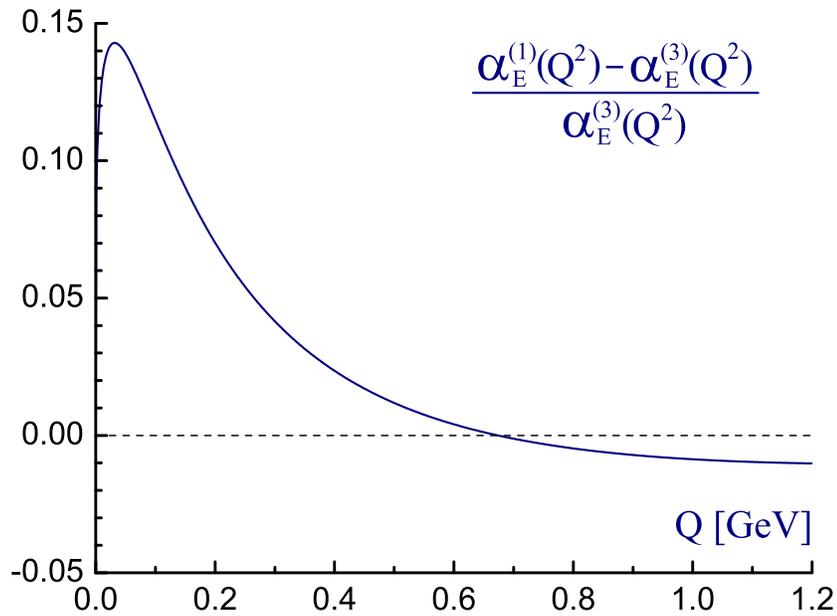}}
\vskip7.5mm
\caption{\footnotesize Relative difference between the one-loop analytic running
coupling $\alpha^{(1)}_{\rm E}(Q^2)$ with $\Lambda^{(1,{\rm eff})}_{n_f=3}
= 193\,$MeV and three-loop $\alpha^{(3)}_{\rm E}(Q^2)$ with
$\Lambda^{(3)}_{n_f=3}=417\,$MeV in the range $0<Q<1.2\,$ GeV.}
\label{dif}
\end{figure}

In the present work~\cite{Baldicchi:2007ic,Baldicchi:2007zn} 
a similar calculation with the input
(\ref{ARC1L}) and a slight different choice of the parameters
is made (preliminary results were given in~\cite{Baldicchi:2006az}).
First, the string tension has been fixed a priori to the value $\sigma=0.18\,\,{\rm
GeV^2}\,$ (consistent with hadron phenomenology and lattice
simulations) and the scale constant to
$\Lambda_{n_f=3}^{(1,{\rm eff})}=193\,$MeV. The whole set of remaining
parameters, all the quark masses, are then determined by fitting
the $\pi\,$, $\phi\,$, $J/\psi\,$ and $\Upsilon\,$
masses. It turns out $m_u=m_d=196\,$MeV, $m_s=352\,$MeV,
$m_c=1.516\,$GeV and $m_b=4.854\,$GeV. The results for the meson
spectrum are given in the fourth column of the tables in App.~B.\\
The effective value for the QCD scale $\Lambda_{n_f=3}^{(1,{\rm eff})}$ has been
dictated by the comparison with the 3-loop analytic coupling
normalized at the Z boson mass (see Eqs.\ (\ref{sp3},\ref{sp3t}) and (\ref{ARC3L})) 
according to the world average. As displayed in Fig.~\ref{dif}, the relative difference
between the two curves is no more than 1$\%$ in the region of momentum transfer 
$0.5<Q<1.2\,$GeV, to which the bulk of the states used as an input
in the calculation belongs.\\
Furthermore, as already noted, the set of equations (\ref{upt}-\ref{eq:uconf1}) 
refer to a single definite
quark-antiquark channels. So, having correct relativistic
kinematics, they do not include coupling with
other channels like any potential model (see App.~A). Therefore, one can not
expect to have any insight into the splitting of
over-threshold complicated multiplets which involve mixture of different
states. Even the position of the c.o.g.\ mass is expected to be reproduced
only within one-half of the width of the state. This has been taken into
account in the estimate of the theoretical error (see Sec.~4).

The resolution method of the eigenvalue equation for the operator
(\ref{eq:M2}, \ref{upt}-\ref{eq:uconf1}) used 
in~\cite{BP,Baldicchi:2004wj} and in the present work can be summarized
as follows.

a) In the static limit the problem can be reduced to the corresponding
one for the center of mass Hamiltonian (see App.~A)
\be
H_{\rm CM} = w_1 + w_2  - {4 \over 3} {\alpha_s \over r } +  \sigma
r \, .
\label{eq:static}
\ee
b) The eigenvalue equation for (\ref{eq:static}) is solved
for a convenient fixed $\alpha_s$ by the
Rayleigh-Ritz method, using the three dimensional harmonic oscillator
basis and diagonalizing a $30 \times 30$ matrix.

c) The square of the meson mass is evaluated as $\langle \phi_a |
M^2 |\phi_a\rangle $, $ \phi_a $ being the eigenfunction obtained
in step b) (with $a$ the whole set of quantum numbers) and the
operator $M^2$ given by Eq.\ (\ref{eq:M2}).

d) Prescription c) is equivalent to treat $ M^2-H_{\rm CM}^2 $ as a first
order perturbation. Consistently the hyperfine separation should be given
by
\begin{eqnarray}
& & (^3m_{nl})^2-(^1m_{nl})^2 =
 {32 \over 9\pi} \int_0^\infty \!
     dk \, k^2  \int_0^\infty \! dk^{\prime} \,
     k^{\prime 2} \varphi_{nl}^* (k)  \varphi_{nl} (k^{\prime})\,\times
\nonumber \\
& &  \qquad \qquad \qquad \sqrt {{w_1+w_2 \over w_1 w_2}}
     \sqrt {{w_1^{\prime}+w_2^{\prime}
     \over w_1^{\prime} w_2^{\prime}}} \int_{-1}^1 \! d\xi \,
     \alpha_s({\bf Q}^2) P_l(\xi)\, ,
\label{eq:hyper}
\end{eqnarray}
where $\varphi_{nl}$ is the radial part of the complete eigenfunction
$\phi_a\,$.\\
For the quark masses and string tension $\sigma$ in~(\ref{eq:static})
the same values listed above have been used, and as far as 
$\alpha_{s}\,$ is concerned, that is supposed to be a constant
in~(\ref{eq:static}), the value $\alpha_s=0.35\,$ has been used, which is
the one typically used in non-relativistic calculations, and is also the
freezing value adopted in~\cite{BP}.

\section{Extracting $\alpha_s^{\rm exp}(Q^2)$ from the data}

One focus now on the reversed problem, i.e., the determination
of the $\alpha_s^{\rm exp}(Q^2)$ values at the characteristic
scales of a selected number of ground and excited states.\\
In order to estimate $\alpha_s^{\rm exp}(Q^2)$ at low energy 
one needs first to assign an effective $Q$-value to each state.
To this end one first rewrites the squared mass, as given by point c)
in Sec.\ 3.2, more explicitly as the sum of the unperturbed part, the
perturbative and the confinement ones respectively
\be
m^2_{a}
=\langle\phi_a|M_0^2|\phi_a\rangle+
\langle\phi_a|U_{\rm OGE}|\phi_a\rangle+
\langle\phi_a|U_{\rm Conf}|\phi_a\rangle\,.
\lb{m_th}
\ee
Here $U_{\rm OGE}$ is given by the second line of (\ref{upt}) and $U_{\rm Conf}$
by Eq.\ (\ref{ucf}) and the first two lines of (\ref{eq:uconf1}).
From the OGE contribution one then extracts for each state the
fixed coupling value $\alpha_{a}^{\rm th}\,$ that leads
to the same theoretical mass as by using $\alpha_{\rm E}^{(1)}(Q^2)$
given by Eq.\ (\ref{ARC1L}). This can be done by means of the
relation
\be
\langle\phi_a|U_{\rm OGE}|\phi_a\rangle
\equiv\langle\phi_a|\alpha_{\rm E}^{(1)}({\bf Q} ^2){\cal O}
({\bf  q};{\bf Q})
|\phi_a\rangle=\alpha_{a}^{\rm th}\langle\phi_a|{\cal O}
({\bf q};{\bf Q})
|\phi_a\rangle,
\lb{a_th}
\ee
where ${\cal O}({\bf q};{\bf Q})$ can be drawn again by the
second line of Eq.~(\ref{upt}). The effective momentum transfer $Q_a$
associated to each bound state is then identified by the equation
\be
\alpha_{\rm E}^{(1)}(Q_a^2)=\alpha_{a}^{\rm th}\,.
\lb{Qeff}
\ee
The next step is to search for the correct (fixed) value of the
coupling that exactly reproduces the experimental mass of
each state. This is defined by the relation
\be
\langle\phi_a|M_0^2|\phi_a\rangle+
\alpha_s^{\rm exp}(Q^2_a)\langle\phi_a|{\cal O}({\bf q};{\bf Q})
|\phi_a\rangle +
\langle\phi_a|U_{\rm Conf}|\phi_a\rangle=m^2_{\rm exp}\,,
\lb{m_exp}
\ee
so that, by combining Eqs.\ (\ref{m_th}), (\ref{a_th}) and (\ref{m_exp})
one finally obtains
\be
\alpha_s^{\rm exp}(Q^2_a)=\alpha_{a}^{\rm th}+\frac{m^2_{\rm exp}-m^2_{a}}
{\langle\phi_a|{\cal O}({\bf q};{\bf Q})|\phi_a\rangle}\,.
\lb{a_exp}
\ee
This procedure has been applied to a number of light-light,
light-heavy and heavy-heavy ground as well as excited states.

Before discussing the uncertainties related to the extracted 
$\alpha_s^{\rm exp}(Q^2)\,$, few comments are in order.\\
First note that in the evaluation of $Q_a$ in (\ref{a_th})
the hyperfine splitting has been neglected, whereas it has been taken
into account in (\ref{a_exp}), bringing possibly to different values of
$\alpha_s^{\rm exp}$ for the singlet and the triplet states
(when there are reliable data for both). \\
Furthermore, the sensitivity of the effective $Q$'s derived as above 
from the specific coupling (\ref{ARC1L}), has been
checked by analyzing their deviation for a $25\%\,$ shift of
$\Lambda_{n_f=3}^{(1,{\rm eff})}$ around the value 193\ MeV, and one finds
that the average change in the momentum scale amounts to~$3\%\,$.
This makes the resulting
$\alpha_s^{\rm exp}(Q^2)$ reliable, at least qualitatively, even in
the deep IR region ($Q<0.2\,$GeV), where the discrepancy with respect
to massless $\alpha_{E}^{(1)}(Q^2)\,$ is sizable.\\
Obviously the theoretical meson masses are sensitive to a variation 
of the quark mass parameters (particularly in the case of the $\pi$ meson), 
whereas $\alpha_s^{\rm exp}\,$ and the relative 
$Q_a$ turn out to be much more stable. For instance, in the 
light-light sector, an increase in the light quark mass of 5$\%$ amounts to 
a change of about 
$2\%$ in the value of $\alpha_s^{\rm exp}\,$ and $0.2\%$ in the relative
$Q_a\,$.\\ 
Finally, a subtle point concerns the choice of
the ``unperturbed'' $\alpha_s$ involved in the static
Hamiltonian (\ref{eq:static}). Actually, the value adopted is very near
to the $\alpha_{a}^{\rm th}$ pertaining to the $b\bar{b}(1S)$ state,
but definitively smaller than the typical
$\alpha_{a}^{\rm  th}\,$'s. 
The point is that the hyperfine splitting is much
more sensible than the c.o.g.\ mass to the behaviour of the unperturbed
wave function at small distance (large momentum), which is specifically
controlled by the value of the unperturbed~$\alpha_s\,$. As a result,
the effective fixed value $ \alpha_s $ in
Eq.\ (\ref{eq:static}) that reproduces the same splitting as by using the coupling
$ \alpha_{\rm E}^{(1)}(Q^2) $ turns out to
be significantly smaller than $\alpha_{a}^{\rm th}$ calculated from
the c.o.g.\ mass. Essentially, it was chosen a phenomenological value
for the unperturbed $\alpha_s$ in order to have a good reproduction
of the hyperfine splitting so as to reasonably reconstruct the
c.o.g.\ of the doublet when one component is missing. It was then used
the position of the c.o.g.\ (which is rather stable w.r.t.\
the unperturbed $\alpha_{s}$) to extract the final result 
$\alpha_{s}^{\rm exp}(Q_a^2)\,$.

\section{Theoretical uncertainties}

First of all, it should be clearly addressed the unavoidable model dependence of the 
above results. Indeed, the formalism 
exploited in their derivation is based on the  
ansatz~(\ref{eq:wilson}), consisting of 
the sum of two contributions that one knows to be asymptotically correct for
small and large quark-antiquark distances. More sophisticated ansatz also 
exist (see e.g.~\cite{Billo:2006zg}, \cite{Dosch:1987sk} or 
\cite{Baker:1994nq}), but they turn out to be difficult to implement
within BS formalism. \\
In the context of the model in hand the sources of error are the istantaneous
approximation implied by the three-dimensional reduction 
of the BS equation, the approximations introduced into the resolution of
the eigenvalue equation, the inclusion of only the leading perturbative
contribution in the BS kernel $I$, and finally having neglected the coupling
between different
quark antiquark channels.\\ 
Thus, aside from minor effects due to approximations in the resolution
of the eigenvalue equation and from the three-dimensional reduction, the main 
sources of theoretical error in the
whole procedure are expected to arise from neglecting the NLO
contribution to the BS kernel as well as the coupling with other
channels, and these have been explicitly estimates.
\begin{figure}[!t]
\vspace{-3.5truecm}
\centerline{\includegraphics[width=0.8\textwidth,height=0.8\textheight]{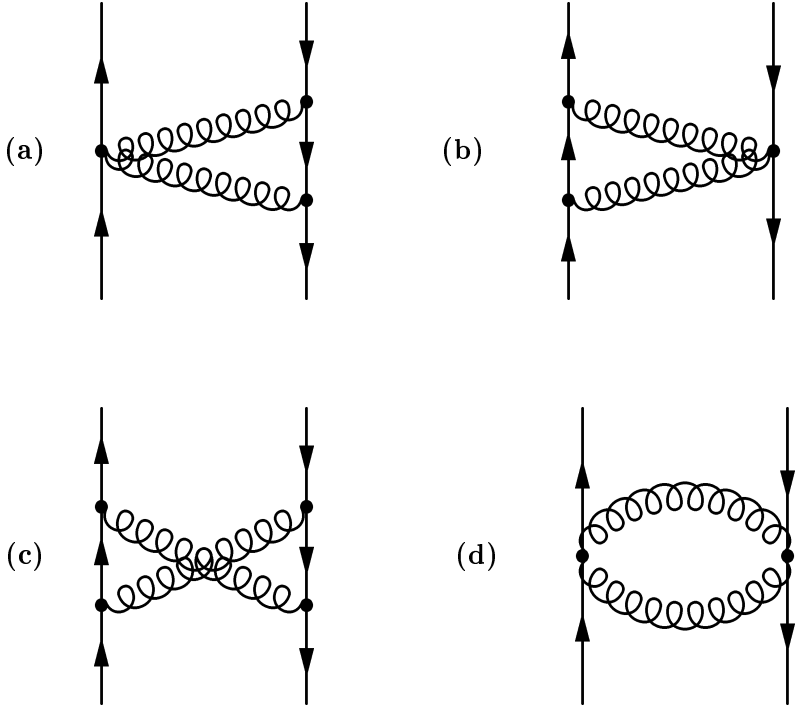}
}\caption{{\footnotesize NLO contributions to the second order BS kernel $I$.}} 
\lb{nlo}
\end{figure}

The NLO contribution to the perturbative
part of the BS-kernel comes from four diagrams with two-gluon
exchange; two triangular graphs containing a four-line vertex of
the type $g^2\phi^*\phi A_{\mu}A^{\mu}$ and two three-line
vertices $g\phi^*\partial_{\mu}\phi A^{\mu}$ (the spin independent
part of the second order BS formalism is quite similar to scalar
QED), one {\it fish diagram} with two four-line vertices, and a
crossing box with four three line vertices (see Fig.\ \ref{nlo}). If the renormalization
scale is identified with the momentum transfer $Q$ the fish graphs
contribution is completely reabsorbed in the renormalization. On
the other hand, a somewhat crude estimate of the contribution of
each of the two triangular graphs yields
\begin{equation}
        I_{\rm triang} \sim 4\left({4 \over 3}\,
        \alpha_s\right)^2 {9m^2 \over 4Q^2 + 2m^2}
\end{equation}
and for the crossing box graph, similarly
\begin{equation}
     I_{\rm crsbox} \sim  {64 \over 3}
     \left({4 \over 3}\,\alpha_s\right)^2
{m^4 \over (Q^2 + m^2 + k^2)^2}.
\end{equation}
These expressions have to be compared with the leading one-gluon
term used in the present calculation 
\begin{equation}
      I_{\rm OGE} \sim 16\pi {4 \over 3 }\,\alpha_s {m^2
          \over Q^2}\,.
\end{equation}
Putting all things together, the overall error due to the omission
of such NLO contributions to the BS kernel is then
\begin{equation}
{\Delta I\over I} = \sqrt{\left(2\,{I_{\rm triang}\over I_{\rm
OGE}}\right)^2+\left({I_{\rm crsbox}\over I_{\rm
OGE}}\right)^2}\,,
\end{equation}
and this produces
\begin{equation}
{{\Delta\cal O}\over{\cal O}}\sim{\Delta I\over I}\,.
\end{equation}
By using Eqs.~(\ref{a_th}-\ref{a_exp}), after some algebra it is easy
to recognize that the NLO effects on $\alpha_s^{\rm exp}$ turn
out to be of the same order, that is
\begin{equation}
\Delta_{\rm NLO}\alpha_s\sim\alpha_{a}^{\rm th}\,{\Delta
I\over I}\,,
\lb{errnlo}
\end{equation}
which is what is assumed in the foregoing.
The NLO errors do not exceed $5\%$ for heavy
quark states whereas they are enhanced up to $20\%$ when light and
strange quarks are involved.\\
Finally, since the strength of the neglected coupling with other
channels (OC) is obviously measured  by the width $\Gamma_a$ of
the state, one roughly estimates an error of the order of $\Delta
m_a\sim \Gamma_a/2\,$ in the evaluation of the theoretical meson mass $m_a\,$. On this
basis, for each determination of
$\alpha_s^{\rm exp}\,(Q^2_a)$
the related theoretical error is given by
\begin{equation}
\Delta_{\rm \Gamma}\alpha_s=\frac{m_a}
{\langle\phi_a|{\cal O}({\bf q};{\bf Q})|\phi_a\rangle}\,\Gamma_a\,.
\label{err}
\end{equation}
Usually the error $\Delta m_{\rm exp}$ on the experimental mass $ m_{\rm exp}$
is much smaller than $\Gamma_a/2\,$. When, however, this is not the case
one has to consider also the experimental error
$\Delta_{\rm exp}\alpha_s\,$, obtained from (\ref{err}) by
replacing $m_a\,\Gamma_a$ with $2m_{\rm exp}\Delta m_{\rm exp}\,$.

All other sources of uncertainties aformentioned, that is, the errors implied
by the three dimensional reduction, and by the approximations introduced in the 
resolution of the eigenvalue equation for $M^2\,$, and the one due to the model 
dependence, though difficult to be explicitly estimated, 
can be globally taken into account by means of a comparison between the calculated and the
experimental spectrum.  To this aim, one can restrict the considerations to a
sample of better established data which exclude high orbital excitations
(D and F states), and introduce an additional conventional error
$\overline{\Delta}m$ on all the theoretical masses, defined by
\begin{equation}
\chi_m^2=\frac{1}{N_{SP}} \sum_{a = 1}^{N_{SP}}
(m_{a} - m_{\rm exp} )^{2}/[(\Delta_{\rm tot} m_{a})^{2}
+ (\overline{\Delta}m)^{2}] \sim 1,
\end{equation}
where $\Delta_{\rm tot} m_{a} $ is the total error resulting from all the 
sources explicitly evaluated, i.e., $( \Delta_{\rm tot} m_{a} )^{2} =
m_{a}^{2} \Delta I/I + (\Gamma_{a}/2)^{2} + ( \Delta m_{\rm exp}
)^{2}$, and the sum is restricted to the selected sample of data
(namely, S and P states only). It turns out $\overline{\Delta}m\simeq
20\,$MeV, and then it can be set
$\overline{\Delta}\alpha_{s}=(2m_a\overline{\Delta}m)/
\langle\phi_a|{\cal O}({\bf q};{\bf Q})|\phi_a\rangle\,$. This
uncertainty, that turns out to be about $5\%$ on average, has not
been included in the tables, but is used in the analysis of the
following section.

\section{BS-model results: concert of low and high energy data via APT}

All results are displayed in details in tables I-VII of App.~B, and
pictorially in Fig.\ \ref{low}.  
The first three columns specify the
state and its experimental mass as given by~\cite{pdg}. The fourth
column gives theoretical results for the  meson masses, and the
last three give the effective $Q\,$'s, the relative 3-loop APT coupling
$\alpha_{E}^{(3)}(Q^2)$ and the experimental coupling with errors (both 
theoretical and experimental).\\
In Fig.\ \ref{low} values of $\alpha_s^{\rm exp}$ at the same Q
from triplet and singlet states have been combined through a weighted
average according to their errors.

As one may infer from Fig.~\ref{low}, the experimental points exhibit
a remarkable evolution from 500~MeV down to 200~MeV, where only the
safer S and P states are involved, in good agreement with the 3-loop
analytic coupling $\alpha_{\rm E}^{(3)}(Q^2)$ properly normalized,
i.e., with $\Lambda_{n_f=3}^{(3)}=417\,$MeV, and discussed in Sec.~2
(cf. Eqs.~(\ref{ARC3L}) and~(\ref{sp3}), (\ref{sp3t})).\\ 
 Specifically, defining
$\Delta_{\rm tot}\alpha_{s}$ as the total error explicitly evaluated
by means of Eqs.~(\ref{errnlo}) and~(\ref{err}), one finds 
\begin{equation}
 \chi^{2}_{\alpha} = \frac{1}{N_{SP}} \sum_{a = 1}^{N_{SP}}
( \alpha^{\rm exp}_{s}(Q_{a}^{2}) - \alpha_{E}^{(3)}(Q_{a}^{2}))^{2}/
[(\Delta_{\rm tot} \alpha_{s} )^{2} +
( \overline{\Delta} \alpha_{s} )^{2} ] \sim 0.8\,.
\end{equation}
The agreement quantitatively supports the relevance of the APT
approach to IR phenomena down to a few hundred~MeV.

On the other hand, at energies below 200~MeV there seems to exist a general tendency of
$\alpha_{\rm s}^{\rm exp}(Q^2)$ to deviate from the APT curve and to
approach zero, or at least a finite limit, which is less than the
universal APT freezing value. Note, however, that the analysis of
such an extreme IR behaviour is based on high orbital excitations (D
and F states), that lie well above the strong decay thresholds and
possess large widths. As a consequence, the theoretical reliability
of the method is lower at these scales, as apparent from the large
estimated errors. Moreover, also the discrepancy between $\alpha_{\rm
E}^{(1)}(Q^2)$ (used in the calculation) and $\alpha_{\rm
E}^{(3)}(Q^2)$ (used as a reference term) rises above 10$\%$ at these
scales. In fact, only two states $\pi_2$(1670) (interpreted as $s\bar
s\,(1^1D_2)$) and $f_2$(2150) ($s\bar s\,(1^3F_2)$), corresponding to
$Q\simeq 120\,$MeV, generate $\alpha_{\rm s}^{\rm exp}(Q^2)$
(marginally) out of the error bands, and the state $f_2$(2150), which
has been observed only once, has never been confirmed.\\
Nevertheless, as already noted, such a deep IR behaviour could
correlate with some lattice results~\cite{Lattice}, and could be
discussed within theoretical models, in particular within the recently developed
``massive'' modification of APT, briefly 
discussed in Sec.~2.4. Specifically, as displayed
in Fig.~\ref{low}, the 1-loop coupling $\alpha(Q^2,
m^2)$~(\ref{MARC}) with an effective mass $m\ind{}{eff} \simeq (38
\pm 10)\,$MeV reasonably fits all experimental points down to the deep
infrared region.
%
\begin{figure}[!t]
\centerline{\includegraphics[width=125mm]{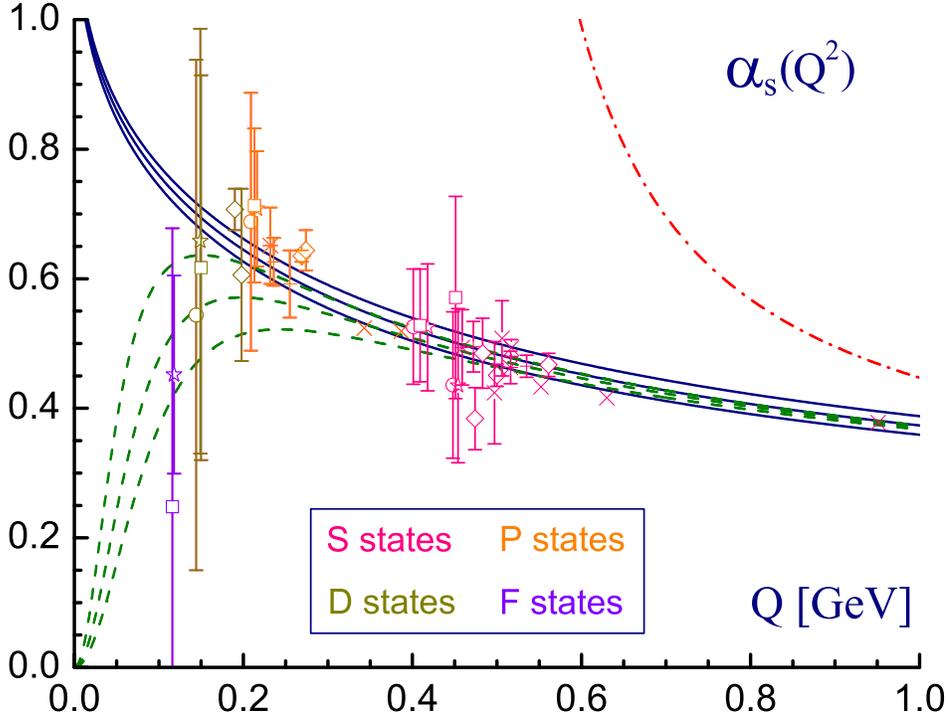}}
\caption{{\footnotesize Extracted values of $\alpha_{s}^{\rm exp}(Q^2)$
against the 3-loop APT coupling~(\ref{AAPT}) with
$\Lambda^{(3)}_{n_f=3}=(417\pm 42)\,$MeV (solid), and its
perturbative counterpart (dot-dashed). The ``massive'' 1-loop APT
coupling ($n=1$ in~(\ref{AMAPT})) refers to $\Lambda^{(1,{\rm
eff})}_{n_f=3}=204\,$MeV and $m\ind{}{eff}=(38 \pm 10)\,$MeV
(dashed). Circles, stars and squares refer respectively to $q\bar
q\,$, $s\bar s\,$ and $q\bar s\,$ with $q=u,d\,$, diamonds and
crosses to $c\bar c\,$ and $b\bar b\,$; asterisks stay for $q\bar
c\,$ and $q\bar b\,$, whereas plus signs for $s\bar c\,$ and $s\bar
b\,$.}}
\lb{low}
\end{figure}

Notice that in the selection of states irregular
and incomplete multiplets have been excluded as a rule. Of this type, e.g., 
in the light quark sector, are the $3S$ states ($ m_{3\,^3S_1}-m_{3\,^1S_0} $
is anomalously large and about twice as
$ m_{2\,^3S_1}-m_{2\,^1S_0} $), $1\,^3P$ ($ m_{1\,^3P_0} $ being larger than
$ m_{1\,^3P_1} $), $1\,^3D$, $F$, $G$, $H$ (incomplete).
If however included in the analysis, all these states would bring the results 
in agreement with the general tendency outlined.

As far as the renormalization scheme dependence is concerned, note that 
the coupling definition is implicitly embodied in ansatz~(\ref{eq:wilson}). 
Specifically, here one assumes that after the area term subtraction, $i\ln W$ 
is dominated by the OGE term, with the fixed
coupling~$\alpha_{s}$ replaced by the running coupling ~$\alpha_{s}(Q^2)\,$. 
This amounts to including all the dressing effects into $\alpha_{s}(Q^2)$.
It is
worth noting that the coupling defined in such way is free of
unphysical singularities by construction.\\
At the same time, the 1-loop effective APT coupling
$\al{}{E}(Q^2)$, involved in the calculation of the meson spectrum, is remarkably
stable with respect to both the higher loop corrections and the
choice of renormalization scheme (see Sec.~2. and a detailed
discussion of this issue in Ref.~\cite{APT-07}). Thus one might expect that
the same situation should also occur for~$\al{}{exp}(Q^2)$, with the possible
exception for the deep infrared region, where other
nonperturbative effects could be relevant.\\
Quark self-energy effects have been taken into account by a recursive 
resolution of the Dyson-Schwinger equation. In the second order BS 
formalism this simply amounts to replacing the current quark masses with 
the constituent masses \cite{Baldicchi:2004wj}.
\begin{figure}[!hb]
\centerline{\includegraphics[width=125mm]{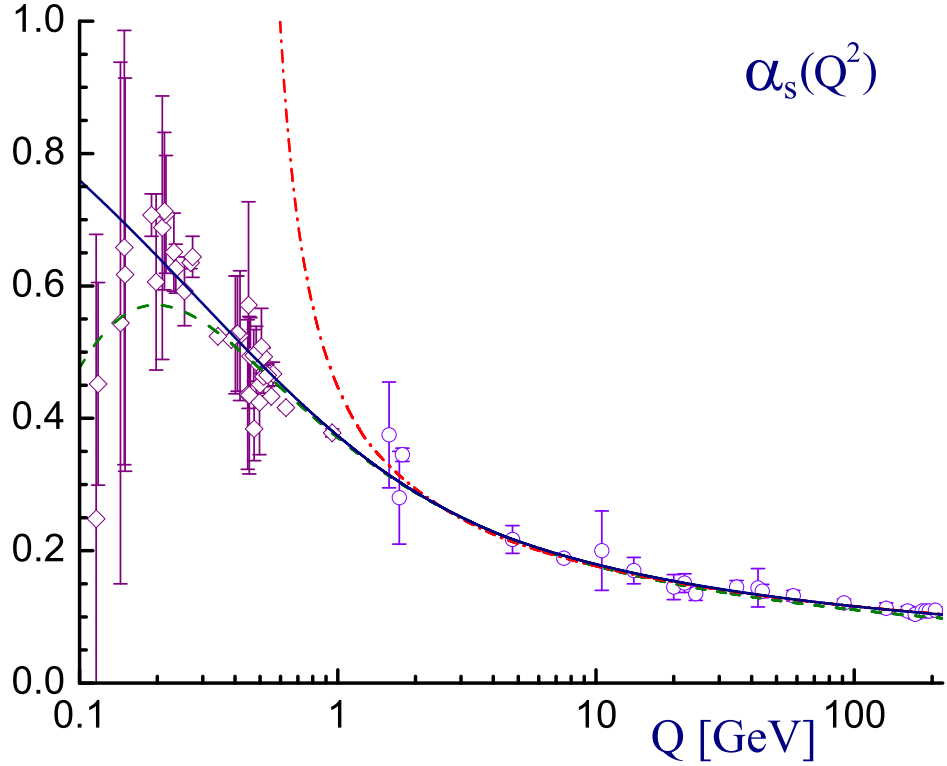}}
\vskip7.5mm
\caption{\footnotesize Summary of low ($\diamond$) and high energy ($\circ$) data
against the 3-loop analytic coupling~(\ref{ARC}) (solid curve) and its
perturbative counterpart (dot-dashed curve) both normalized at the Z~boson
mass. Also shown is the ``massive'' 1-loop analytic
coupling~(\ref{MARC}) (dashed curve) same as in Fig.~\ref{low}.}
\lb{tot}
\end{figure}

Finally, Fig.~\ref{tot}
displays a synthesis of the results for
$\alpha_{s}(Q^2)$ extracted from the bound states calculation in the BS framework, 
with a sample of high energy data as given by S.~Bethke~\cite{bethke},
against the 3-loop APT coupling $\alpha^{(3)}_{E}(Q^2)$~(\ref{ARC})
and its massive modification~(\ref{MARC}).
Also shown in the figure is the common perturbative 3-loop coupling with IR 
singular behaviour that is ruled out by the data. As can be seen, the
BS-APT theoretical scheme allows a rather satisfactory correlated
understanding of very high and rather low energy phenomena.

\chapter{Conclusive remarks}

To summarize, as discussed in Sec.~1, straightforward application 
of the RG method to perturbative expansions in QCD eventually
gives rise to unphysical singularities in both the running coupling
and the physical observables, in the low energy domain. The Landau singularities
severely complicate theoretical analyses of hadron dynamics.\\
Then, an useful prescription to eliminate unphysical singularities is
provided by the APT approach, outlined in Sec.~2, which is 
based upon the analycity requirement of the QCD coupling. 
This prescription has been used in previous works in the framework
of a Bethe-Salpeter like formalism to compute the meson spectrum.\\ 
In this work, the point of view has been then somewhat reversed,   
namely, as discussed in Sec.~3, the comparison between the
calculated meson spectrum within the BS framework and the experimental 
data is used, in order to extract information on the infrared behavior 
of the QCD running coupling. 

The method consists in solving the eigenvalue equation for the
squared mass operator as given by Eq.~(\ref{eq:M2}), obtained by a
three dimensional reduction of the original BS equation. The
relativistic potential $U$ then follows from a proper
ansatz~(\ref{eq:wilson}) on the Wilson loop to encode confinement,
and is the sum~(\ref{eq:M2}) of a one-gluon-exchange term $U_{\rm
OGE}$ and a confining term $U_{\rm Conf}\,$. The strong coupling occurring
in the perturbative part of the potential needs to be IR finite since
its argument has to be identified with the momentum transfer in the
$q\bar{q}$ interaction, and this typically takes values down to few
hundreds MeV.\\ 
As a first step, the common perturbative running coupling
$\alpha_{s}$ has then been replaced by its 1-loop analytized 
counterpart $\alpha_{E}^{(1)}$ Eq.~(\ref{ARC1L}) as derived by the 
APT algorithm, which avoids the hurdle of the spurious singularities in the 
IR region~\cite{ShSol96-7}, the whole theoretical scheme being
extensively discussed in Sec.~2.\\
An effective QCD scale $\Lambda_{n_f=3}^{(1, \rm{eff})}=193$~MeV
has been imposed in order to reasonably reproduces the 3-loop APT coupling, 
normalized at the Z~boson mass along with world average~\cite{pdg} (i.e.,
$\Lambda_{n_f=5}^{(3)}=236\,$MeV and $\Lambda_{n_f=3}^{(3)}=417\,$MeV by 
continuous threshold matching), specifically in the neighbohood of 0.5~GeV, 
which is the typical scale of most of the imput states.\\
Then, it has been taken advantage of the BS results for the meson
spectrum, both in the light and heavy quark sector, to deduce within
this framework the fixed coupling value for each state that exactly
matches the theoretical and experimental mass.\\
A key point is finally the comparison of such experimental determinations
$\alpha_{s}^{\rm exp}$ with the ``Euclidean'' APT coupling $\alpha_{E}^{(3)}(Q^2)\, $
and its further ``massive'' development.

The results are twofold. On the one hand, as expected on the basis of
the reasonable agreement between the theoretical and the experimental meson data, the 3-loop
analytic coupling remarkably fits the extracted points $\alpha_{s}^{\rm exp}(Q^2)$
from 1~GeV down to 200~MeV, within the evaluated errors, both theoretical and
experimental. This confirms and
yields a quantitative estimate of the relevance of the APT
to IR phenomena down to 200~MeV.\\
Besides, below this scale, the experimental points
exhibit a tendency to deviate with respect to the APT curve.
Despite the sizable errors, this could give a hint 
on the vanishing of $\alpha_{\rm s}(Q^2)$ as $Q\to0\,$ in
concert with some results from lattice simulations~\cite{Lattice}. 
It should be emphasized, however, that the existence of a finite IR limit 
of $\alpha_{s}(Q^2)\,$, lower than the universal APT freezing value,
can not be ruled out by these data.\\
Neverthless, a vanishing QCD coupling at zero momentum transfer
can be theoretically discussed in the framework of a recent ``massive'' 
modification~\cite{MAPT} of the APT formalism, which takes into account 
effects of a finite threshold in the dispersion relations. Since in the 
extremely low $Q$ region confinement forces play the dominant role, the
overall reasonable agreement between the ``massive'' APT model and the 
results of the BS formalism would suggest a relation between the linear
potential, arising from the area term in the ansatz (\ref{eq:wilson}),
and the thresholds effects in the analytic properties of the QCD coupling, 
that needs to be further investigated.
\\
\newpage


\section*{Appendix A: Second order Bethe-Salpeter formalism}

In the QCD framework a {\it second order} four point quark-antiquark
function and the full quark propagator can be defined as
\begin{equation}
H_{4}(x_1,x_2;y_1,y_2)= -{1\over 3} {\rm Tr _{color}}
\langle \Delta_1 (x_1,y_1;A)
{\Delta}_2(y_2,x_2;A)\rangle
\label{eq:so4point}
\end{equation}
\noindent
and
\begin{equation}
H_{2}(x-y) = {i \over \sqrt{3}}{\rm Tr_{color}}
\langle \Delta(x,y;A)\rangle \,,
\label{eq:so2point}
\end{equation}
where
\begin{equation}
\langle f[A] \rangle = \int  DA\, M_F [A]\, e^{iS_G[A]} f[A]  \,,
\label{eq:expt}
\end{equation}
\noindent
$
M_F[A] = {\rm Det} \, \Pi_{j=1}^2 [1 + g\gamma^\mu A_\mu
( i\gamma_j^\nu \partial_{j\nu} - m_j^{\rm curr})^{-1}]
$
and $\Delta (x,y;A)$ is the {\it second order} quark
propagator in an external gauge field.

The quantity $ \Delta $ is defined by
the second order differential equation
\begin{equation}
(D_\mu D^\mu +m^2_{\rm curr} -{1\over 2} g \, \sigma^{\mu \nu} F_{\mu \nu})
\Delta (x,y;A) = -\delta^4(x-y) \, ,
\label{eq:soprop}
\end{equation}
($\sigma^{\mu \nu} = {i\over 2} [\gamma^\mu, \gamma^\nu]$ and
$D_\mu=\partial_\mu + ig A_\mu$) and it is related to
the corresponding first order propagator by
$
S(x,y;A) = (i \gamma^\nu D_\nu + m_{\rm curr}) \Delta (x,y;A) \,
$, $m_{\rm curr}$ being the so-called current mass of the quark.

The advantage of considering second order quantities is that the spin
terms are more clearly separated and it is possible to write for
$\Delta$ a generalized Feynman-Schwinger representation, i.e., to
solve Eq.~(\ref{eq:soprop}) in terms of a quark path integral
\cite{BMP,BP}. Using the latter in (\ref{eq:so4point}) or
(\ref{eq:so2point}) a similar representation can be obtained for
$H_{4}$ and $H_{2}$.

The interesting aspect of this final representation is that the gauge field
appears in it only through a Wilson line correlator $W$.
In the limit $x_2 \to x_1$, $y_2 \to y_1$ or $y \to x$ the Wilson lines
close in a single Wilson loop $\Gamma$
and if $\Gamma$ stays on a plane,
$i\ln W$ can be written according to (\ref{eq:wilson}) as

\begin{eqnarray}
&& i\ln W = {16\pi\over 3}\alpha_{\rm s} \oint dz^\mu \oint dz^{\nu \prime}
D_{\mu \nu}(z-z^\prime) +
\label{eq:wils} \\
&& \sigma \oint dz^0 \oint dz^{0 \prime} \delta (z^0-z^{0\prime})
|{\bf z} - {\bf z}^\prime| \int_0^1 d\lambda
 \Big \{ 1 -  [\lambda {d{\bf z}_\perp \over dz^0}
 + (1-\lambda) {d{\bf z}_\perp ^\prime \over dz^{0 \prime}} ]^2
\Big \}^{1\over 2} \, . \nonumber
\end{eqnarray}
The area term here is written as the algebraic sum of successive
equal time strips and $ d{\bf z}_\perp = d{\bf z} -
(d{\bf z}\cdot {\bf r}){\bf r}/r^2 $ denotes the transversal component of
$ d{\bf z} $. The basic assumption now
is that, in the center of mass frame, 
(\ref{eq:wils}) remains a good approximation even
in the general case, i.e.,
for non flat curves and  when $x_2 \ne x_1$,
$y_2 \ne y_1$ or $y \ne x$.
Then, by appropriate manipulations on the resulting expressions,
an inhomogeneous
Bethe-Salpeter equation for the 4-point function
$H_{4}(x_1,x_2;y_1,y_2)$ and a Dyson-Schwinger equation for
$H_{2}(x-y)$ can be derived in a kind of generalized ladder and rainbow
approximation respectively. This should appear plausible, even from the point of
view of graph resummation, for the analogy between the perturbative
and the confinement terms in (\ref{eq:wils}).
In momentum representation, the corresponding homogeneous
BS-equation becomes
\begin{eqnarray}
\Phi_P (k) &=& -i \int {d^4u \over (2 \pi)^4} \;
   \hat I_{ab} \left( k-u; \, {1 \over 2}P
   +{k+u \over 2}, \,
   {1 \over 2}P-{k+u \over 2} \right)\,\,\times \nonumber \\
    & & \qquad\times\,\,
   \hat H_{2}^{(1)}   \left({1 \over 2} P  + k \right)
      \sigma^a  \, \Phi_P (u) \, \sigma^b \,
   \hat H_{2}^{(2)} \left(-{1 \over 2} P + k \right) \, ,
\label{eq:bshom}
\end{eqnarray}
\noindent
where $\sigma^0=1$; $a, \, b = 0, \, \mu\nu$;
the c.m.\ frame has to be understood, $P=(m_B, {\bf 0})$;
$\Phi_P (k)$ denotes the appropriate {\it second order} wave function,
that in terms of the second order field $\phi (x) = (i\gamma^\mu
D_\mu + m_{\rm curr})^{-1}\psi(x)$ can be defined as the Fourier
transform of
$
  \langle 0|\phi({\xi \over 2}) \bar\psi(-{\xi \over 2}) |P\rangle\,
$.

Similarly, in terms of the irreducible self-energy, defined by
$\hat H_{2}(k) = i (k^2-m_{\rm curr}^2)^{-1} +
i ( k^2-m_{\rm curr}^2)^{-1} \, i \,
\Gamma (k) \, \hat H_{2}(k) \,$,
the Dyson-Schwinger equation can be written
\begin{equation}
\hat \Gamma(k) =  \int {d^4 l \over (2 \pi)^4}  \,
\hat I_{ab} \Big ( k-l;{k+l \over 2},{k+l \over 2} \Big )
\sigma^a \hat H_{2}(l) \, \sigma^b \ .
\label{eq:dshom}
\end{equation}

The kernels are the same in the two Eqs.\
(\ref{eq:bshom}) and (\ref{eq:dshom}), consistently with the
requirement of chiral symmetry limit \cite{Hecht:2000xa}, being given by
\begin{eqnarray}
& & \hat I_{0;0} (Q; p, p^\prime)  =
   16 \pi {4 \over 3} \alpha_{\rm s} p^\alpha p^{\prime \beta}
  \hat D_{\alpha \beta} (Q)  + \nonumber \\
& &  \quad + 4 \sigma  \int \! d^3 {\bf \zeta} e^{-i{\bf Q}
   \cdot {\bf \zeta}}
    \vert {\bf \zeta} \vert \epsilon (p_0) \epsilon ( p_0^\prime )
   \int_0^1 \! d \lambda \{ p_0^2 p_0^{\prime 2} -
   [\lambda p_0^\prime {\bf p}_{\rm T} +
   (1-\lambda) p_0 {\bf p}_{\rm T}^\prime ]^2 \} ^{1 \over 2} \nonumber \\
& & \hat I_{\mu \nu ; 0}(Q;p,p^\prime) = 4\pi i {4 \over 3} \alpha_{\rm s}
   (\delta_\mu^\alpha Q_\nu - \delta_\nu^\alpha Q_\mu) p_\beta^\prime
   \hat D_{\alpha \beta}(Q)  - \nonumber \\
& & \qquad \qquad \qquad  - \sigma  \int d^3 {\bf \zeta} \, e^{-i {\bf Q}
\cdot {\bf \zeta}} \epsilon (p_0)
   {\zeta_\mu p_\nu -\zeta_\nu p_\mu \over
   \vert {\bf \zeta} \vert \sqrt{p_0^2-{\bf p}_{\rm T}^2}}
   p_0^\prime  \nonumber \\
& & \hat I_{0; \rho \sigma}(Q;p,p^\prime) =
   -4 \pi i{4 \over 3} \alpha_{\rm s}
   p^\alpha (\delta_\rho^\beta Q_\sigma - \delta_\sigma^\beta Q_\rho)
   \hat D_{\alpha \beta}(Q) + \nonumber \\
& & \qquad \qquad  \qquad  + \sigma  \int d^3 {\bf \zeta} \, e^{-i{\bf Q}
  \cdot {\bf \zeta}} p_0
  {\zeta_\rho p_\sigma^\prime - \zeta_\sigma p_\rho^\prime \over
  \vert {\bf \zeta} \vert \sqrt{p_0^{\prime 2}
   -{\bf p}_{\rm T}^{\prime 2}} }
  \epsilon (p_0^\prime)  \nonumber \\
& & \hat I_{\mu \nu ; \rho \sigma}(Q;p,p^\prime) =
   \pi {4\over 3} \alpha_{\rm s}
  (\delta_\mu^\alpha Q_\nu - \delta_\nu^\alpha Q_\mu)
  (\delta_\rho^\alpha Q_\sigma - \delta_\sigma^\alpha Q_\rho)
  \hat D_{\alpha \beta}(Q) \, ,
\label{eq:imom}
\end{eqnarray}
\noindent
where in the second and in the third equation $\zeta_0 = 0$ has to be
understood. Notice that, due to the privileged role given to the c.m.\ frame,
the terms proportional to $\sigma$ in (\ref{eq:imom}) formally are not
covariant.\\
In fact, it can be checked that $\Gamma(k)$ can
be consistently assumed to be spin independent and
Eq.~(\ref{eq:dshom}) can be rewritten in the simpler form
\begin{equation}
\Gamma (k) = i \int {d^4l \over (2 \pi)^4} \,
  {R(k,l) \over l^2-m^2+ \Gamma(l)} ,
\label{eq:dssc}
\end{equation}
\noindent
with
\begin{eqnarray}
R(k,l)&=& 4\pi {4 \over 3} \alpha_{\rm s}
\left [(k+l)^{\mu}(k+l)^{\nu}D_{\mu\nu}(k-l)+\right.
\nonumber\\
& &\left.(k-l)_{\nu}(k-l)^{\nu}D_{\mu}^{\,\,\mu}(k-l)
-(k-l)^{\mu}(k-l)^{\nu}D_{\mu\nu}(k-l)\right]
+\nonumber \\
  & & + \sigma \int d^3 {\bf r} e^{-i({\bf k}-{\bf l})
      \cdot{\bf r}} r (k_0 + l_0)^2 \sqrt{1-{({\bf k_\perp}+{\bf
      l_\perp})^2 \over (k_0 + l_0)^2}} \,,
\label{eq:dskernel}
\end{eqnarray}
\noindent
${\bf k_\perp}$ and ${\bf l_\perp}$ denoting as above the transversal
part of ${\bf k}$ and ${\bf l}$. Eq.\ (\ref{eq:dssc}) can be solved
by iteration resulting in an expression of the form $\Gamma(k^2,\bf{k}^2)$,
since~(\ref{eq:dskernel}) is not formally covariant.  Then
the constituent (pole) mass $m$ is defined by the
equation
\be
m^2 -m^2_{\rm curr}+\Gamma(m^2,{\bf k}^2)=0
\lb{mcurr}
\ee
and the dependence on ${\bf k}^2\,$,
being an artifact of the ansatz~(\ref{eq:wilson}),
is eliminated by extremizing $m(\bf{k}^2)$ in $\bf{k}^2\,$.

The 3-dimensional reduction of Eq.~(\ref{eq:bshom}) can be
obtained by a usual procedure of replacing $H_2(k)$ with
$i (k^2-m^2)^{-1} $
and $\hat{I}_{ab}$ with its so-called instantaneous approximation
$ \hat{I}_{ab}^{\rm inst}({\bf k}, {\bf u})\,$.
In this way, one can explicitly integrate over $u_0$ and arrive to
a 3-dimensional equation in the form of the eigenvalue equation
for a squared mass operator Eq.~(\ref{eq:M2}), with \cite{BMP}
\begin{equation}
   \langle {\bf k} \vert U \vert {\bf k}^\prime \rangle =
        {1\over (2 \pi)^3 }
        \sqrt{ w_1 + w_2 \over 2  w_1  w_2} \; \hat I_{ab}^{\rm \; inst}
        ({\bf k} , {\bf k}^\prime) \; \sqrt{ w_1^\prime + w_2^\prime \over 2
         w_1^\prime w_2^\prime}\; \sigma_1^a \sigma_2^b \,.
\label{eq:quadrrel}
\end{equation}
Finally by using Eq.~(\ref{eq:imom}) one obtains
Eqs.~(\ref{upt}-\ref{eq:uconf1}).

  Alternatively, in more usual terms, one could look for the eigenvalue of the
mass operator or center of mass Hamiltonian
$          H_{\rm CM} \equiv M = M_0 + V  $
with $V$ defined by $M_0V+VM_0+V^2=U$. Neglecting term $V^2$
the linear form potential $V$ can be
obtained from $U$ by the replacement
$
\sqrt{ (w_1+w_2) (w_1^\prime +w_2^\prime)\over w_1w_2w_1^\prime w_2^\prime}
\to {1\over 2\sqrt{w_1 w_2 w_1^\prime w_2^\prime}}
$.
The resulting
expression is particularly useful for a comparison with models based
on potential. In particular, in the static limit $V$ reduces to the Cornell
potential
\begin{equation}
V_{\rm stat} = - {4 \over 3} {\alpha_{\rm s} \over r } + \sigma r \, .
    \label{eq:static2}
\end{equation}
Note that it is necessary to introduce a cut-off $B$ in
Eq.~(\ref{eq:dssc}). As a consequence the constituent mass turns out
to be a function of the current mass and of $B$, $m=m\,(m_{\rm curr},
B)$. Then if one uses a running current mass $m_{\rm curr}(Q^2)$, 
a running constituent mass $m(Q^2)$ is obtained, as it has been done
in~\cite{Baldicchi:2004wj}. However the
singular expression used there
\begin{equation}
 m_{\rm curr}(Q^2) = {\hat m} \left( \frac{Q^{2}/ \Lambda^{2} -1}
{Q^2/ \Lambda^2 \ln ( Q^2/ \Lambda^2 )} \right)^{ \gamma_0/
2 \beta_0}
\end{equation}
is not consistent with Eq.~(\ref{ARC1L}),
and if a more consistent assumption is taken, e.g.,
\begin{equation}
 m_{\rm curr}(Q^2) = {\hat m} \left(
\alpha_{\rm E}^{(1)} (Q^2) \right)^{ \gamma_0/
2 \beta_0}\,,
\end{equation}
or the other resulting from the analytization of a similar expression
with $\alpha_{\rm E}^{(1)}(Q^2)$ replaced with the ordinary
perturbative $\alpha_{\rm s}^{(1)}(Q^2)$, the dependence of $m$ on
$Q^2$ is strongly reduced. For this reason even the light quark mass
is here treated as a constants to be adjusted with the the
data\footnote{In this way the only role that is left to the DS
equation is to justify the difference between the constituent and the
current masses.}.


\newpage
\section*{Appendix B: Numerical results}

The tables below display the complete set of results as explained in
Sec.~3.5. The values of all the parameters are 
$\sigma=0.18\;{\rm GeV}^{2}\,$, $\Lambda_{n_f=3}^{(1,{\rm
eff})}=193\;{\rm MeV}\,$, $m_{q}=196\;{\rm MeV}\;(q=u,d)\,$,
$m_{s}=352\;{\rm MeV}\,$, $m_{c}=1.516\;{\rm GeV}\,$ and
$m_{b}=4.854\; {\rm GeV}\,$. Meson masses are given in MeV. The last
column displays the experimental coupling $\alpha_{\rm s}^{\rm
exp}(Q_a^2)$ with the theoretical error $\Delta_{\rm NLO}$ due to 
the next-to-leading order terms neglected, the
theoretical error $\Delta_{\rm \Gamma}$ from the half width $
\Gamma/2 $ and the experimental error $\Delta_{\rm exp}$
respectively.\\
\\
{\footnotesize \dag~Center of gravity masses of the incomplete
multiplets estimated in analogy with other multiplets.}

\begin{table} [!ht]
\footnotesize{
\caption{$q \bar{q} \;\, (q = u,d) $
\vspace{0.7cm} } \hspace{-2.truecm}\begin{tabular}{ccccccc}
\hline\hline
  &  &  &  &  &  & \\
States & (MeV) & $m_{\rm exp}$ & $m_{\rm th}$ & $Q$ &
$\alpha_{\rm E}^{(3)}$ &
$\alpha_{\rm s}^{\rm exp}\pm\Delta_{\rm NLO}\pm\Delta_{\rm \Gamma}\pm
 \Delta_{\rm exp}$
\\
\hline
$ 1 \, {^{1} {\rm S}_{0}} $ &
$
\left\{
\begin{array}{c}
\pi^{0}   \\
\pi^{\pm}
\end{array}
\right.
$
&
$
\left.
\begin{array}{c}
134.9766 \pm 0.0006  \\
139.57018 \pm 0.00035
\end{array}
\right\}
138
$ & 136 & 401 & 0.522 & $ 0.534 \pm 0.122 \pm - \pm - $  \\
$ 1 \, {^{3} {\rm S}_{1}} $ & $ \rho (770) $ & 775.5 $ \pm $ 0.4 &
749 & &  & $ 0.517 \pm 0.122 \pm 0.048 \pm - $ \\
$ 1 \Delta {\rm SS} $ &  & 638 & 613 &   &  &  \\
$ 2 \, {^{1} {\rm S}_{0}} $ & $ \pi (1300) $ & 1300 $ \pm $ 100 &
1223 & 448 & 0.502 & $ 0.451 \pm 0.114 \pm 0.152 \pm 0.081 $  \\
$ 2 \, {^{3} {\rm S}_{1}} $ & $ \rho (1450) $ & 1459 $ \pm $ 11 &
1363 &  &  & $ 0.427 \pm 0.114 \pm 0.062 \pm 0.010 $  \\
$ 2 \Delta {\rm SS} $ &  & 159 & 139 &  &  &   \\
\hline
$ 1 \, {^{1} {\rm P}_{1}} $ & $ b_{1} (1235) $
& 1229.5 $ \pm $ 3.2 & 1234 & 209 & 0.637 & $ 0.688 \pm 0.155
\pm 0.124 \pm 0.006 $ \\
\hline
$ 1 \, {^{1} {\rm D}_{2}} $ & $ \pi_{2} (1670) $ & $ 1672.4
\pm 3.2 $ & 1595 & 144 &
 0.701 & $ 0.544 \pm 0.151 \pm 0.364 \pm 0.009 $ \\
\hline\hline
\end{tabular}
\vspace{-0.2truecm}}
\end{table}
\begin{table}[!ht] 
\footnotesize{
\caption{ $ s \bar{s} $
\vspace{3.5cm}}
\hspace{-2.truecm}\begin{tabular}{ccccccc}
\hline\hline
  &  &  &  &  &  &  \\
States & (MeV) & $m_{\rm exp}$ & $m_{\rm th}$ & $Q$ & $
\alpha_{\rm E}^{(3)}$ &
$\alpha_{\rm s}^{\rm exp}\pm\Delta_{\rm NLO}\pm\Delta_{\rm \Gamma}\pm
 \Delta_{\rm exp}$
\\
\hline
$ 1 \, {^{3} {\rm S}_{1}} $ & $ \phi (1020) $
& 1019.460 $ \pm $ 0.019 & 1019 & 418 & 0.514 &
$ 0.525 \pm 0.098 \pm 0.002 \pm - $  \\
$ 2 \, {^{3} {\rm S}_{1}} $ & $ \phi (1680) $ & 1680 $ \pm $ 20
& 1602 & 454 & 0.500 & $ 0.435 \pm 0.096 \pm 0.068 \pm 0.019 $ \\
\hline
$ 1 \, {^{1} {\rm P}_{1}} $ & $ h_{1} (1380) $
& 1386 $ \pm $ 19 & 1472 & 216 & 0.631 & $ 0.824 \pm 0.098
\pm 0.083 \pm 0.032 $ \\
$
\begin{array}{c}
1 \, {^{3} {\rm P}_{2}} \\
1 \, {^{3} {\rm P}_{1}} \\
1 \, {^{3} {\rm P}_{0}}
\end{array}
$ &
$
\begin{array}{c}
f_{2}^{\prime} (1525) \\
f_{1} (1510) \\
f_{0} (1500)
\end{array}
$ &
$
\left.
\begin{array}{c}
1525 \pm 5 \\
1518 \pm 5 \\
1507 \pm 5
\end{array}
\right\}
1521
$ & 1484 &  &  & $ 0.603 \pm 0.098 \pm 0.070 \pm 0.009 $  \\
\hline
$ 1 \, {^{1} {\rm D}_{2}} $ & $ \eta_{2} (1870) $ &
$ 1842 \pm 8 $ & 1807 & 149 & 0.695 &
$ 0.658 \pm 0.079 \pm 0.318 \pm 0.023 $ \\
\hline
$
\begin{array}{c}
1 \, {^{3} {\rm F}_{4}} \\
1 \, {^{3} {\rm F}_{3}} \\
1 \, {^{3} {\rm F}_{2}}
\end{array}
$ &
$
\begin{array}{c}
  \\  \\ f_{2} (2150)  \\
\end{array}
$ &
$
\left.
\begin{array}{c}
    \\   \\ 2156 \pm 11  \\
\end{array}
\right\} 2165^{\dag} $ &
2070 & 118 & 0.733 & $ 0.452 \pm 0.064 \pm 0.137 \pm 0.024 $ \\
\hline\hline
\end{tabular}
}
\end{table}
\begin{table}[!ht] 
\footnotesize{
\caption{ $ q \bar{s} \; (q = u,d) $
\vspace{0.5cm}} \hspace{-2.truecm}\begin{tabular}{ccccccc}
\hline\hline
& & & & & & \\
States & (MeV) & $m_{\rm exp}$ & $m_{\rm th}$ & $ Q $ & $
\alpha_{\rm E}^{(3)} $ &
$ \alpha_{\rm s}^{\rm exp}\pm\Delta_{\rm NLO}\pm\Delta_{\rm \Gamma}\pm
 \Delta_{\rm exp} $
\\
\hline
$ 1 \, {^{1} {\rm S}_{0}} $ &
$
\left\{
\begin{array}{c}
K^{0}   \\
K^{\pm}
\end{array}
\right.
$
&
$
\left.
\begin{array}{c}
497.648 \pm 0.022  \\
493.677 \pm 0.016
\end{array}
\right\}
495
$ & 491 & 409 & 0.518 & $ 0.529 \pm 0.122 \pm - \pm - $  \\
$ 1 \, {^{3} {\rm S}_{1}} $ &
$
\left\{
\begin{array}{c}
 K^{\ast} (892)^{0}   \\
 K^{\ast} (892)^{\pm}
\end{array}
\right.
$
&
$
\left.
\begin{array}{c}
 896.00 \pm 0.25 \\
 891.66 \pm 0.26
\end{array}
\right\}
893.11 $ & 887 &  &  & $ 0.526 \pm 0.122 \pm 0.017 \pm - $ \\
$ 1 \Delta {\rm SS} $ &  & 398 & 396 &  &  &  \\
$ 2 \, {^{3} {\rm S}_{1}} $ & $ K^{\ast} (1410) $ & 1414 $ \pm $ 15 &
1485 & 451 & 0.501 & $ 0.571 \pm 0.117 \pm 0.102 \pm 0.013 $ \\
\hline
$ 1 \, {^{1} {\rm P}_{1}} $ & $ K_{1} (1270) $ & 1272 $ \pm $ 7
& 1355 & 213 & 0.634 & $ 0.820 \pm 0.129 \pm 0.081 \pm 0.012 $ \\
$
\begin{array}{c}
1 \, {^{3} {\rm P}_{2}}
\vspace{3mm}
 \\
1 \, {^{3} {\rm P}_{1}} \\
1 \, {^{3} {\rm P}_{0}}
\end{array}
$ &
$
\begin{array}{c}
\left\{
\begin{array}{c}
K_{2}^{\ast} (1430)^{0} \\ K_{2}^{\ast} (1430)^{\pm}
\end{array}
\right. \\
K_{1} (1400) \\
K_{0}^{\ast} (1430)
\end{array}
$ &
$
\left.
\begin{array}{c}
1432.4 \pm 1.3 \\
1425.6 \pm 1.5 \\
1402 \pm 7 \\
1414 \pm 6
\end{array}
\right\} 1417.7 $ & 1367 &  &  & $ 0.583 \pm 0.129 \pm 0.133 \pm 0.007 $ \\
\hline
$
\begin{array}{c}
1 \, {^{3} {\rm D}_{3}} \\
1 \, {^{3} {\rm D}_{2}} \\
1 \, {^{3} {\rm D}_{1}}
\end{array}
$ &
$
\begin{array}{c}
K_{3}^{\ast} (1780) \\
K_{2} (1770) \\
K^{\ast} (1680)
\end{array}
$ &
$
\left.
\begin{array}{c}
1776 \pm 7  \\
1773 \pm 8  \\
1717 \pm 27
\end{array}
\right\}
1763
$ & 1712 & 150 & 0.694 & $ 0.617 \pm 0.113 \pm 0.273 \pm 0.031 $ \\
\hline
$
\begin{array}{c}
1 \, {^{3} {\rm F}_{4}} \\
1 \, {^{3} {\rm F}_{3}} \\
1 \, {^{3} {\rm F}_{2}}
\end{array}
$ &
$
\begin{array}{c}
K_{4}^{\ast} (2045) \\
K_{3} (2320) \\
K_{2}^{\ast} (1980)
\end{array}
$ &
$
\left.
\begin{array}{c}
2045 \pm 9  \\
2324 \pm 24  \\
1973 \pm 25
\end{array}
\right\} 2121 $ &
1973 & 116 & 0.736 & $ 0.248 \pm 0.095 \pm 0.413 \pm 0.071 $ \\
\hline\hline
\end{tabular}
}
\vspace{9.truecm}
\end{table}

\newpage
\begin{table}[!ht] 
\footnotesize{
\caption{ $ c \bar{c} $
\vspace{0.05cm}}
\hspace{-2.truecm}\begin{tabular}{ccccccc}
 \hline\hline
& & & & & & \\
States & (MeV) & $m_{\rm exp}$ & $m_{\rm th}$ &
$ Q $ & $ \alpha_{\rm E}^{(3)} $&
$ \alpha_{\rm s}^{\rm exp}\pm\Delta_{\rm NLO}\pm\Delta_{\rm \Gamma}\pm
 \Delta_{\rm exp} $ \\
\hline
$ 1 \, {^{1} {\rm S}_{0}} $ & $ \eta_{c} (1S) $
& 2980.4 $ \pm $ 1.2 & 2980 & 561 & 0.464 &
$ 0.467 \pm 0.025 \pm 0.008 \pm 0.001 $ \\
$ 1 \, {^{3} {\rm S}_{1}} $ & $ J/\psi (1S) $ & 3096.916 $ \pm $ 0.011
& 3097 &  &  & $ 0.467 \pm0.025 \pm - \pm - $ \\
$ 1 \Delta {\rm SS} $ &  & 117 & 118 &  &  &  \\
$ 2 \, {^{1} {\rm S}_{0}} $ & $ \eta_{c} (2S) $ & 3638 $ \pm $ 4
& 3595 & 500 & 0.483 & $ 0.446 \pm 0.023 \pm 0.007 \pm 0.004 $ \\
$ 2 \, {^{3} {\rm S}_{1}} $ & $ \psi (2S) $
& 3686.093 $ \pm $ 0.034 & 3653 &  &  & $ 0.455 \pm 0.023 \pm - \pm - $ \\
$ 2 \Delta {\rm SS} $ &  & 48 & 58 &  &  &   \\
$ 3 \, {^{3} {\rm S}_{1}} $ & $ \psi (4040) $ & 4039 $ \pm $ 1
& 4030 & 483 & 0.489 & $ 0.485 \pm 0.022 \pm 0.049 \pm 0.001 $ \\
$ 4 \, {^{3} {\rm S}_{1}} $ & $ \psi (4415) $ & 4421 $ \pm $ 4 &
4337 & 474 & 0.492 & $ 0.384 \pm 0.022 \pm 0.042 \pm 0.006 $ \\
\hline
$ 1 \, {^{1} {\rm P}_{1}} $ & $ h_{c} (1P) $ & 3525.93 $ \pm $ 0.27 &
3532 & 269 & 0.592 & $ 0.631 \pm 0.012 \pm - \pm - $  \\
$
\begin{array}{c}
1 \, {^{3} {\rm P}_{2}} \\
1 \, {^{3} {\rm P}_{1}} \\
1 \, {^{3} {\rm P}_{0}}
\end{array}
$ &
$
\begin{array}{c}
\chi_{c2} (1P) \\
\chi_{c1} (1P) \\
\chi_{c0} (1P)
\end{array}
$ &
$
\left.
\begin{array}{c}
3556.20 \pm 0.09 \\
3510.66 \pm 0.07 \\
3414.76 \pm 0.35
\end{array}
\right\} 3525.3 $ & 3537 &  &  & $ 0.640 \pm 0.012 \pm 0.002 \pm - $ \\
$
\begin{array}{c}
2 \, {^{3} {\rm P}_{2}} \\
2 \, {^{3} {\rm P}_{1}} \\
2 \, {^{3} {\rm P}_{0}}
\end{array}
$ &
$
\begin{array}{c}
\chi_{c2} (2P) \\
               \\
X(3872)
\end{array}
$ &
$
\left.
\begin{array}{c}
3929 \pm 5   \\
             \\
3871.2 \pm 0.5
\end{array}
\right\} 3915^{\dag} $ & 3929 & 274 & 0.589 &
$ 0.644 \pm 0.013 \pm 0.027 \pm 0.009 $ \\
\hline
$
\begin{array}{c}
1 \, {^{3} {\rm D}_{3}} \\
1 \, {^{3} {\rm D}_{2}} \\
1 \, {^{3} {\rm D}_{1}}
\end{array}
$ &
$
\begin{array}{c}
   \\
   \\
 \psi (3770)
\end{array}
$ &
$
\left.
\begin{array}{c}
   \\
   \\
3771.1 \pm 2.4
\end{array}
\right\} 3820^{\dag} $ &
3822 & 190 & 0.654 & $ 0.707 \pm 0.008 \pm 0.030 \pm 0.006 $ \\
$
\begin{array}{c}
2 \, {^{3} {\rm D}_{3}} \\
2 \, {^{3} {\rm D}_{2}} \\
2 \, {^{3} {\rm D}_{1}}
\end{array}
$ &
$
\begin{array}{c}
   \\    \\   \psi (4160)
\end{array}
$ &
$
\left.
\begin{array}{c}
   \\    \\
4153 \pm 3
\end{array}
\right\} 4183^{\dag} $ &
4150 & 198 & 0.646 & $ 0.606 \pm 0.009 \pm 0.132 \pm 0.008 $ \\
\hline\hline
\end{tabular}
Quantum numbers of $ h_{c} (1P) $ and $ X(3872) $ mesons
are not well established.
}
\end{table}
\begin{table}[!ht] 
\vspace{-0.08truecm}
\footnotesize{
\caption{ $ b \bar{b} $
\vspace{0.07cm}}
\hspace{-2.6truecm}\begin{tabular}{ccccccc}
 \hline\hline
  & & & & & &  \\
States & (MeV) & $m_{\rm exp}$ & $m_{\rm th}$ & $ Q $ & $
\alpha_{\rm E}^{(3)} $ &
$ \alpha_{\rm s}^{\rm exp}\pm\Delta_{\rm NLO}\pm\Delta_{\rm \Gamma} \pm
 \Delta_{\rm exp} $
\\
\hline
$ 1 \, {^{3} {\rm S}_{1}} $ & $ \Upsilon (1S) $
& 9460.30 $ \pm $ 0.26 & 9461 & 951 & 0.381
& $ 0.378 \pm 0.006 \pm - \pm - $ \\
$ 2 \, {^{3} {\rm S}_{1}} $ & $ \Upsilon (2S) $
& 10023.26 $ \pm $ 0.31 & 9987 & 630 & 0.445 &
$ 0.416 \pm 0.004 \pm - \pm - $ \\
$ 3 \, {^{3} {\rm S}_{1}} $ & $ \Upsilon (3S) $
& 10355.2 $ \pm $ 0.5 & 10321 & 552 & 0.466 &
$ 0.433 \pm 0.003 \pm - \pm - $  \\
$ 4 \, {^{3} {\rm S}_{1}} $ & $ \Upsilon (4S) $
& 10579.4 $ \pm $ 1.2 & 10588 & 517 & 0.478 &
$ 0.493 \pm 0.003 \pm 0.013 \pm 0.002 $  \\
$ 5 \, {^{3} {\rm S}_{1}} $ & $ \Upsilon (10860) $
& 10865 $ \pm $ 8 & 10820 & 497 & 0.484 &
$ 0.424 \pm 0.003 \pm 0.078 \pm 0.011 $  \\
$ 6 \, {^{3} {\rm S}_{1}} $ & $ \Upsilon (11020) $
& 11019 $ \pm $ 8 & 11034 & 506 & 0.481 &
$ 0.508 \pm 0.003 \pm 0.057 \pm 0.012 $  \\
\hline
$
\begin{array}{c}
1 \, {^{3} {\rm P}_{2}} \\
1 \, {^{3} {\rm P}_{1}} \\
1 \, {^{3} {\rm P}_{0}}
\end{array}
$ &
$
\begin{array}{c}
\chi_{b2} (1P) \\
\chi_{b1} (1P) \\
\chi_{b0} (1P)
\end{array}
$ &
$
\left.
\begin{array}{c}
9912.21 \pm 0.26\pm 0.31 \\
9892.78 \pm 0.26\pm 0.31 \\
9859.44 \pm 0.42 \pm 0.31
\end{array}
\right\} 9899.87
$ & 9880 & 387 & 0.528 & $ 0.519 \pm 0.002 \pm - \pm - $  \\
$
\begin{array}{c}
2 \, {^{3} {\rm P}_{2}} \\
2 \, {^{3} {\rm P}_{1}} \\
2 \, {^{3} {\rm P}_{0}}
\end{array}
$ &
$
\begin{array}{c}
\chi_{b2} (2P) \\
\chi_{b1} (2P) \\
\chi_{b0} (2P)
\end{array}
$ &
$
\left.
\begin{array}{c}
10268.65 \pm 0.22 \pm 0.50 \\
10255.46 \pm 0.22 \pm 0.50 \\
10232.5 \pm 0.4 \pm 0.50
\end{array}
\right\}
10260.24
$ & 10231 & 343 & 0.549 & $ 0.524 \pm 0.002 \pm - \pm 0.001 $  \\
\hline\hline
\end{tabular}
}
\end{table}

\newpage
\begin{table}[!ht] 
\footnotesize{
\caption{Light-heavy quarkonium systems
\vspace{0.3cm}}
\hspace{-2.truecm}\begin{tabular}{ccccccc}
 \hline\hline
  & & & & & &  \\
States & (MeV) & $m_{\rm exp}$ & $m_{\rm th}$ & $ Q $ & $
\alpha_{\rm E}^{(3)} $ &
$ \alpha_{\rm s}^{\rm exp}\pm\Delta_{\rm NLO}\pm\Delta_{\rm \Gamma}\pm
 \Delta_{\rm exp} $
\\
\hline
$ q \bar{c}$ &  &  &  &  &  &   \\
$ 1 \, {^{1} {\rm S}_{0}} $ &
$
\left\{
\begin{array}{c}
D^{\pm} \\
D^{0}
\end{array}
\right.
$
&
$
\left.
\begin{array}{c}
1869.3 \pm 0.4  \\
1864.5 \pm 0.4
\end{array}
\right\}
1867.7
$
& 1843 & 459 & 0.498   & $ 0.488 \pm 0.082 \pm - \pm - $  \\
$ 1 \, {^{3} {\rm S}_{1}} $ &
$
\left\{
\begin{array}{c}
D^{\ast}(2010)^{\pm} \\
D^{\ast}(2007)^{0}
\end{array}
\right.
$
&
$
\left.
\begin{array}{c}
2010.0 \pm 0.4  \\
2006.7 \pm 0.4
\end{array}
\right\}
2008.9
$
& 2000 &  &  & $ 0.499 \pm0.082 \pm - \pm - $  \\
$ 1 \Delta {\rm SS} $ &  & 141 $ \pm $ 1 & 157 &  &  &   \\
\hline
$
\begin{array}{c}
\\
1 \, {\rm P}_{2} \\  \\
1 \, {\rm P}_{1} \\  \\
\end{array}
$ &
$
\begin{array}{c}
\left\{
  \begin{array}{c}
    D_{2}^{\ast} (2460)^{\pm}  \\
    D_{2}^{\ast} (2460)^{0}
  \end{array}
\right.
\\
\left\{
  \begin{array}{c}
    D_{1} (2420)^{\pm}  \\
    D_{1} (2420)^{0}
  \end{array}
\right.
\\
\end{array}
$ &
$
\left.
\begin{array}{c}
  \begin{array}{c}
    2459 \pm 4   \\
    2461.1 \pm 1.6
  \end{array}
\\
  \begin{array}{c}
    2423.4 \pm 3.1   \\
    2422.3 \pm 1.3
  \end{array}
\\
\end{array}
\right\}
$ & 2443 & 232 & 0.619 & $ 0.651 \pm 0.051 \pm 0.028 \pm 0.005 $ \\
\hline
\hline \\[-3.5mm]
$ q \bar{b}$ &  &  &  &  &  &  \\
$ 1 \, {^{1} {\rm S}_{0}} $ &
$
\left\{
\begin{array}{c}
B^{\pm} \\
B^{0}
\end{array}
\right.
$
&
$
\left.
\begin{array}{c}
5279.0 \pm 0.5  \\
5279.4 \pm 0.5
\end{array}
\right\}
5279.1
$
& 5246 & 516 & 0.478 & $ 0.456 \pm 0.036 \pm - \pm - $  \\
$ 1 \, {^{3} {\rm S}_{1}} $ &
$ B^{\ast} $ & 5325.0 $ \pm $ 0.6 & 5311 &  &  &
$ 0.471 \pm 0.036 \pm - \pm - $  \\
$ 1 \Delta {\rm SS} $ &  & 46 $ \pm $ 3 & 64  &  &   &   \\
\hline\hline
\end{tabular}
}
\end{table}
\begin{table} [!ht]
\vspace{-0.1truecm}
\footnotesize{
\caption{Light-heavy quarkonium systems
\vspace{0.3cm}}
\hspace{-2.truecm}\begin{tabular}{ccccccc}
 \hline\hline
  & & & & & & \\
States & (MeV) & $m_{\rm exp}$ & $m_{\rm th}$  & $ Q $ & $
\alpha_{\rm E}^{(3)} $ &
$ \alpha_{\rm s}^{\rm exp}\pm\Delta_{\rm NLO}\pm\Delta_{\rm \Gamma}\pm
 \Delta_{\rm exp} $
\\
\hline
$ s \bar{c} $ &  &  &  &   &  &  \\
$ 1 \, {^{1} {\rm S}_{0}} $ &
$ D_{s}^{\pm} $ & 1968.2 $ \pm $ 0.5 & 1959 & 472 & 0.493 &
$ 0.494 \pm 0.055 \pm - \pm - $  \\
$ 1 \, {^{3} {\rm S}_{1}} $ &
$ D_{s}^{\ast \pm} $ & 2112.0 $ \pm $ 0.6 & 2109 & &  &
$ 0.497 \pm 0.055 \pm - \pm - $  \\
$ 1 \Delta {\rm SS} $ &  & 144 $ \pm $ 1 & 149 &  &  &   \\
\hline
$
\begin{array}{c}
1 \, {\rm P}_{2} \\
1 \, {\rm P}_{1}
\end{array}
$ &
$
\begin{array}{c}
D_{s2} (2573)^{\pm} \\
D_{s1} (2536)^{\pm} \\
\end{array}
$ &
$
\left.
\begin{array}{c}
2573.5 \pm 1.7 \\
2535.35 \pm 0.34 \pm 0.5 \\
\end{array}
\right\}
$ & 2545 & 236 & 0.616 & $ 0.626 \pm 0.036 \pm 0.009 \pm 0.002 $ \\
\hline
\hline \\[-3.5mm]
$ s \bar{b} $ &  &  &  &  &  &  \\
$ 1 \, {^{1} {\rm S}_{0}} $ &
$ B_{s}^{0} $ & 5367.5 $ \pm $ 1.8 & 5343 & 535 & 0.472 &
$ 0.457 \pm 0.024 \pm - \pm 0.001 $ \\
$ 1 \, {^{3} {\rm S}_{1}} $ &
$ B_{s}^{\ast} $ & 5412.8 $ \pm $ 1.7 & 5408 &  &  &
$ 0.473 \pm 0.024 \pm - \pm 0.001 $  \\
$ 1 \Delta {\rm SS} $ &  & 47 $ \pm $ 4 & 65  &  &  &   \\
\hline $ 1 \, {\rm P} $ & $ B_{sJ}^{\ast} (5850) $ & 5853 $ \pm $
15 & 5830 & 255 & 0.602 & $ 0.592 \pm 0.013 \pm 0.042 \pm 0.027 $ \\
\hline\hline
\end{tabular}
}
\end{table}



\part{Progress in the determination of the gluon polarization}

\chapter{Polarized parton distributions from DIS}

One of the fundamental properties of the nucleon structure is the
spin distribution among its quark and gluon constituents. The spin 
structure of the nucleon has been extensively
investigated by inclusive polarized lepton scattering off polarized 
protons and neutrons, since the discovery of the EMC spin 
effect~\cite{Ashman:1987hv}.
Despite the naive expectation that the nucleon spin is carried by quarks,
the experimental results revealed that only a small fraction is
actually due to quarks (the so-called ``proton spin crisis'').
Considerable efforts, both experimentally and theoretically, have 
subsequentely gone into understanding where the remaining fraction of
the nucleon spin resides (see e.g. Ref.~\cite{Lampe:1998eu} for a recent 
review).\\
To this end, the determination of the first moments of the polarized 
parton distribution functions (pdfs) is necessary. 
Attention is currently devoted to the reconstruction of pdfs
at all values of the Bjorken $x$ and the momentum transfer $Q^2$. Of  
special interest is the determination of the gluon polarization, in
order to evaluate the total contribution carried by the gluons to the nucleon 
spin.

The possibility of inferring the polarized gluon density from scaling 
violations has been extensively studied. However, phenomenological
analyses of the current inclusive DIS data only yield weak constraints 
on the gluon first moment, and leave the $x$-shape completely
undetermined. This issue has been thoroughly investigated in the present
work, by performing a NLO analysis in perturbative QCD of world data from 
inclusive DIS experiments, as discussed in detail in Sec.~6.

Actually, some light may be shed on the 
gluon content of the nucleon by observing specific events that receive 
leading contributions from gluon initiated subprocesses, that is, by 
directly measuring the polarized gluon density.
In particular, as discussed in Sec.~7, in polarized fixed target experiments,
the gluon polarization can be probed through the photon-gluon fusion process, 
by selecting open-charm events or hadron pairs with high transverse momentum, 
and both strategies are currently used at the COMPASS experiment (CERN). Indeed
charmed mesons in the final states provide a clean tag of photon-gluon fusion,
since there are essentially no competing processes and a clean perturbative analysis
is possible. 
Therefore, a phenomenological study 
of open-charm photoproduction from deep-inelastic muon scattering off
nucleons, with longitudinally polarized beam and target, 
has been also performed, with reference to upcoming results on experimental
asymmetries from
COMPASS. The results are presented 
in Sec.~7 and, specifically, the constraints that can be obtained on the 
polarized gluon density at the COMPASS kinematics are discussed.

In what follows the main theoretical tools are briefly reviewed. Specifically 
Sec.~5.1
recalls the basic ingredients for the study of polarized lepton-nucleon 
scattering, i.e. the polarized structure functions $g_1$ and $g_2$. 
In Sec.~5.2 the expectations of the naive quark-parton model are summarized, 
whereas Sec.~5.3 is devoted to the main aspects of the perturbative QCD
analysis, that is, the relations between structure functions and polarized
parton distributions. In Sec.~5.4 the way experimental information
on the relevant structure function $g_1$ are extracted from the experimentally
measured quantities is described, and, finally Sec~5.5 sketches how 
target mass corrections are taken into account in the present analysis.

\section{Structure functions and their moments}

Polarized structure functions are the form factors which parametrize the
cross section for deep-inelastic scattering of polarized leptons off a 
polarized hadronic target. If the momentum transfer $Q^2$ involved is much
smaller than the $Z$ boson mass, the antisymmetric part of 
the hadronic tensor can be expressed in terms of two spin-dependent structure
functions, $g_1$ and $g_2$, i.e.~\cite{Kodaira,Forte:1994dw}
\bea
&& iW_A^{\mu\nu}\equiv\frac{1}{4\pi}\int d^4x e^{iqx}<p,s|J^{[\mu}(x)J^{\nu]}(0)|p,s>\nn\\
&& =im\varepsilon^{\mu\nu\rho\sigma}q_{\rho}\left[\frac{s_{\sigma}}{pq}\,g_1(x,Q^2)+
\frac{s_{\sigma}\,pq-p_{\sigma}\,qs}{(pq)^2}\,g_2(x,Q^2)\right]
\lb{Wa}
 \eea
where $p^{\mu},\,m$ and $s^{\mu}$ are, respectively, the target four-momentum, mass, 
and spin (normalized as $s^{\mu}s_{\mu}=-1$). Neglecting weak interaction effects,  
$J^{\mu}$ in Eq.~(\ref{Wa}) is the electric current. The deeply inelastic region is 
identified by the Bjorken limit 
\be
Q^2=-q^2\to\infty\,,\quad\textrm{with}\quad x=\frac{Q^2}{2pq}\quad\textrm{fixed}\,,
\lb{kin}
\ee
in which the invariant mass of the hadronic system in the final state is much larger
than the nucleon mass, namely
\be
W^2=m^2+Q^2\,\frac{1-x}{x}\,\gg\, m^2\,.
\lb{wm}
\ee
In the case when the nucleon spin $s$ is purely longitudinal, that is, 
\be
s=s_L=\lambda\left(p-\frac{m^2}{pq}q\right)\,,\quad 
\lambda=\pm\frac{1}{m\sqrt{1+4m^2x^2/Q^2}}\,,
\lb{sl}
\ee
the antysimmetric tensor Eq.~(\ref{Wa}) can be written as
\be
W_A^{\mu\nu}=\lambda m\varepsilon^{\mu\nu\rho\sigma}\frac{q_{\rho}p_{\sigma}}{pq}
\left[g_1(x,Q^2)-\frac{4m^2x^2}{Q^2}\,g_2(x,Q^2)\right]\,.
\lb{g1l}
\ee
In the Bjorken limit the product $\lambda m\to 1$, thus leaving a finite residue 
as $m\to 0$. Furthermore, the factor $4m^2x^2/Q^2$ vanishes, and the structure
function $g_2$ decouples. As a result only $g_1$ is asimptotically relevant.\\
In case of transverse polarization both structure functions equally contribute, 
but the whole cross section is strongly suppressed by an overall target mass
factor.

The light-cone expansion of the current product in Eq.~(\ref{Wa}) implies
that the first moment of the relevant structure function $g_1$ for a nucleon
target, at leading twist, is given by~\cite{ope}
\be
\Gamma_1(Q^2)\equiv\int_0^1 dx\,g_1(x,Q^2)=\frac{1}{2}\left[\sum_{i=1}^{n_f}e_i^2
C_i(Q^2)a_i\right]\,,
\lb{fm}
\ee
where $a_i$ are given by matrix elements of the axial current 
for the $i$-th flavor, namely 
\be
m a_i s_{\mu}\equiv <p,s|\bar\psi_i\gamma_{\mu}\gamma_5\psi_i|p,s>\,,
\lb{ai}
\ee
$e_i$ is the electric charge, and $C_i(Q^2)$ are perturbatively calculable
coefficient functions. Then, assuming that only the three lightest flavors 
are activated, Eq.~(\ref{fm}) can be recast as
\be
\Gamma_1^{p,n}(Q^2)=\frac{1}{12}\left[C_{\rm NS}(Q^2)\left(\pm a_3+\frac{1}{3} a_8\right)+
\frac{4}{3}C_{\rm S}(Q^2)a_0\right]\,,
\lb{fm38}
\ee
where the plus (minus) sign refers to a proton (neutron) target, and the singlet
and nonsinglet matrix elements are defined by
\bea
&& a_0=a_u+a_d+a_s\nn\\
&& a_3=a_u-a_d\,,\quad a_8=a_u+a_d-2a_s
\lb{mtx}
\eea
respectively. $C_{\rm S}(Q^2)$ and $C_{\rm NS}(Q^2)$ are $Q^2$-dependent singlet 
and nonsinglet QCD coefficient functions; to order $\alpha_s$, in the $\overline{\rm MS}$
scheme e.g., one has  
\be
C_{\rm S}(Q^2)=C_{\rm NS}(Q^2)=1-\frac{\alpha_s}{\pi}+\mathcal{O}(\alpha_s^2)\,,
\lb{cfs}
\ee
whereas $C_{\rm S}\neq C_{\rm NS}$ at higher orders~\cite{Mertig:1995ny}. QCD 
corrections turn out to be sizable in the $Q^2$-range of the present experiments 
and are thus relevant for the comparison with the data~\cite{Karliner}.\\
The singlet and nonsinglet components of $\Gamma_1$ can be then extracted using 
Eq.~(\ref{fm38}). 
Indeed, the triplet and octet currents are conserved and therefore scale invariant,
and the related matrix elements can be derived from any other process.
The triplet axial charge is equal to the axial coupling measured in nucleon
$\beta$-decay 
\be
a_3=\frac{g_A}{g_V}=F+D=1.2601\pm0.0025
\lb{a3}
\ee
where $F$ and $D$ are weak hyperon decay constants in flavor SU(3) 
symmetry~\cite{Yao:2006px,FD}.
The octet matrix element is then given by
\be
a_8=3F-D=0.588\pm0.033\,.
\lb{a8}
\ee
On the other hand, the singlet axial charge $a_0$ is scale dependent due 
to the anomalous nonconservation of the singlet axial current~\cite{axanom},
and can be determined from a measurement of the proton first moment, by 
rewriting Eq.~(\ref{fm38}) as
\be
C_{\rm S}(Q^2)a_0(Q^2)=9\Gamma_1^p(Q^2)-\frac{1}{2}C_{\rm NS}(Q^2)(3F+D)\,.
\lb{a0ex}
\ee
Since the two terms on the right hand side are roughly of the same order
(typically $\Gamma_1\sim 0.1$), the value of $a_0$ arises from a large cancellation
between them (see e.g.~\cite{Forte:1994dw}). 
Present data indeed indicate that $a_0$ is compatible with zero,
but values as large as $a_0(10\,{\rm GeV}^2)\sim 0.3$ are not excluded.

Finally, a fundamental prediction of the theory is the Bjorken sum 
rule
\be
\Gamma_1^p(Q^2)-\Gamma_1^n(Q^2)=\frac{1}{6}C_{\rm NS}(Q^2)\,a_3\,,
\lb{bj}
\ee
that was originally derived from current algebra and isospin 
symmetry~\cite{Bjorken}. Eq.~(\ref{bj}) relates the integral over all $x$
of the nonsinglet polarized structure function $g_1^p-g_1^n$, at fixed $Q^2$,  
to the well-measured $\beta$-decay coupling $a_3$. A comparison with the data,   
using the computed $C_{\rm NS}(Q^2)$ thus allows a direct
test of isospin in this channel, as well as the predicted scale dependence.
In addition, it is possible to use the measurements to extract a relatively
accurate determination of the strong coupling $\alpha_s(Q^2)$~\cite{Altarelli:1998nb}.

\section{The parton model predictions}

In the naive quark-parton model the spin-dependent structure function $g_1$ 
can be simply expressed as the charge-weighted difference between momentum
distributions for quark helicities aligned parallel $(q^{\uparrow})$ and
antiparallel $(q^{\downarrow})$ to the longitudinally polarized parent nucleon
\be
g_1(x)=\frac{1}{2}\sum_{i=1}^{n_f}e_i^2\Delta q^+(x)\,,\quad
\Delta q^+(x)=\Delta q_i(x)+\Delta\bar q_i(x)\,.
\lb{pm}
\ee
The polarized quark distributions are defined as
\be
\Delta q_i(x)=q_i^{\uparrow}(x)-q_i^{\downarrow}(x)\,,
\lb{qdef}
\ee
and similarly for antiquarks. Assuming $n_f=3$, one can define the fla\-vor 
sin\-glet and nonsinglet combinations of polarized quark densities, namely 
\bea
&& \Delta \Sigma=\Delta u^+ +\Delta d^+ +\Delta s^+\nn\\
&& \Delta q_3=\Delta u^+ -\Delta d^+\nn\\
&& \Delta q_8=\Delta u^+ +\Delta d^+ -2\Delta s^+\,.
\lb{plusd}
\eea
Then the structure function $g_1$ can be cast into the form
\be
g_1^{p,n}(x)=\frac{<e^2>}{2}\left[\Delta \Sigma(x)\pm\frac{3}{4}\Delta q_3(x)+
\frac{1}{4}\Delta q_8(x)\right]\,,
\lb{pm2}
\ee
where the plus (minus) sign stays for a proton (neutron) target, and the 
average charge is $<e^2>=2/9$ for $n_f=3$.\\ 
In the naive parton model gluons do not contribute to the nucleon spin, and 
the first moments of singlet and nonsinglet quark densities are 
straightforwardly related to the matrix elements of axial currents, such 
that
\be
a_0=\int_0^1 dx \Delta \Sigma(x)\,,\quad a_3=\int_0^1 dx \Delta q_3(x)\quad 
\textrm{and}\quad a_8=\int_0^1 dx \Delta q_8(x)\,.
\lb{aipm}
\ee
In this context, the first moments of $\Delta q_i^+(x)$ can be interpreted
as the contribution of the quark flavor $i$ to the nucleon spin, and thus
the singlet axial charge $a_0$ as the net quark helicity.\\ 
Then, under the assumption 
that the strange sea in the nucleon is unpolarized, namely
\be
\left|\int_0^1 dx \Delta s^+(x)\right|\ll \left|\int_0^1 dx \Delta q^+(x)\right|_{q=u,d}\,,
\lb{sea}
\ee
the quark helicity coincides with the nonsinglet (octet) axial charge, $a_0\simeq a_8$.
Thus, from Eq.~(\ref{a8}) one roughly has 
\be
a_0\simeq 0.6\,,
\lb{a08}
\ee
whereas the value extracted from the data (see Eq.~(\ref{a0ex})) is compatible 
with zero. The Ellis-Jaffe sum rules~\cite{Ellis:1973kp} then  
give the parton model predictions for the proton and neutron first moments
\be
\Gamma_1^{p,n}=\pm \frac{1}{12}a_3+\frac{5}{36}a_8\,,
\lb{ej}
\ee
that are not consistent with the measured values, at the level of several standard
deviations. Thus, against the Ellis-Jaffe prediction and intuition, quarks seem
to carry a very small fraction of the total nucleon helicity. Actually, it is only
at the naive parton level that the first moment of the singlet part of the structure
function $g_1$ corresponds to the total helicity fraction carried by quarks.

\section{Polarized structure functions and pdfs in QCD}

When including QCD corrections, also the gluon density contributes and the  
structure function $g_1$ is related to the scale-dependent 
polarized quark and gluon distributions by the convolution with appropriate
coefficient functions~\cite{Zijlstra:1993sh,Altarelli:1981ax}
\be
g_1(x,Q^2)=\frac{<e^2>}{2}\left[C_{\rm NS}\otimes \Delta q_{\rm NS}+
C_{\rm S}\otimes \Delta \Sigma + 2n_f C_g\otimes \Delta g\right],
\lb{g1}
\ee
where $<e^2>=n_f^{-1}\sum_{i=1}^{n_f}e_i^2$, $n_f$ being the number of active flavors 
with electric charge $e_i$, and $\otimes$ denotes the convolution product 
with respect to $x$
\be
f\otimes g=\int_x^1 \frac{dy}{y}f\left(\frac{x}{y}\right)g(y)\,.
\lb{conv}
\ee
In Eq.~(\ref{g1}) $\Delta g(x,Q^2)$ is the polarized gluon distribution, whereas 
the singlet and nonsinglet quark distributions are defined as
%
%
%
\bea
&&\Delta \Sigma(x,Q^2)\equiv\sum_{i=1}^{n_f}
\left(\Delta q_i+\Delta \bar q_i\right)\\
&&\Delta q_{\rm NS}(x,Q^2)\equiv \sum_{i=1}^{n_f}\left(\frac{e_i^2}{<e^2>}-1\right)\,
\left(\Delta q_i+\Delta \bar q_i\right)\,,\nn
\lb{nss}
\eea
where $\Delta q_i$ and $\Delta \bar q_i$ are the scale-dependent quark and 
antiquark polarized densities of flavor $i$, defined as in Eq.~(\ref{qdef}). 

The perturbative part of the $Q^2$-dependence of the polarized parton 
densities is given by the Altarelli-Parisi evolution equations~\cite{Altarelli:1977zs}.
The singlet quark and gluon distributions mix according to
\be
\frac{\partial}{\partial\ln t}
\left( \begin{array}{c}
\Delta\Sigma\\
\Delta g
\end{array} \right)
=\frac{\alpha_s(t)}{2\pi}
\left( \begin{array}{cc}
\Delta P_{qq}^S & 2 n_f\Delta P_{qg}^S\\
\Delta P_{gq}^S & \Delta P_{gg}^S
\end{array} \right)
\otimes\left( \begin{array}{c}
\Delta\Sigma\\
\Delta g
\end{array} \right)\,,
\lb{AP1}
\ee
while the nonsinglet quark evolves independently as
\be
\frac{\partial}{\partial\ln t}\Delta q_{\rm NS}=\frac{\alpha_s(t)}{2\pi}
\Delta P_{qq}^{\rm NS}\otimes\Delta q_{\rm NS}\,.
\lb{AP2}
\ee
In Eqs.~(\ref{AP1}) and~(\ref{AP2}) $\Delta P(x,Q^2)$ denote the spin-dependent
splitting functions, and $t=\ln(Q^2/\Lambda^2)$ with $\Lambda$ the QCD scale
parameter. The factorization and renormalization 
scales are both taken equal to $Q^2$, so that all the scale dependence appears 
through $t$.\\ 
Eqs.~(\ref{AP1}) and~(\ref{AP2}) are valid
for all orders of perturbative QCD; the coefficient functions $C$ and the 
evolution kernels $\Delta P$ may each be expanded in 
powers of $\alpha_s$. At NLO
\be
C(x,\alpha_s)=C^{(0)}(x)+\frac{\alpha_s(t)}{2\pi}C^{(1)}(x)+\mathcal{O}(\alpha_s^2)
\lb{nlocf}
\ee
\be
\Delta P(x,\alpha_s)=\Delta P^{(0)}(x)+\frac{\alpha_s(t)}{2\pi}\Delta P^{(1)}(x)+
\mathcal{O}(\alpha_s^2)\,.
\lb{nlodp}
\ee
It is also convenient to introduce anomalous dimensions
\be 
\gamma(n,\alpha_s)\equiv\int_0^1dx\, x^{n-1}\Delta P(x,\alpha_s)
\lb{anom}
\ee
i.e. the Mellin transforms of the splitting functions. One can analogously
define moment-space coefficient functions $C(n,\alpha_s)$ and parton distributions
$\Delta q_{\rm NS}(n,Q^2)$, $\Delta\Sigma(n,Q^2)$ and $\Delta g(n,Q^2)$.\\
In accordance with the partonic picture, in Eq.~(\ref{nlocf}) 
$C^{(0)}_{\rm NS}(x)=C^{(0)}_{\rm S}(x)=\delta(1-x)$, while $C^{(0)}_g(x)=0$, so that 
at order $\alpha_s^0$ $g_1$ decouples from $\Delta g$, and
is just a linear combination of polarized quark distributions, whose $Q^2$ dependence 
is entirely specified by Eqs.~(\ref{AP1}) and~(\ref{AP2}) (the leading order 
splitting functions are calculated in Ref.~\cite{Altarelli:1977zs}).

Beyond leading order, splitting functions and coefficient functions are no
longer universal, hence even though the scale dependence of the observable 
structure function $g_1$ is determined uniquely, at least up to higher order
corrections, its separation into contributions due to quarks and gluons is scheme 
dependent and thus essentially arbitrary. The NLO coefficient functions may be
modified by a change of the factorization scheme which is partially compensated 
by a corresponding change in the NLO anomalous dimensions, hence both are required
in a full NLO order computation. The complete set of NLO coefficient functions and 
splitting functions can be found in Ref.~\cite{Mertig:1995ny}.\\
The modified minimal subtraction ($\overline{\rm MS}$) factorization scheme is 
commonly used in the analysis of polarized parton distributions. 
In the $\overline{\rm MS}$ scheme 
the first moment of the gluon coefficient function vanishes, the gluon density does not
contribute to the first moment of $g_1$ and the scale dependent singlet axial
charge is thus equal to the singlet quark first moment
\be
a_0(Q^2)=\Delta\Sigma_{\overline{\rm MS}}(1,Q^2)\,. 
\lb{a0MS}
\ee
 Alternatively,
one can define a different scheme, the Adler-Bardeen (AB) scheme~\cite{axanom}, such that 
the singlet quark first moment is conserved 
at all orders and can be identified with the total quark helicity, whereas the
polarized gluon density is defined as in the $\overline{\rm MS}$
scheme.   
In the AB scheme the gluon polarization directly contributes 
to the singlet axial charge (and thus to the first moment of $g_1$), which is now 
written as
\be
a_0(Q^2)=\Delta\Sigma_{\rm AB}(1)-n_f\frac{\alpha_s(Q^2)}{2\pi}\Delta g(1,Q^2)\,.
\lb{a0AB}
\ee
Eqs.~(\ref{a0MS}) and~(\ref{a0AB}) then 
yield the relation between the first moments of the singlet quark distributions in the two
schemes. They differ by the product $\alpha_s(Q^2)\Delta g(1,Q^2)$, which is due to the 
anomalous nonconservation of the singlet axial current~\cite{axanom}, and is scale 
invariant at leading order; this implies that $\Delta g(1,Q^2)$ increases as 
$1/\alpha_s(Q^2)$ with $Q^2$, and the gluon contribution in Eq.~(\ref{a0AB}) is not 
asymptotically suppressed by powers of $\alpha_s$. As a result, this scheme dependence 
does not vanish at large $Q^2$, and the definition of the singlet quark first moment 
is therefore maximally ambiguous.

In the AB scheme, which will be adopted in the following phenomenological analysis 
(see Sec.~6), the first moment of the gluon 
coefficient function at NLO reads~\cite{axanom}
\be
C_{g}(1,\alpha_s)=-\frac{\alpha_s}{4\pi}+\mathcal{O}(\alpha_s^2)\,,
\lb{cg}
\ee
and the first moment of the structure function $g_1$ is given at NLO
by 
\bea
&&\Gamma_{1}(Q^2)\equiv\int_0^1dx\,g_1(x,Q^2)=\lb{Ga1}\\
&&\frac{<e^2>}{2}\left[\left(1-\frac{\alpha_s}{\pi}
\right)\left(\Delta q_{\rm NS}(1,Q^2)+\Delta\Sigma(1,Q^2)\right)
- n_f\frac{\alpha_s}{2\pi}
\Delta g(1,Q^2)\right]\,.\nn
\eea
Different factorization schemes were also discussed in 
Refs.~\cite{Ball:1995td}, and the dependence of the results of phenomenological
analyses of spin-dependent pdfs on the choice of the scheme was studied. 

It should be finally noted that a number of theoretical models attempt to 
explain the quark helicity distribution 
within the nucleon and the unexpected smallness of the singlet axial charge. A 
possible interpretation is to assume a large and negative contribution from the 
strange sea polarization $\Delta s^+$. This suppression of the axial charge might 
be explained by invoking nonperturbative mechanisms based on instanton-like vacuum
configuration~\cite{forte}. In this case $\Delta s=\Delta \bar s$. Another scenario
is possible, where the smallness of the singlet axial charge is due to the intrinsic 
strangeness, i.e. the C-even strange combination is large, but  $\Delta s$ differs
significantly from $\Delta \bar s$. Specifically, in the Skyrme models of the
nucleon~\cite{Skyrme} the strange distribution (in particular its first moment) is 
large, while the antistrange 
distribution is much smaller and does not sizably contribute to the axial 
charge~\cite{Ma}.\\
Finally, another possible mechanism has been proposed to
understand the nucleon spin structure (see~\cite{Altarelli:1998nb} and references 
therein), which includes the gluon contribution. Indeed, in the Adler-Bardeen scheme,
one can assume, on the basis of Eq.~(\ref{a0AB}), that a cancellation 
between a large (scale-indepedent) singlet quark distribution and a large gluon 
polarization takes place. In this case $\Delta s^+\ll \Delta u^+,\,\Delta d^+$,
as expected in the parton model.

\section{Phenomenology of $g_1$}

Experimental information on the structure functions $g_1(x,Q^2)$ and 
$g_2(x,Q^2)$ are extracted from measurements of spin-asymmetries. 
Longitudinally polarized leptons are scattered
off a hadronic target that is polarized either longitudinally or transversely.
The longitudinal $(A_{||})$ and tranverse $(A_{\perp})$ asymmetries are
formed by combining data taken with opposite beam helicity
\be
A_{||}=\frac{\sigma^{\uparrow\downarrow}-\sigma^{\uparrow\uparrow}}
{\sigma^{\uparrow\downarrow}+\sigma^{\uparrow\uparrow}}\,,\quad\quad
A_{\perp}=\frac{\sigma^{\downarrow\rightarrow}-\sigma^{\uparrow\rightarrow}}
{\sigma^{\downarrow\rightarrow}+\sigma^{\uparrow\rightarrow}}\,.
\lb{asy}
\ee
The symbols $\sigma^{\uparrow\uparrow}$ and $\sigma^{\uparrow\downarrow}$ denote
the cross sections for the lepton-nucleon scattering with their parallel
and antiparallel helicity states, respectively, whereas  
$\sigma^{\uparrow\rightarrow}$ and $\sigma^{\downarrow\rightarrow}$ represent 
the scattering cross sections for transversely polarized nucleon targets.\\
The asymmetries $A_{||}$ and $A_{\perp}$ are related to the photon 
absorption cross section asymmetries $A_1$ and $A_2$ by the relations 
\be
A_{||}=D\left(A_1+\eta A_2\right)\,,\qquad A_{\perp}=d\left(A_2-\zeta A_1\right)\,,
\lb{Alt}
\ee
where 
\be
A_1(x,Q^2)=\frac{\sigma^T_{1/2}-\sigma^T_{3/2}}{\sigma^T_{1/2}+\sigma^T_{3/2}}\,,\qquad
A_2(x,Q^2)=\frac{2\sigma^{LT}}{\sigma^T_{1/2}+\sigma^T_{3/2}}\,.
\lb{A12}
\ee
Here, $\sigma^T_{1/2}$ and $\sigma^T_{3/2}$ represent the absorption cross sections
of virtual transverse photons for the total helicity of the photon-nucleon system
of $1/2$ and $3/2$ respectively; $\sigma^{LT}$ denotes the interference term between the
transverse and longitudinal photon-nucleon amplitudes.\\
The factor $D$ in the first of Eqs.~(\ref{Alt}) is interpreted as the depolarization 
of the photon with respect to the primary lepton beam
\be
D=\frac{y(1+\gamma^2y/2)(2-y)}{y^2(1+\gamma^2)+2(1+R)(1-y-\gamma^2y^2/4)}\,,
\lb{D}
\ee
and depends upon the kinematic factors
\be
y=1-\frac{E'}{E}\quad\textrm{and}\quad  \gamma=\frac{2mx}{\sqrt{Q^2}}
\lb{yge}
\ee
where $y$ is the fractional energy lost by the lepton. The depolarization factor
also depends on the ratio $R(x,Q^2)$ of the cross sections for the longitudinally
polarized photon to the transverse one, and is experimentally measured. 
Moreover,
\be
\eta=\frac{\gamma(1-y-\gamma^2y^2/4)}{(1+\gamma^2y/2)(1-y/2)}\,.
\lb{et}
\ee
The kinematic factors in the second of Eqs.~(\ref{Alt}) are given respectively
by
\be
d=\frac{D\sqrt{1-y-\gamma^2y^2/4}}{1-y/2}\,,\qquad
\zeta=\frac{\gamma(1-y/2)}{1+\gamma^2y/2}\,.
\lb{dz}
\ee
Then, the asymmetries $A_1$ and $A_2$ can be related to 
the polarized structure functions $g_1$ and $g_2$ by
\be
A_1(x,Q^2)=\frac{g_1(x,Q^2)-\gamma^2 g_2(x,Q^2)}{F_1(x,Q^2)}\,,\quad
A_2(x,Q^2)=\frac{\gamma (g_1(x,Q^2)+g_2(x,Q^2))}{F_1(x,Q^2)}\,,
\lb{ga12}
\ee
where $F_1(x,Q^2)$ is the unpolarized structure function of the nucleon.\\
If both the longitudinal and transverse asymmetries, $A_{||}$ and $A_{\perp}$, 
are measured, one can extract both structure functions, $g_1(x,Q^2)$ and 
$g_2(x,Q^2)$ from experimental data with minimal assumptions, using the relations
\bea
&& g_1(x,Q^2)=\frac{F_1(x,Q^2)}{(1+\gamma^2)(1+\eta\zeta)}\left[(1+\gamma\zeta)
\frac{A_{||}}{D}-(\eta-\gamma)\frac{A_{\perp}}{d}\right]\,,\\
&& g_2(x,Q^2)=\frac{F_1(x,Q^2)}{(1+\gamma^2)(1+\eta\zeta)}
\left[\left(\frac{\zeta}{\gamma}-1\right)\frac{A_{||}}{D}+
\left(\eta+\frac{1}{\gamma}\right)\frac{A_{\perp}}{d}\right]\,.
\lb{g1g2}
\eea
Actually, the transverse asymmetry is generally poorly determined, and in most
cases only the longitudinal asymmetry $A_{||}$ is measured. In this case,
the relevant structure function $g_1$ can be expressed via the observable $A_{||}$ 
and the unknown structure function $g_2$, namely
\be
g_1(x,Q^2)=\frac{F_1(x,Q^2)}{(1+\gamma\eta)}\frac{A_{||}}{D}+
\frac{\gamma(\gamma-\eta)}{\gamma\eta+1}\,g_2(x,Q^2)\,.
\lb{g1ex}
\ee
An experimental determination of $g_1$ thus relies on a theoretical 
assumption on $g_2$. In Eq.~(\ref{g1ex}) $F_1$ is expressed in 
terms of the unpolarized structure function $F_2$, usually extracted from unpolarized 
DIS experiments, i.e. 
\be
F_1(x,Q^2)=\frac{F_2(x,Q^2)}{2x(1+R(x,Q^2))}(1+\gamma^2)\,.
\lb{fr12}
\ee
Then, as discussed in Sec.~6, parametrizations for the unpolarized structure 
function $F_2(x,Q^2)$ and the ratio $R(x,Q^2)$ of 
longitudinal and transverse virtual photo-absorption cross sections must be used
in order to extract experimental information on the structure function $g_1(x,Q^2)$.\\
Finally, the experimentally measured counting rate asymmetry $A_{\rm exp}$ 
is related to the cross section asymmetry $A_{||}$ by
\be
A_{\rm exp}=f_t P_t P_b A_{||}\,,
\lb{aex}
\ee
where $P_b$ is the beam polarization, $P_t$ the polarization of the target
nucleon, and $f_t$ the target dilution factor, i.e. the fraction of polarized
nucleons in the target material.

Relations (\ref{g1ex}) and (\ref{fr12}) simplify if the nucleon mass $m$ is
neglected. Indeed, in this case the kinematic factor $\gamma\simeq 0$, and the structure 
function $g_2$ decouples. This is generally a reasonable approximation 
in the small-$x$ or large-$Q^2$ region, where the factor $\gamma^2$ is of the
order of $10^{-2}-10^{-3}$. Then, one simply has
\be
g_1(x,Q^2)\simeq \frac{F_2(x,Q^2)}{2x(1+R(x,Q^2))}\,A_1\,,\quad\textrm{with}\quad
A_1\simeq\frac{A_{||}}{D}\,.
\lb{g1apx}
\ee

It should be finally noted that some of the experimental measurements concern  
the asymmetry $A_1$ (or $A_{||}$), while others concern 
values of the combination of $A_{||}$ and $A_{\perp}$ which corresponds
to $g_1/F_1$. The two quantities coincide for $m=0$ (aside from the 
depolarization factor in the case of $A_{||}$), but they do not when
mass corrections are included. Then, in the case $A_1$ is measured, the 
related theoretical asymmetry 
is given by $[g_1-(4m^2x^2/Q^2)g_2]/F_1$ (or by the similar relation 
obtained by Eq.~(\ref{g1ex}) if $A_{||}$ is measured), in the second case 
simply by $g_1/F_1$.

\section{Target mass corrections}

A large part of experimental data in polarized deep-inelastic scattering
are taken at relatively low values of $Q^2$. Specifically, $Q^2$ is usually 
around few GeV$^2$ for data points in the small-$x$ region (data at 
$Q^2< 1$GeV$^2$ are also available, but they are 
usually not included in perturbative analyses). In this kinematical region,  
contributions suppressed by inverse powers of $Q^2$ arise when taking
into account the finite value of the nucleon mass $m$, and could play 
a relevant role.\\ 
Another source of power-suppressed terms originate from the 
operator product expansion of the hadronic tensor $W_A^{\mu\nu}$ in 
Eq.~(\ref{Wa}), i.e. from matrix elements of operators of non-leading 
twist (see e.g.~\cite{ope}). 
These contributions are usually referred to as dynamical higher twist, and their 
effect can not be calculated in perturbation theory, so that it is difficult 
to assess their relevance in any phenomenological analysis of polarized pdfs.

When considering target mass corrections, in order to obtain experimental 
information on $g_1$ from the measured asymmetry $A_{||}$, the complete 
relation Eq.~(\ref{g1ex}) must be used (or the first of Eq.~(\ref{ga12}) 
if the data concern $A_1$), that is, all the kinematical factors must be
retained. Moreover, also the structure function $g_2$ now contributes 
when comparing the theoretically predicted observable $g_1$ to the data. 
However, experimental data on the structure function $g_2$ are restricted
to a limited range in the $(x,Q^2)$-plane and are affected by large 
uncertainties (see e.g.~\cite{Abe:1995dc}). Therefore, a model for $g_2$ 
is usually invoked in phenomenological analyses.\\ 
To take target mass corrections into account, following 
Ref.~\cite{Piccione:1997zh}, the moments of the polarized structure 
functions $g_1$ and $g_2$ can be expressed in terms of the matrix elements of the 
twist-2 and 3 operators appearing in the light-cone expansion of the forward scattering 
amplitude. Specifically, at the first order in $m^2/Q^2$,   
the $n$-th moments of
$g_1$ and $g_2$ read~\cite{Piccione:1997zh}
\bea
&& g_1^n(Q^2)=a_n+\frac{m^2}{Q^2}\frac{n(n+1)}{(n+2)^2}(n\,a_{n+2}+4\,d_{n+2})+
\mathcal{O}\left(\frac{m^4}{Q^4}\right)\lb{nm12}\\
&& g_2^n(Q^2)=\frac{n-1}{n}(d_n-a_n)+\frac{m^2}{Q^2}\frac{n(n-1)}{(n+2)^2}
[n\,d_{n+2}-(n+1)a_{n+2}]+\mathcal{O}\left(\frac{m^4}{Q^4}\right)\nn\,.
\eea
Here, $a_n$ and $d_n$ are given by matrix elements of the twist-2 and 3 polarized 
operators
\bea
&& <p,s|i^{n-1}\left[\bar\psi\gamma_5\gamma^{\sigma}D^{\mu_1}\dots 
D^{\mu_{n-1}}\lambda_i\psi\right]_S|p,s>=-m\,a_n M_1^{\sigma\mu_1\dots\mu_{n-1}}\quad\\
&& <p,s|i^{n-1}\left[\bar\psi\gamma_5\gamma^{\lambda}D^{\sigma}D^{\mu_1}\dots 
D^{\mu_{n-2}}\lambda_i\psi\right]_{S'}|p,s>=m\,d_n M_2^{\lambda\sigma\mu_1\dots\mu_{n-2}}
\quad\nn
\lb{ope}
\eea
where $m,p$ and $s$ are mass, four-momentum and spin of the target nucleon, 
the symbol $[\dots]_{S}$ means complete symmetrization in the indices 
$\sigma,\mu_1\dots\mu_{n-1}$, whereas $[\dots]_{S'}$ denotes antisymmetrization
with respect to $\lambda$ and $\sigma$ and symmetrization over the other indices.
The symbol $M_1^{\sigma\mu_1\dots\mu_{n-1}}$ denotes the most general rank-$n$ tensor
which can be formed with one spin four-vector $s$ and $n-1$ momentum four-vectors $p$; 
similarly $M_2^{\lambda\sigma\mu_1\dots\mu_{n-2}}$ is antisymmetric in $\lambda$
and $\sigma$ and symmetric in all other indices.\\ 
If $m=0$ is taken in Eqs.~(\ref{nm12}) the $n$-th moments of the structure functions
(aside from QCD perturbative corrections) reduce to the simple form~\cite{Kodaira:1979ib}
\bea
&& g_{1\,0}^n(Q^2)=a_n+\mathcal{O}\left(\frac{m^2}{Q^2}\right)\lb{nm012}\\
&& g_{2\,0}^n(Q^2)=\frac{n-1}{n}(d_n-a_n)+\mathcal{O}\left(\frac{m^2}{Q^2}\right)\nn\,.
\eea
One can then use Eq.s~(\ref{nm012}) to eliminate the matrix elements $a_n$ and 
$d_n$ from Eqs.~(\ref{nm12}) in favour of the $n$-th moments of the structure
functions at zero nucleon mass, i.e. $g_{1\,0}^n$ and $g_{2\,0}^n$. Then, to order
$\mathcal{O}(m^2/Q^2)$ one has~\cite{Piccione:1997zh}
\bea
&& g_1^n(Q^2)=g_{1\,0}^n(Q^2)+\frac{m^2}{Q^2}\frac{n(n+1)}{(n+2)^2}
\left[(n+4)g_{1\,0}^{n+2}(Q^2)+\frac{4(n+2)}{(n+1)}g_{2\,0}^{n+2}(Q^2)\right]\nn\\
&& g_2^n(Q^2)=g_{2\,0}^n(Q^2)+\frac{m^2}{Q^2}\frac{n(n-1)}{(n+2)^2}
\left[\frac{n(n+2)}{(n+1)}g_{2\,0}^{n+2}(Q^2)-g_{1\,0}^{n+2}(Q^2)\right]\,.
\lb{nmf12}
\eea
If $d_n=0$, the structure functions obey the so-called 
Wandzura-Wilczek relation~\cite{Wandzura:1977qf}
\be
g_{2}^{n}(Q^2)=-\frac{n-1}{n}\,g_{1}^{n}(Q^2)\,.
\lb{wr}
\ee
Actually, twist-3 contributions are not power-suppressed with respect to the leading 
twist; however, if $d_n\ll a_n$ is assumed, one can then use Eq.~(\ref{wr}) to express 
$g_1^n$ and $g_2^n$ in 
Eqs.~(\ref{nmf12}) in terms of the zero-mass structure function $g_{1\,0}^n$ alone, namely
\be
g_{2\,0}^{n+2}(Q^2)=-\frac{n+1}{n+2}\,g_{1\,0}^{n+2}(Q^2)\,,
\lb{wr2}
\ee
with $g_{1\,0}$ given by Eq.~(\ref{g1}). Thus, when including mass
corrections in the analysis, the theoretically predicted structure function
$g_1$ (or its $n$-th moment) as given by Eq.~(\ref{g1}) is corrected by the
first of Eqs.~(\ref{nmf12}), together with the hypothesis (\ref{wr2}) on 
the structure function $g_2$ at zero mass.\\


\chapter{Global analysis of inclusive DIS data}

The problem of extracting relevant physical quantities from the existing 
inclusive DIS data is considered here in detail. The analysis is mainly aimed 
at assessing the relevance of inclusive measurements to the determination
of the spin-dependent gluon density, its first moment and the $x$-shape.

To begin with, some details on the analysis method are given 
in Sec.~6.1, and the boundary conditions for parton distribution functions 
are discussed in Sec.~6.2. Sec.~6.3 is devoted to the study of the
impact of the most recent inclusive DIS data on the extraction of polarized
pdfs. In particular, the effect of very low-energy
data is studied in Sec.~6.4, in connection with higher-twist corrections.\\ 
The results on the gluon distribution function are given in Sec.~6.5., and their
dependence upon input densities extensively investigated. Finally, the main consequences of
the above analysis on the structure function $g_1$ and moments of quark and
gluon polarizations are summarized in Sec.~6.6.

\section{Analysis method}

The nucleon structure function $g_1$ is constructed at the 
measured points in the $(x,Q^2)$ plane by convoluting the hard coefficient 
functions with the solutions of the NLO evolution equations. A parametrization 
for the polarized pdfs in the singlet, nonsinglet and gluon sectors defined in 
Sec.~5, namely $\Delta\Sigma,\,\Delta q_{\rm NS}$ and $\Delta g$, as a function 
of a number of parameters is thus assigned at a given initial scale $Q_0^2$. The 
whole set of parameters is then optimized by minimizing~\footnote{The $\chi^2$ 
minimization procedure is performed by the CERN program library MINUIT~\cite{minuit}.} 
the total $\chi^2$, namely
\be
\chi^2=\sum_i\left(\frac{g_1(x,Q^2)-g_1^{\rm data}(x,Q^2)}
{\delta g_1^{\rm data}(x,Q^2)}\right)^2\,,
\lb{chi2}
\ee
where the sum runs over the available data points. Experimental data for 
$g_1$ are derived from the measured longitudinal spin asymmetry as discussed 
in Sec.~5.4, and are usually given along curves $Q^2(x)$ in a restricted region of
the final-state invariant mass for each experimental set of data. The uncertainty 
$\delta g_1^{\rm data}$ includes both statistical and systematic errors added 
in quadrature, and $g_1(x,Q^2)$ denotes the theoretically predicted observable.\\ 
A full next-to-leading order analysis of the available experimental data on
the polarized structure function $g_1$ of the nucleon has been performed in
the Adler-Bardeen factorization scheme, as discussed in Sec.~5.3.\\ 
The AP evolution is performed in the Mellin space of the $n$-th 
moments, and, by the inverse transformation, the structure function $g_1$ in 
terms of the Bjorken-$x$ is finally recovered for the fitting procedure, 
i.e. the $\chi^2$-minimization.

In order to extract $g_1$ data from the experimental asymmetries, as 
discussed in Sec.~5.4, one needs a parametrization for both the ratio 
$R=\sigma_{\rm L}/\sigma_{\rm T}$ of the photoabsorption cross sections,
and the unpolarized structure function $F_2$.\\
Because of the slightly better coverage of the small-$x$ region, due to
the COMPASS data (see Sec.~6.3 below), the parametrization R1990~\cite{Whitlow:1990gk} 
used in previous analyses~\cite{Altarelli:1998nb,Forte:2001ph} for the ratio 
$R=\sigma_{\rm L}/\sigma_{\rm T}$ has been replaced by the 
most recent one R1998~\cite{Abe:1998ym}.
The latter parametrization indeed extends the kinematic range to lower and higher values
of $x$ ($0.005\lsim x \lsim 0.86$), by adding to the fit a considerable number of 
data~\cite{Tao:1995uh} on $R(x,Q^2)$.\\ 
Furthermore, the commonly used parametrization of the unpolarized structure function $F_2$, 
which also enters into the experimental determination of $g_1$, given by NMC~\cite{nmc} has 
been replaced by the unbiased one~\cite{F2nn}, obtained by training sets of neural networks 
on the experimental data (see~\cite{Del Debbio:2004qj}).\\ 
However, no sizable effects are observed as a result of both these implementations, that 
is, all the parameters of pdfs turned out to be always largely consistent within the 
errors, and the quality of the fit essentially unchanged.

Finally, it should be noted that target mass corrections have been included as a rule  
up to the first order in $m^2/Q^2$, as explained in Sec.~5.5, throughout this
analysis. Indeed, when considering the finite size of the nucleon
mass $m$, also the structure function $g_2$ enters into the theoretical determination of
$g_1$. Then, due to the lack of precise experimental data on $g_2$, the Wandzura-Wilczek 
hypotesis has been invoked in the calculations (see Sec.~5.5). 

\section{Parametrization of pdfs}

The functional form of spin-dependent parton distribution functions at the initial scale 
$Q_0^2$ is fixed according to the conventional parametrization
\be
\Delta f(x,Q_0^2)=N_f\eta_f x^{\alpha_f}(1-x)^{\beta_f}(1+\ga_f x^{\delta_f})
\qqd\textrm{(type-A)}
\lb{fitA}
\ee
where $\Delta f$ denotes $\Delta\Sigma,$ $\Delta q_{\rm NS},$ and 
$\Delta g$. The overall factor $N_f$ is defined by the normalization 
condition
\be
\int_0^1 dx\,\Delta f(x,Q_0^2)=\eta_f\,,
\lb{ndf}
\ee
i.e. such that the parameters $\eta_f$, fitted to the data, directly reflect 
the value of the first moment of pdfs at the initial scale, which is taken as
a rule at $Q_0^2=1\,$GeV$^2$.\\ 
The nonsinglet quark density at $Q_0^2=1\,$GeV$^2$ is defined by the linear 
combination of the triplet and octect distributions
\be
\Delta q_{\rm NS}(x,Q_0^2)=\pm\frac{3}{4}\Delta q_3(x,Q_0^2)+
\frac{1}{4}\Delta q_8(x,Q_0^2)\,,
\ee
where the plus (minus) sign stays for a proton (neutron) target.
The distributions $\Delta q_3$ and $\Delta q_8$ are assumed to have the same 
$x$-dependence, while the parameter $\eta_8$, corresponding to the first moment 
of $\Delta q_8$, is fixed to the value $\eta_8=0.588\pm0.025$ 
by using SU(3) symmetry and octet baryon decay constants~\cite{Yao:2006px}. 
The parameter $\eta_3$ is instead fitted to the data. 
At higher scales, new contributions are generated dynamically as heavy quark 
thresholds are crossed.\\
Since the large-$x$ behavior of the polarized gluon distribution 
can not be determined by inclusive DIS data, i.e. by scaling violations, 
the value of $\beta_g$ needs to be fixed a priori.  
Typical values of $\beta_g$ examined in the following analysis are 
restricted to the range $4\leq\beta_g\leq10$, and a correlation with 
the gluon first moment at the initial scale $\eta_g$ is observed, in that
increasing the (fixed) value of $\beta_g$ leads to a slight decrease of 
$\eta_g$. Then, both parameters have been followed out as new data are 
included and the whole fit improved.\\ 
Moreover, all the exponents $\delta_f$ in the last factor of Eq.~(\ref{fitA})
have been fixed, as in Ref.~\cite{Altarelli:1998nb}, respectively to the 
values~\footnote{The choice for $\delta_{\rm NS}$ in particular is discussed 
in~\cite{Altarelli:1997}, and is made in order to avoid asymptotic 
behaviors in the nonsinglet channel more singular than 
$x^{-0.5}$ as $x\to0$, corresponding to the maximum saturation of the partonic 
constraint $|\Delta q_{\rm NS}(x,Q^2)|\lsim q_{\rm NS}(x,Q^2)$. 
Indeed, a correlation between the parameters $\delta_{\rm NS}$ and 
$\alpha_{\rm NS}$ has been pointed out in~\cite{Altarelli:1997}, such that 
$\alpha_{\rm NS}$ approaches zero as $\delta_{\rm NS}$ decreases from unity.}
\be
\delta_{\Sigma}=\delta_g=1\qd\rm{and}\qd\delta_{\rm NS}=0.75\,.
\ee

A different functional form has been also adopted for the initial 
parametrizations, to evaluate the impact of the input densities on the final results.
Specifically, while keeping the initial scale at $Q_0^2=1\,$GeV$^2$, the less 
singular input
\bea
&&\Delta\Sigma=N_{\Sigma}\eta_{\Sigma}x^{\alpha_{\Sigma}}\left(
\ln 1/x\right)^{\beta_{\Sigma}}\qqd\qqd\qqd\qqd\qd
\textrm{(type-B)}\lb{fitB}\\
&&\Delta f=N_f\eta_f \left[\left(\ln 1/x\right)^{\alpha_f}+
\ga_f\,x^{\delta_f}\,\left(\ln 1/x\right)^{\beta_f}\right]\qd
\Delta f=\Delta q_{\rm NS},\,\Delta g\nn
\eea
is used for an alternative fit, with the same normalization condition
(\ref{ndf}). Similarly to the standard parametrization, the exponents
$\delta_f$ are fixed to the values $\delta_g=1$ and $\delta_{\rm NS}=0.75$. 
At variance with the initial densities~(\ref{fitA}), in Eq.~(\ref{fitB}) 
the rise at small $x$ is at most logarithmic, and then softer than any power. 
Since $\ln 1/x\sim(1-x)$ as $x\to1$, the large-$x$ 
behavior in Eq.~(\ref{fitB}) is similar to Eq.~(\ref{fitA}) for the singlet 
distribution, whereas both terms contribute in the gluon and 
nonsinglet cases.

The total number of parameters determined from the fit amounts, in 
both cases, to at most eleven, and no limits are imposed on their variation. 
Also the sign of all parameters are left free, including the first moment at
the initial scale, although the data always choose $\eta_f$ to
be positive.\\ 
Finally note that, in order to avoid any bias produced by further constraints,
the input densities Eqs.\ (\ref{fitA}) and (\ref{fitB}) are 
not explicitly forced by the positivity condition $|\Delta f(x,Q^2)|\leq f(x,Q^2)$, at 
variance with many recent analyses (see e.g.~\cite{Hirai:2006sr,Leader:2005ci}), 
where, on the other hand, the initial parametrizations for the polarized gluon, 
valence and sea quark distributions are taken proportional to the related 
unpolarized quantities, with proportionality factors given by powers of $x$. 
However, the positivity bound has been checked a posteriori, and, in the case 
of fits of type-A, turns out to be 
always largely fulfilled (see also~\cite{Altarelli:1998gn}).

\section{Update}

The main purpose of the following analysis is to investigate the impact of 
the recent experimental data on the polarized parton distributions, with a  
special focus on the gluon polarization. The whole analysis is performed at NLO in 
the Adler-Bardeen factorization scheme, and the normalization of the strong 
coupling $\alpha_s(M^2_Z)=0.118$ is assumed throughout.\\
The starting point is a global fit of type-A (Eq.~(\ref{fitA})), of inclusive 
DIS data, on the basis of 176 experimental points used in the previous 
analyses~\cite{Altarelli:1998nb,Forte:2001ph}, 
as summarized in Tab.~1. 

\vspace{0.7truecm}
\begin{tabular}[t]{|l|l|l|l|l|} 
\hline
Data set (target)  & $x$-range & $Q^2$-range (GeV$^2$) & No. data & Ref.  \\
\hline\hline
SMC (p)  & 0.005 - 0.479 & 1.3 - 58  & 12 & \cite{Adeva:1998vv} \\
E143 (p,d)  & 0.031 - 0.749 & 1.27 - 9.52 & 56  & \cite{Abe:1998wq} \\
E155 (p) & 0.015 - 0.75 & 1.22 - 34.72 & 24 & \cite{Anthony:2000fn} \\
HERMES (p) & 0.023 - 0.66 & 1.01 - 7.36 & 20 & \cite{Airapetian:1998wi}\\
SMC (d)  & 0.005 - 0.480 & 1.3 - 54.8  & 12 & \cite{Adeva:1998vv} \\
E155 (d) & 0.015 - 0.75 & 1.22 - 34.79 & 24 & \cite{Anthony:1999rm}\\
E142 (n) & 0.035 - 0.466 & 1.1 - 5.5 & 8 & \cite{Anthony:1996mw}  \\
HERMES (n) & 0.033 - 0.464 & 1.22 - 5.25 & 9 & \cite{Ackerstaff:1997ws} \\
E154 (n) & 0.014 - 0.564 & 1.2 - 15 & 11 & \cite{Abe:1997cx}\\
\hline
\end{tabular}

{\footnotesize Tab.~1 Initial set of inclusive DIS data (176 experimental points).}
\vspace{0.7truecm}

\begin{figure}[!ht]
\centerline{\includegraphics[width=.6\linewidth,angle=-90]{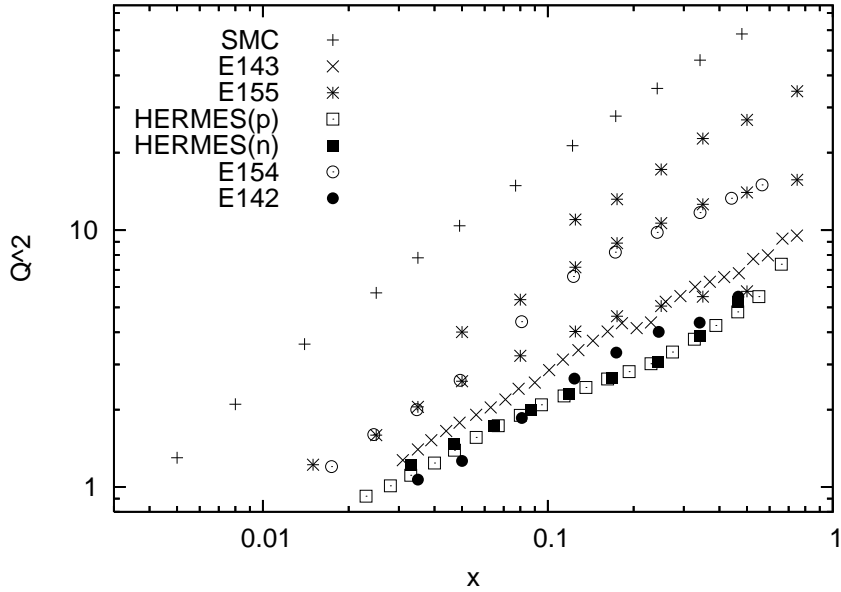}}
\vspace{0.5truecm}
\caption{\footnotesize Distribution in the $(x,Q^2)$-plane of the set of 176 
DIS data up to 2000, as summarized in Tab.~1.}
\label{176}
\end{figure}

Note that the effects of the E155 proton data~\cite{Anthony:2000fn} alone have 
been analyzed in recent works~\cite{Hirai:2003pm}, and the conclusion was reached 
that their accuracy significantly contributes to reducing the statistical errors 
of pdfs.\\
As can be seen by the scatter plot in Fig.~\ref{176}, the covered region of the 
$(x,Q^2)$-plane is 
roughly restricted to $0.01\lsim x \lsim 0.8$ for the Bjorken $x$ (with the 
exception of very few SMC points at lower $x$), and to $1\leq Q^2 \lsim 58\,$GeV$^2$  
for the momentum transfer. Indeed the cut $Q^2\geq 1\,$GeV$^2$ is imposed as a rule to 
all the data.

\vspace{0.5truecm}
\vspace{0.5truecm}

The results of a fit of type-A on this set of data are shown in the second column 
of Tab.~3. By comparing such results with those of the previous 
analyses~\cite{Altarelli:1998nb,Forte:2001ph}, globally the same picture arises, 
and, in particular, the $x$-shapes of pdfs are similar to those of 
Refs.~\cite{Altarelli:1998nb,Forte:2001ph}, and likewise not very precisely determined.\\ 
A sizable first moment of the polarized gluon density at the initial scale is found, 
and, both the gluon and the nonsinglet quark distributions rise as $x\to0$, whereas 
a flat behavior is observed in the singlet sector.\\ 
Note that the parameters $\gamma_g$ and $\gamma_{\Sigma}$, which control the shape of 
the gluon and singlet distributions at intermediate $x$, can not be disentangled in this 
analysis. As observed in~\cite{Ball:1995td}, this is due to the mixing of $\Delta g$ 
and $\Delta\Sigma$ in the AP evolution. 
Therefore $\gamma_g=\gamma_{\Sigma}$ is taken, thus leaving ten parameters to be determined 
by the $\chi^2$ minimization. 
Finally, the errors given here are statistical errors from the fit.

A bulk of more precise data has become recently available. These are in particular
15 points at rather high energy from COMPASS~\footnote{Note that the latest COMPASS 
data~\cite{comp07} at very low $x$ and $Q^2$ have been excluded altogether in this analysis, all
lying below the cut $Q^2\geq 1\,$GeV$^2$ imposed to the data set.}, and a number of lower 
energy points from HERMES and JLAB as summarized in Tab.~2. 
Furthermore, a set of complementary data from SMC has been also included, 
that was not considered in the previous analyses~\cite{Altarelli:1998nb,Forte:2001ph}. 
A large amount of very low-energy data from CLAS~\cite{Fatemi:2003yh,Dharmawardane:2006zd} 
will be considered separately, in connection with higher-twist effects (see Sec.~6.3).

\begin{tabular}[t]{|l|l|l|l|l|} 
\hline
Data set (target) & $x$-range & $Q^2$-range (GeV$^2$) & No. data & Ref.  \\
\hline\hline
COMPASS (d) & 0.0046 - 0.566 & 1.10 - 55.3 & 15 & \cite{Alexakhin:2006vx} \\
HERMES (p,d) & 0.0264 - 0.7311 & 1.12 - 14.29 & 74 & \cite{Airapetian:2007mh} \\
JLAB & 0.33 - 0.60 & 2.71 - 4.83 & 3 & \cite{Zheng:2004ce} \\
SMC (p,d) & 0.0043 - 0.121 & 1.09 - 23.1 & 16 & \cite{Adeva:1999pa} \\
\hline
\end{tabular}\\

{\footnotesize Tab.~2. New set of inclusive DIS data up to 2006.}
\vspace{0.5truecm}

\begin{figure}
\centerline{\includegraphics[width=.6\linewidth,angle=-90]{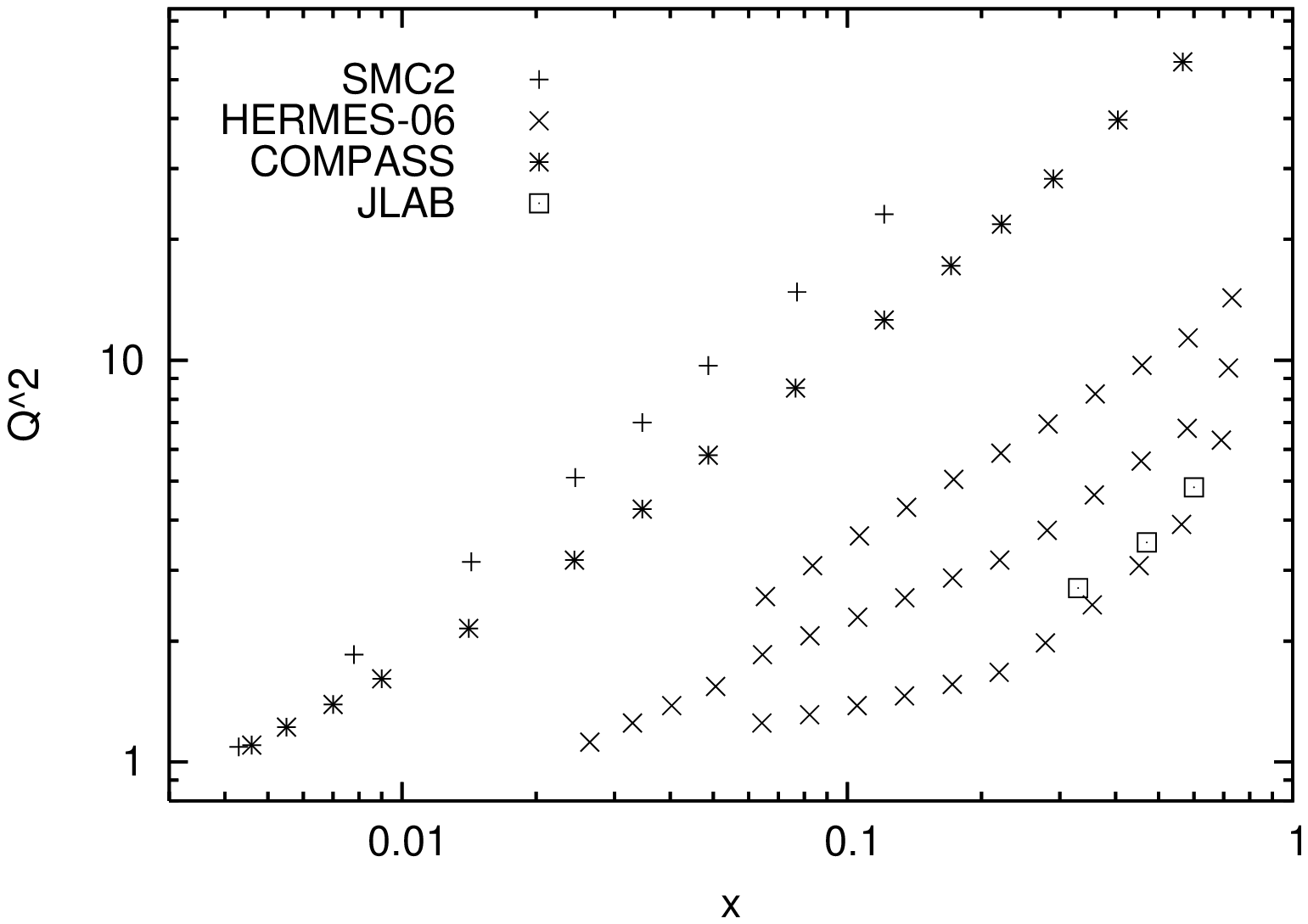}}
\vspace{0.4truecm}
\caption{\footnotesize Distribution in the $(x,Q^2)$-plane of the new set of 
DIS data up to 2006, as summarized in Tab.~2.}
\label{xqnew}
\vspace{0.6truecm}
\centerline{\includegraphics[width=.6\linewidth,angle=-90]{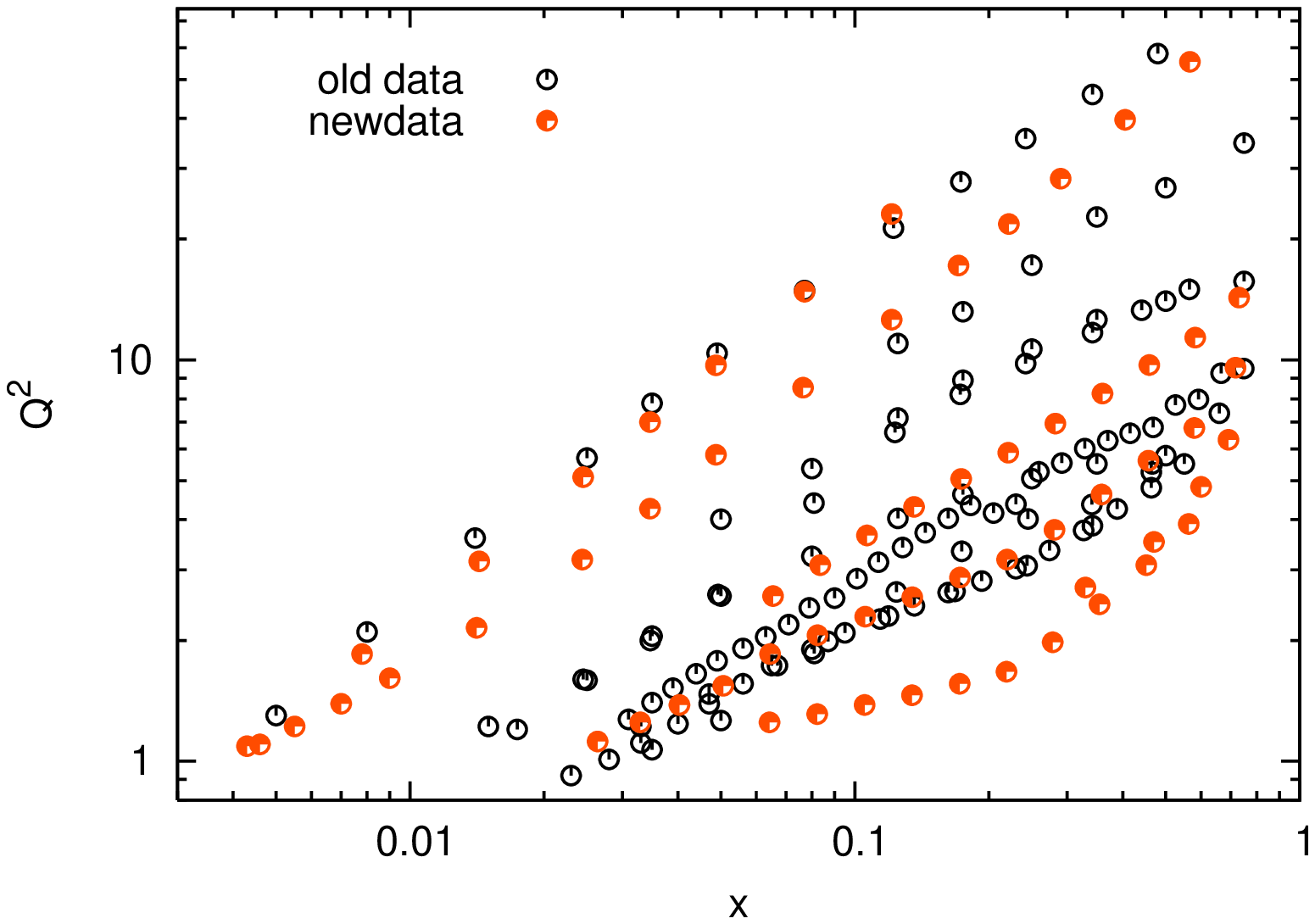}}
\vspace{0.4truecm}
\caption{\footnotesize Distribution in the $(x,Q^2)$-plane of the complete 
set of data.}
\label{xqtot}
\end{figure}

\begin{figure}
\vspace{-0.8truecm}
\centerline{\includegraphics[width=.8\linewidth]{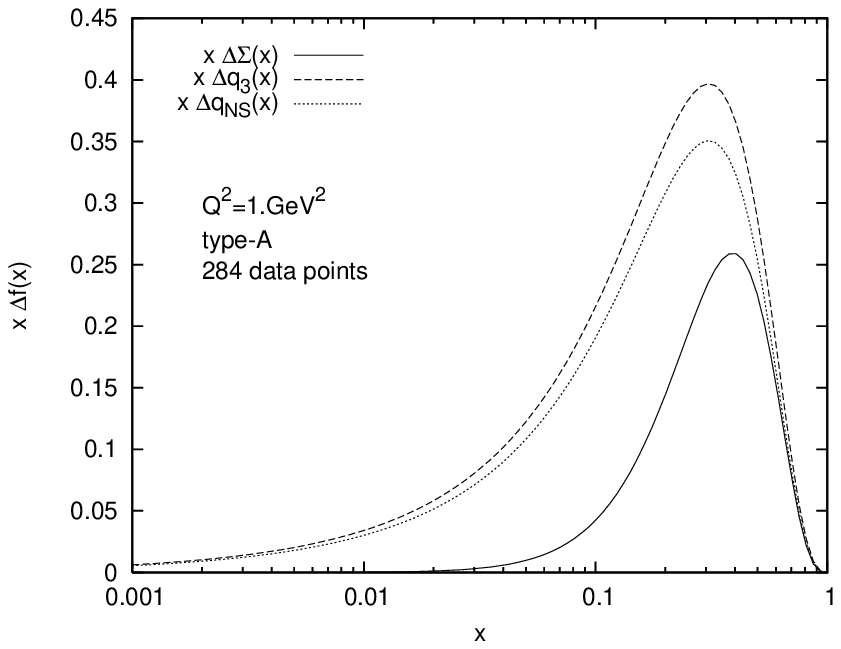}}
\vspace{-0.6truecm}
\caption{\footnotesize Singlet and nonsinglet distributions (proton target)
at $Q^2=1\,$GeV$^2$ from a fit of type-A on the whole set of available data.}
\label{sns284}
\centerline{\includegraphics[width=.8\linewidth]{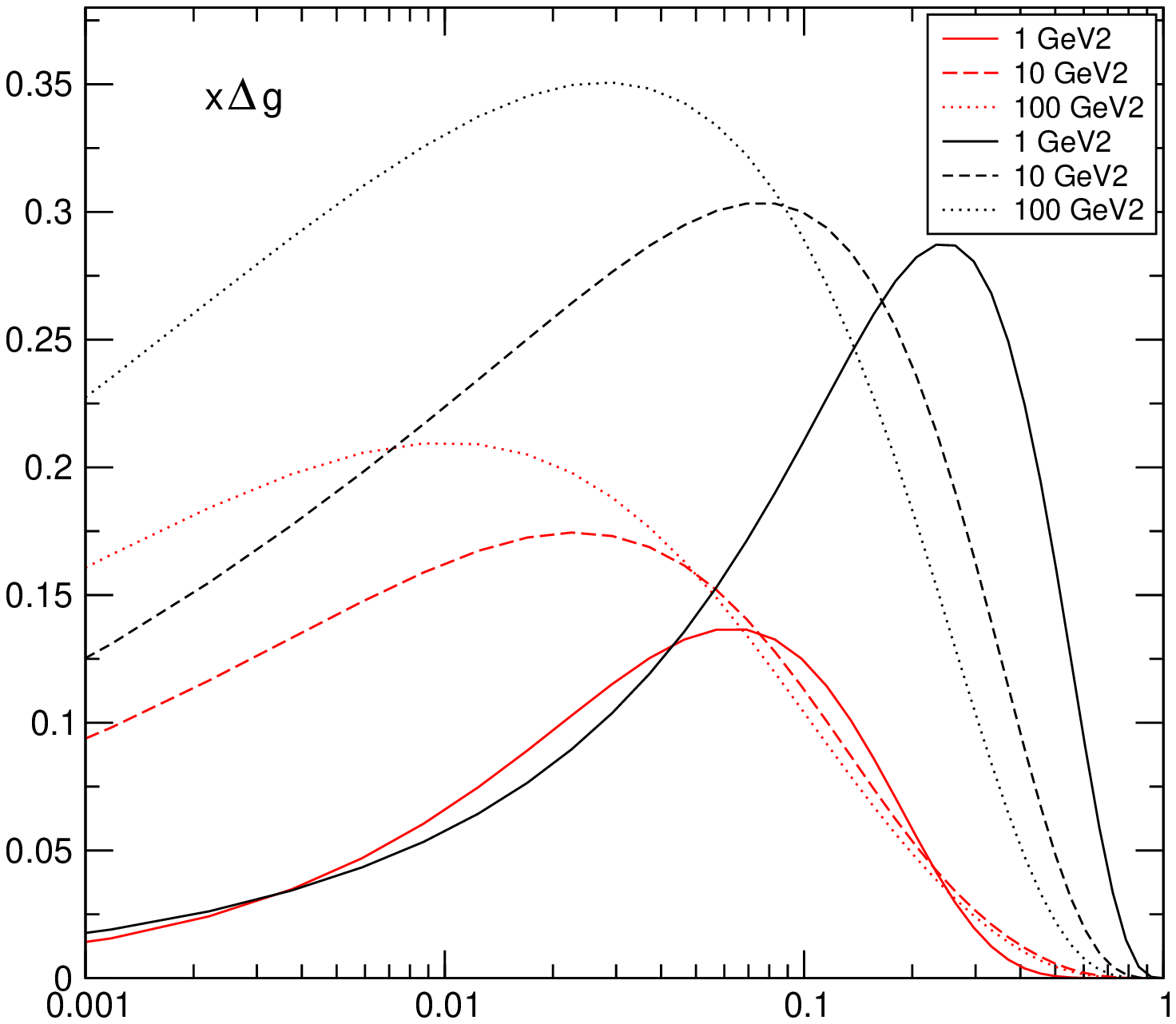}}
\vspace{-1.truecm}
\caption{\footnotesize Best-fit forms (type-A) of $x\Delta g(x)$ in both fits, 
respectively to the set of 176 data points (black curves) and to the full set including 
data of Tab.~2 (red curves), with incresing $Q^2$.}
\label{oldgluon}
\end{figure}

As shown in Figs.~\ref{xqnew} and~\ref{xqtot}, such data slightly improve the 
coverage of the $(x,Q^2)$-plane, and in particular the low-$x$ region down to 
$x\gsim 0.0046$ (COMPASS).\\
The results of a fit of type-A, including all these most recent data is displayed in 
the third column of Tab.~3. Due to the increased number of experimental points and 
their accuracy, an overall reduction of the statistical errors from the fit is observed. 
Aside from a somewhat flatter behavior at small-$x$ of the polarized singlet distribution, 
essentially the same shape of pdfs turns out in both the singlet and nonsinglet sectors, 
which are displayed at the initial scale in Fig.~\ref{sns284}. In particular,  
the new small-$x$ data points confirm the rise of the nonsinglet distribution 
$\Delta q_{\rm NS}$, already determined in the previous 
analyses~\cite{Altarelli:1998nb,Altarelli:1997}. The singlet and nonsinglet first moments 
are fairly unaffected by the new experimental data.

\begin{center}
\begin{tabular}[t]{|l|l|l|} 
\hline
Parameters ($Q^2_0=1\,$GeV$^2$) & 176 data & 284 data \\
\hline\hline
$\eta_{\Sigma}$  & 0.347 $\pm$ 0.025 & 0.344 $\pm$ 0.011 \\
$\alpha_{\Sigma}$   &0.732 $\pm$ 0.504 & 1.421 $\pm$ 0.167\\
$\beta_{\Sigma}$   & 4.307$\pm$0.626 & 3.212 $\pm$ 0.918 \\
$\gamma_{\Sigma}$   & 10.044 $\pm$16.356 & -0.673 $\pm$ 0.805\\
$\textcolor{red}{\eta_g}$  & \textcolor{red}{0.856$\pm$0.666} 
& \textcolor{red}{0.402 $\pm$ 0.062} \\
$\alpha_g$  & -0.772 $\pm$ 0.443 & -0.300 $\pm$ 0.194\\
$\beta_g$  & 4. (fixed) & 10. (fixed) \\
$\gamma_g$  &10.044 $\pm$16.356 & -0.673 $\pm$ 0.805\\
$\eta_3$ & 1.119 $\pm$0.035 & 1.095 $\pm$ 0.024\\
$\alpha_{NS}$ &-0.364  $\pm$ 0.307 & -0.276 $\pm$ 0.165\\
$\beta_{NS}$  & 2.982$\pm$ 0.291 & 3.186 $\pm$ 0.157\\
$\gamma_{NS}$  & 9.749$\pm$ 15.426 & 7.356 $\pm$ 6.271\\
\hline\hline
 $\chi^2/{\rm d.o.f}$ &  0.936 & 0.879\\
\hline\hline
$\textcolor{red}{a_0}(10\,$GeV$^2)$  &\textcolor{red}{ 0.142$\pm$0.023} 
&\textcolor{red}{0.234 $\pm$ 0.010} \\
$\Gamma_1^p(10\,$GeV$^2)$  & 0.113$\pm$0.002 & 0.121 $\pm$ 0.001 \\
\hline
\end{tabular}\\
\end{center}
{\footnotesize Tab.~3. Fits of type-A, respectively to the set of 176 data points 
of Tab.~1 (second column) and to the full set including data of Tab.~2 (third column).}
\vspace{0.5truecm}
 
The most relevant feature here is the sizable reduction of the gluon first moment and 
of its statistical error.  
Actually, such a low value of $\eta_g$ is also correlated to the large fixed value 
of $\beta_g$, that seems, however, the value preferred by the data. Indeed, the 
same fit has been repeated by exploring lower (fixed) values of $\beta_g$, and the minimum 
$\chi^2$ is found, for this set of data, with $\beta_g=10$. The correlation
between the parameter that controls the large-$x$ shape of $\Delta g$ and the first moment  
turns out to be a distinctive feature of the type-A parametrization, and is not observed 
in the case of the type-B fit (see Sec.~6.5.2). A deeper discussion of the impact of each
set of experimental points on $\eta_g$ is given in Sec.~6.5.1.\\
The resulting polarized gluon density (type-A) is 
displayed in Fig.~\ref{oldgluon} with increasing scale $Q^2$ (red lines), and compared 
with the analogous result of the reference fit to 176 data points (black lines).\\
Finally, the value of the singlet axial charge $a_0(Q^2)$ is considerably 
increased, and is given in Tab.~3 at $Q^2=10\,$GeV$^2$ for both fits. Also shown in 
Tab.~3, as an example, is the first moment of the polarized structure function $g_1$ for 
a proton target at the same scale and computed over the full $x$-range. Complete
results for $g_1$ are summarized in Sec.~6.6.1.

\section{Low-energy data}

A large amount of low-energy and very precise data points from the CLAS 
collaboration~\cite{Fatemi:2003yh,Dharmawardane:2006zd}, 
both on proton and deuteron targets, has been left aside in the previous analysis
in order to disentangle the effects of high and low energy data on polarized pdfs. 
Indeed, CLAS data are restricted to a limited region of the final-state 
hadronic invariant mass, roughly 1~GeV$\lsim W\lsim 3\,$GeV, where non-perturbative 
effects could be relevant.
This kinematical region involves large Bjorken-$x$ at moderate values of the 
squared momentum transfer $Q^2$, and is 
characterized by the presence of nucleon resonances which contribute to 
higher-twist effects in the structure functions.\\
In order to perform a consistent perturbative analysis of world data, a 
quantitative criterion for selecting experimental points from CLAS has been studied.
To this end, a lower bound on the invariant mass $W$ has been preliminarly  
estimated on the basis of the relevance of higher-twist contributions to the
moments of the polarized structure function $g_1$.\\
As a next step, the precise location of the kinematic cut to be imposed on the 
whole set of data has been determined by explicitly evaluating residual higher-twist
effects.

\subsection{Selection of CLAS data}

In order to fix a lower cut on the $(x,Q^2)$ plane to  
the full set of CLAS data, a recent analysis~\cite{Osipenko:2005nx} of 
the higher moments of the proton structure function $g_1^p$ has been 
exploited. All the available data, both in the DIS and resonance 
region~\cite{Fatemi:2003yh}, down to very low $Q^2$, are 
used in~\cite{Osipenko:2005nx} to estimate the higher-twist contributions (twist-4 
and 6) to the $n$-th moments of $g_1^p$
\be
M_n(Q^2)=\delta\eta_n(Q^2)+HT_n(Q^2)\,.
\lb{Mn}
\ee
with $n=3,5,7$.  
Here the leading and higher-twist terms, $\delta\eta_n(Q^2)$ and $HT_n(Q^2)$
respectively (see Eqs.~(32) and (45) in Ref.~\cite{Osipenko:2005nx}), have been
determined by fitting a number of parameters to the data. 
As expected, the total higher-twist term turned out~\cite{Osipenko:2005nx} to be sizable 
mainly for $Q^2\sim$ few GeV$^2$ and is still non-negligible even at 
$Q^2\simeq 10\,$GeV$^2$ for the higher moments. Moreover, it has
been pointed out in~\cite{Osipenko:2005nx} that 
the total higher-twist contribution is significantly larger in the polarized case 
than in the unpolarized one (see also~\cite{Simula:2001iy}).\\
On the basis of the above analysis, the ratio $HT_n(Q^2)/\delta\eta_n(Q^2)$ can be 
easily evaluated for each $n$ on a wide range of $Q^2$.
This fact has been used here to select a lower bound on 
the momentum transfer, $Q^2_0$, for each $n$, by 
requiring $HT_n(Q^2)/\delta\eta_n(Q^2)\lsim10\%$ for $Q^2 \gsim Q^2_0$.\\ 
Then, 
in order to define a proper cut to CLAS data in the $(x,Q^2)$-plane, to each $n$ a  
corresponding value of $x$ must be also assigned. To this end, the integrand 
$x^{n-1}\,g_1^p(x,Q_0^2)$ of the $n$-th moment of the 
proton structure function $g_1^p$ has been evaluated 
at the quoted $Q^2_0$, and the position of the peak, $x_0$, has been taken as a 
characteristic value of $x$. 

\begin{figure}[!h]
\vspace{0.3truecm}
\centerline{\includegraphics[width=.6\linewidth,angle=-90]{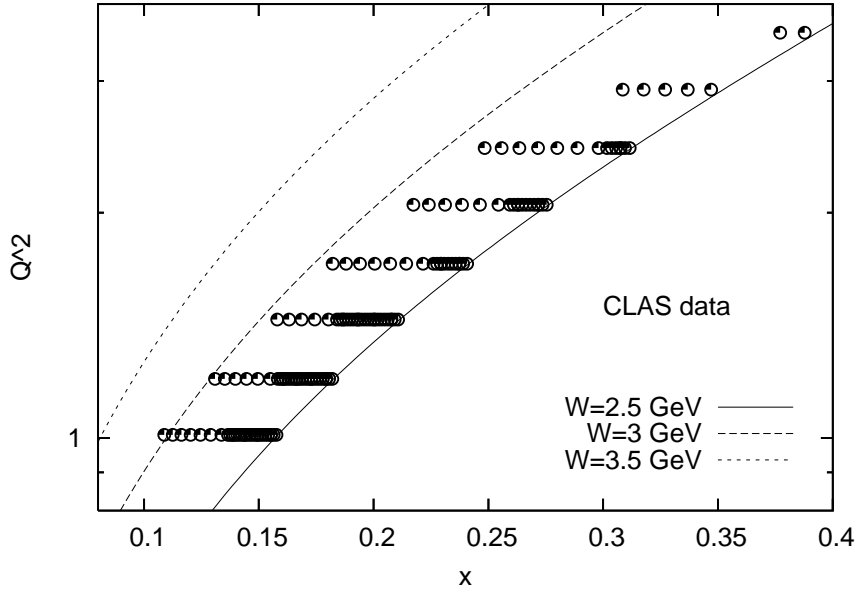}}
\vspace{0.3truecm}
\caption{\footnotesize Set of CLAS data cut at $W\geq 2.5\,$GeV used in the present
analysis.}
\label{clas}
\vspace{0.3truecm}
\end{figure}

By combining these values of $x_0$ and $Q^2_0$ 
for each $n=3,5,7$, the corresponding invariant mass $W_0$ has been computed 
according to Eq.~(\ref{wm}).
It has been found $W_0=1.93,\,2.48,\,2.96\,$GeV for the moments $n=3,5,7,$ respectively.\\  
Then, taking an average over the three values of $W_0$, one roughly finds 
$W_0\simeq 2.5\,$GeV as a lower bound on the final-state invariant mass. 
One may thus conclude that data points below this threshold
can not be safely included in the analysis without taking
into account higher-twist corrections, whose contribution amounts to at least $10\%$
of the leading twist term in this region of the $(x,Q^2)$-plane.\\
As a first result, this leads to a preliminary selection on the full set of CLAS data. 
Indeed, only 148 experimental points lying above the cut $W_0=2.5\,$GeV  
are retained in what follows, and their distribution in the $(x,Q^2)$-plane is displayed in
Fig.~\ref{clas}. 

\subsection{Higher-twist effects}

\begin{figure}[!ht]
\centerline{\includegraphics[width=.61\linewidth,angle=-90]{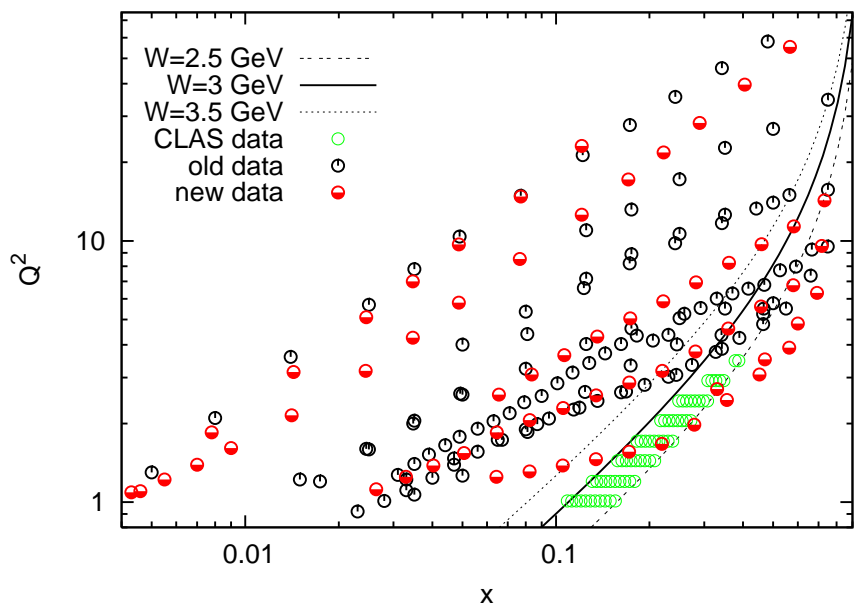}}
\vspace{0.3truecm}
\caption{\footnotesize Distribution in the $(x,Q^2)$-plane of the whole set of data including
selected CLAS data, compared with lines of fixed final-state invariant mass $W=2.5,3$ and $3.5\,$GeV.}
\label{clas2}
\vspace{-0.2truecm}
\end{figure}

As a result of the above analysis, a lower bound should be consistently imposed on the whole 
set of available data. Indeed, all the data below $W_0=2.5\,$GeV must be certainly
discarded, and the precise position of the global cut has been investigated by studying the
related effects of residual higher twist corrections on the results.\\ 
Three possible cuts are compared in Fig.~\ref{clas2} with the whole set of data, including 
the selected CLAS data (cut at $W\geq 2.5\,$GeV). 
As can be seen, a cut at $W\geq 3\,$GeV would entail the exclusion of all the experimental 
points from CLAS (with the exception of only one point on proton target), and of
a number of new HERMES data as well. Higher cuts would mean a 
sizeable loss of information, whereas imposing the lower bound at $W=2.5\,$GeV does not 
significantly reduce the set of data but higher twist effects could still play a relevant role
in this region.

\begin{figure}[!h]
\centerline{\includegraphics[width=.55\linewidth,angle=-90]{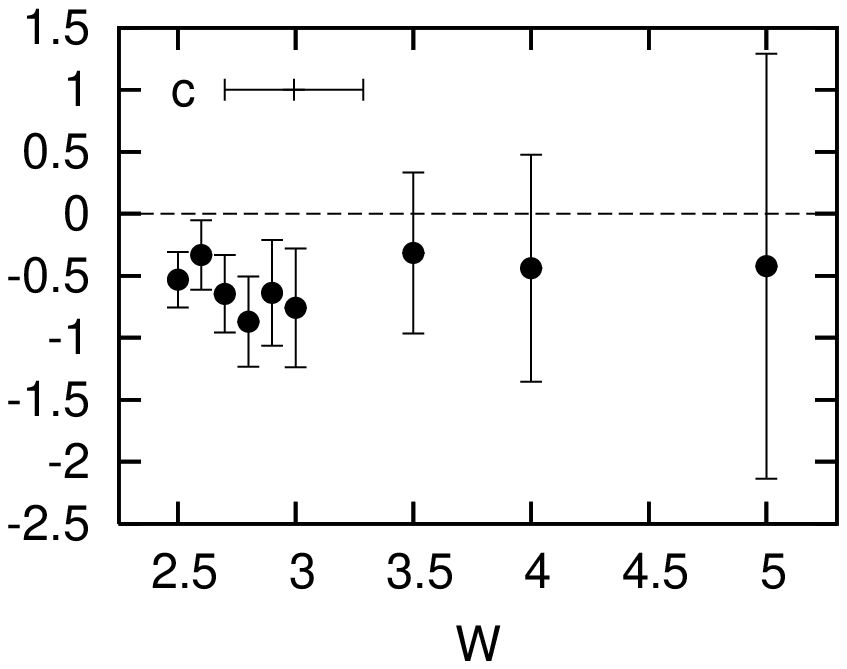}}
\caption{\footnotesize Values with errors of the parameter $c$ fitted to the set
of data gradually cut at higher $W$.}
\label{ht}
\end{figure}

This issue has been quantitatively evaluated by 
introducing a phenomenological term for the structure function 
$g_1$, so as to take into account higher-twist contributions in the Bjorken-$x$ space,
namely
\be
g_1(x,Q^2)\,\to\, g_1(x,Q^2)\left[1+\frac{c}{W^2}\right]\,.
\lb{cht}
\ee
The multiplicative factor in Eq.~(\ref{cht}) depends on one dimensional parameter $c$ 
to be fitted to the data, in order to study the role of lower energy points in the
determination of higher-twist corrections. A broad range of possible cuts $W_0$ has 
been investigated; specifically,  
the same fit of type-A, with one more fitting parameter $c$, has been performed on   
the whole set of data (including the selected subset of CLAS data), progressively 
cut at higher values $W$ of
the invariant mass.\\ 
The values of the HT parameter $c$ with their errors, as determined by these fits, are 
displayed in Fig.~\ref{ht} as a function of the cut $W_0$ imposed. As a result, the higher-twist 
contributions turn out to be gradually less relevant as the full set of data is restricted to 
higher values of $W$. Indeed, as can be seen by Fig.~\ref{ht}, the largeness of the
error bar increases with the value of the cut $W_0$, making the higher-twist parameter
more consistent with zero.\\ 
A glance to Figs.~\ref{clas2} and~\ref{ht} reveals that the choice $W_0=3\,$GeV, where 
the higher-twist parameter $c$ differs from zero by no more than 1.5 standard deviations, 
seems a reasonable one, in that it justifies a purely perturbative analysis while 
reducing the loss of experimental information.

In Tab.~4 are shown, as an example, the results of two fits of type-A with 11 fitted 
parameters and 
different cuts on $W$, i.e., no cut at all (except for CLAS data cut at $W\gsim 2.5\,$GeV)
and $W\gsim 3.5\,$GeV over all the available points (which excludes CLAS data altogether). 

\begin{center}
\begin{tabular}[t]{|l|l|l|} 
\hline
Parameters ($Q^2_0=1\,$GeV$^2$) & no cut & $W\geq 3.5\,$GeV \\
\hline\hline
$\eta_{\Sigma}$  & 0.401 $\pm$ 0.014 & 0.397 $\pm$ 0.021\\
$\alpha_{\Sigma}$   & 1.357 $\pm$ 0.273 & 1.347 $\pm$ 0.304\\
$\beta_{\Sigma}$   & 2.655 $\pm$ 1.233 & 2.435 $\pm$ 1.198\\
$\gamma_{\Sigma}$   & -1.015 $\pm$ 0.448 & -1.191 $\pm$ 0.320\\
$\textcolor{red}{\eta_g}$  & \textcolor{red}{0.444 $\pm$ 0.066} & 
\textcolor{red}{0.494 $\pm$ 0.075}  \\
$\alpha_g$  & -0.388 $\pm$ 0.184 & -0.451 $\pm$ 0.179\\
$\beta_g$  & 8. (fixed) & 6. (fixed) \\
$\gamma_g$  & -1.015 $\pm$ 0.448 & -1.191 $\pm$ 0.320 \\
$\eta_3$ & 1.129 $\pm$ 0.027 & 1.137 $\pm$ 0.041\\
$\alpha_{NS}$ & -0.236 $\pm$ 0.158 & -0.327 $\pm$ 0.218\\
$\beta_{NS}$  & 3.119 $\pm$ 0.163 & 2.903 $\pm$ 0.300\\
$\gamma_{NS}$  & 7.114 $\pm$ 5.630 & 8.979 $\pm$ 10.051\\
\textcolor{red}{$c$} & \textcolor{red}{-0.561 $\pm$ 0.210} &
\textcolor{red}{-0.315 $\pm$ 0.650}\\
\hline\hline
 $\chi^2/{\rm d.o.f}$ &  0.937 & 1.056\\
\hline
\end{tabular}\\
\end{center}

{\footnotesize Tab.~4. Fits of type-A, respectively to the full set of data with no
cut except for CLAS ($W\geq 2.5\,$GeV) (second column), and with a global cut at 
$W\geq 3.5\,$GeV (third column).}
\vspace{0.5truecm}

Comparing the second column of Tab.~4 with the third one of Tab.~3 it can be seen 
that CLAS data do
not sizably affect the shape of quark and gluon densities, whereas the main effects of
the lower energy points, all cut in the third column of Tab.~4, is essentially the 
improvement in the determination of higher-twist contributions to $g_1$. Similar
conclusions were also reached in Ref.~\cite{Leader:2006xc} where the impact of a much
extended subset of CLAS data~\cite{Dharmawardane:2006zd} (i.e. $W\gsim 2\,$GeV) on 
polarized pdfs has been studied.

\section{Gluon polarization from $g_1$ data}

\begin{figure}
\centerline{\includegraphics[width=.65\linewidth,angle=-90]{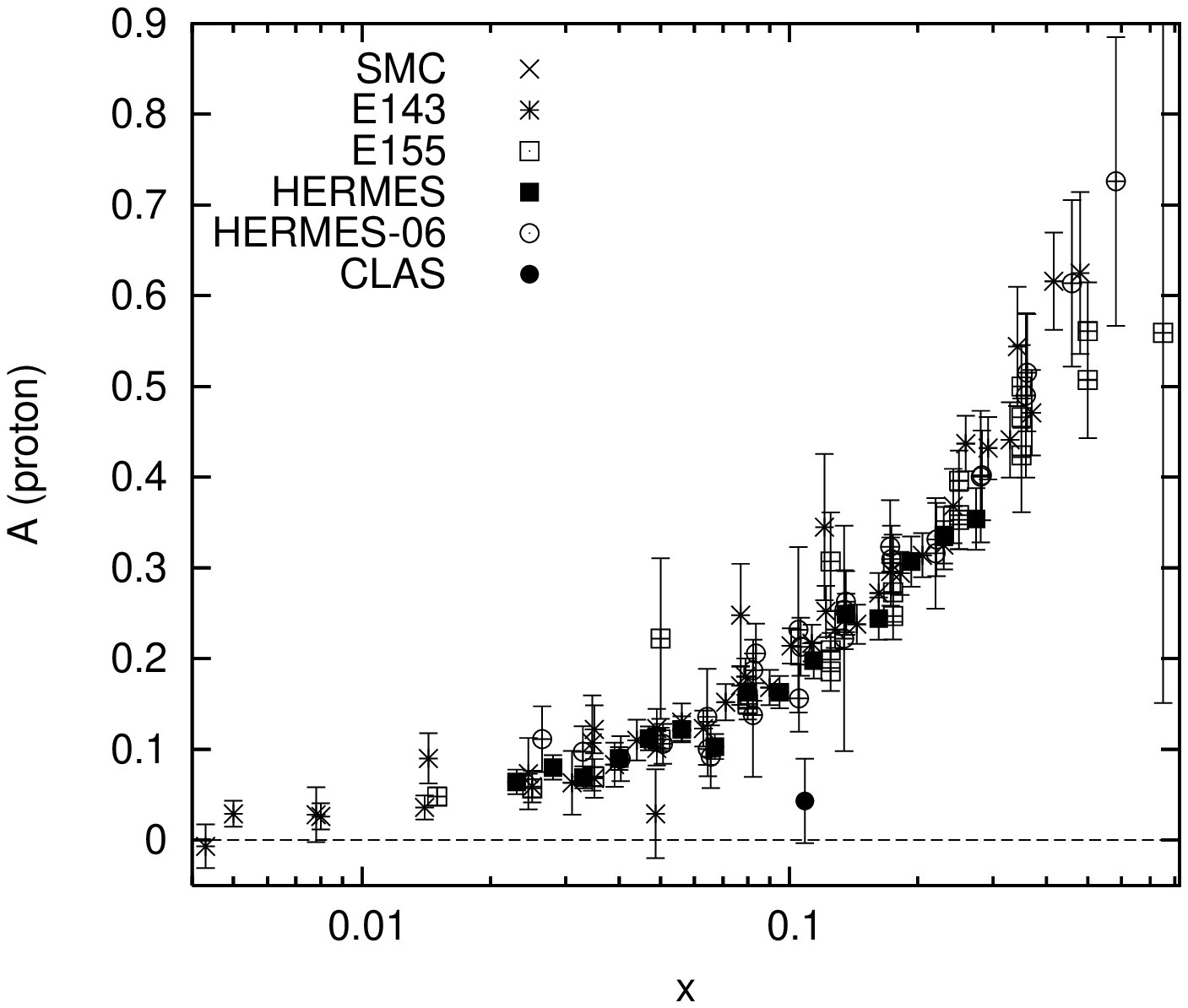}}
\caption{\footnotesize Asymmetry data on proton target summarized in Tabs.~1 and~2,
cut at $W\geq 3\,$GeV.}
\label{a1P}
\centerline{\includegraphics[width=.65\linewidth,angle=-90]{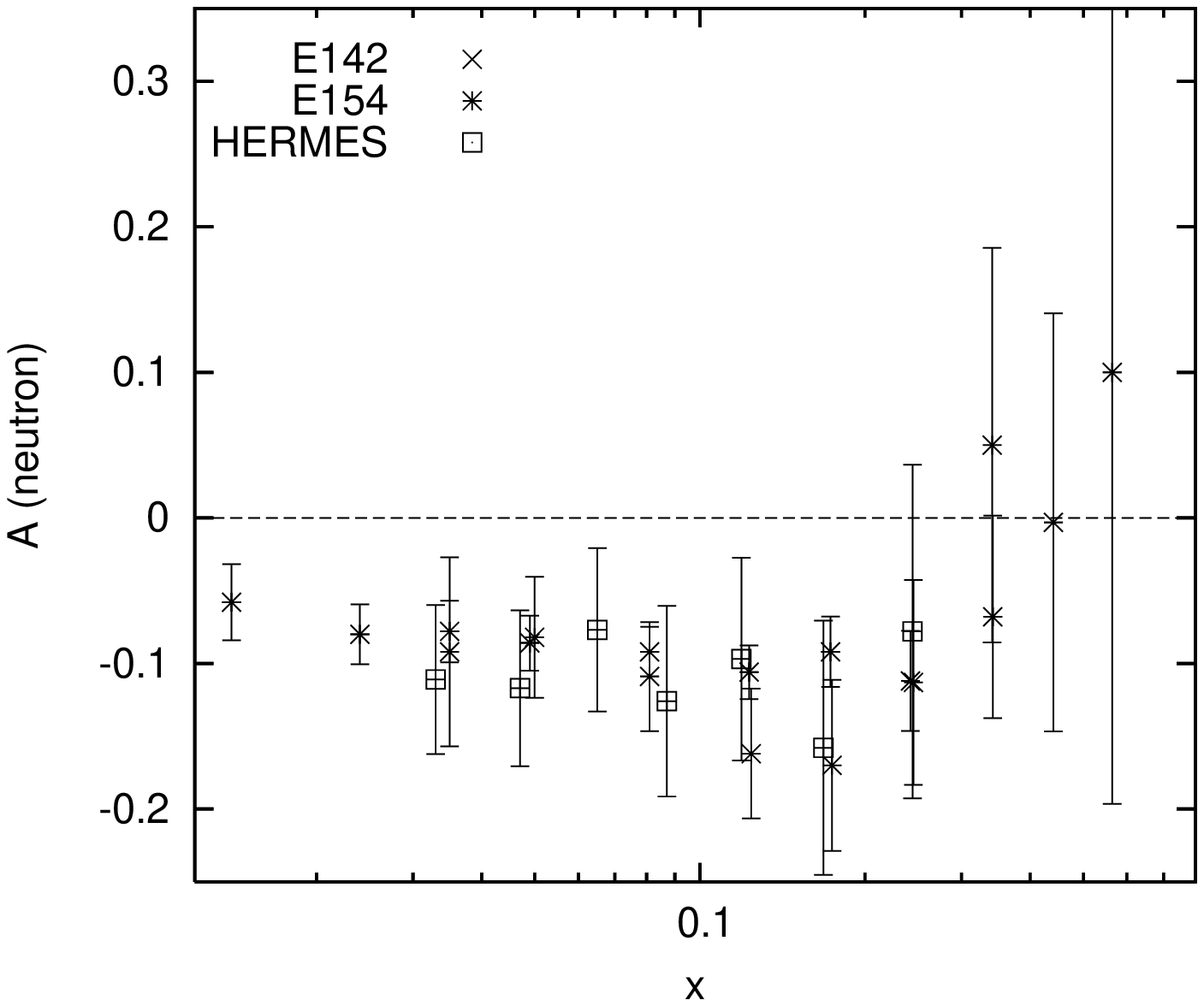}}
\caption{\footnotesize Asymmetry data on neutron target summarized in Tabs.~1 and~2,
cut at $W\geq 3\,$GeV.}
\label{a1N}
\end{figure}

\begin{figure}[!ht]
\centerline{\includegraphics[width=.65\linewidth,angle=-90]{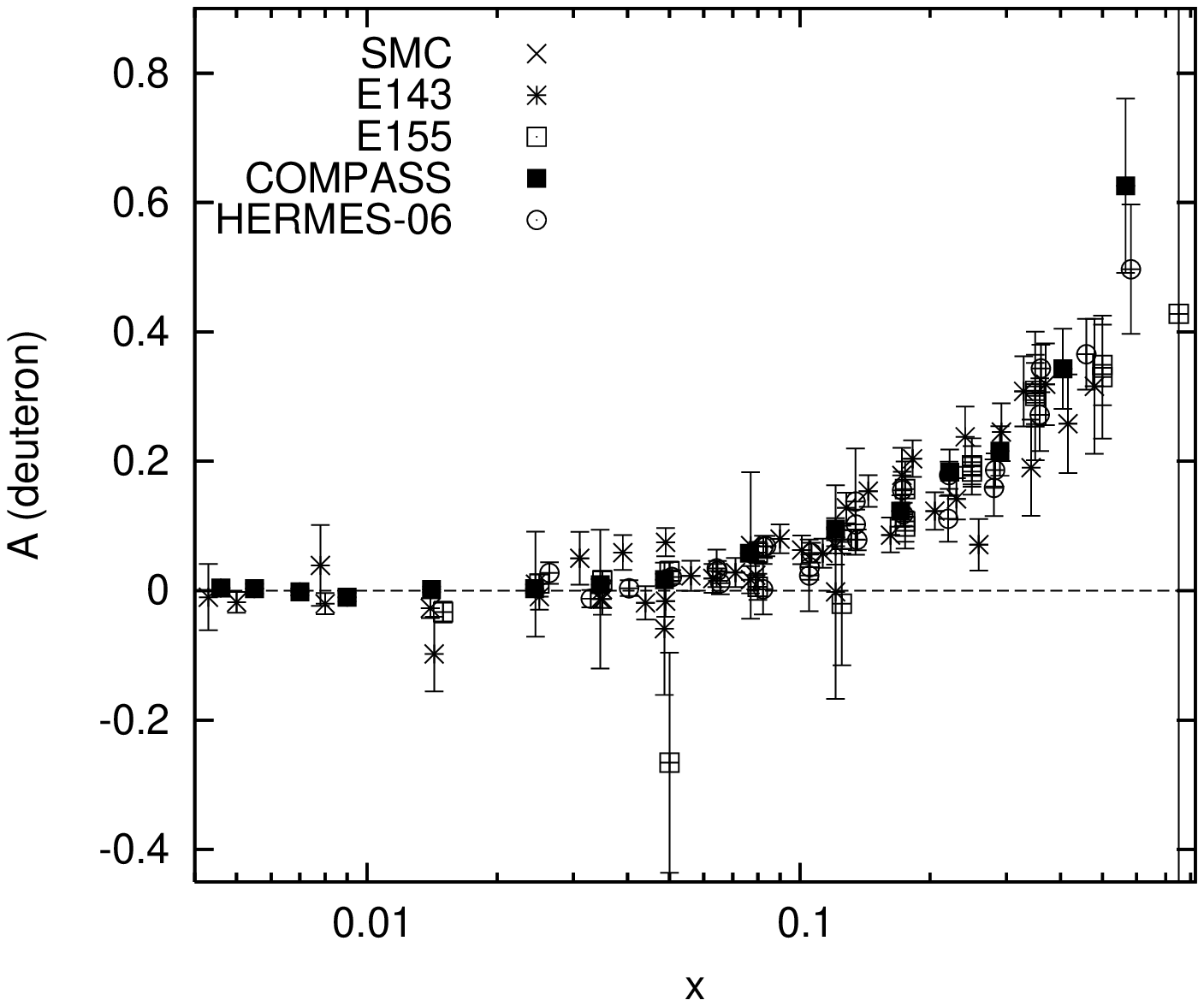}}
\caption{\footnotesize Asymmetry data on deuteron target summarized in Tabs.~1 and~2,
cut at $W\geq 3\,$GeV.}
\label{a1D}
\end{figure}

As discussed in the previous section, in order to perform a consistent 
perturbative treatment of world data,
avoiding systematic errors induced by higher-twist effects, 
the final set of experimental points retained is the one 
selected by the cut on the invariant mass at $W\geq 3\,$GeV, which  
amounts to 238 data points and covers the region above the solid curve 
in Fig.~\ref{clas2}.
Indeed, in this kinematical region the higher-twist contributions can be
safely neglected.\\ 
The corresponding experimental asymmetries on proton, neutron and deu\-te\-ron targets 
included in the analysis are given as a function of $x$ in Figs.~\ref{a1P}, \ref{a1N} 
and \ref{a1D} respectively.

\vspace{-0.25truecm}

As already noted, all the CLAS data (except for one point) have been excluded, together
with the three points from JLAB and also a number of lower energy HERMES data. Such a 
cut has obviously no effect on the higher energy data from COMPASS and SMC.\\
The information that can be extracted specifically on the polarized gluon density
by a global fit of the present inclusive DIS data are discussed in the following
sections.

\subsection{Best-fit results and impact of the data on the gluon first moment}

The results of a type-A fit to this set of experimental points is shown in Tab.~5. 
Here, higher-twist corrections are switched off ($c=0$ fixed), whereas target mass
corrections are included as a rule. As a cross check, it has been verified that no 
sizable differences can be traced 
with respect to the results of a similar fit performed on the same set of data where 
the higher-twist parameter $c$ has been fitted ($c=-0.76 \pm 0.48$).\\ 
Comparing the results of Tab.~5 with those of the third column of Tab.~3 (including 
the whole set of data except for CLAS) makes clear that 
the main effect of the cut on the quark and gluon polarizations amounts to an overall 
increase of the statistical errors due to the loss of a number of experimental points, 
mainly from HERMES. On the other hand, the systematic error due to higher-twist effects 
is reduced by excluding low-energy data.\\
The $x$-shapes of the quark singlet and nonsinglet distributions are 
very similar to those displayed in Figs.~\ref{sns284}, and slightly differ only
for the normalization. 
Specifically, the nonsinglet quark distribution rises at small-$x$, whereas the
Adler-Bardeen singlet quark turns out to be flat at the initial scale. The best-fit
form of quark distributions is displayed in Fig.~\ref{snB} (black lines).\\ 
The first moment of the polarized gluon density (and its statistical error) at the 
initial scale is slightly 
increased, but is neverthless much smaller than the previous determinations (see 
Tab.~3, second column), in agreement with many current analyses (see Sec.~6.5).\\
The parameter $\gamma_g$ in Tab.~5 has been 
disentangled from $\gamma_{\Sigma}$, but clearly the present data can not give precise 
constrains on its value. Again, the behavior 
of $\Delta g$ as $x\to 1$ can not be determined by inclusive DIS measurements 
and the parameter $\beta_g$ has to be fixed. 
The range $4\leq\beta_g\leq 10$ has been explored and the minimum $\chi^2$ 
is found for $\beta_g=6$.
  
\vspace{0.2truecm}
\begin{center}
\begin{tabular}[b]{|l|l|} 
\hline
Parameters ($Q^2_0=1\,$GeV$^2$) & type-A ($W\geq 3\,$GeV, 238 data)  \\
\hline\hline
$\eta_{\Sigma}$  & 0.391 $\pm$ 0.021  \\
$\alpha_{\Sigma}$   &1.154 $\pm$ 0.334\\
$\beta_{\Sigma}$   & 1.918$\pm$0.972  \\
$\gamma_{\Sigma}$   & -1.232  $\pm$  0.192\\
$\textcolor{red}{\eta_g}$  & \textcolor{red}{0.555 $\pm$0.146}  \\
$\alpha_g$  & -0.484  $\pm$0.293 \\
$\beta_g$  & 6. (fixed)  \\
$\gamma_g$  &3.829  $\pm$10.443\\
$\eta_3$ & 1.127  $\pm$ 0.034 \\
$\alpha_{NS}$ &  -0.321 $\pm$ 0.271 \\
$\beta_{NS}$  &  2.970 $\pm$ 0.336 \\
$\gamma_{NS}$  &  9.475 $\pm$ 13.376 \\
\hline\hline
 $\chi^2/{\rm d.o.f}$ &  0.932  \\
\hline\hline
$\textcolor{red}{a_0}(10\,$GeV$^2)$  & \textcolor{red}{ 0.246 $\pm$0.025 } \\
$\Gamma_1^p(10\,$GeV$^2)$  &0.125  $\pm$  0.003 \\
\hline
\end{tabular}\\
\end{center}

{\footnotesize Tab.~5. Best-fit result of type-A on the selected data set, and 
$c=0$ fixed.}
\vspace{0.2truecm}

\vspace{0.2truecm}
\begin{center}
\begin{tabular}[b]{|l|l|l|l|}
\hline
data set  & $\eta_g$ ($Q_0^2=1\,$GeV$^2$) & $\beta_g$ (fixed) & $\chi^2/{\rm d.o.f.}$ \\
\hline\hline
176 set & 0.86 $\pm$ 0.67 & 4. & 0.94\\
 + SMC2 & 0.69 $\pm$ 0.52 & 6. & 0.96 \\
 + COMPASS & 0.53 $\pm$ 0.14 & 8. & 0.93 \\
 + HERMES-06 & 0.43 $\pm$ 0.07 & 8. & 0.88\\
 + JLAB & 0.40 $\pm$ 0.06 & 10. & 0.88\\
 238 set ($W\geq 3\,$GeV) & 0.56 $\pm$ 0.15 & 6. & 0.93\\
\hline
\end{tabular}\\
\end{center}

{\footnotesize Tab.~6. Impact of each set of experimental data on the value of the
first moment $\eta_g$ of the gluon distribution. Also shown are the corresponding 
value of $\beta_g$ and the related $\chi^2$.}
\vspace{0.3truecm}

The $x$-shape of the best-fit result of type-A (black 
curves) for the polarized gluon distribution is displayed in Fig.~\ref{AB} with 
increasing momentum transfer.\\
Finally, the impact of each set of data on the first moment of 
the gluon polarization is summarized in Tab.~6, 
together with the corresponding best value of $\beta_g$ and the related $\chi^2$. 
As can be seen, all the measurements contribute to decreasing $\eta_g$, the much 
relevant effect being that from the high-energy 
data by COMPASS, lying well above the cut imposed on the $(x,Q^2)$-plane.
 
\subsection{Dependence of $\Delta g$ on input densities}

The assumed functional form of input densities sizably affects the determination of 
relevant quantities, such as the $x$-shape of the gluon polarization. 
A way of evaluating this effect is to perform the whole analysis by changing the 
initial parametrization of pdfs. This has been done here by using the functional 
form given by Eq.~(\ref{fitB}) (type-B) for an alternative fit on the selected 
set of data cut at 
$W\geq 3\,$GeV used in the type-A fit (Tab.~5).  

\vspace{0.3truecm}
\begin{center}
\begin{tabular}[b]{|l|l|} 
\hline
Parameters ($Q^2_0=1\,$GeV$^2$)& type-B ($W\geq \,$GeV, 238 data)  \\
\hline\hline
$\eta_{\Sigma}$  & 0.417 $\pm$ 0.023  \\
$\alpha_{\Sigma}$   &2.496 $\pm$ 0.425\\
$\beta_{\Sigma}$   & 3.512$\pm$0.467  \\
$\textcolor{red}{\eta_g}$  & \textcolor{red}{0.801$\pm$0.190}  \\
$\alpha_g$  & 1.050 $\pm$ 0.486\\
$\beta_g$  & 3.113$\pm$0.868 \\
$\gamma_g$  & -1.447$\pm$0.647\\
$\eta_3$ & 1.219 $\pm$0.038 \\
$\alpha_{NS}$ & 1.733 $\pm$ 0.121 \\
$\beta_{NS}$  & 4.655$\pm$ 0.138 \\
$\gamma_{NS}$  & -0.269$\pm$ 0.060 \\
\hline\hline
 $\chi^2/{\rm d.o.f}$ &  0.926 \\
\hline\hline
$\textcolor{red}{a_0}(10\,$GeV$^2)$  & \textcolor{red}{0.218$\pm$0.027}  \\
$\Gamma_1^p(10\,$GeV$^2)$  & 0.129$\pm$0.004  \\
\hline
\end{tabular}\\
\end{center}

{\footnotesize Tab.~7. Best-fit result of type-B on the selected data set, and
$c=0$ fixed.}
\vspace{0.4truecm}

\begin{figure}
\vspace{-0.9truecm}
\centerline{\includegraphics[width=.8\linewidth,angle=-90.]{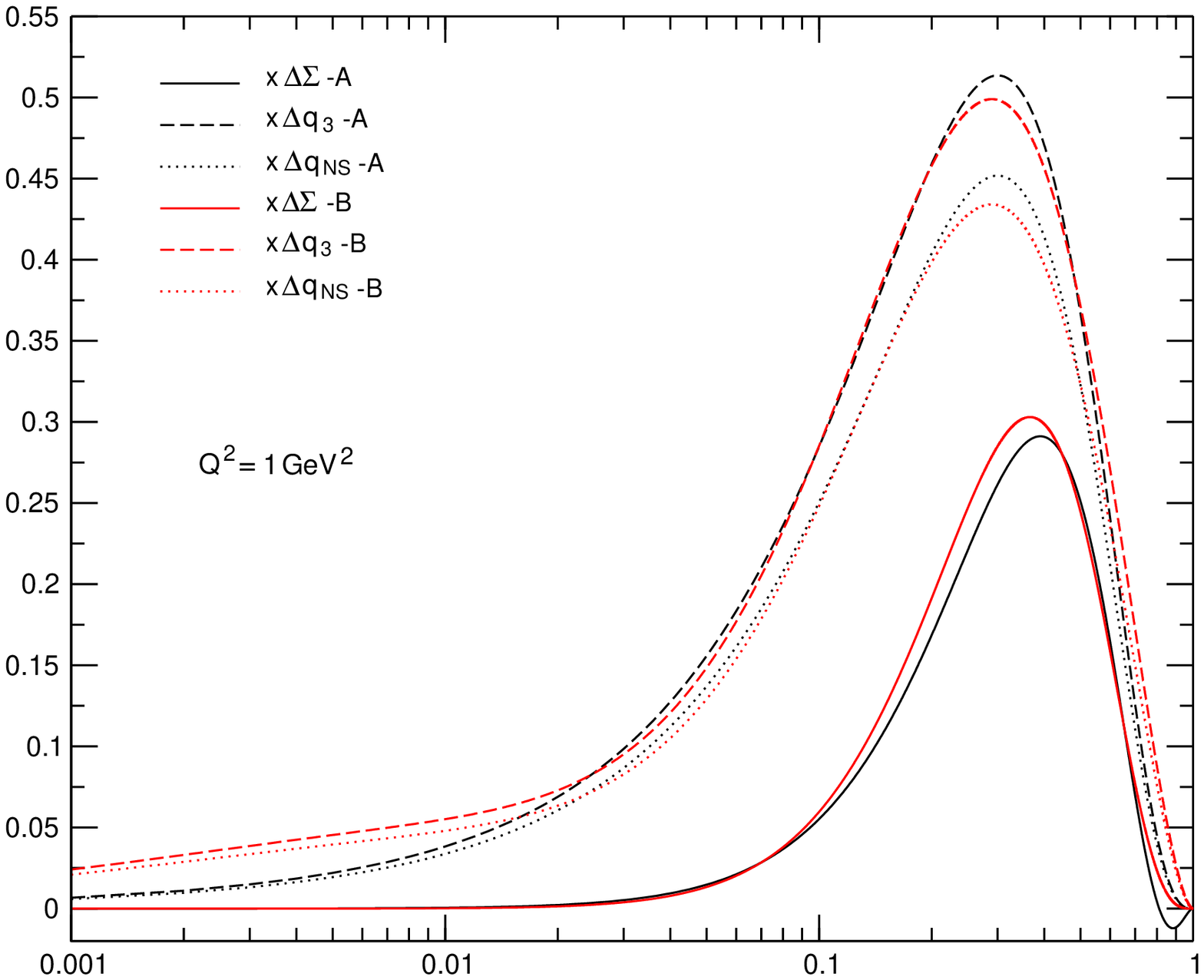}}
\vspace{-0.85truecm}
\caption{\footnotesize Best-fit results of type-A and B (black and red lines respectively) 
for the polarized quark singlet 
and nonsinglet distributions (proton target) at $Q^2=1\,$GeV$^2$, over the selected 
set of available data ($W\geq 3\,$GeV).} 
\label{snB}
\vspace{-0.2truecm}
\centerline{\includegraphics[width=.8\linewidth,angle=-90.]{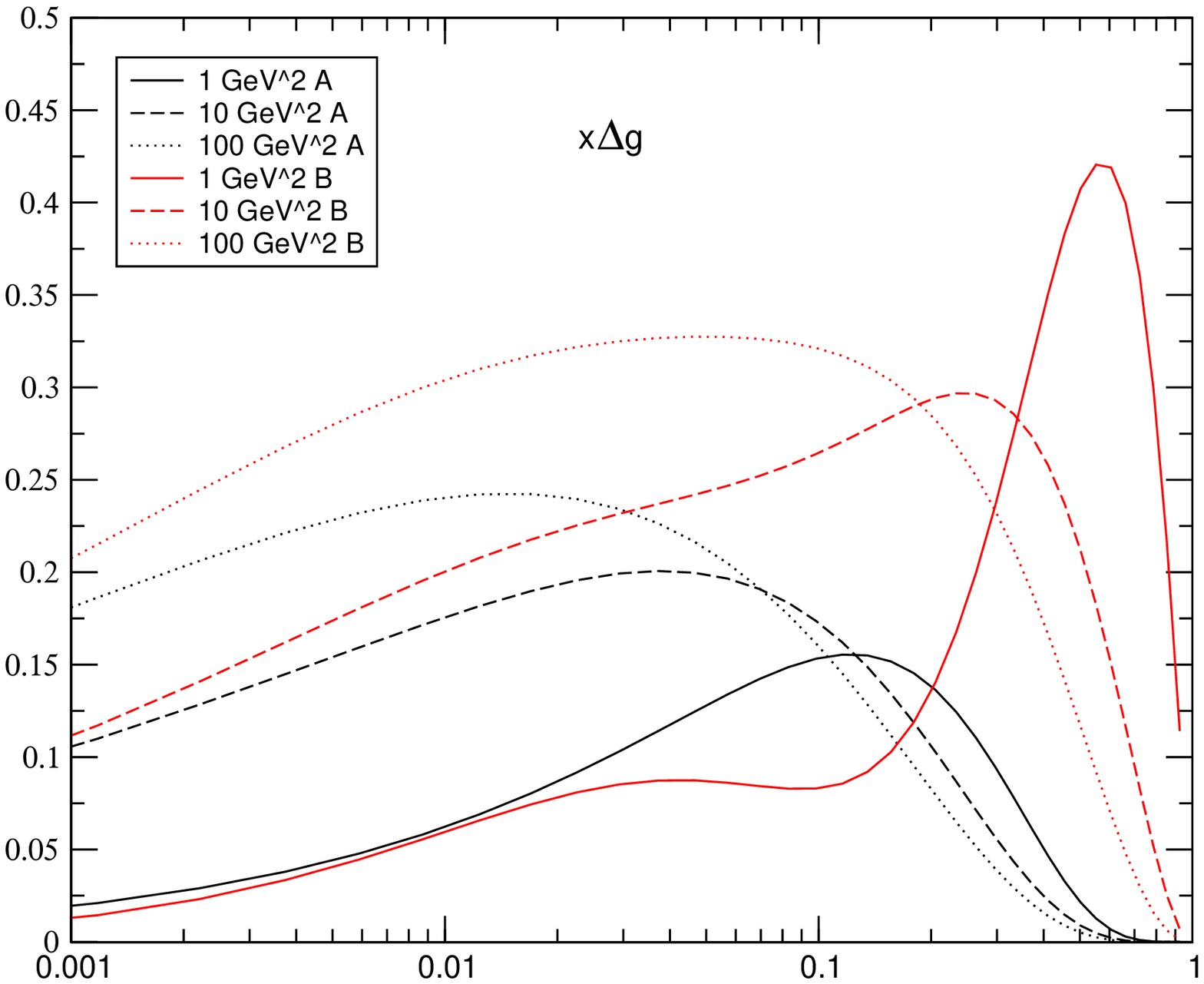}}
\vspace{-1.truecm}
\caption{\footnotesize Best-fit results for the polarized gluon distribution with 
increasing $Q^2$, over the selected set of available data ($W\geq 3\,$GeV). The 
black curves denote the fit of type-A, and the red ones the fit of type-B.}
\label{AB}
\end{figure}

The analysis is performed in the 
Adler-Bardeen scheme, and the results for the pdf parameters,  
the singlet axial charge $a_0(Q^2)$ and the first moment $\Gamma_1^p(Q^2)$ of 
the proton structure function at $Q^2=10\,$GeV$^2$ are given in Tab.~7.  
These results may be compared with those of the related fit of type-A (Tab.~5).\\
In the quark sector first moments turns out to be compatible within the errors
with those from the type-A fit. The resulting $x$-shapes of singlet and nonsinglet  
distributions are 
also very similar to the best-fit forms  
of type-A, and are displayed in Fig.~\ref{snB} at the initial scale (red lines).\\
A stronger sensitivity to the initial functional form is observed in the gluon 
sector. The central value of the gluon first moment turns out to be larger
than that of type-A, even though 
still compatible within the errors.\\ 
On the other hand, the $x$-shape of $\Delta g$ is considerably changed with respect 
to type-A, and both are displayed in Fig.~\ref{AB} with increasing momentum transfer
$Q^2$. 
Note that, at variance with the fits of type-A, the behavior
of $\Delta g$ as $x\to 1$ has not been fixed here a priori.\\
A quantitative estimate of the systematic uncertainty produced by the choice
of input densities may be performed by analyzing distinctive features of the $x$-shape 
in both type-A and B gluon distributions, such as the position of the peak and the 
corresponding value of the function. This study has been performed in Sec~7,   
in connection with pseudo-data analyses for charm photoproduction at the COMPASS 
experiment that aim at measuring $\Delta g$ directly (see Sec.~7).

Finally, being the quality of the fit fairly the same in both cases, no one of the 
assumed functional forms is neatly preferred by the data, that is both choices of
input densities give equally good fits of inclusive world data, although the gluon
$x$-shapes are significantly different. 
This result makes clear the large bias produced by the initial parametrization  
on the spin-dependent gluon distribution. This fact, sometimes hidden
within most of the available phenomenological studies, mainly performed by
imposing standard input densities (and constrained by the positivity
condition), has been here clearly stated and taken into
account in the estimation of the theoretical errors (see
Sec.~6.6.2 below, and also Sec.~7.3).
One may thus conclude 
that, even though the better coverage of the $(x,Q^2)$-plane, mainly due to the 
precise COMPASS data, allows a more reliable determination of the first moment 
of the gluon polarization by scaling violations, its $x$-shape is still largely 
unconstrained by the present inclusive DIS data.

\section{Phenomenological implications}

The main results of the above analysis as far as phisically relevant quantities are
concerned are summarized in the following sections. Specifically, the best-fit forms 
of structure functions
are presented, and the final estimate for the singlet axial charge and the polarization 
of the gluon and the quark flavor singlet combination in the nucleon are discussed.

\subsection{Best-fit results for $g_1$}

\begin{figure}
\centerline{\includegraphics[width=.8\linewidth]{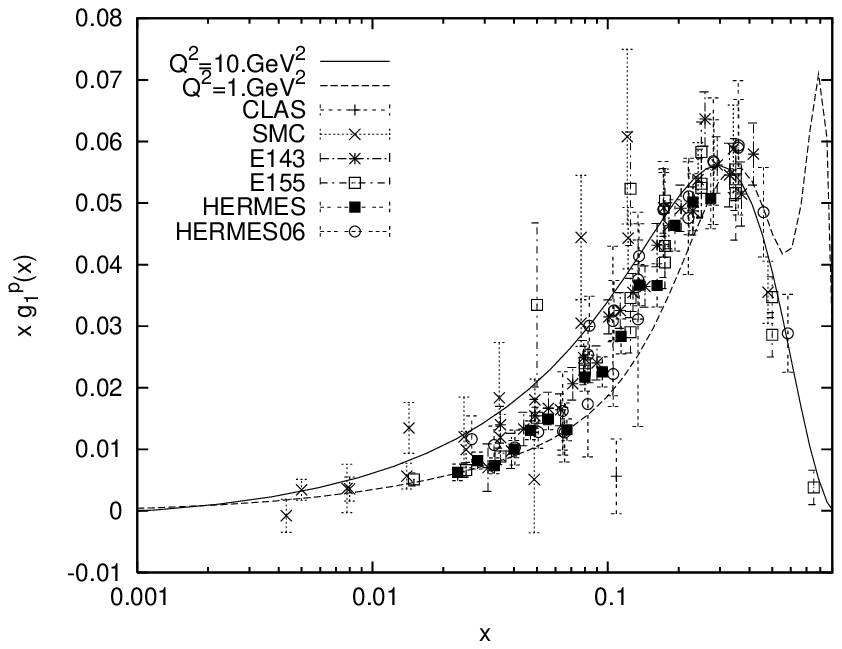}}
\caption{\footnotesize Best-fit result (type-A) for the proton structure function 
compared
with the data points used in the analyis ($W\geq 3\,$GeV).}
\label{xg1p}
\vspace{0.5truecm}
\centerline{\includegraphics[width=.8\linewidth]{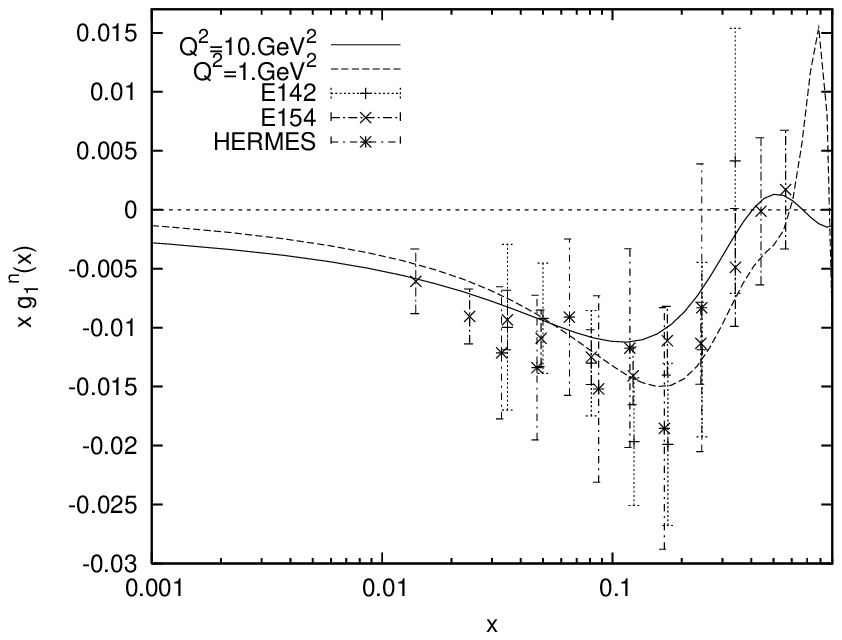}}
\caption{\footnotesize Best-fit result (type-A) for the neutron structure function 
compared
with the data points used in the analyis ($W\geq 3\,$GeV).}
\label{xg1n}
\end{figure}

\begin{figure}
\centerline{\includegraphics[width=.8\linewidth]{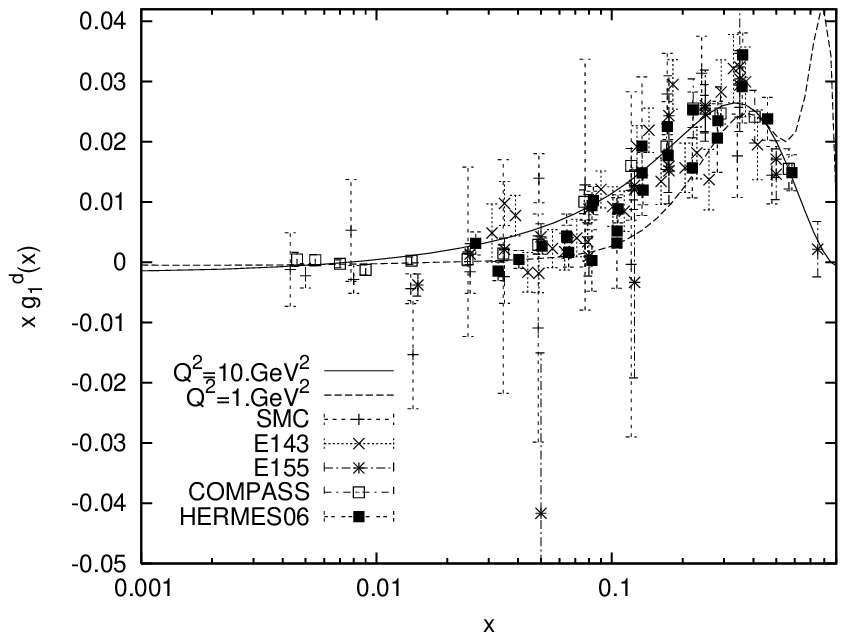}}
\caption{\footnotesize Best-fit result (type-A) for the deuteron structure function 
compared
with the data points used in the analyis ($W\geq 3\,$GeV).}
\label{xg1d}
\vspace{0.5truecm}
\centerline{\includegraphics[width=.8\linewidth]{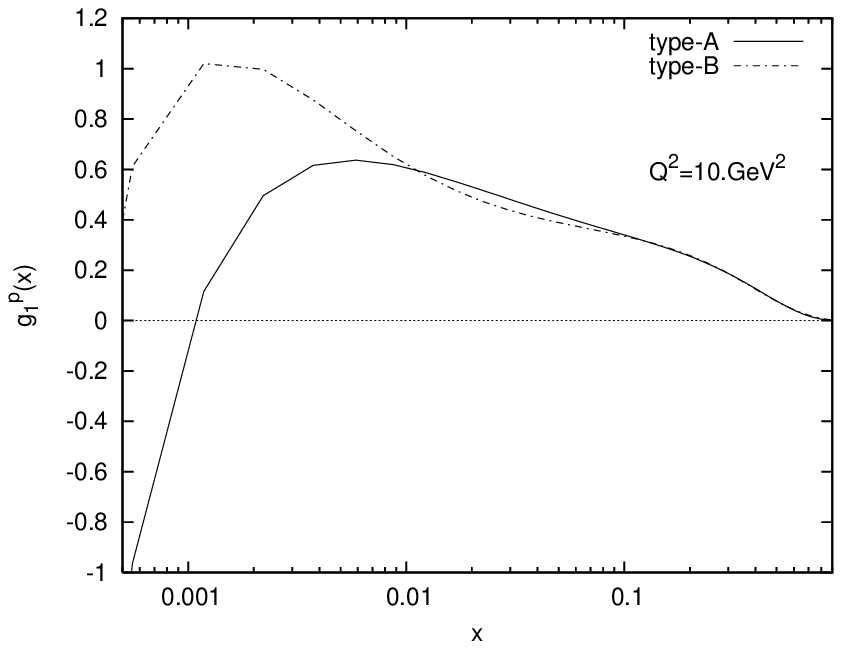}}
\caption{\footnotesize Best-fit results of type-A (solid) and type-B (dot-dashed) 
for the proton structure function.}
\label{g1p10}
\end{figure}

The best-fit structure functions $g_1$ (type-A) for proton, neutron and deuteron 
targets,
are displayed in Figs.~\ref{xg1p},~\ref{xg1n} and~\ref{xg1d} respectively, in the range 
of momentum transfer covered by the data (i.e. $Q^2=1,10\,$GeV$^2$), and compared with 
the experimental points with their errors used in the analysis.\\ 
Note that the sharp rise at large-$x$ 
observed in all the three curves at the initial scale ($Q^2=1\,$GeV$^2$) is due to 
the inclusion of target mass corrections, and it disappears if $m^2=0$ is taken 
exactly for the 
nucleon mass. Such an effect, however, is not physical since the approximation
adopted for the calculation of target mass corrections becomes 
unreliable at very large $x$ (see~\cite{Piccione:1997zh}). Indeed, because of 
the $m^2/Q^2$ 
suppression factor, this unphysical peak is washed out at higher $Q^2$, which is 
actually the experimentally interesting region for large-$x$.  
\vspace{0.4truecm}

\begin{figure}
\centerline{\includegraphics[width=.8\linewidth]{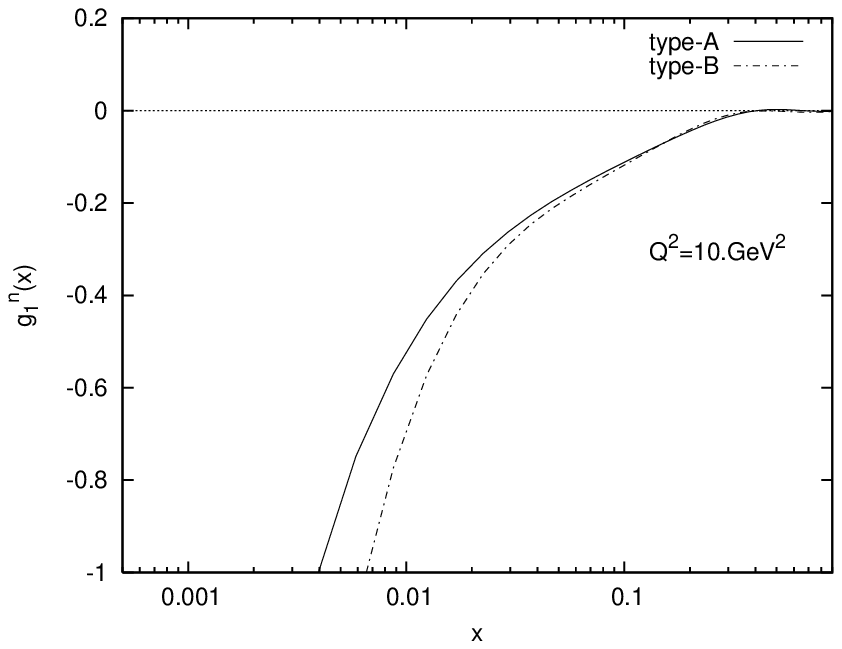}}
\caption{\footnotesize Best-fit results of type-A (solid) and type-B (dot-dashed) 
for the neutron structure function.}
\label{g1n10}
\vspace{0.5truecm}
\centerline{\includegraphics[width=.8\linewidth]{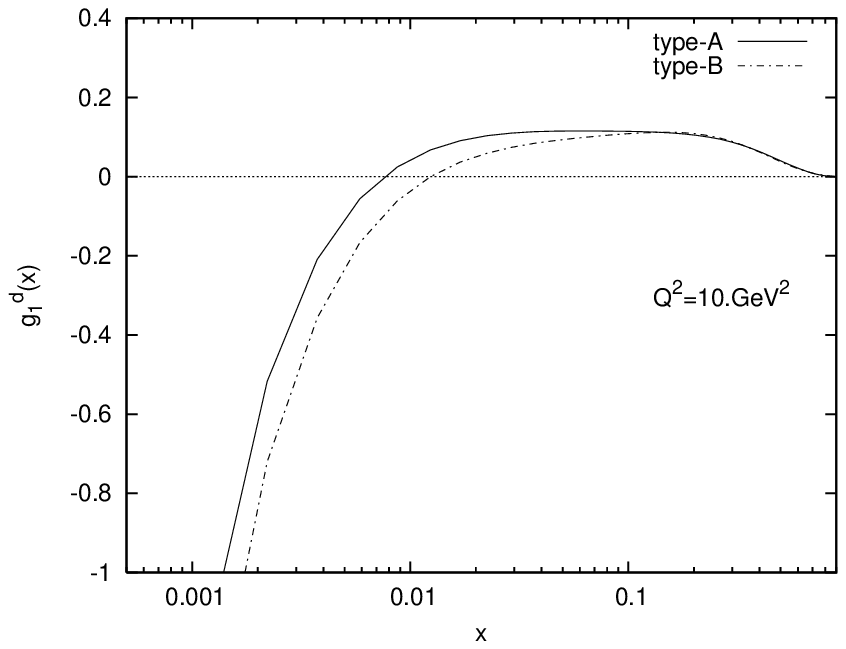}}
\caption{\footnotesize Best-fit results of type-A (solid) and type-B (dot-dashed) 
for the deuteron structure function.}
\label{g1d10}
\end{figure}
\vspace{-0.25truecm}

Figs.~\ref{g1p10},~\ref{g1n10} and~\ref{g1d10} compare the best-fit forms of $g_1$ 
for each target, corresponding to both fits, type-A and B respectively, 
at $Q^2=10\,$GeV$^2$. Sizable differences can be traced in the small-$x$ region in the two
fits. This fact is due to the way the functional forms of pdfs extrapolate the behavior
observed in the last measured points, emphasizing the impact of the initial parametrization
on the small-$x$ behavior of the structure functions. In all cases $g_1$ is driven negative 
at small-$x$ by the rise of the polarized gluon density in this region.\\
The values of the first moment of $g_1$ obtained by integration over both the full $x$-range
and only over the measured range, are given in Tab.~8, at typical values of $Q^2$ for each 
target. Both 
fits, type-A and B respectively, are considered. The truncated moments turn out to be 
generally quite close to each other.  

\vspace{0.3truecm}
\begin{center}
\begin{tabular}[b]{|l|l|l|l|} 
\hline
  & $\Gamma_1^p(10\,{\rm GeV}^2)$ & $\Gamma_1^d(10\,{\rm GeV}^2)$ & 
$\Gamma_1^n(5\,{\rm GeV}^2)$ \\
\hline\hline
type-A (full range) & 0.125 $\pm$ 0.003 & 0.038 $\pm$ 0.003 & -0.047 $\pm$ 0.004\\
type-A (meas. range) & 0.129 & 0.047 & -0.029\\
\hline\hline
type-B (full range) & 0.129 $\pm$ 0.004 & 0.035 $\pm$ 0.003 & -0.058 $\pm$ 0.005\\
type-B (meas. range) & 0.128 & 0.044 & -0.032\\
\hline
\end{tabular}\\
\end{center}

{\footnotesize Tab.~8. First moment of the structure function $g_1^p$ for proton,
deuteron and neutron targets.}
\vspace{0.4truecm}

\subsection{Polarized quark and gluon densities}

Moments of quark singlet and gluon polarizations as well as the value of the 
singlet axial charge may be evaluated by taking the average over the central values 
in the two fits, type-A and B respectively, and the final estimates thus read 
\bea
&&\Delta\Sigma(1)=0.40\,\pm\,0.02\,({\rm exp})\pm\,0.06\,({\rm th})\nn\\
&&\Delta g(1,1\,{\rm GeV}^2)=0.68\,\pm\,0.12\,({\rm exp})\pm\,0.29\,({\rm th})\lb{qga0}\\
&&a_0(10\,{\rm GeV}^2)=0.23\,\pm\,0.02\,({\rm exp})\,^{+0.35}_{-0.16}\,({\rm th}).\nn
\eea
The first error quoted in Eqs.~(\ref{qga0}) is the one related to the statistical errors 
from the fit, the second error is the theoretical one.\\
Indeed, as discussed in~\cite{Altarelli:1998nb,Altarelli:1997}, the main sources of 
theoretical uncertainty in the determination of quark and gluon polarizations,   
and of the related physical quantities, stem from the truncation of 
the perturbative expansions, as well as from the choice of the assumed initial 
parametrization of pdfs as emphasized in Sec.~6.5.2.\\
The error on the gluon first moment $\Delta g(1,1\,{\rm GeV}^2)$ due 
to the choice of input densities can be evaluated by comparing the results
of the type-A and B fits, i.e. by taking the half-difference of the two
central values as an estimate of the related uncertainty. It turns out roughly 
$\simeq\pm 0.12$. Similarly for the singlet quark and the axial charge one has
$\,\simeq\pm 0.05$ and $\,\simeq\pm 0.01$ respectively.\\  
The uncertainty induced by the truncation of the perturbative series for the 
coefficient functions and the evolution kernels is reflected by the dependence 
of the results on the renormalization scale $\mu_R$ and on the 
factorization scale $\mu_F$ respectively, that are usually both fixed  
to the value of $Q^2$ of each data point. A way of studying the effects of higher orders 
is by changing the value of $\mu_R$ and $\mu_F$ around the chosen values. 
On the basis of the previous 
analyses~\cite{Ridolfi:2003di,Altarelli:1998nb,Altarelli:1997}, 
the theoretical error on $\Delta g(1,1\,{\rm GeV}^2)$ due to higher order corrections 
has been estimated to be roughly $\simeq\pm 0.26$. Similarly for $\Delta\Sigma(1)$ and 
$a_0(10\,{\rm GeV}^2)$ one has $\simeq\pm 0.03$ and $\simeq^{+0.35}_{-0.16}$ respectively.\\  
Another source of theoretical uncertainty is due to higher-twist corrections, and, as 
extensively discussed in Sec.~2.4, the related systematic error has been strongly reduced
in the present analysis by discarding very low energy data, i.e. below the cut at the
final-state invariant mass $W=3\,$GeV. However, by exploting results of Sec.~2.4.2,  
an estimate of higher-twist effects may be attained by comparing the 
results of the same fit of type-A to the whole set of available data (Tab.~4, 
second column) with the one performed on the selected subset cut at $W\geq 3\,$GeV 
(Tab.~5). The related error on the gluon first moment $\Delta g(1,1\,{\rm GeV}^2)$ 
amounts to $\simeq\pm 0.06$, and it is thus
negligibly small if compared to the theoretical errors from the higher orders and
the initial parametrization. For 
$\Delta\Sigma(1)$ and $a_0(10\,{\rm GeV}^2)$ one finds $\simeq\pm 0.01$ and 
$\simeq\pm 0.02$ respectively.\\
The overall theoretical uncertainties in Eqs.~(\ref{qga0}) is then obtained
by combining the errors quoted above. 
Other sources of theoretical error, such as the uncertainty on the value of 
$\alpha_s(M_Z^2)$ and $\eta_8$, and the one 
related to the positions of heavy quark thresholds, all entail very small 
effects~\cite{Ridolfi:2003di} and have been neglected in the final estimates. 
Also the impact of 
target mass corrections is found to be small (see~\cite{Ridolfi:2003di,Blumlein:1998nv}).\\  

In order to compare these results with those of current analyses, it should be recalled
that in the Adler-Bardeen factorization scheme the 
polarized singlet quark density $\Delta\Sigma(1)$ is scale independent, whereas in the 
$\overline{\rm MS}$ scheme, adopted in most of the recent 
works~\cite{Hirai:2006sr,Leader:2005ci,Bluemlein:2002be,Gluck:2000dy}, 
it coincides with the (non conserved) singlet axial charge $a_0(Q^2)$ (see Eq.~(\ref{a0AB})). 
The polarized gluon 
density $\Delta g$ should be the same in the two schemes at NLO accuracy. A reasonable 
agreement between the above results and the most recent 
estimates is found. As an example, in Ref.~\cite{Hirai:2006sr} the quoted values read
\be
\Delta\Sigma(1,1\,{\rm GeV}^2)=0.25\,\pm\,0.10\,,\quad\,
\Delta g(1,1\,{\rm GeV}^2)=0.47\,\pm\,1.08\,,
\lb{aac06}
\ee
that are very similar to their previous estimates~\cite{Hirai:2003pm}.\\
Furthermore, in Refs.~\cite{Alexakhin:2006vx,Leader:2006xc} two solutions with either
$\Delta g>0$ or $\Delta g<0$ have been found; the first moment of $\Delta g$ at 
the initial scale turns out to be small in absolute value, i.e. 
$|\eta_g|<0.3$~\cite{Alexakhin:2006vx,Leader:2006xc} for both solutions, but the
shapes of the distributions are very different.  
It is finally worth noting that in the analysis of Ref.~\cite{Alexakhin:2006vx} the positive
solution for $\Delta g$ seems to be favored by the data.

\begin{figure}[!ht]
\centerline{\includegraphics[width=.8\linewidth,angle=-90]{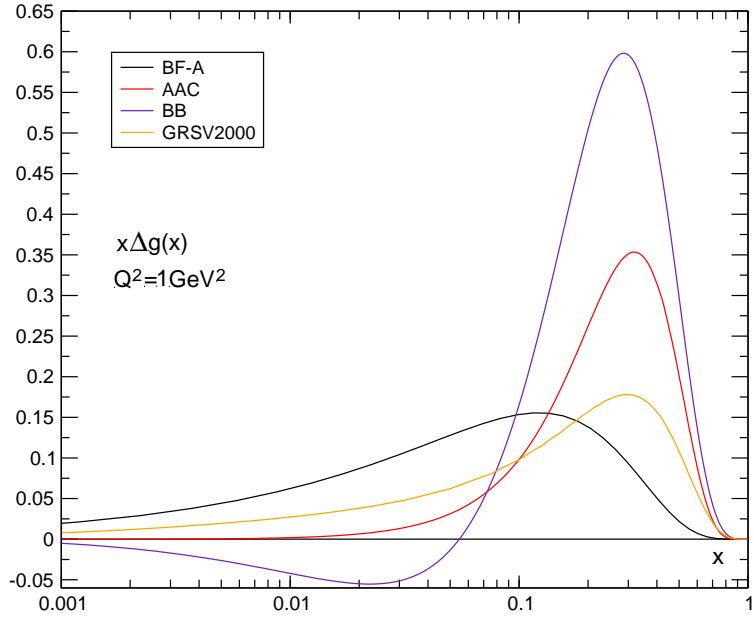}}
\caption{\footnotesize Best-fit result (type-A) for $x\Delta g(x)$ compared
at $Q^2=1\,$GeV$^2$ with the other parametrizations, i.e. AAC~\cite{Hirai:2003pm},
BB~\cite{Bluemlein:2002be} and GRSV~\cite{Gluck:2000dy}.}
\label{alldg}
\end{figure}

Then, even though the most recent global 
fits to world data fairly agree within the errors as far as  
the first moment of the polarized gluon density is concerned, they produce very 
different shapes of $\Delta g$, as shown in Fig.~\ref{alldg}. Here, the best-fit result
of type-A (Tab.~5) is compared at the scale $Q^2=1\,$GeV$^2$ with other available 
parametrizations~\cite{Hirai:2003pm,Bluemlein:2002be,Gluck:2000dy} of the spin-dependent 
gluon distribution. 
One may thus concludes that to the present accuracy inclusive  
DIS data are not capable to discriminate among the existing different scenarios, and thus
to pin down the $x$-shape of the gluon polarization. As discussed extensively 
in the next section, 
some light may be shed on this issue by direct measurements of $\Delta g$, rather than
by scaling violations, that is, by including in the analysis experimental data from
exclusive reactions in which gluons contribute at leading order.


\chapter{Open-charm photoproduction at COMPASS}


The spin-dependent gluon distribution $\Delta g$ in the nucleon seems 
to be hardly constrained by the inclusive DIS measurements. Actually, 
as discussed in Sec.~6, a global fit to world data including the most recent and precise 
experimental results on the polarized structure function $g_1$, 
allows a better determination of its first moment, that turns out to be 
much smaller than previous estimates (see e.g.\cite{Altarelli:1998nb,Gluck:1995yr}). 
However, the polarized gluon density 
exhibits a sizable dependence on the assumed functional form (see Sec.\ 6.4.2), and, as a 
result, the $x$-shape is still largely unknown.

At present, experimental constraints on the gluon polarization are 
expected to come from exclusive reactions, aimed at
measuring $\Delta g$ directly, rather than through scaling
violations. Such measurements entail the observation of specific events  
that receive leading contributions from gluon initiated subprocesses,
such as, for example, production of heavy-flavored hadrons and jets with large 
transverse momentum. 
In particular, as discussed below, the open-charm 
photoproduction represents a promising approach for extracting 
$\Delta g(x)$. This channel is currently experimentally
studied at COMPASS, and first results on the 
spin asymmetry are upcoming. Therefore, a complete analysis of polarized
photoproduction of open-charm at the COMPASS kinematics has been performed
in the present work, in order to assess the relevance of such measurements
in constraining the $x$-shape of $\Delta g(x)$.

A brief overview of the current scenario on direct measurements of the
gluon polarization is sketched in Sec.~7.1. Sec.~7.2 is devoted to the
theoretical study of the COMPASS experiment, and finally phenomenological
results are discussed in Sec.~7.3.

\section{Direct determination of $\Delta g$}

Reactions dominated by gluon initiated processes at the partonic level 
can be investigated with both electromagnetic and strong probes, 
that is, in lepto- and photo-production reactions and hadro-production 
respectively.

\subsection{Experiments at polarized proton-proton colliders}

Inelastic proton-proton collisions with polarized beams at high energies
provide the possibility to investigate the gluon spin distribution 
by using strong interaction among partons from the colliding hadrons.\\
New perspectives have been opened up by the RHIC-Spin (Relativistic Heavy 
Ion Collider) machine at Brookhaven National Laboratories, where
a number of high-$p_T$ processes can be generated by proton beam  
collisions at c.m.s. energies of $\sqrt{S}=200-500\,$GeV, which are 
sensitive to $\Delta g$, such as prompt photon and heavy flavor production, 
jet, single-inclusive and di-hadron production (see e.g.~\cite{Bunce:2000uv}).
Indeed, in all cases, the polarized gluon density plays a prominent role 
already at LO of QCD, through the underlying hard processes of gluon-gluon 
fusion and gluon-quark scattering.\\
Complete NLO corrections to polarized QCD hard-scattering processes, such as 
high-$p_T$ pion and jet production by polarized proton collisions,
are also available~\cite{Jager:2002xm,deFlorian:2002az,Jager:2004jh}.\\ 
In particular, in the NLO analysis~\cite{Jager:2002xm} of high-$p_T$ pion 
hadroproduction, the predicted asymmetry $A_{LL}^{\pi_0}$, i.e. the ratio of polarized 
and  unpolarized cross sections for the specific reaction, exhibits sizable 
differences for two distinct sets of spin-dependent pdfs~\cite{Gluck:2000dy}, 
which both provide good fit to inclusive DIS data, but differ significantly in 
the polarized gluon density (see also~\cite{Jager:2004jh}).
The gluon polarizations considered in Ref.~\cite{Jager:2002xm} are the maximal 
gluon polarization ``GRSV-max'' -- based on the maximal 
saturation of the positivity contraint, i.e. the assumption $\Delta g=g$ at the 
input scale -- and the ``standard'' distribution of the GRSV analysis~\cite{Gluck:2000dy}.  
The latter in particular is compared in Fig.~\ref{alldg} with the best fit result (type-A)
of Sec.~6 and with other available parametrizations. 
As a result, observed spin asymmetries should be capable of discriminating 
among different scenarios. 

First experimental results from PHENIX and STAR detectors at RHIC on double 
helicity asymmetry at c.m.s. energy $\sqrt{S}=200\,$ GeV, in 
pion~\cite{Adler:2004ps} and jet~\cite{Kiryluk:2005mh} production, 
probing the kinematic region in the parton momentum fraction $0.03<x<0.3$,
turn out to be consistent with a moderate gluon polarization, such as the
``standard'' GRSV distribution~\cite{Gluck:2000dy}. 
Specifically, the observed asymmetries are small, and a large positive 
gluon contribution to the nucleon spin, as given by the maximally polarized gluon  
``GRSV-max''~\cite{Gluck:2000dy}, is clearly disfavored by the 
data~\cite{Adler:2004ps,Kiryluk:2005mh}.\\
Besides, the PHENIX $\pi_0$ data points have been also included in a recent 
analysis of inclusive DIS world data by the AAC Collaboration~\cite{Hirai:2006sr}, 
and the conclusion was reached that the ensuing uncertainty on $\Delta g$ is 
significantly reduced (by about 60$\%$). 
Nonethless, within this analysis, along with a positive gluon polarization, 
a negative gluon distribution in the small-$x$ region -- type-1 and 
-3 respectively~\cite{Hirai:2006sr} -- is still consistent with the data, and 
resides outside the estimated error band of type-1 $\Delta g$. 
Further insights into the low-$x$ shape of the polarized gluon density may be 
supplied by higher energy runs ($\sqrt{S}=500\,$ GeV) at RHIC.\\

\subsection{Lepton-nucleon scattering in polarized fixed target experiments}

Polarized lepton-nucleon interactions still play a crucial
role for investigating the spin structure of the nucleon.
In order to obtain observables which are sensitive to the polarized gluon 
density, one has to consider reactions that are less inclusive than DIS,
in which one measures one or more outgoing final-state particles resulting from
spin-dependent gluon interactions. 
The largest cross sections typically stem from the photoproduction
regime, which is characterized by scattered leptons at very small angles, 
and thus by exchanged photons almost on-shell.\\
A crucial point here is the clear-cut sensitivity of cross sections and 
spin asymmetries in photoproduction reactions, such as for example heavy 
flavored hadron production, at both collider and fixed target energies, 
to the shape and size of the spin-dependent gluon distribution, as was shown 
in earlier studies (see e.g.~\cite{Frixione:1996ym,Stratmann:1996xy}).

At present, experimental results on lepton-nucleon interactions are supplied 
by low-energy fixed target experiments, where a beam of longitudinally  
polarized leptons is scattered off longitudinally polarized nucleon targets at
c.m.s energies $\sqrt{S}$ of at most few tens GeV.\\ 
In the framework of polarized DIS, the gluon polarization can be 
directly accessed via the underlying photon-gluon fusion (PGF) mechanism, 
resulting in a quark-antiquark pair. Experimental signatures to tag this 
subprocess are hadron pairs with high-$p_T$ in the final states, and 
open-charm events where the $q\bar{q}$ pair is required to be a $c\bar{c}$ 
pair and an outgoing charmed meson is reconstructed.\\
First measurements of longitudinal spin asymmetries in high-$p_T$ hadron-pair 
production are given by HERMES~\cite{Airapetian:1999ib} at DESY and more recently
by COMPASS~\cite{Ageev:2005pq} at CERN, both in the photoproduction regime, 
and by SMC~\cite{Adeva:2004dh} at CERN in the DIS region ($Q^2\ge1\,$GeV$^2$).
However, in such a reaction, the measured asymmetries receive contributions not 
only by pure PGF events, but also by a significant fraction of background events, 
mainly due to the two competing processes of gluon radiation by QCD Compton 
scattering ($\ga^* q\to q g$) and photon absorption at the lowest order of DIS 
($\ga^* q\to q$). High-$p_T$ hadron pair photoproduction is affected by large 
higher order corrections, and a full perturbative understanding is lacking.  
Estimates of the relative contributions from the background 
processes then rely on Monte-Carlo simulations, which are tuned to describe 
the data (see e.g.~\cite{Ageev:2005pq}).
The average gluon polarization is probed within a restricted range of momentum 
fraction, i.e. roughly $x\sim0.1$, and turned out, on the whole, to be consistent 
with a moderate gluon polarization (for a summary of such measurements  
see~\cite{Ageev:2005pq}). Phenomenological studies of hadron-pair polarized 
photoproduction at HERMES and COMPASS can be found
in~\cite{Hendlmeier:2006pd,Hendlmeier:2007pq,Bravar:1997kb}.

\subsection{The open-charm method at COMPASS}

At variance with hadron-pair production, the open-charm approach is free of 
background, since the PGF subprocess is the main mechanism for producing  
charm quarks in polarized DIS. A clean perturbative description is thus possible,
and higher order corrections have been found to be reasonably small~\cite{Bojak:1998bd}.  
This approach is currently used at COMPASS~\cite{COMP} to extract the ratio
of the polarized and unpolarized gluon distribution $\Delta g/g$. Here, 
polarized muons with a beam energy 
of $E_{\mu}=160\,$GeV scatter off deuterons in a polarized $^6$LiD target, 
corresponding to a c.m.s. energy of roughly $\sqrt{S}\simeq18\,$GeV, in the 
photoproduction regime where the photon virtuality is $Q^2\simeq 0$ (a description 
of the experimental setup can be found in~\cite{Abbon:2007pq}). $D$ mesons 
in the final states, originated from charm quark fragmentation, are reconstructed 
from their decay products in the channel $D^0\to K\pi$, with a branching
ratio of only $\sim4\%$, which results in low statistics. In addition, the channel 
$D^*\to D^0\pi_s$, where $\pi_s$ is a slow pion, is studied, increasing the 
signal-to-brackground ratio~\cite{LeGoff:2007zz}.\\
From the measured spin asymmetry $A_{LL}$ a preliminary determination of $\Delta g/g$ has 
been obtained by a LO analysis in QCD, that relies on the approximate 
relation~\cite{Kurek:2006fw}
\be
\vspace{-0.3truecm}
A_{LL}=R_{\rm PGF}\,\hat{a}_{LL}\frac{\Delta g}{g}+A_{\rm BG}\,,
\lb{all}
\ee
where $\hat{a}_{LL}$ is the partonic asymmetry, i.e. the ratio of spin-dependent
and spin-averaged hard cross sections, $R_{\rm PGF}$ is the fraction of PGF 
events in the selected sample and $A_{\rm BG}$ is the background asymmetry. An
estimation of $\hat{a}_{LL}$ on a event-by-event basis is then provided as a 
function of the measured kinematical variables, and is based on the comparison 
of the data sample with Monte-Carlo studies (see~\cite{Koblitz:2007zt} for details 
on the analysis method).\\ 
The extracted value~\cite{Koblitz:2007zt} of $\Delta g/g$ for data collected from
2002 up to 2004 turned out to be negative, i.e. 
$<\Delta g/g>=-0.57\pm0.41\,$(stat)$\pm0.17\,$(syst), 
but still consistent with zero, and represents an average over the probed 
$x$-range, $<x>\simeq0.15$. The hard scale is approximatively given by the charm 
quark mass $\mu^2=4(m_c^2+p_{tD}^2)\simeq13\,$GeV$^2$, where $p_{tD}$ is the $D^0$ meson
transverse momentum with respect to the photon direction.\\
However, the above Monte-Carlo approach entails a number of approximations, thus
missing a clean theoretical description of the process. On the other hand, as already
noted, since the photon-gluon fusion is the main underlying mechanism for
producing charm quarks by polarized lepton-nucleon scattering, the open-charm reaction can 
be reliably calculated in perturbation theory, with the charm quark mass as the
hard scale required. 
\newpage

\section{Theoretical analysis of charm asymmetries at COMPASS}

A thorough analysis of open-charm production from deep inelastic
muon scattering off nucleons, with longitudinally polarized beam and
target, has been performed here in the 
photoproduction approximation, where only quasi-real photons ($Q^2\simeq0$), 
radiated from the incoming leptons, are considered. The analysis is 
indeed tailored to the COMPASS experiment~\cite{COMP,Koblitz:2007zt} with relation to 
upcoming results on spin asymmetries in open-charm photoproduction, and is 
aimed at assessing the sensitivity of the actually observed 
quantities to the polarized gluon distribution.\\
As already noted, many earlier studies of heavy quark polarized 
photoproduction, at both high and low energy up to NLO, are also 
available~\cite{Frixione:1996ym,Stratmann:1996xy,Bojak:1998bd}. This 
issue is here re-analysed by constructing a close theoretical 
description of the experimental spin asymmetry 
at COMPASS, as a function of the measured kinematic variables. 

\subsection{General framework}

The spin-dependent and spin-averaged cross sections are defined by
\be
\Delta\sigma=\frac{1}{2}\left[\sigma^{\uparrow\uparrow}-\sigma^{\uparrow\downarrow}
\right]\qd\textrm{and}\qd
\sigma=\frac{1}{2}\left[\sigma^{\uparrow\uparrow}+\sigma^{\uparrow\downarrow}\right]
\lb{sda}
\ee
respectively, where the arrows denote parallel and opposite helicities of the
scattering particles. The spin-dependent photoproduction cross 
section for the process $l N\to l' H X$, where a longitudinally polarized lepton  
beam scatters off a longitudinally polarized nucleon target producing one observed 
hadron $H$ in the final state, can be written, on the basis of factorization 
theorem of QCD, as
\bea
&& \hspace{-1.truecm} d\Delta\sigma_{l N}=\sum_{ijk}\int dx_i\,dx_j\,dz_k\,
\Delta f_i^{(l)}(x_i,\mu_F)\,\Delta f_j^{(N)}(x_j,\mu_F)\lb{fcs}\\
&& \times\,\,d\Delta\hat{\sigma}_{ij}(x_i,x_j,S,P_H/z_k,\mu_R,\mu_F,\mu'_F)
 \,D_k^H(z_k,\mu'_F)\,.\nn
\eea
Here, $\Delta f_j^{(N)}(x_i,\mu_F)$ is the usual polarized parton density  
in the nucleon at the factorization scale $\mu_F$, and the non-pertubative 
function $D_k^H(z_k,\mu'_F)$ describes the fragmentation process of the produced   
parton $k$ into the observed hadron $H$ at the scale $\mu'_F$, with momentum 
fraction $z_k$. The polarized hard scattering cross section 
$d\Delta\hat{\sigma}_{ij}$ can be calculated in QCD order-by-order in the 
strong coupling $\alpha_s(\mu_R^2)$, $\mu_R$ being the renormalization scale, 
and the sum in Eq.\ (\ref{fcs}) is taken over all partonic channels. Finally, 
the function $\Delta f_i^{(l)}(x_i,\mu_F)$ represents the parton density in 
the lepton.

In the framework of the photoproduction approximation, which is the physically
interesting case in connection with the COMPASS experiment, 
the initial-state lepton is considered as a source of quasi-real photons 
(equivalent-photon approximation), with energy distribution given by the 
Weizsacker-Williams spectrum~\cite{von Weizsacker:1934sx} 
that can be theoretically computed. The function $\Delta f_i^{(l)}$ is given by 
the polarized Weizsacker-Williams function~\cite{deFlorian:1999ge}
\be
\Delta f_{\ga}^{(l)}(y)=\frac{\alpha_{\rm em}}{2\pi}\left[
\frac{1-(1-y)^2}{y}\ln\frac{Q_{max}^2(1-y)}{m_l^2y^2}+2m_l^2 y^2\left(
\frac{1}{Q_{max}^2}-\frac{1-y}{m_l^2y^2}\right)\right]\,,
\lb{wwf}
\ee
where $m_l$ is the lepton mass, $y$ the lepton momentum fraction carried by
the radiated photon and $Q_{max}$ is the allowed upper value to 
the photon's virtuality, to be fixed according to the experimental 
conditions~\cite{Frixione:1993yw}. The unpolarized counterpart of Eq.\ (\ref{wwf})
reads~\cite{Frixione:1993yw}
\be
f_{\ga}^{(l)}(y)=\frac{\alpha_{\rm em}}{2\pi}\left[
\frac{1+(1-y)^2}{y}\ln\frac{Q_{max}^2(1-y)}{m_l^2y^2}+2m_l^2 y\left(
\frac{1}{Q_{max}^2}-\frac{1-y}{m_l^2y^2}\right)\right]\,.
\lb{wwu}
\ee

Actually, quasi-real photons undergo interactions with partons in  
the nucleon either directly (``point-like'' component) or via their partonic 
structure (``resolved'' or ``hadronic'' component). In the first case photons
simply interact as elementary point-like particles, whereas in the second case 
photons fluctuate into a hadronic state, before the hard 
QCD interaction takes place, and the so-called ``resolved'' contributions to the
cross section compete with the ``direct'' part.  
Therefore, the experimentally measurable cross section is the sum of the 
point-like and the resolved photon contributions (see for 
example~\cite{Klasen:2002xb,Jager:2003vy,Drees:1988uv} and references therein)
\be
d\Delta\sigma_{l N}=d\Delta\sigma^{\rm dir}_{l N}+d\Delta\sigma^{\rm res}_{l N}\,.
\lb{dr}
\ee
Here, the direct component is given by Eqs.\ (\ref{fcs}) and (\ref{wwf}),
and the resolved component can also be cast in the form of Eq.\ (\ref{fcs}),
by properly defining the parton density of the lepton 
as the convolution product~\cite{Stratmann:1996xy}
\be
\Delta f_i^{(l)}(x_i,\mu_F)=\int_{x_i}^1\,\frac{dy}{y}\Delta f_{\ga}^{(l)}(y)
\Delta f_i^{(\ga)}\left(\frac{x_i}{y},\mu_F\right)\,,
\lb{res}
\ee
with $\Delta f_{\ga}^{(l)}(y)$ given by Eq.\ (\ref{wwf}). The partonic
content of the polarized photon is unmeasured so far and models are usually 
invoked~\cite{Stratmann:1996an} in actual calculations. Furthermore,
the direct photon contribution can be viewed as a particular case of
Eq.~(\ref{res}), by defining
\be
\Delta f_i^{(\ga)}=\delta\left(1-\frac{x_i}{y}\right)\,.
\ee
The corresponding spin-averaged cross sections are finally given by Eq.~(\ref{fcs}) 
replacing polarized parton densities and hard cross sections by the related
unpolarized quantities. The experimentally relevant observable is the spin
asymmetry, i.e. the ratio of polarized and unpolarized cross sections 
\be
A_{LL}=\frac{d\Delta\sigma_{l N}}{d\sigma_{l N}}\,.
\lb{as}
\ee

Focussing on the determination of the gluon polarization from photoproduction
reactions, such resolved contributions act as a background. 
Extensive analyses of their relevance with relation to the kinematic 
regime have been performed in studies on high-energy lepton-proton
collisions (see e.g.~\cite{Stratmann:1996xy,Jager:2003vy,deFlorian:1998fq}), 
in view of the polarized mode of HERA at DESY and the planned eRHIC project at 
BNL~\cite{erich}. 
Lower energies (fixed-target) experiments are generally expected to be 
less sensitive to the unmeasured partonic content of the polarized 
photon (see~\cite{Jager:2003vy,deFlorian:1998fq} and references therein), and 
LO estimates~\cite{Stratmann:1996xy} have revealed that the resolved contributions  
are negligibly small in the case of polarized photoproduction
of charm quarks at COMPASS energies. In the analysis of this experiment, one 
can thus safely neglect the resolved piece in Eq.\ ({\ref{dr}}).

In the case of 
open-charm photoproduction, as already noted, the only partonic 
subprocess contributing in LO to the spin-dependent cross section 
Eq.\ (\ref{fcs}), is photon-gluon fusion $\ga g\to c\bar{c}$. However, 
in the NLO approximation of QCD, along with 1-loop virtual corrections to 
PGF and real corrections with an additional gluon in the final state, new 
subprocesses, induced by a light quark (antiquark) replacing the gluon in 
the initial-state, i.e. $\ga q\,(\bar{q})\to c\bar{c}q\,(\bar{q})$, can also contribute,  
diluting the dependence of the asymmetry on $\Delta g(x)$.\\ 
All NLO contributions have been explicitly calculated in~\cite{Bojak:1998bd},  
for the polarized case~\footnote{The NLO QCD corrections in
the unpolarized case have been calculated for the first time in 
Refs.~\cite{Smith:1991pw,Ellis:1988sb}. Photoproduction of heavy quarks in 
the unpolarized case at fixed target experiments has been extensively 
studied~\cite{Ridolfi:1998dm}.}, and found to be reasonably small 
(around 10$\%$) 
in the energy range accessible at COMPASS (roughly 
$\sqrt{S_{\ga N}}\simeq 12\,$GeV), 
and should be 
outside of the present level of experimental accuracy at COMPASS~\cite{COMP}. 
Furthermore, the background due to the light quark induced subprocesses
turned out~\cite{Bojak:1998bd} to be fairly negligible. In particular, the 
PGF mechanism dominates in the COMPASS kinematical region, and a clear 
determination of $\Delta g$ may be performed by detecting open-charm 
events.\\
On this basis, the theoretical analysis of the spin asymmetries in open-charm 
photoproduction is performed below at the leading order of QCD, and by retaining only the 
point-like photon channel. The calculation is then specialized to the COMPASS
kinematics in order to reconstruct the actual experimentally observed quantities.

\vspace{-0.1truecm}
\subsection{Leading order cross section}

Open-charm events in the final states are a clear signature of the LO PGF
channel, namely

\be
\ga(q,\lambda_{\ga})+g(p,\lambda_g)\to c(k)+\bar c(k')\,,
\ee
where, $\lambda_{\ga},\,q$ and $\lambda_g,\,p$ denote photon and gluon 
helicities and four-momenta respectively, such that $q^2=p^2=0$. 
For the outgoing heavy quark pair $k^2=k'^2=m_c^2$, $m_c$ being the charm
quark mass. The corresponding Mandelstam variables 
are then defined by
\be
s=(q+p)^2,\qd t_1=(p-k)^2-m_c^2,\qd\textrm{and}\qd u_1=(q-k)^2-m_c^2, 
\lb{Mv}
\ee
with $s+t_1+u_1=0$. To order ${\mathcal O}(\alpha_{\rm em}\alpha_s)\,$ the 
differential partonic cross section can be written as~\cite{Frixione:1996ym,G1991}
\be
\frac{d^2\hat{\sigma}_{\ga g}}{dt_1du_1}(s,t_1,u_1,\lambda_{\ga},\lambda_g)=
\frac{2\pi\alpha_{\rm em}\alpha_s(\mu_R^2)}{9s^2}\,\left[\Sigma+
\lambda_{\ga}\lambda_g\Delta\right]\,\delta(s+t_1+u_1)
\lb{hard}
\ee
with
\be
\Sigma=-\frac{8m_c^4s^2}{t_1^2u_1^2}+2\frac{t_1^2+u_1^2+4m_c^2s}{t_1u_1}\,,
\lb{hS}
\ee
\be
\Delta=\frac{4m_c^2(t_1^3+u_1^3)}{t_1^2u_1^2}+2\frac{t_1^2+u_1^2-2m_c^2s}{t_1u_1}\,.
\lb{hD}
\ee

The spin-dependent and spin-averaged hard cross sections are 
defined as usual by
\be
d\Delta\hat{\sigma}_{\ga g}=\frac{1}{2}
\left[d\hat{\sigma}_{\ga g}^{\uparrow\uparrow}-
d\hat{\sigma}_{\ga g}^{\uparrow\downarrow}\right]\,,\quad
d\hat{\sigma}_{\ga g}=\frac{1}{2}
\left[d\hat{\sigma}_{\ga g}^{\uparrow\uparrow}+
d\hat{\sigma}_{\ga g}^{\uparrow\downarrow}\right]\,
\lb{xsects}
\ee
respectively, where the arrows denote parallel and antiparallel polarizations of
incoming photons and gluon. The Born-level cross sections thus explicitly
read

\be
\frac{d^2\Delta\hat{\sigma}_{\ga g}}{dt_1du_1}=\frac{4\pi\alpha_{\rm em}\alpha_s}
{9s^2}\left(\frac{t_1}{u_1}+\frac{u_1}{t_1}\right)\left(1-\frac{2m_c^2s}{t_1u_1}
\right)\,\delta(s+t_1+u_1)\,,
\lb{hP}
\ee
\be
\frac{d^2\hat{\sigma}_{\ga g}}{dt_1du_1}=\frac{4\pi\alpha_{\rm em}\alpha_s}{9s^2}
\left[\frac{t_1}{u_1}+\frac{u_1}{t_1}+\frac{4m_c^2s}{t_1u_1}\left(1-\frac{m_c^2s}
{t_1u_1}\right)\right]\,\delta(s+t_1+u_1)\,.
\lb{hU}
\ee
\vspace{0.7truecm}

Physical cross sections for the leptoproduction process $\mu N\to \mu'H X$ 
where $H$ is a charmed meson in the final state, can be related to the 
photoproduction analogue by the Weizsacker-Williams function Eq.~(\ref{wwf}). 
To order ${\mathcal O}(\alpha_{\rm em}\alpha_s)\,$, 
at the partonic level only the gluon distribution is relevant. Then 
Eq.\ (\ref{fcs}), in the unpolarized (polarized) case, can be now rewritten
schematically as
\be
d(\Delta)\sigma_{\mu N}=\int_{y_{\rm min}}^{y_{\rm max}} dy\,(\Delta)f_{\ga}^{(\mu)}(y)\,
d(\Delta)\sigma_{\ga N}
\lb{wP}
\ee
where $(\Delta)f_{\ga}^{(\mu)}(y)$ is given by Eq.\ (\ref{wwu}) 
(Eq.\ (\ref{wwf})), $q=y P_\mu$ with $P_{\mu}$ the incoming muon 
four-momentum, and the integration range over $y$ is fixed by the
experimental conditions. The photoproduction cross section in Eq.~(\ref{wP})  
at LO is given by 
\be
d(\Delta)\sigma_{\ga N}=\int\,dx\,(\Delta)g(x,\mu_F^2)\int\,dz\,D_c^H(z)\,
d(\Delta)\hat{\sigma}_{\ga g}(x,z,y,S,T,U,\mu_R^2)\,,
\lb{phcs}
\ee
where the (photoproduction) hadronic invariants $S,T,U$ are related to their
partonic counterparts by 
\be
s=xS,\quad t_1=\frac{xT}{z}\quad\textrm{and}\quad u_1=\frac{U}{z}\,.
\lb{hmandl}
\ee
The only parton distribution involved at LO is the unpolarized (polarized) 
gluon density $(\Delta)g(x,\mu_F^2)$, whereas $D_c^H(z)$ is the 
non-perturbative fragmentation function of a produced charm quark into 
an observed charmed meson.\\
Using the expressions Eqs.\ (\ref{hP}) and (\ref{hU}) of the hard scattering  
cross sections, Eq.\ (\ref{phcs}) explicitly reads

\bea
&&\hspace{-1.truecm}\frac{d^2\Delta\sigma_{\ga N}}{dTdU}=
\frac{4\pi\alpha_{\rm em}\alpha_s(\mu_R^2)}{9S^2}\int\,\frac{dx}{x}\,
\Delta g(x,\mu_F^2)\int\,\frac{dz}{z^2}\,D_c^H(z)\,\times\nn\\
&&\left(\frac{U}{xT}+\frac{xT}{U}\right)\,\left(1-\frac{2m_c^2z^2S}{TU}\right)\,
\delta\left(xS+\frac{xT}{z}+\frac{U}{z}\right)\,,
\lb{phP}
\eea
\bea
&&\hspace{-1.truecm}\frac{d^2\sigma_{\ga N}}{dTdU}=
\frac{4\pi\alpha_{\rm em}\alpha_s(\mu_R^2)}{9S^2}
\int\,\frac{dx}{x}\,g(x,\mu_F^2)\int\,\frac{dz}{z^2}\,D_c^H(z)\,\times\nn\\
&&\left[\frac{U}{xT}+\frac{xT}{U}+\frac{4m_c^2z^2S}{TU}\left(1-\frac{m_c^2z^2S}{TU}
\right)\right]\,\delta\left(xS+\frac{xT}{z}+\frac{U}{z}\right)\,,
\lb{phU}
\eea
in the spin-dependent and spin-averaged case respectively.

Actually, if the momentum fraction $y$ of the incoming muon carried by the 
photon is reconstructed event-by-event, the open-charm asymmetries do depend on 
$y$, as well as on the measured variables of the 
outgoing $D$ meson.\\
Within this framework, no average over the Weizsacker-Williams spectrum, 
Eq.\ (\ref{wP}), is required, and the observed quantity is consistently 
described by the photoproduction asymmetry
\be
A_{LL}^D=\frac{d\Delta\sigma_{\ga N}}{d\sigma_{\ga N}}
\lb{sa}
\ee
with $d(\Delta)\sigma_{\ga N}$ given by Eqs.\ (\ref{phP}) and~(\ref{phU}).

\subsection{COMPASS kinematics}

Specifying the above general discussion to the case of the (fixed target) 
open-charm experiment at COMPASS, the muon momentum fraction $y$ carried by
the photon is indeed reconstructed event-by-event with no experimental cut.
Thus, the polarized and unpolarized cross 
sections in Eq.\ (\ref{sa}) are double differential distributions, parametrized 
by the measured $D$ meson variables in the nucleon rest frame, that is, the 
transverse momentum $p_{tD}$ with respect to the photon direction, and the
energy fraction $z_D=E_D/E_{\ga}\,$ with respect to the photon energy $E_{\ga}$. 
They also depend on $y$ through the incoming photon energy, given as a 
function of the muon beam energy $E_{\ga}=y E_{\mu}\,$, with $E_{\mu}=160$~GeV. 
The hadronic invariant system in the laboratory
frame then reads
\be
T=-Sz_D,\qd U=-\frac{S^2}{2M^2}\left(z_D-\chi_D\right)\qd\textrm{and}
\qd S=2ME_{\ga}
\lb{TUS}
\ee
\be
\textrm{with}\qqd\chi_D=\sqrt{z_D^2-\frac{4M^2m_{tD}^2}{S^2}}\,,
\lb{chD}
\ee
and $m_{tD}=\sqrt{p_{tD}^2+m_D^2}$, i.e. the $D$ meson transverse mass.   
Using the Jacobian $dTdU=(2p_{tD}S/\chi_D)dz_Ddp_{tD}$
to express Eqs.\ (\ref{phP}) and~(\ref{phU}) in the nucleon rest 
frame variables Eq.\ (\ref{TUS}), the relevant spin-dependent and spin-averaged 
physical cross sections become

\bea
&&\frac{d^2\Delta\sigma_{\ga N}}{dp_{tD}\,dz_D}=\frac{16\pi\alpha_{\rm em}
\alpha_s(\mu^2) M^2}{9S^3}\frac{p_{tD}}{z_D\chi_D(z_D-\chi_D)}\int_{x_{\rm min}}^1
\frac{dx}{x}\Delta g(x,\mu^2)\left(z_D+w_D\right)\nn\\
&&\qd\times\,\left[1-\frac{2z_Dw_D}{\left(z_D+w_D\right)^2}\right]\left[1-
\frac{4m_c^2M^2}{z_D(z_D-\chi_D)S^2}\left(z_D+w_D\right)^2\right]D
\left(z_D+w_D\right)\nn\\
&&\lb{dsig}
\eea
and 
\bea
&&\frac{d^2\sigma_{\ga N}}{dp_{tD}\,dz_D}=\frac{16\pi\alpha_{\rm em}
\alpha_s(\mu^2) M^2}{9S^3}\frac{p_{tD}}{z_D\chi_D(z_D-\chi_D)}\int_{x_{\rm min}}^1
\frac{dx}{x} g(x,\mu^2)\left(z_D+w_D\right)\times\nn\\
&&\hspace{-0.8truecm}\,\left[1-\frac{2z_Dw_D}{\left(z_D+w_D\right)^2}+
\frac{4m_c^2}{xS}\left(1-\frac{2m_c^2M^2}{z_D(z_D-\chi_D)S^2}\left(z_D+w_D\right)^2
\right)\right] D\left(z_D+w_D\right)\,,
\nn\\
&&
\lb{sig}
\eea
respectively, where the shorthand 
\be
w_D=\frac{(z_D-\chi_D)S}{2xM^2}
\lb{wD}
\ee
has been used. The actual observable at COMPASS is finally the ratio of Eqs.~(\ref{dsig}) 
and~(\ref{sig}), according to Eq.~(\ref{sa}).\\
It is worth noting that in Eqs.\ (\ref{dsig}) and (\ref{sig})
integration over the fragmentation variable $z$ has been explicitly performed,
being constrained by the $\delta$-function. Moreover, the integration range
over the gluon momentum fraction $x$ depends on the measured variables through
the relation 
\be
x_{\rm min}(p_{tD},z_D)=\frac{E_{\ga}}{M}\left(\frac{z_D-\chi_D}{1-z_D}\right)\,.
\lb{xmin}
\ee

On the other hand, if one explicitly integrates over $x$, the integration 
threshold over $z$ is constrained by the relation 
\be
z_{\rm min}(p_{tD},z_D)=z_D+\frac{E_{\ga}}{M}(z_D-\chi_D)\,.
\lb{zmin}
\ee
This actually means that the gluon momentum fraction $x$ and the fragmentation
variable $z$ are not independent variables.

The strong coupling $\alpha_s(\mu_R^2)$, as well as the polarized 
gluon density $\Delta g(x,\mu_F^2)$ are evaluated in NLO at the common fixed
value $\mu_R=\mu_F=2m_c$ for the renormalization and factorization scales
respectively. For the charm quark mass $m_c=1.5$~GeV is taken, $M\simeq0.938\,$GeV  
for the nucleon mass, and the $D$ mass is taken as an average of the experimentally 
detected $D^0$ and $D^*$ masses~\cite{Yao:2006px}, roughly $m_D\simeq 1.94\,$GeV. 
The strong coupling is normalized as 
$\alpha_s(M_Z^2)\simeq0.118$ at the $Z$ boson mass~\cite{Yao:2006px}.\\
Note that the experimental cut $z_D\gsim0.2$ is imposed on the fractional $D$ energy, 
and the most populated bins are estimated~\cite{COMP} as $0.32\lsim z_D\lsim 0.66$ and 
$0.35\lsim p_{tD}\lsim 1\,$GeV.

\begin{figure}
\centerline{\includegraphics[width=.7\linewidth]{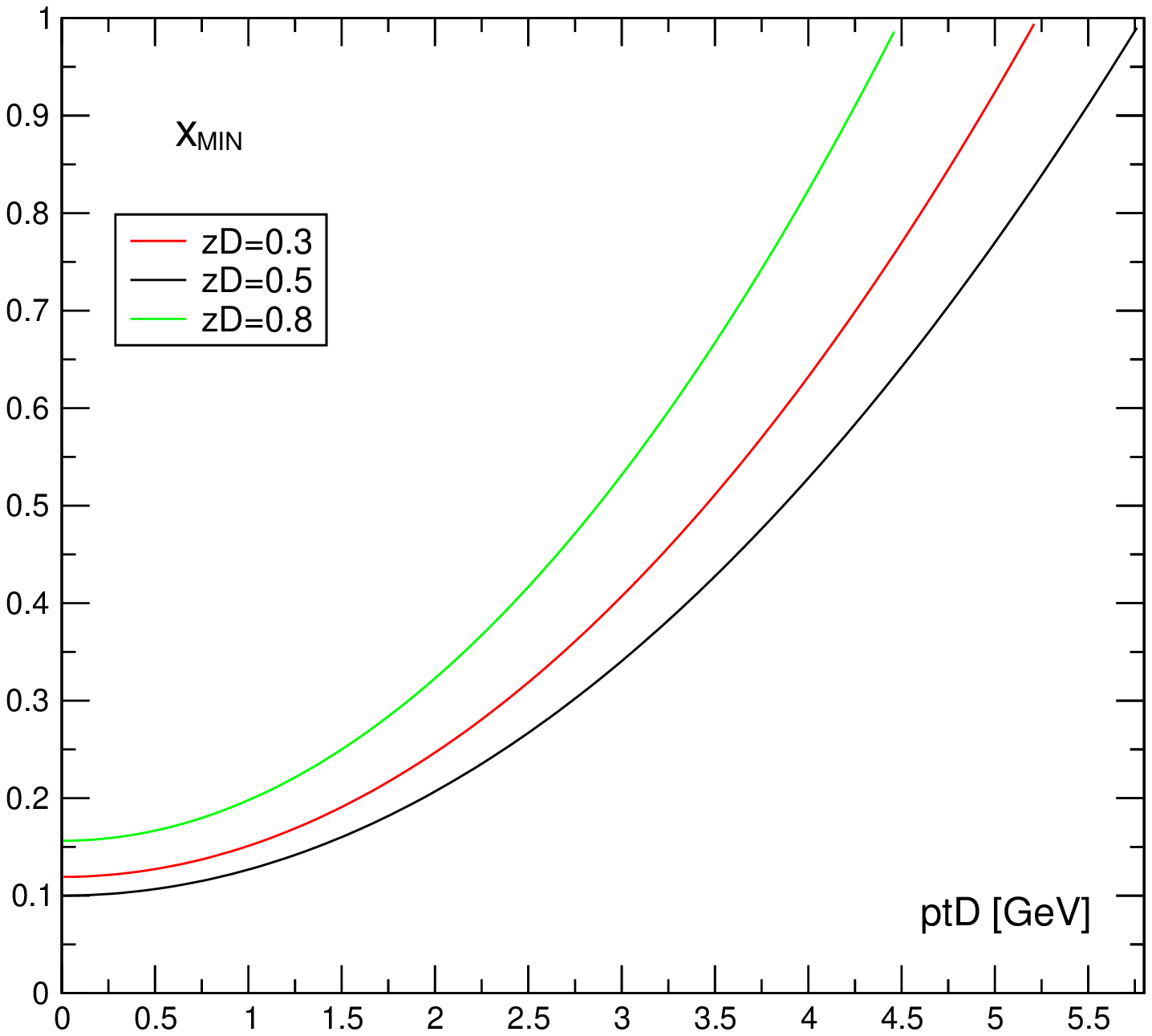}}
\vspace{-0.8truecm}\caption{\footnotesize Variation of the lower bound 
$x_{\rm min}$ in the integration over the gluon momentum fraction $x$, as 
a function of $p_{tD}$ at three different values of $z_D$.}
\label{xmpt}
\centerline{\includegraphics[width=.7\linewidth]{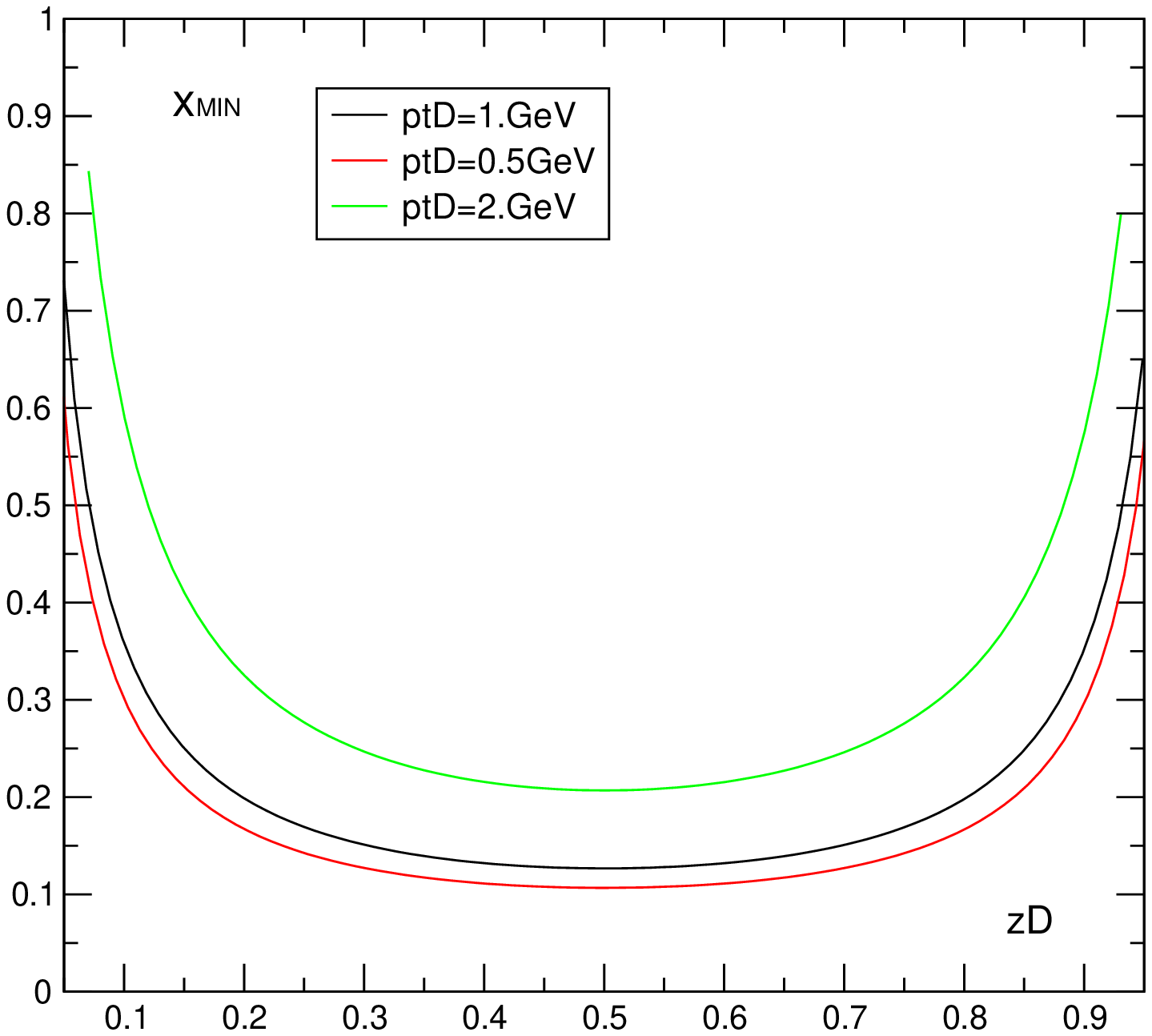}}
\vspace{-0.8truecm}\caption{\footnotesize Variation of the lower bound 
$x_{\rm min}$ in the integration over the gluon momentum fraction $x$, 
as a function of $z_D$ at three different values of $p_{tD}$.}
\label{xmzd}
\end{figure}

\begin{figure}
\centerline{\includegraphics[width=.7\linewidth]{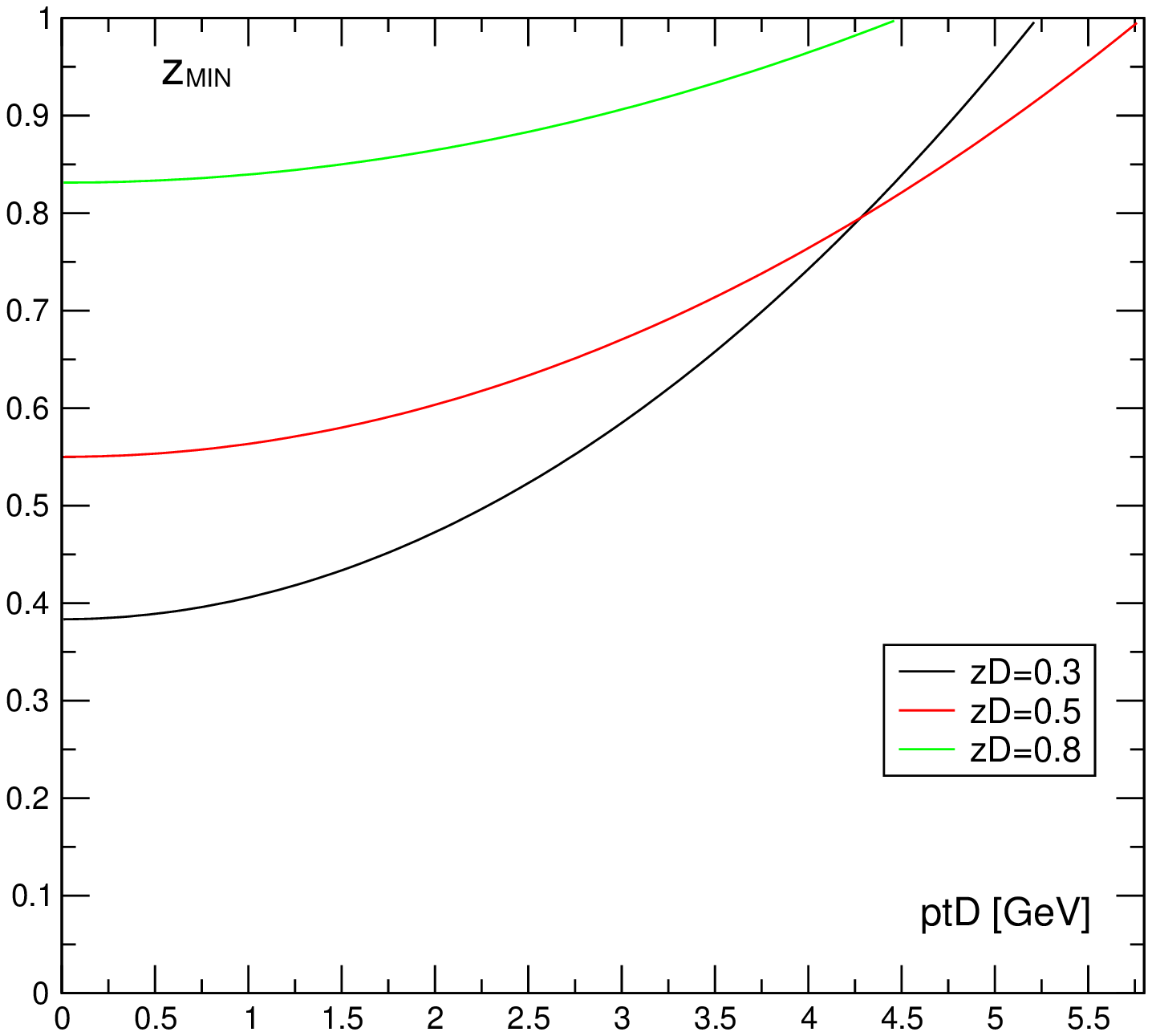}}
\vspace{-0.8truecm}\caption{\footnotesize Variation of the lower bound 
$z_{\rm min}$ in the integration over the fragmentation variable $z$, 
as a function of $p_{tD}$ at three different values of $z_D$.}
\label{zmpt}
\centerline{\includegraphics[width=.7\linewidth]{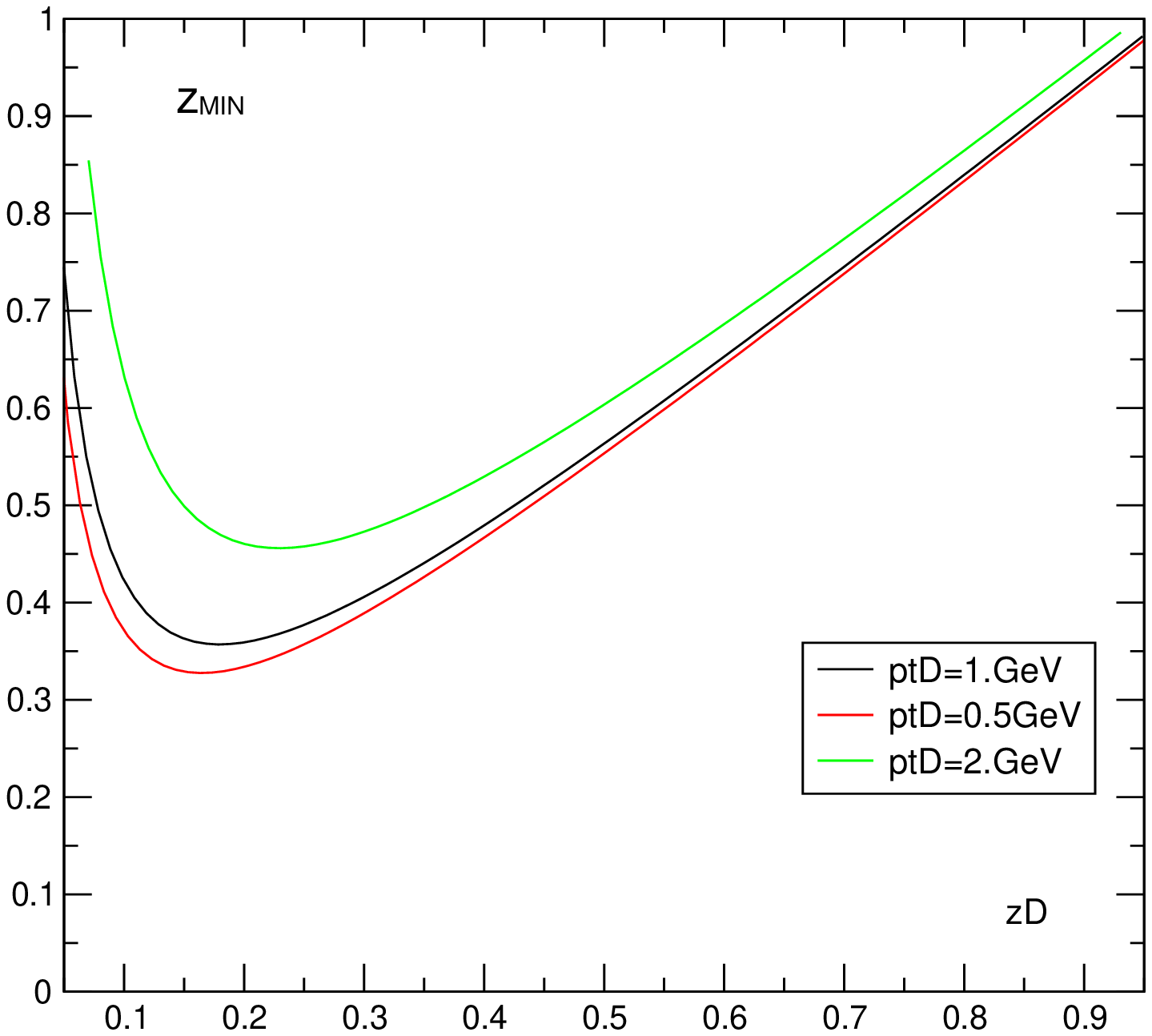}}
\vspace{-0.8truecm}\caption{\footnotesize Variation of the lower bound 
$z_{\rm min}$ in the integration over the fragmentation variable $z$, 
as a function of $z_D$ at three different values of $p_{tD}$.}
\label{zmzd}
\end{figure}

Finally, as far as Eqs.\ (\ref{xmin}) and (\ref{zmin}) are concerned some comments
are in order. Indeed, for fixed values of the 
observed kinematic variables $z_D,\,p_{tD}$, the fragmentation variable 
$z$ is completely fixed by the momentum fraction $x$, and thus only the 
integral over $x$ remains as shown by the expression of the cross sections
Eqs.\ (\ref{dsig}) and (\ref{sig}). 
Furthermore, the integration threshold over $x$ is determined by the values
of $z_D$ and $p_{tD}$. This bound, given by Eq.\ (\ref{xmin}), is displayed in Figs.\
\ref{xmpt}~and~\ref{xmzd} as a function of each of the two variables.
As expected, the low $x$ region ($x\lsim0.1$) is unaccesible at 
COMPASS energies, and the integration region involved is stretched
towards larger $x$ with increasing $p_{tD}\,$~(Fig.~\ref{xmpt}). The same 
situation is observed as the $D$ meson energy approaches the kinematical boundaries 
(Fig.~\ref{xmzd}), whereas in the physically meaningful region of $z_D$  
$x_{\rm min}$ is nearly costant ($x_{\rm min}\sim0.1$).\\
As already noted, one could equivalently determine $x$ for given $z$ and then
only the integration over $z$ remains. In such a case, it is the fragmentation 
variable $z$ which is bounded from
below with $z_{\rm min}$ determined by $z_D$ and $p_{tD}$ according to Eq.\
(\ref{zmin}). The bound on $z$ is displayed in Figs.~\ref{zmpt}~and~\ref{zmzd}
versus $z_D$ and $p_{tD}$ respectively. Both figures show that a sizable  
amount of the charm quark momentum is carried by the $D$ meson in the whole
kinematical region.\\ 
\subsection{Pdfs and fragmentation function}

\begin{figure}
\centerline{\includegraphics[width=0.7\linewidth]{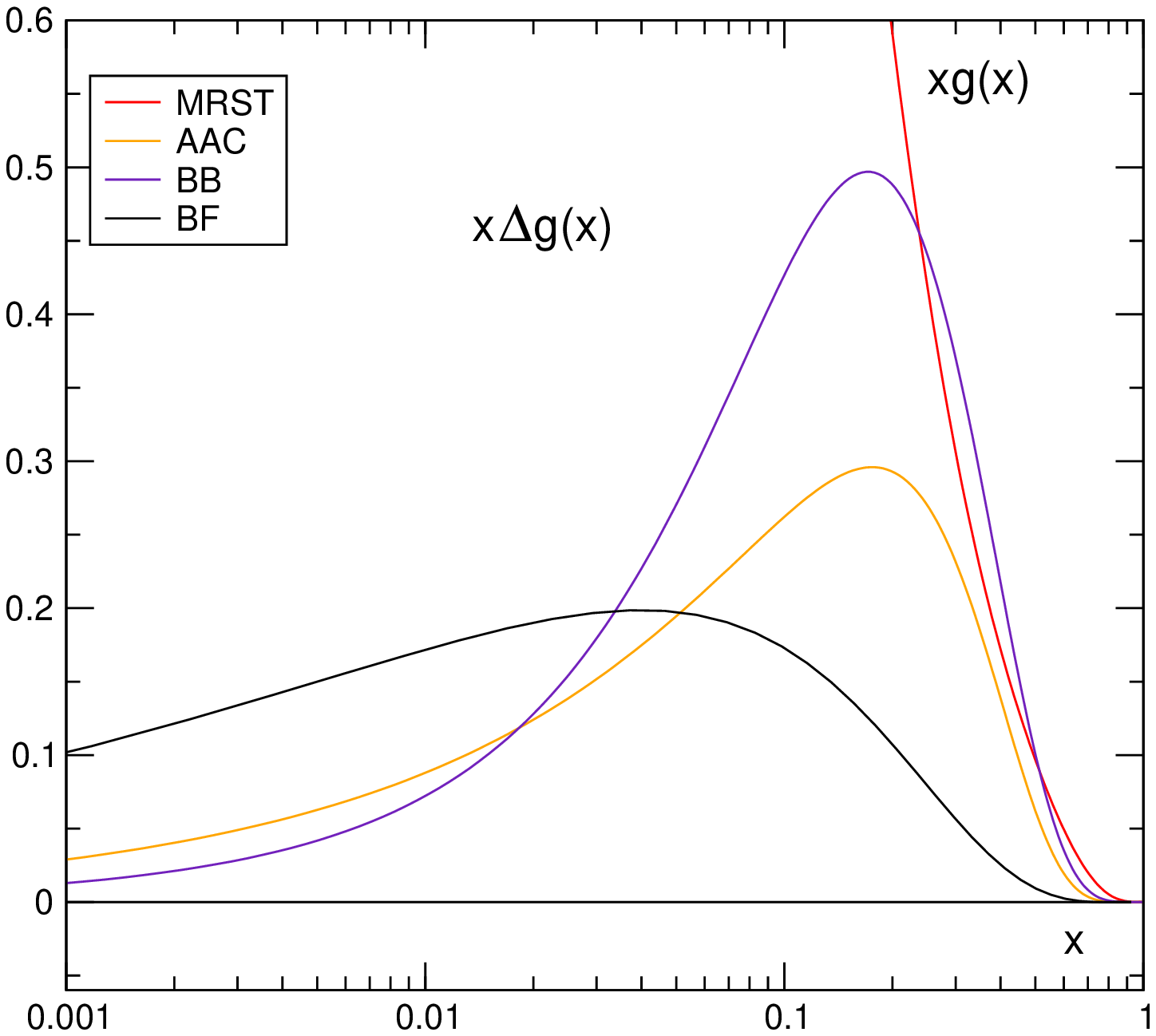}}
\vspace{-0.8truecm}\caption{\footnotesize NLO parametrizations of gluon 
distribution functions used in the analysis, namely 
MRST2004f4~\cite{Martin:2006qz} for $g(x)$, AAC~\cite{Hirai:2003pm}, 
BB~\cite{Bluemlein:2002be} and the most recent best-fit result BF-A for 
$\Delta g(x)$, at the scale $\mu_F^2=4m_c^2$.}
\label{xdg}
\centerline{\includegraphics[width=.7\linewidth]{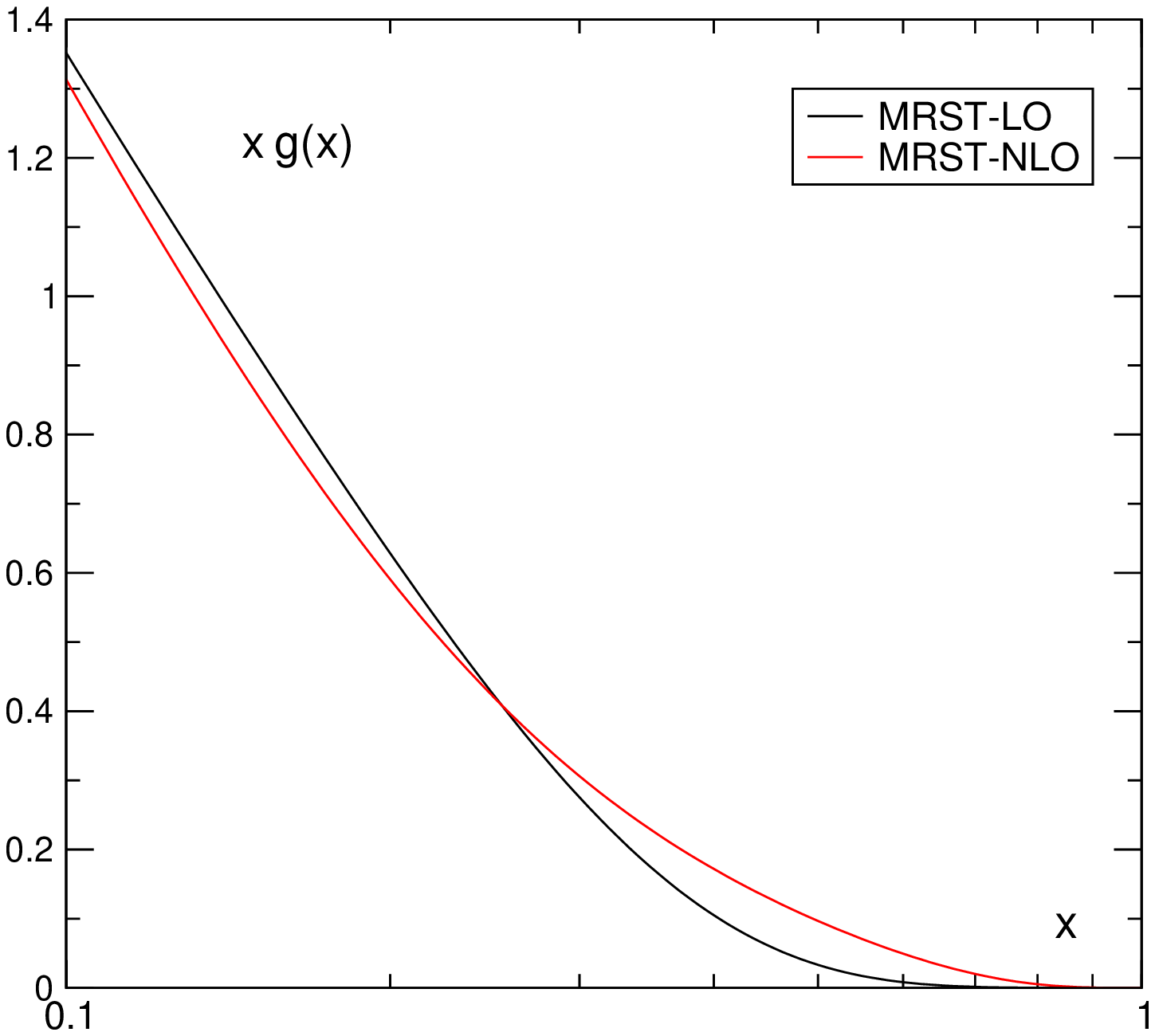}}
\vspace{-0.8truecm}\caption{\footnotesize LO and NLO MRST2004f4 
parametrizations~\cite{Martin:2006qz} of the unpolarized gluon distribution 
in the region of interest, at the scale $\mu_F^2=4m_c^2$.}
\label{xg}
\end{figure}

The sensitivity of the asymmetry on the spin-dependent gluon distribution  
$\Delta g(x,\mu_F^2)$ is investigated here by using three (NLO) parametrizations, 
obtained from fit to inclusive DIS data, namely AAC~\cite{Hirai:2003pm}, 
BB~\cite{Bluemlein:2002be}, and the most recent best-fit result BF-A (i.e. type-A) 
discussed in Sec.~6.5.1.\\ 
The use of NLO parton distributions in a LO calculation has obviously
no effect on the accuracy of the calculation which remains LO. Note also
that the AAC and BB polarized gluons are given in the $\overline{\rm MS}$ 
factorization scheme, whereas the BF-A gluon is given in the Adler-Bardeen 
scheme. However, as already noted, the polarized gluon densities are the same
in the two schemes at NLO accuracy. 
The three parametrizations used for $\Delta g(x,\mu_F^2)$ are displayed at 
the common scale $\mu_F=2m_c$ in Fig.~\ref{xdg}, together with the 
MRST2004f4~\cite{Martin:2006qz} unpolarized gluon distribution, which is
used as a rule in the spin-averaged cross section Eq.\ (\ref{sig}).\\
In Fig.~\ref{xg} the LO and NLO MRST2004f4 parametrizations for the
unpolarized gluon distribution $g(x)$ are compared in the region of 
interest for $x$ ($x\gsim0.1$), at the scale $\mu=2m_c$. This will be used 
to estimate the dependence of the results on the unpolarized gluon distribution.\\ 
Note also that some of the available LO parametrizations of $\Delta g(x)$ 
(e.g. BB and LSS~\cite{Leader:2005ci}) violate the positivity constraint 
imposed by the LO MRST determination of $g(x)$ at intermediate and large  
$x\,$.  This problem is also marginally present for the NLO BB polarized 
gluon as seen in Fig.~\ref{xdg}, but does not affect significantly the analysis.
Indeed, the uncertainty on the unpolarized gluon
at large-$x$ makes violation of the positivity bound less serious.
 
\begin{figure}[!ht]
\centerline{\includegraphics[width=.7\linewidth]{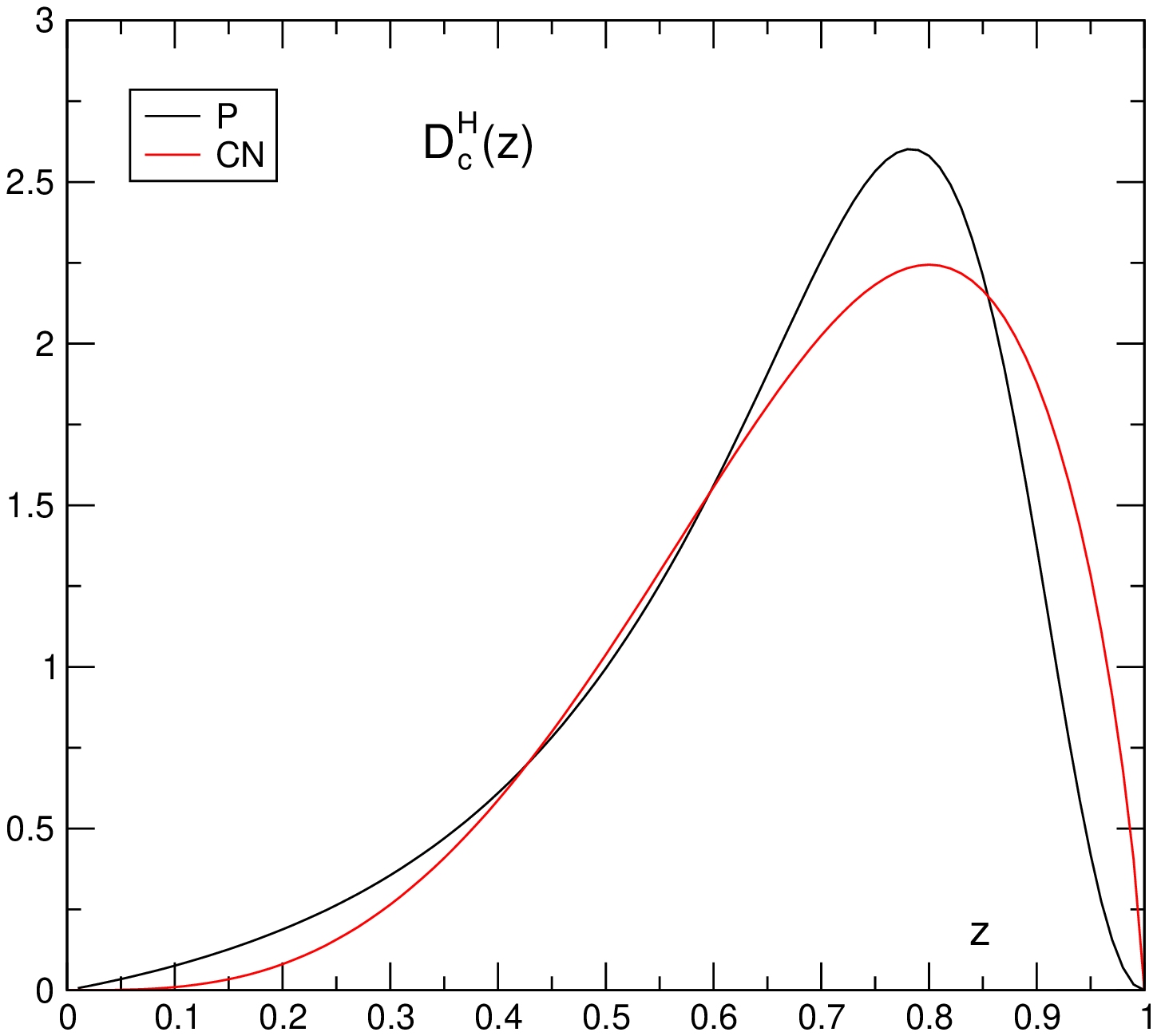}}
\vspace{-0.8truecm}\caption{\footnotesize Fragmentation functions $D_c^H(z)$ 
used in the analysis, namely Eq.~(\ref{Pfr}) (P)~\cite{Peterson:1982ak,Chrin:1987yd} 
and Eq.~(\ref{CNfr}) (CN)~\cite{Colangelo:1992kh}.}
\label{dzp}
\end{figure}

Finally, for the fragmentation function of a charm quark into a $D$ meson
($D^0$ and $D^*$) the non-pertubative Peterson (P) 
function~\cite{Peterson:1982ak} 
\be
D_c^H(z)=\frac{N}{z\left[1-1/z-\varepsilon/(1-z)\right]^2}
\lb{Pfr}
\ee
is used as a baseline, with $N$ the normalization factor. The 
phenomenological parameter $\varepsilon$, that  
is predicted to scale as $\varepsilon\sim\Lambda_{\rm QCD}^2/m_Q^2$ in 
the fragmentation of a heavy quark $Q$, is fixed to the usual value 
$\varepsilon=0.06$~\cite{Chrin:1987yd} (see also~\cite{Cacciari:1997ad}). 
In Fig.~\ref{dzp} the Peterson fragmentation function Eq.\ (\ref{Pfr}) is
compared to the slightly different parametrization (CN) given 
in Ref.~\cite{Colangelo:1992kh}, namely
\be
D_c^H(z)=N(1-z)^{a_1}z^{a_2},\qd\qd a_1=0.8,\,\,a_2=3.2.
\lb{CNfr}
\ee
The latter is also used in the present analysis in order to estimate 
the impact of the precise shape of the fragmentation function on the results.

\section{Phenomenological results}

The theoretically predicted asymmetries $A_{LL}^D$, given by 
Eqs.\ (\ref{sa}), (\ref{dsig}) and (\ref{sig}), are analysed here as
a function of the relevant kinematic variables $y,z_D,p_{tD}$, so as 
to determine their sensitivity to the polarized gluon density.\\
Next, the effect of including COMPASS results in a global fit of pdfs,
along with inclusive DIS data, is discussed by using pseudo-data generated
on the basis of the theoretical prediction and the expected statistical
error.

\subsection{Asymmetries as a function of the measured variables}

\begin{figure}[!hb]
\centerline{\includegraphics[width=.7\linewidth]{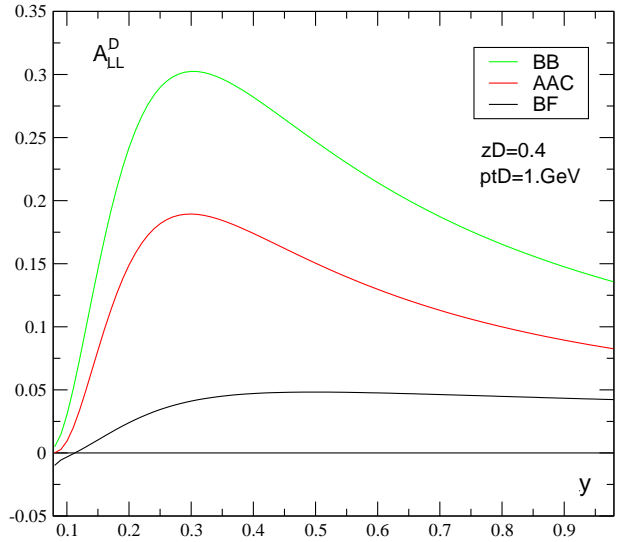}}
\vspace{-0.8truecm}\caption{\footnotesize The spin asymmetry as a 
function of the photon energy $y=E_{\gamma}/E_{\mu}$, at $p_{tD}=1$~GeV 
and $z_D=E_D/E_{\gamma}=0.4\,$.
Three (NLO) parametrizations for the $\Delta g$ are used,
namely AAC, BB and BF-A, whereas $g$ is given by MRST2004f4 (NLO). 
The Peterson fragmentation function is used with $\varepsilon=0.06\,$.}
\label{yas}
\end{figure}

The asymmetries $A_{LL}^D$ as a function of the incoming photon energy
$E_{\gamma}=y E_{\mu}$ are compared in Fig.~\ref{yas} for the aforementioned 
parametrizations of $\Delta g(x)$. The $D$ meson trasverse momentum and
fractional energy are fixed to the values $p_{tD}=1$~GeV and 
$z_D=E_D/E_{\gamma}=0.4\,$, corresponding roughly to a maximum in both 
the polarized and unpolarized cross sections at the COMPASS kinematics. 
With increasing photon energy, the asymmetry slowly decreases due to the 
faster growth of the unpolarized cross section, similarly to the case of 
total charm photoproduction~\cite{Stratmann:1996xy}. The 
gluon polarization has a significant impact on the size of the asymmetry, 
in that the normalization is mostly determined by the value of $\Delta g(x)$ for 
$x> x_{\rm min}\approx 0.1$. Because for all the given parametrizations
$\Delta g(x)$ is rapidly decreasing in this region, the asymmetry is
in fact essentially determined by the value of  $\Delta g(x_{\rm min})$. 
This is indeed in substantial agreement with the estimated~\cite{Koblitz:2007zt} 
average $x=<0.15>$ over the probed region at COMPASS.
The shape of the asymmetry is instead dominated by that of the 
underlying hard process.

\begin{figure}[!ht]
\centerline{\includegraphics[width=.7\linewidth]{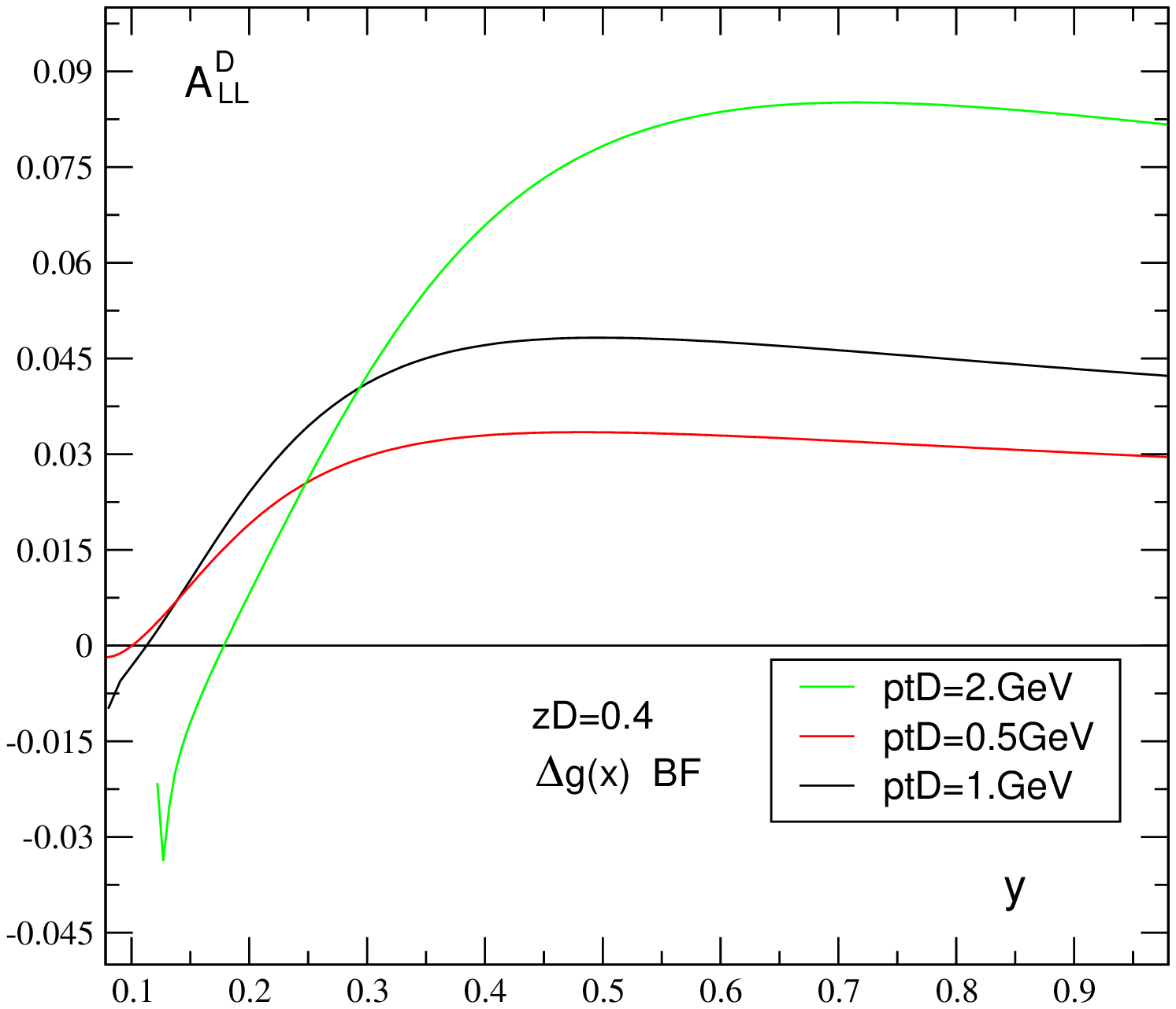}}
\vspace{-0.8truecm}\caption{\footnotesize The spin asymmetry as a function 
of $y$ at different $p_{tD}$ and $\Delta g$ given by BF-A ($z_D$, $g(x)$ and 
$D_c^H(z)$ as in Fig.~\ref{yas})}
\label{y2as}
\end{figure}
The $y$-shape of $A_{LL}^D$ is also displayed in Fig.~\ref{y2as} at different 
fixed values of $p_{tD}$, and for the same polarized gluon density BF-A, used 
as a benchmark in what follows.  
As the measured transverse momentum $p_{tD}$ is increased the large $x$ 
region of $g(x)$ and $\Delta g(x)$ is probed. Because the ratio $\Delta
g(x)/g(x)$ increases with $x$, the asymmetry correspondingly increases
with growing  $p_{tD}$.

\begin{figure}
\centerline{\includegraphics[width=0.7\linewidth]{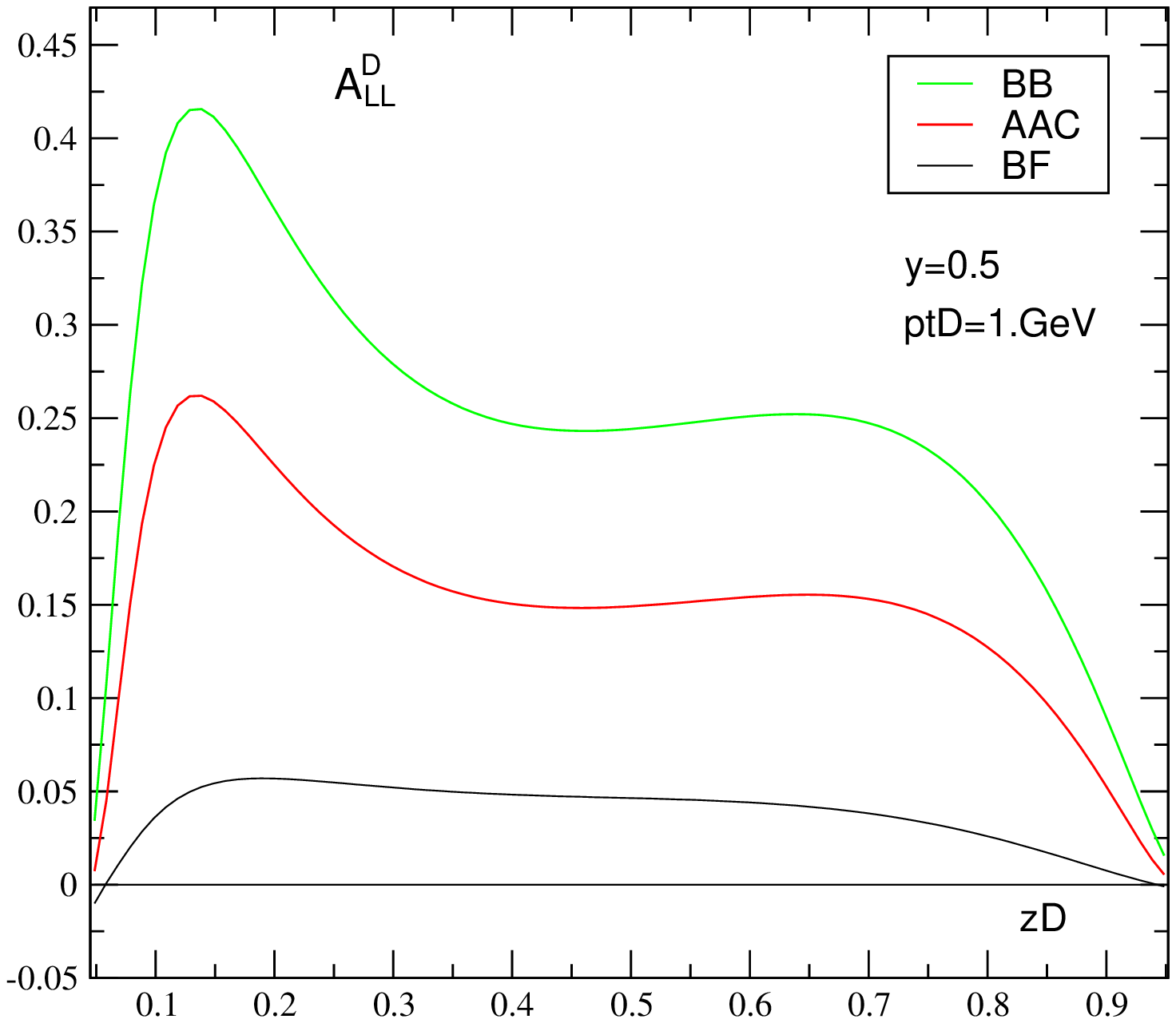}}
\vspace{-0.8truecm}\caption{\footnotesize The spin asymmetry as a function 
of the fractional energy $z_D=E_D/ E_{\ga}$, at fixed photon energy $y=0.5\,$ 
($p_{tD}$, $g(x)$, $\Delta g(x)$ and $D_c^H(z)$ as in Fig.~\ref{yas}).}
\label{zas}
%
%
\centerline{\includegraphics[width=.7\linewidth]{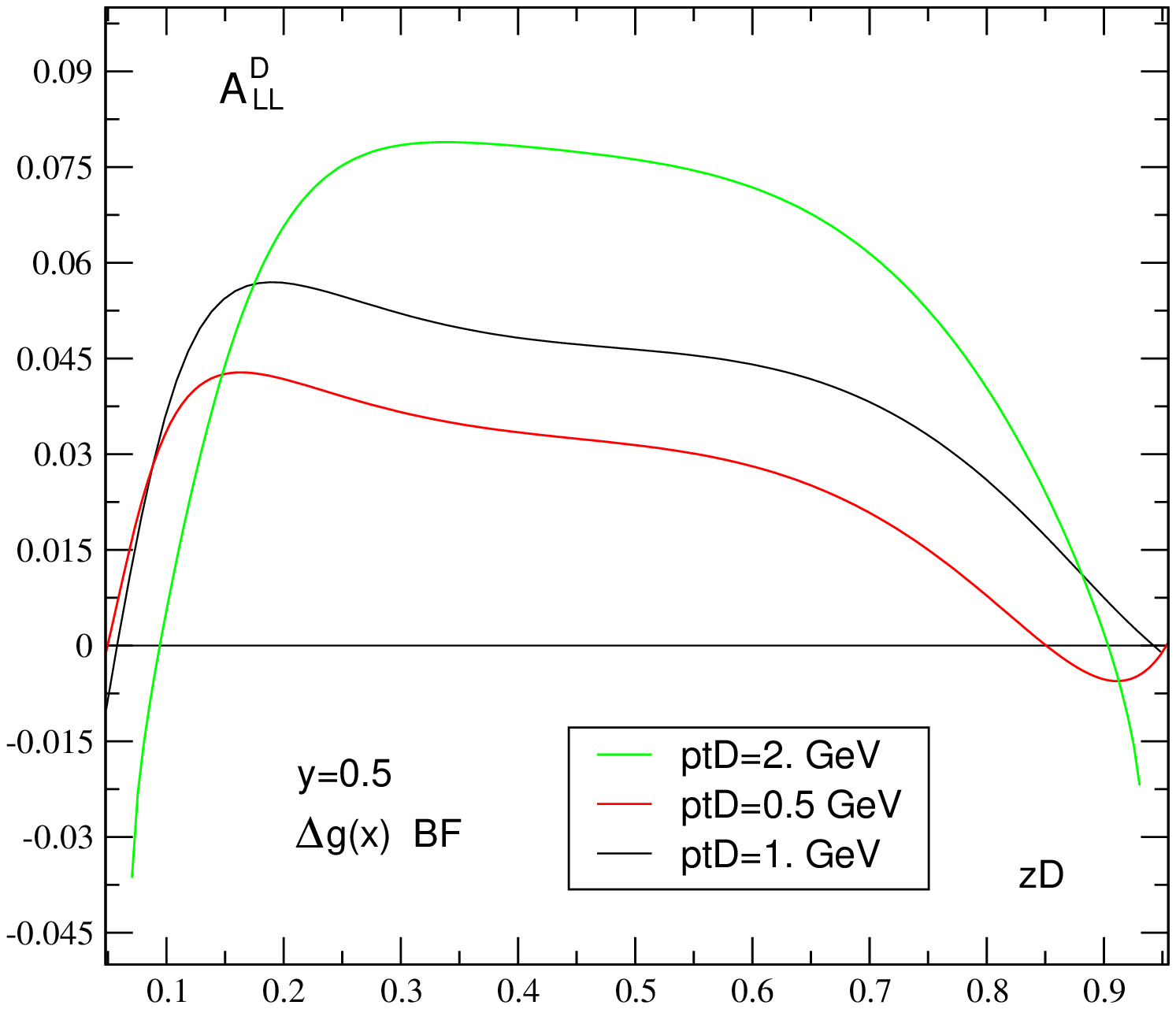}}
\vspace{-0.8truecm}\caption{\footnotesize The spin asymmetry as a function 
of $z_D$ at different $p_{tD}$ and $\Delta g$ given by BF-A ($y$, $g(x)$ and 
$D_c^H(z)$ as in Fig.~\ref{zas})}
\label{z2as}
\end{figure}

The asymmetry as a function of the fractional energy $z_D$ is given in 
Fig.~\ref{zas} at fixed photon energy $y=0.5\,$ and $p_{tD}=1\,$GeV, for
the three different parametrizations of $\Delta g(x)$. Once again the
normalization is determined mostly by the value of $\Delta g(x_{\rm min})$ 
and the shape by that of the underlying hard process. 
The whole kinematically allowed region is displayed, even though the
COMPASS data are  cut at $z_D>0.2$. 
Note that in the small $z_D\lsim0.2$ region $x_{\rm min}$ grows as
$z_D$ decreases. Therefore, this kinematical region probes $\Delta g$ at increasingly large
values of $x$, and it is thus sensitive to the
poorly known polarized gluon at large $x$. The asymmetry (and even the individual cross sections)
in this region remains sizable despite the fact that the pdf is very small. \\
Fig.~\ref{z2as} displays again the effects of increasing $p_{tD}\,$,
that is, of sampling only events with larger $x$, with $\Delta g(x)$ again
given by BF-A.

\begin{figure}[!ht]
\centerline{\includegraphics[width=0.7\linewidth]{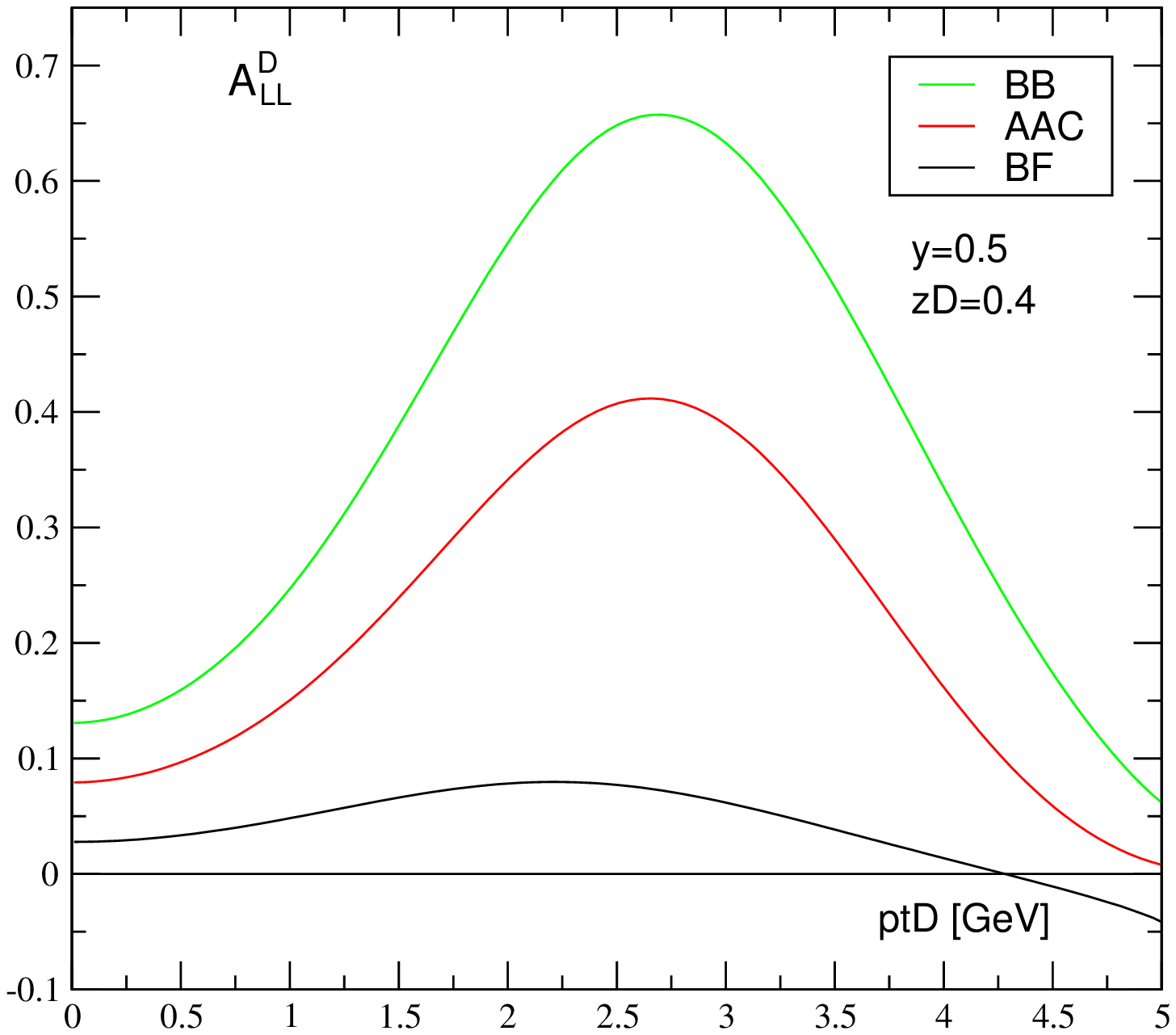}}
\vspace{-0.8truecm}\caption{\footnotesize The spin asymmetry as a function 
of the D meson transverse momentum $p_{tD}$, at fixed photon energy $y=0.5\,$ 
($z_D$, $g(x)$, $\Delta g(x)$ and $D_c^H(z)$ as in Fig.~\ref{yas}).}
\label{pas}
\end{figure}
\begin{figure}
\centerline{\includegraphics[width=0.7\linewidth]{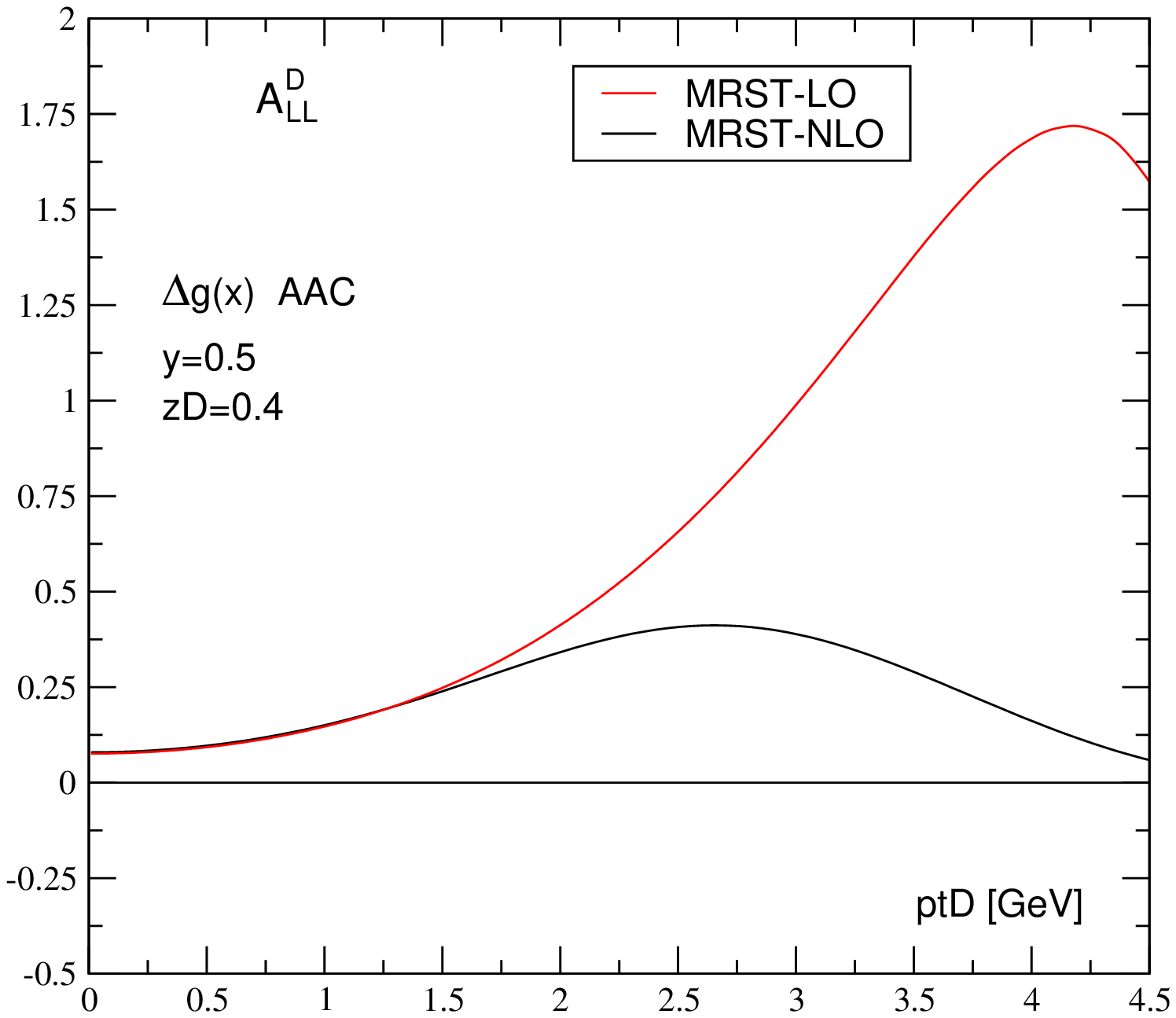}}
\vspace{-0.8truecm}\caption{\footnotesize Effect of the uncertainty in the 
unpolarized gluon distribution at large $x$ on the asymmetry as a function 
of $p_{tD}$. LO and NLO MRST parametrizations of $g$ are compared, with 
$\Delta g$ given by AAC and $D_c^H(z)$, $y,\,z_D$ as in Fig.~\ref{pas}.}
\label{p3as}
%
\centerline{\includegraphics[width=0.7\linewidth]{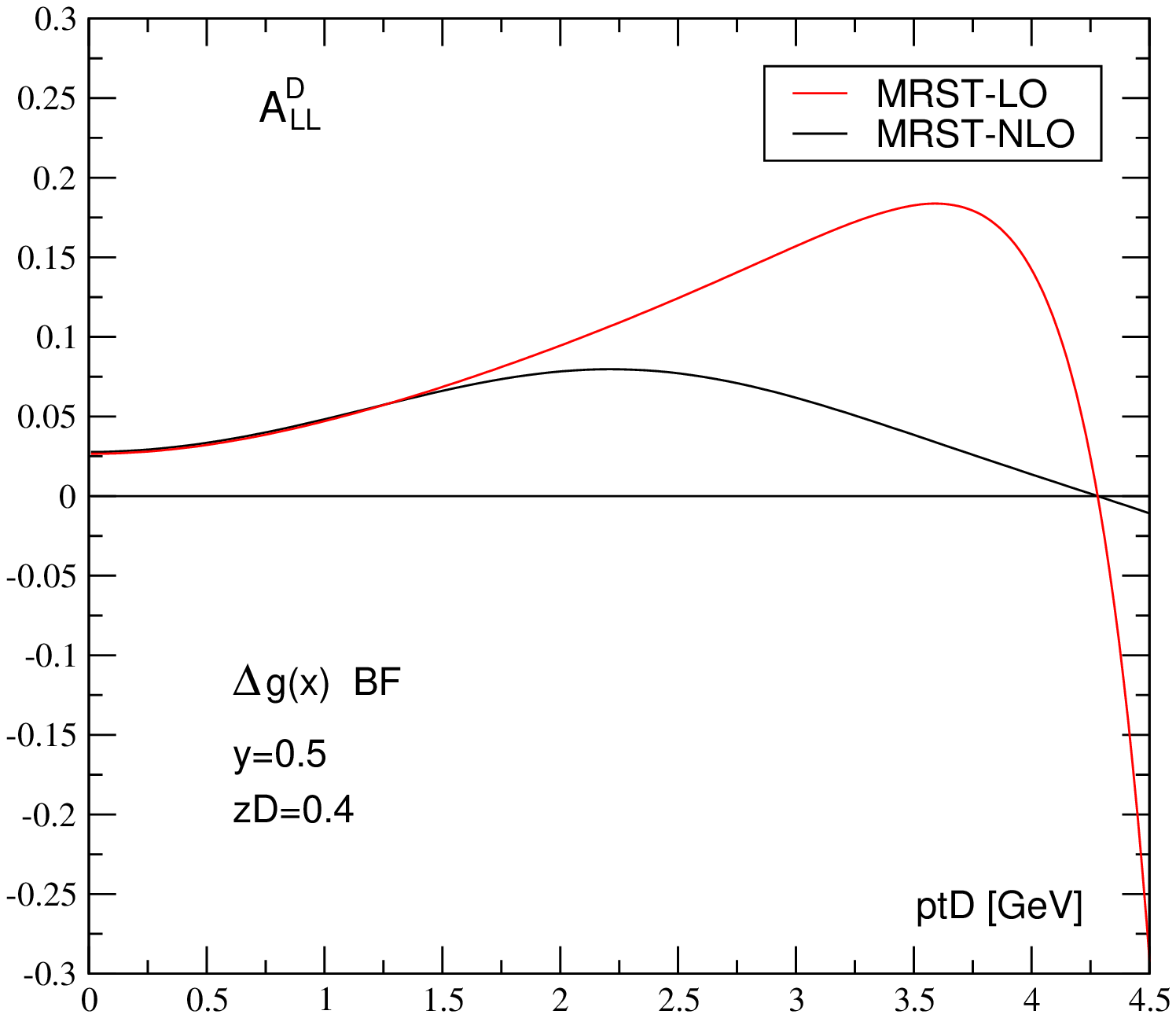}}
\vspace{-0.8truecm}\caption{\footnotesize Effect of the uncertainty in the 
unpolarized gluon distribution $g$ as in Fig.~\ref{p3as}, with $\Delta g$ 
given by BF-A.}
\label{p2as}
\end{figure}
The $p_{tD}$ distribution is shown in Fig.~\ref{pas}, as before for the three
parametrizations of $\Delta g(x)$, at fixed values of $y_D$ and $z_D$. 
The observed $p_{tD}$, being the minimum 
transverse momentum of the parent charm quark, is directly related to the 
minimum gluon momentum required to initiate the process. Generally speaking, 
at low $p_{tD}$ (less than 1~GeV) a larger sample of allowed partonic events 
contributes, but the measurable asymmetry is small. 
On the other hand, the high $p_{tD}$ region (greater than 3~GeV) corresponds to a smaller 
sample of subprocesses restricted to larger $x$, where the uncertainty in the 
unpolarized gluon distribution dominates.

Figs.~\ref{p3as} and~\ref{p2as} show the dependence of the results on the 
unpolarized gluon distribution by comparing the $p_{tD}$-shape of $A_{LL}^D$ 
obtained using the LO and NLO MRST parametrizations of $g(x)$ and two different 
polarized gluon distributions, AAC and BF-A respectively. In both cases, at LO the 
positivity constraint is violated by the gluon distributions at large $x$. 
Specifically, the asymmetry exceeds~1 at $p_{tD}\ge3\,$GeV if $\Delta g$ is given 
by AAC (Fig.~\ref{p3as}), and at $p_{tD}\ge4.7\,$GeV with BF-A (Fig.~\ref{p2as}).
This is due to the fact that both polarized and
unpolarized cross sections become very small in this region. 
Actually, the large difference in the asymmetries in the two cases, that is,
with the LO and the NLO parametrizations of $g(x)$, in both Figs.~\ref{p3as}
and~\ref{p2as},  
indicates that for $p_{tD}\gsim 2$~GeV the uncertainty in the
unpolarized gluon distribution makes the result 
completely unreliable. 

\begin{figure}
\centerline{\includegraphics[width=0.7\linewidth]{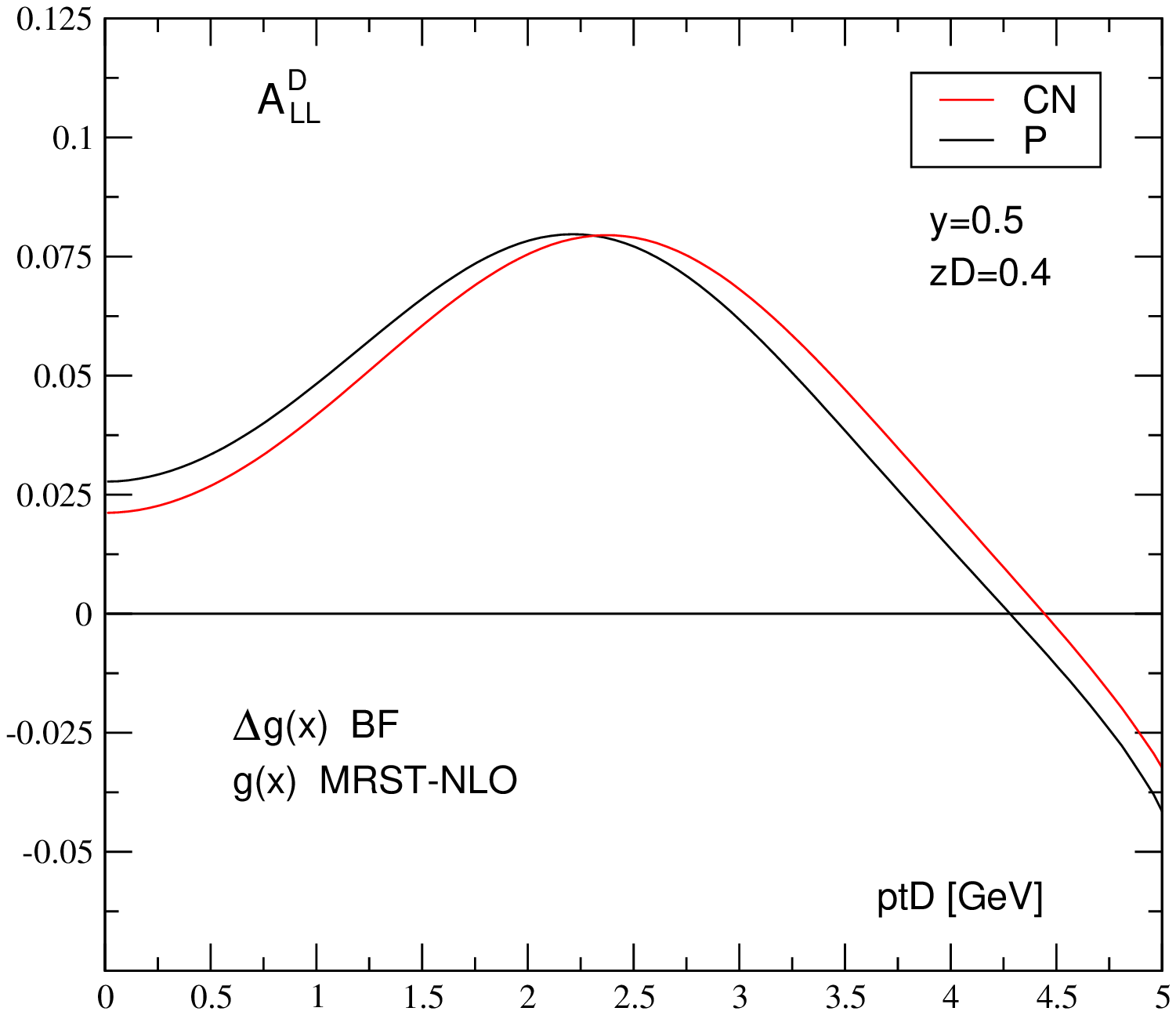}}
\vspace{-0.8truecm}\caption{\footnotesize The asymmetry as a function of 
$p_{tD}$ with two different fragmentation functions, P and CN. $\Delta g,\,y$ 
and $z_D$ are as in Fig.~\ref{p2as} and $g$ is given by MRST at NLO.}
\label{pfas}
%
\centerline{\includegraphics[width=0.7\linewidth]{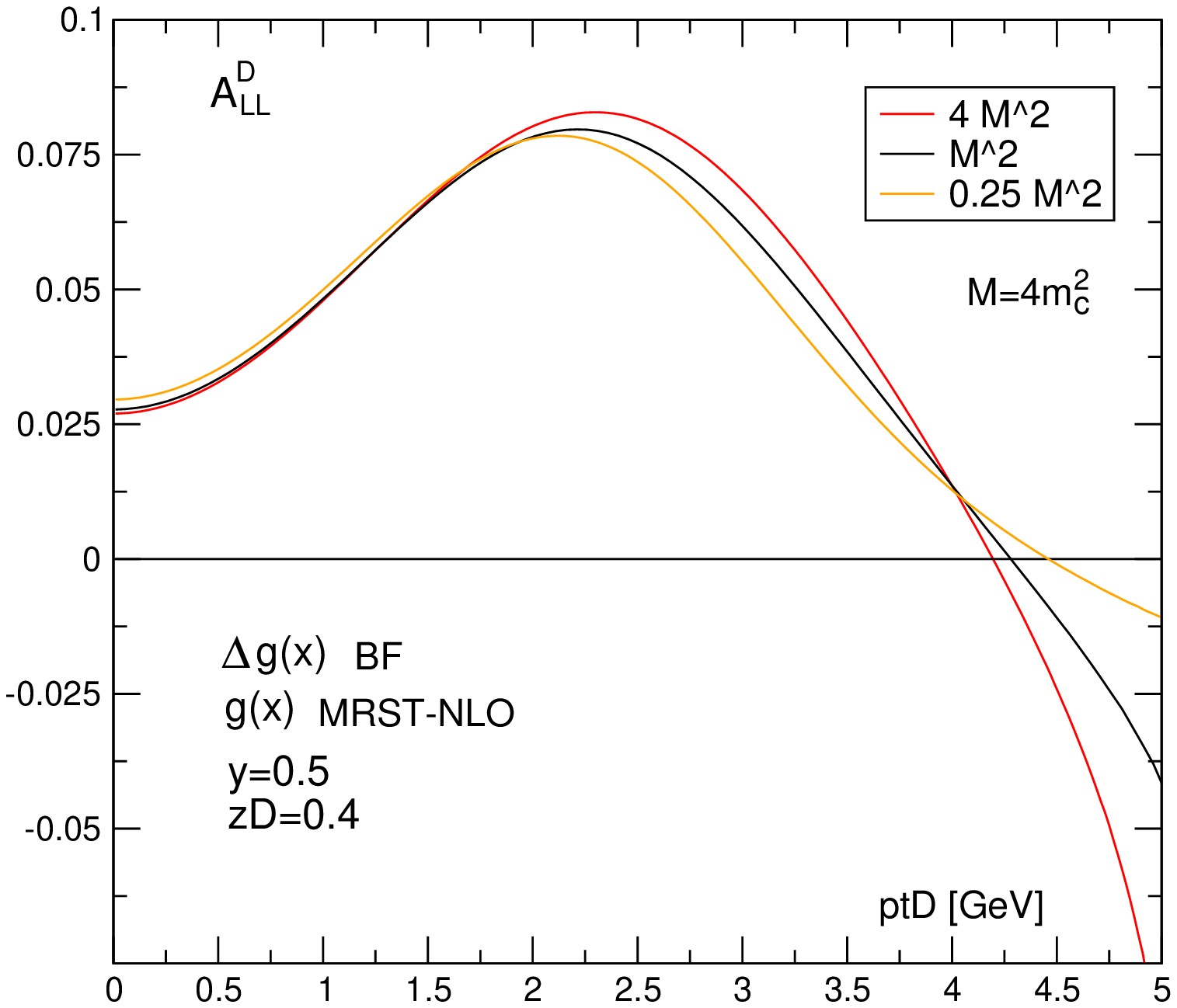}}
\vspace{-0.8truecm}\caption{\footnotesize The asymmetry as a function of 
$p_{tD}$ at different scales $\mu_F^2=\mu_R^2=a\,4m_c^2$, $a=0.25,1,4$. 
$\Delta g,\,g,\,y$ and $z_D$ are as in Fig.~\ref{p2as} and $D_c^H$ is given by 
the Peterson function.}
\label{pscale}
\end{figure}

As an example, the dependence of the asymmetry on the precise shape 
of the fragmentation function is shown as a function of $p_{tD}$ in 
Fig.~\ref{pfas}, at fixed $y$ and $z_D$. The BF-A and NLO MRST
parametrizations are used for the polarized and unpolarized gluon
respectively. The choice of the fragmentation function has no significant
impact on the $p_{tD}$-distribution over the entire kinematical region. 
Similar results are found for the asymmetries as a function of the other
relevant variables.\\
Finally,  as already 
noted in~\cite{Stratmann:1996xy}, also a moderate scale-dependence 
of the asymmetry as a function of $p_{tD}$ is observed in 
the region of interest, i.e. for $p_{tD}\lsim 2\,$GeV. This is displayed 
in Fig.~\ref{pscale} with $\Delta g(x)$ and $g(x)$ as in Fig.~\ref{pfas},
for a simultaneous variation of the factorization and renormalization scales 
by a factor $a=0.25,1,4$ around the value $\mu_F^2=\mu_R^2=4m_c^2$.

\newpage
\subsection{Results from pseudo-data fit}

As a last step, pseudo-data for open-charm asymmetries $A_{LL}^{D}$ have been generated 
by a Gaussian-distributed random shift around the theoretical values. The 
latter are based upon two different assumptions on the polarized gluon density, 
namely the AAC parametrization~\cite{Hirai:2003pm} and the best-fit result
of type-A (BF-A) discussed in Sec.~6. Both are displayed in Fig.~\ref{xdg}
at the scale of the process at hand.\\ 
The primary goal is to determine to what extent such data can pin down the $x$-shape 
of $\Delta g$, so far unconstrained by inclusive DIS measurements, and potentially 
reduce the systematic error induced by the choice of the initial parametrization.

To this end, twelve bins have been selected for the measured kinematic 
variables, with central values $y=0.35,0.65$,$\,z_D=0.45,0.65$ and 
$p_{tD}=0.5,1,2\,$GeV. These are chosen such that they uniformly cover the 
experimentally relevant kinematical region.\\
The related asymmetries $A_{LL}^D$ are 
calculated in each bin, for the chosen polarized gluon. On the basis of the total 
number of the reconstructed $D$ mesons and their distribution 
with respect to each relevant variable~\cite{Koblitz:2007zt}, 
a statistical error is assigned to the predicted asymmetries, i.e.
$\delta A_{LL}^D=\sqrt{(1-(A_{LL}^D)^2)/N}$, where $N$ is the estimated total
number of events in each bin. The theoretical values 
are then 
randomly shifted with Gaussian distribution to produce the final set of 12 
pseudo-data for the assumed polarized gluon parametrization.\\
As expected, due to the sensitivity of $A_{LL}^D$ to $\Delta g$, very
different values of pseudo-asymmetries are found by using the two distinct polarized gluons, 
i.e. AAC and BF-A; 
\begin{figure}[!ht]
\centerline{\includegraphics[width=0.75\linewidth, angle=-90]{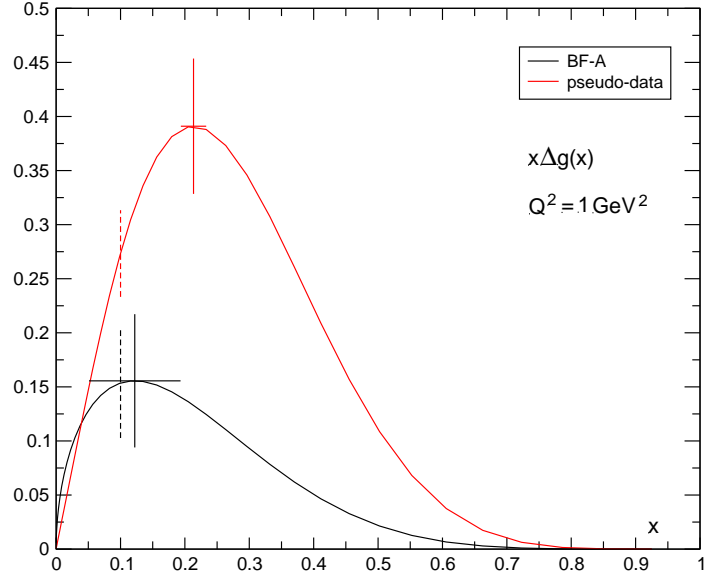}}
\vspace{-0.8truecm}\caption{\footnotesize Best fit result (type-A) for $\Delta g$ 
at $Q^2=1\,$GeV$^2$ including $\tilde{A}_{\rm AAC}$ pseudo-data, compared to  
BF-A (only inclusive data).}
\label{ch1}
\end{figure}
the two sets of related pseudo-data, when included in the global fit of pdfs
along with inclusive DIS data (see Sec.\ 6), 
also produce very different results for the fitted gluon polarizations.

Indeed, a first set of 12 pseudo-data has been generated by using the AAC
parametrization for $\Delta g(x)$ ($\tilde{A}_{\rm AAC}$ in what follows), and 
included in the global fit of type-A (Sec.~6).
A glance to Fig.~\ref{ch1} reveals the significant impact of the $\tilde{A}_{\rm AAC}$ 
pseudo-data on the $x$-shape of $\Delta g$. Here, the original best fit result 
BF-A (only inclusive data) for $\Delta g$ at the initial scale $Q^2=1\,$GeV$^2$ 
is compared with that obtained by including $\tilde{A}_{\rm AAC}$ pseudo-data.\\ 
Also shown in Fig.~\ref{ch1} are the statistical errors propagated to characteristic points 
of $x\Delta g(x)$, namely $x=0.1$ (roughly corresponding to the region probed 
by COMPASS), the position and the value of the peak in $x\Delta g(x)$.
\begin{figure}[!hb]
\centerline{\includegraphics[width=0.75\linewidth, angle=-90]{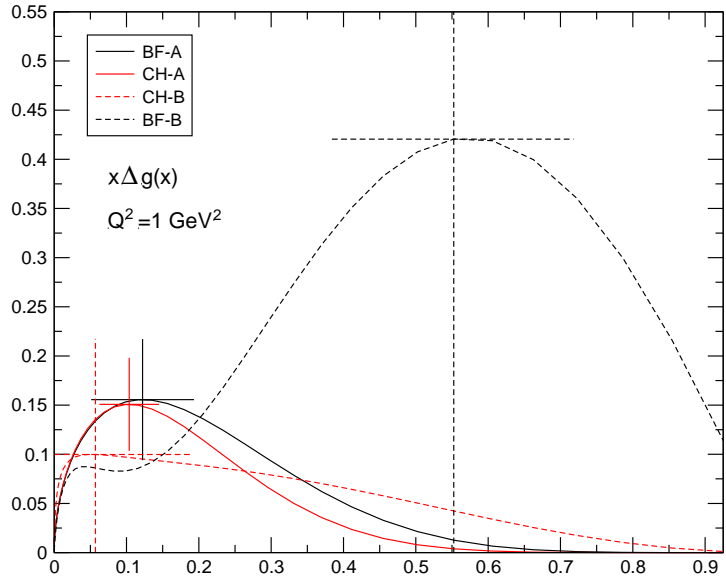}}
\vspace{-0.8truecm}\caption{\footnotesize Best fit results for $\Delta g$ 
at $Q^2=1\,$GeV$^2$ for inclusive DIS data alone (black curves) and
including $\tilde{A}_{\rm BF-A}$ pseudo-data (red curves).}
\label{ch2}
\end{figure}
As can be seen, in particular the peak values in the two cases are not consistent within
several standard deviations.

Then, in order to perform a quantitative study of the impact of open-charm asymmetries
on the $x$-shape of $\Delta g$, a second set of pseudo-data has been generated
by using the best-fit result of type-A for the polarized gluon ($\tilde{A}_{\rm BF-A}$).\\ 
$\tilde{A}_{\rm BF-A}$ pseudo-data are then included in two types of fit
to world data, type-A and B respectively (see Sec.~6.5.2), that differ by the 
assumed functional form of pdfs at the initial scale. The results for the gluon polarization
are summarized in Fig.~\ref{ch2}.\\ 
As already pointed out, the two original best-fit gluons, type-A and B
respectively, both provide equally good 
fit to inclusive DIS data, even though they significantly differ in shape. These are
represented in Fig.~\ref{ch2} by black-solid (BF-A) and black-dashed (BF-B) curves 
respectively. 
The red-solid and red-dashed curves in Fig.~\ref{ch2} denote the type-A and B 
results respectively 
(CH-A and CH-B), obtained by adding the $\tilde{A}_{\rm BF-A}$ pseudo-data to the
fit.\\ 
As a measure of the systematic error induced by the choice of the input
gluon density, one may compare the values of characteristic points (and  
the related errors) of the $x$-shape for 
the two types of best-fit gluons (type-A and B), with and without charm data. 
Specifically, for only inclusive data fits, the position of the peak of 
$x\Delta g(x)$ and the corresponding function value in the case of type-A parametrization
turn out to be
\be
x_{\rm max}=0.122\pm 0.071,\quad x_{\rm max}\Delta g(x_{\rm max})=0.156\pm 0.062 
\quad\textrm{(BF-A)},
\lb{pointsA}
\ee
whereas for the type-B fit they read
\be
x_{\rm max}=0.552\pm 0.169\quad x_{\rm max}\Delta g(x_{\rm max})=0.421\pm 0.871
\quad\textrm{(BF-B)}\,.
\lb{pointsB}
\ee
The quoted uncertainties are obtained by propagating the related statistical 
errors of the parameters from the fit.\\
On the other hand, by including charm pseudo-data $\tilde{A}_{\rm BF-A}$, one has 
\be
x_{\rm max}=0.104\pm 0.041,\quad x_{\rm max}\Delta g(x_{\rm max})=0.151\pm 0.047
\quad\textrm{(CH-A)}
\lb{points2A}
\ee
for the type-A parametrization, and 
\be
x_{\rm max}=0.057\pm 0.132,\quad x_{\rm max}\Delta g(x_{\rm max})= 0.100\pm 0.118
\quad\textrm{(CH-B)},
\lb{points2B}
\ee
in the case of type-B, the main effect being obviously on the latter.\\ 
Specifically,
one may take e.g. the half-difference of the central values in the two fits, A and B respectively, 
as an estimate of the uncertainty on the position of the peak in the polarized gluon density; 
in the case of inclusive DIS data alone, from Eqs.~(\ref{pointsA}-\ref{pointsB}) this amounts 
to roughly $\simeq 0.2$, whereas by including charm pseudo-data in the 
two fits it turns out to be strongly reduced, i.e. $\simeq 0.02$ by using 
Eqs.~(\ref{points2A}-\ref{points2B}). Indeed, in the latter case, 
the positions of the peak of $x\Delta g(x)$ for type-A and B are now largely consistent within
the errors. Similar consideration hold also for the corresponding function values.\\
This fact clearly demonstrates the capability of open-charm asymmetries in 
constraining the $x$-shape of the gluon polarization, significantly reducing the bias induced
by the choice of input densities.

Finally, the two best-fit results for all the pdfs parameters, including $\tilde{A}_{\rm BF-A}$ 
pseudo-data are shown in Tab.~9 below. Comparing these results with the type-A and B fits to 
inclusive DIS data alone, 
respectively Tabs.~5 and~7, it can be seen that, as expected, charm pseudo-data have  
no sizable effect on quark distributions. Moreover, the statistical errors are 
generally slightly reduced, and the quality of the fit improved.\\ 
In particular, as far as the gluon first moment is concerned, very close values arise
from the two types of fits if charm pseudo-data are included. Averaging over the two
best-fit values one would obtain $\Delta g(1,1\,{\rm GeV}^2)=0.54\pm 0.08$, to be 
compared with the second of Eqs.~(\ref{qga0}).

One may thus conclude that the much constrained $x$-shape of the spin-dependent 
gluon distribution, beside the more precise determination of the first moment, 
is the most relevant feature of a 
global analysis that includes open-charm asymmetries along with inclusive DIS data. 
Indeed, charm pseudo-data strongly reduce the impact of the initial parametrization on 
the results, that significantly biases inclusive data analyses. Thus, on the basis 
of the above results, real charm asymmetries from the COMPASS experiment should
be capable to discriminate among the existing different best-fit forms of the 
gluon polarization which equally well describe DIS data. 

\begin{center}
\begin{tabular}[t]{|l|l|l|} 
\hline
Parameters ($Q^2_0=1\,$GeV$^2$) & type-A (CH-A) & type-B (CH-B) \\
\hline\hline
$\eta_{\Sigma}$  & 0.388 $\pm$ 0.018 & 0.391 $\pm$ 0.016\\
$\alpha_{\Sigma}$   & 1.184 $\pm$ 0.317 & 2.670 $\pm$ 0.412\\
$\beta_{\Sigma}$   & 1.958 $\pm$ 0.974 & 3.614 $\pm$ 0.456\\
$\gamma_{\Sigma}$   & -1.229 $\pm$ 0.198 & --\\
$\textcolor{red}{\eta_g}$  & \textcolor{red}{0.525 $\pm$ 0.123} & 
\textcolor{red}{0.553 $\pm$ 0.095}  \\
$\alpha_g$  & -0.481 $\pm$ 0.321 & 3.529 $\pm$ 0.257\\
$\beta_g$  & 8. (fixed) & 2.281$\pm$1.360 \\
$\gamma_g$  & 6.620 $\pm$ 14.070 & 11.656 $\pm$ 12.533 \\
$\eta_3$ & 1.127 $\pm$ 0.034 & 1.207 $\pm$ 0.034\\
$\alpha_{NS}$ & -0.315 $\pm$ 0.270 & 1.624 $\pm$ 0.110\\
$\beta_{NS}$  & 2.965 $\pm$ 0.328 & 5.498 $\pm$ 0.170\\
$\gamma_{NS}$  & 8.966 $\pm$ 12.571 & -0.180 $\pm$ 0.050\\
\hline\hline
 $\chi^2/{\rm d.o.f}$ &  0.888 & 0.902\\
\hline
\end{tabular}\\
\end{center}
{\footnotesize Tab.~9. Fits of type-A and B, including 12 charm pseudo-asymmetries
$\tilde{A}_{\rm BF-A}$ to the set of 238 inclusive DIS data points (cut at $W\ge 3\,$GeV, 
see Sec.~6).}
\vspace{0.5truecm}


\chapter{Conclusions}

To summarize, the possibility of inferring the spin-dependent gluon distribution
from scaling violations has been thoroughly studied, by means of a full NLO phenomenological 
analysis of the available experimental inclusive DIS data, performed in the Adler-Bardeen
factorization scheme.

Special attention, in particular, 
has been paid on the selection of the lower-energy data, in order to avoid 
systematic errors induced by higher-twist corrections. 
Indeed, a bulk of data in polarized DIS
are restricted to the region of the final-state invariant mass which involves large
Bjorken-$x$ at moderate values of the momentum transfer $Q^2$, roughly $1\lsim W\lsim 3\,$GeV.\\
It has been shown that such data do not sizably affect the shape of quark and gluon
densities, whereas the main effect of the lower energy points (mainly from the CLAS and the 
HERMES experiments) is essentially the improvement in the determination of the higher-twist
contributions to the structure function $g_1$. Then, for a consistent perturbative treatment  
of world data, lower energy points can
be safely discarded; to this end, a 
quantitative criterion has been established by a phenomenological
study of higher-twist corrections to $g_1$. As a result, a lower bound has been imposed
on DIS data at $W=3\,$GeV, since above this threshold higher-twist corrections are consistent  
with zero within 1.5 standard deviations, thus justifying a pure perturbative analysis.

Furthermore, the bias produced by the choice 
of input densities has been investigated by performing global fits to the same  
subset of DIS data using different boundary conditions for pdfs (namely type-A and B fits).\\
It turns out that the spin-dependent gluon distribution exhibits a sizable dependence 
on the assumed functional form, whereas minor effects are observed in the quark sector. 
Indeed, best-fit gluons obtained with different input densities 
(type-A and B) both provide equally good fits to inclusive DIS data, even though they 
significantly differ in shape. On the other hand, the first
moment of the polarized gluon turns out to be more stable with respect to the initial
parametrization, and, in particular, smaller if compared to previous
estimates. If all the theoretical errors are properly taken into account, the final
estimate reads
\begin{displaymath}
\Delta g(1,1\,{\rm GeV}^2)=0.68\,\pm\,0.12\,({\rm exp})\pm\,0.29\,({\rm th}),
\end{displaymath}
in reasonable agreement with the results of other recent phenomenological studies.\\
Thus, even though the better coverage of the $(x,Q^2)$-plane, mainly due to the recent precise
COMPASS data, allows a more reliable determination of the gluon first moment,
an analyisis based on inclusive DIS data alone is not capable to pin down the $x$-shape
of the gluon distribution, which is still largely unknown.  

At present, further constraints are expected to come from direct measurements of 
$\Delta g(x)$, which are based upon the observation of gluon-initiated processes at the 
partonic level, such as for example production of heavy-flavored hadrons and jets with 
large transverse momentum, from both hadro- and lepto-production.\\ 
Actually, one of the most promising approach is the measurement of open-charm events
from (fixed target) polarized lepton-nucleon scattering, since charmed mesons in the final states are a clean 
signature of the underlying photon-gluon fusion mechanism, with essentially no competing processes
and moderate NLO corrections. Furthermore, asymmetry data of open-charm 
photoproduction from the COMPASS experiment are upcoming. 
Therefore, a LO 
analysis of this process has been 
performed by constructing a close theoretical description of the actually observed 
quantities at COMPASS.
 
As a first result, this study has revealed in particular that, in the range of 
gluon momentum fractions 
accessible at COMPASS energies, that is restricted to $x\gsim 0.1$, the absolute
 value of the gluon 
polarization has a significant impact on the size 
of the asymmetry viewed as a function of each of the kinematic
variables of the outgoing D meson. This is interesting in view of the
fact that available polarized gluon distributions have similar values
of the first moment, but very different shapes and thus very different
values at fixed $x\gsim 0.1$. 
For instance, the best-fit result of type-A of the present analysis, BF-A, and 
the AAC parametrizations both have a 
first moment of $\Delta g$ around 0.5 at $Q^2=1~\rm{GeV}^2$, but at $x=0.2$ they
differ by more than a factor two, as seen in Fig.~\ref{xdg}. 
Open-charm data may thus resolve the gluon $x$-shape in this region.

In order to assess to what extent such data can actually pin down the $x$-shape 
of $\Delta g(x)$, so far unconstrained by scaling violations, and potentially 
reduce the systematic error induced by the choice of input densities,  
a phenomenological study has been performed by using open-charm pseudo-data,
generated by a Gaussian-distributed random shift around the theoretical values.
Adding pseudo-asymmetries to global fits of world data with different input densities,
type-A and B respectively, 
a much constrained $x$-shape for the fitted gluon turns out, as well as a
better determined first moment. Specifically, the two best-fit forms for the gluon polarization become 
largely consistent in shape if charm pseudo-data are included in the analysis.\\
This fact clearly
emphasizes that, at variance with analyses based on scaling violations, the open-charm approach
should be capable to significantly reduce the 
bias induced by the choice of the initial parametrization, which is one of
the most relevant theoretical uncertainties in the determination of the spin-dependent gluon
distribution,  
and thus to discriminate among the available different best-fit forms which
equally well describe DIS data.


\end{document}